# INVESTIGATION OF PEROVSKITE RELATED OXIDES: STRUCTURAL AND MAGNETIC PROPERTIES

**Ph.D THESIS**

*by*

**ANTARA SARKAR**

**DEPARTMENT OF CHEMISTRY**

**INDIAN INSTITUTE OF TECHNOLOGY DELHI**

**SEPTEMBER 2021**

# INVESTIGATION OF PEROVSKITE RELATED OXIDES:

# STRUCTURAL AND MAGNETIC PROPERTIES

*by*

**ANTARA SARKAR (2018CYZ8196)**

Department of Chemistry

Submitted

in partial fulfilment of the requirements of the degree of the

**Doctor of Philosophy**

to the

**DEPARTMENT OF CHEMISTRY**

**INDIAN INSTITUTE OF TECHNOLOGY DELHI**

**SEPTEMBER 2021**

**This thesis is dedicated to my parents and family.**

# Indian Institute of Technology Delhi

## Candidate's Declaration

I hereby certify that the work which is being presented in the thesis entitled, "**Investigation of perovskite related oxides: Structural and magnetic properties**", in partial fulfilment of the requirements for the award of the degree of Doctor of Philosophy and submitted in the Department of Chemistry of the Indian Institute of Technology Delhi. It is an authentic record of my own work carried out during a period of July 2018 to September 2021 under the supervision of Prof. Ashok K. Ganguli, Indian Institute of Technology, Delhi.

The matter presented in the thesis has not been submitted by me for the award of any other degree or in any other institute.

(Antara Sarkar)

This is to certify that the above statement madeby the candidate is correct to the best of my knowledge.

(Prof. Ashok K. Ganguli)

Supervisor

Date 16/9/2021

**Acknowledgements**

First of all, I sincerely express my heartfelt gratitude to my Ph.D. supervisor Prof Ashok K. Ganguli for his invaluable guidance, motivation and encouragement during the entire Ph.D. research. At many stages of my research work, his positive outlook and confidence on my research inspired me and gave me confidence to overcome the obstacles.

I am highly obliged and express my sincere thanks to the present Head of department, Prof. Anil J. Elias as well as other faculty members for their help and kindness. I am grateful to the members of my SRC, Prof. Arunachalam Ramanan (Chairman, department of chemistry), Prof. Shashank Deep (internal member, department of chemistry) and Prof. Pintu Das (external member, department of physics), IIT Delhi.

I am grateful to head of instrumental lab facility (chemistry department), nano research facility, chemistry department M.Tech. labs for providing me easy access to instruments when needed. I wish to place my thanks to the technical and administrative staff of the department of chemistry. Thank you Dimple ma'am and Ashish bhaiya.

I would like to thank my seniors and lab mates for creating an interacting and enthusiastic research atmosphere. I wish to thank Prof. L. C. Gupta for his valuable suggestions and guidance. I would like to specially mention the name of Dr. Nibedita Das, Dr. Soumen Ash, Dr. Anirban Das and Dr. Vaishali Sethi, for playing a major role in completion of this thesis. I would like to thank all my seniors: Dr. Sandeep Dhaka, Dr. Menaka Jha, Dr. Zeba Haque, Dr. Nitin Yadav, Dr. Soumen Saha and Dr. Kasinath Ojha. I would like to thank all lab mates for their help, especially Dr. Preeti Dagar and Moumita Naskar. I would like to thank my all my fellow lab mates and juniors for making the lab atmosphere cordial, Shalini, Kirandeep, Priyanka, Manisha, Prabuddhakant, Nitika, Sourayan, Manoj, Akshita and Shefali.

A special thanks to my family. No words can suffice my feelings and gratitude towards my parents for their unconditional love and motivation which remained as driving force during the entire course of my Ph.D. They have helped me a lot to reach this stage of my Ph.D. My father, Gautamendu Sarkar remained a constant guide and motivator, and my mother, Mita Sarkar showed unmatched patience and confidence in me. I would like to thank my elder brother, Arnab Sarkar for encouragement and support. I would also like to thank my grandparents for showering their blessings on me.

At the end I would like to thank almighty for believing in me and providing me strength.

**List of Publications**

1. Antara Sarkar, Anirban Das, Soumen Ash, Ravi Shankar P. N. and Ashok Kumar Ganguli; Structural and magnetic properties of $R_{0.5}Sr_{0.5}Fe_{0.5}Mn_{0.5}O_3$ (R = Gd, Nd and Pr) perovskites. J. Alloys and Compounds, DOI : 10.1016/j.jallcom.2021.160747.

2. Antara Sarkar, Anirban Das, Soumen Ash, Debjani Karmakar and Ashok K. Ganguli; Investigation of A-site and B-site substitution: structural and magnetic properties of $(La/Ho)_{1-x}Pr_x(Fe/Mn/Cr)O_3$ x = 0, 0.3, 0.5, 0.7 and 1 . (to be communicated).

3. Antara Sarkar, Anirban Das, Soumen Ash, Ravi Shankar P. N. and Ashok Kumar Ganguli; Investigation of structural and magnetic properties of a Ru-deficient perovskite $PrMn_{0.5}Ru_{0.35}O_3$. (to be communicated).

4. Antara Sarkar, Anirban Das, Soumen Ash, Kandalam V. Ramanujachary, Samuel E. Lofland and Ashok K. Ganguli; Investigation of structural and negative magnetic properties of $La_{0.5-x}Sm_xPr_{0.5}CrO_3$ ( x = 0, 0.1, 0.2, 0.3, 0.4 and 0.5). (To be communicated).

5. Antara Sarkar, Anirban Das, Soumen Ash, Kandalam V. Ramanujachary, Samuel E. Lofland and Ashok K. Ganguli; Investigation of Photoelectrocatalytic and Magnetic Properties of $Sr_2YbRu_{1-x}Ta_xO_6$ (x = 0, 0.25, 0.5, 0.75 and 1) (to be communicated).

**Abstract**

Perovskites exist in varied structures and show structure dependent magnetic properties which is of great interest in the field of multiferroics, magnetocaloric and spintronic devices. Here we intend to study the various rare-earth(R) and transition metal (TM) ion interactions depending on their structures. Different Curie temperature, spinglass transition and spin reorientation transitions were observed depending on the degree of interactions and structural properties. It is interesting to study the role of orbital interactions, disorder in magnetic transitions leading to change in superexchange and double exchange interaction.

On substitution at A-site or B-site results in change in magnetic interactions, as well as creates structural distortions. We also obtained results where the spin interaction of the transition metal and rare-earth interaction, magnetic anisotropy, orientations of orbitals, disorder induced interactions were observed.

The structural investigation of various combinations of rare-earth and transition metal ions at A and B-site with La/Pr/Ho and Fe/Mn/Cr respectively were studied. The magnetic susceptibility measured at 1000 Oe, was found to decrease in $Ho_{1-x}Pr_x(Mn_{0.5}Cr_{0.5})O_3$ (x = 0, 0.3, 0.5 and 0.7), while in $Ho_{1-x}Pr_x(Fe_{0.25}Mn_{0.25}Cr_{0.5})O_3$ (x = 0, 0.5 and 1) it increased as the value of x increases, which could be attributed to increased Ho-Mn interaction and competing anisotropy of Pr and Fe orbitals in the first and second case respectively. $Pr(Fe_{0.25}Mn_{0.25}Cr_{0.5})O_3$ showed highest Curie temperature (247 K) and Curie-Weiss temperature (-58 K) and a maximum change in remanent magnetization at 5 K and 20 K amongst all the compositions, which suggests a stable spin structure and rare-earth spin reorientation taking place in this composition.

The investigation on Pr-Mn-Ru-O phase, on effect of structural disorder due to Ru-deficiency at B-site, as calculated from EDAX and XPS measurements by a stoichiometry of 0.15(2), which leads to enhanced spin glass like behavior at temperatures below 31 K, as evident from the bifurcation between the ZFC and FC curves. This pure Pr-Mn-Ru-O phase obtained was refined in the orthorhombic *Pnma* spacegroup, with greater octahedral distortion along the axial direction (157.15(1)˚) than in the equatorial (168.88(1)˚), along with a compression in the *b* lattice parameter, which further suggests greater Ru-vacancy along the axial *b* direction. The magnetic susceptibility increases at 18 K obtained from DC magnetic susceptibility data at 1000 Oe, suggests disorder induced ferromagnetic ordering, and that at 4.5 K which could said due to 'Pr' and Mn – O – Ru ordering. The non-saturation nature of isothermal magnetization at 2 K and 50 K, further suggests short range magnetic ordering.

To further investigate the competing magnetic interactions of rare-earth and transition metal ions, it was observed that the compound $La_{0.5-x}Sm_xPr_{0.5}CrO_3$ (x = 0, 0.1, 0.2, 0.3, 0.4 and 0.5), showed negative magnetization at 100 Oe. The alignment of the rare-earth moment opposite to the net moment generated by the canted antiferromagnetic moment generated by the Cr ion resulted in negative magnetization for x = 0.4 and 0.5 at 100 Oe, ZFC. Pure phase of $La_{0.5-x}Sm_xPr_{0.5}CrO_3$ (x = 0, 0.1, 0.2, 0.3, 0.4 and 0.5) were refined in orthorhombic *Pnma* spacegroup, where the octahedral distortion increased as the value of x increased. This increase of anisotropy as the value of x decreased has resulted in x = 0 and 0.1, to show magnetic reorientation in the temperature range 160 K to 25 K, from the bifurcation and recombination of FCC and FCW at 10,000 Oe. Exchange bias effect in ZFC and ZFW in the composition $La_{0.1}Sm_{0.4}Pr_{0.5}CrO_3$ was investigated, with the observation of negative exchange bias effect at temperature below 230 K, which further suggests canted antiferromagnetic nature in the following compositions.

In order to inspect further the effect of structural distortion with the decrease in rare-earth radii was investigated in orthorhombic $R_{0.5}Sr_{0.5}Fe_{0.5}Mn_{0.5}O_3$ (R: Gd<Nd<Pr). Influence of increased octahedral distortion, primarily due to Jahn-Teller distortion, resulted in decrease of magnetic transition temperature due to increased hybridization between the transition metal and oxygen orbitals. The octahedral distortion decreases as the rare-earth size increases from Gd to Nd to Pr. $Gd_{0.5}Sr_{0.5}Fe_{0.5}Mn_{0.5}O_3$ shows canted antiferromagnetic ordering of $Mn^{3+}$ spins at temperatures below 49 K followed by Gd antiferromagnetic ordering taking place at temperatures below 3 K. The compositions $Nd_{0.5}Sr_{0.5}Fe_{0.5}Mn_{0.5}O_3$ and $Pr_{0.5}Sr_{0.5}Fe_{0.5}Mn_{0.5}O_3$ showed spin glass like nature at 47 K and 58 K respectively, as $Nd_{0.5}Sr_{0.5}Fe_{0.5}Mn_{0.5}O_3$ showed greater octahedral distortion as compared to $Pr_{0.5}Sr_{0.5}Fe_{0.5}Mn_{0.5}O_3$.

To study the effect on disruption of long range ordering of orbitals on increased Curie –Weiss ($\theta_{CW}$) temperature and on photoelecrocatalytic properties were investigated in monoclinic $Sr_2YbRu_{1-x}Ta_xO_6$ (x = 0, 0.25, 0.5, 0.75 and 1) system. The long range ordering of Yb – O –Ru was found to be maximum in $Sr_2YbRuO_6$, which decreased on substituting Ta in place of Ru, as evident from the maximum $\theta_{CW}$ obtained in $Sr_2YbRuO_6$ (-148 K), which decreased substantially for x = 0.75 (-125 K) to x = 0.5 (-118 K), and x = 0.25 (-102 K). On increasing Ta concentration, the bandgap was found to increase linearly from 1.78(1) eV (x = 0) to 2.08(1) eV (x = 0.75) however with a slight decreased deviation (0.04(1) eV) observed in case of $Sr_2YbRuO_6$. Photoelectrocatalytic measurement showed maximum oxygen evolution current density of 17 $\mu A/cm^2$ in light and dark (12 $\mu A/cm^2$), in $Sr_2YbRuO_6$. XPS showed Ru and Ta were majorly in +5 and +4 oxidation states.

**Contents**

**List of Tables**

**Abbreviations**

| | |
|---|---|
| AFM | Antiferromagnetic |
| FM | Ferromagnetic |
| PM | Paramagnetic |
| R | Rare-earth |
| TM | Transition metal |
| nm | Nanometer |
| Å | Angstrom |
| XRD | X-ray diffraction |
| XPS | X-ray photoelectron spectroscopy |
| PPMS | Physical property measurement system |
| UV Vis | Ultraviolet visible spectroscopy |
| DRS | Diffuse reflectance spectroscopy |
| ZFC | Zero field cooled |
| FC | Field cooled |
| PEC | Photoelectrocatalysis |
| OER | Oxygen evolution reaction |
| eV | Electron-Volt |
| Oe | Oersted |
| kV | Kilovolt |
| Kg | Kilogram |
| f.u. | Formula unit |
| B.E. | Binding energy |

**Greek Symbols**

| | |
|---|---|
| $\chi$ | Magnetic susceptibility |
| $\lambda$ | Wavenumber |
| $\mu_B$ | Bohr magnetron |
| $\nu$ | Frequency |

**English Symbols**

| | |
|---|---|
| A | Absorbance |
| $T_C$ | Curie temperature |
| $T_{SR}$ | Spin reorientation temperature |
| H | Magnetic field |
| M | Magnetization |
| $T_N$ | Néel temperature |
| a, b, c | Lattice parameters |
| n | Order of diffraction |
| K | Kelvin |
| I | Intensity |
| Eg | Energy gap |
| h | Planck’s constant |
| g | Gram |
| T | Temperature |

# Chapter I

# Introduction

# Chapter I

# Introduction

## 1.1 Structure of perovskite

Oxides of general formula $ABO_3$, having a perovskite structure, show varying magnetic properties, due to interaction between transition metal ions and interaction between rare-earth and transition metal ions [1]. Perovskites have continued to attract attention over five decades due to their fascinating properties and several applications [1]. The name perovskite was given to $CaTiO_3$ named after Russian mineralogist L. A. Perovski (1792-1856), which is a colourless solid and diamagnetic [1]. Perovskite oxides can have wide-ranging magnetic properties induced by the varying arrangement of transition metal ions and rare-earth ions [2-8]. The perovskite structures have the general formula $ABO_3$ where the A-atom occupies the cuboctahedral position and B-atom is octahedrally surrounded by oxygen atoms (Fig. 1.1). In recent years transition, metal-based perovskites have received considerable attention due to their magnetocaloric as well as multiferroic properties [2-4].

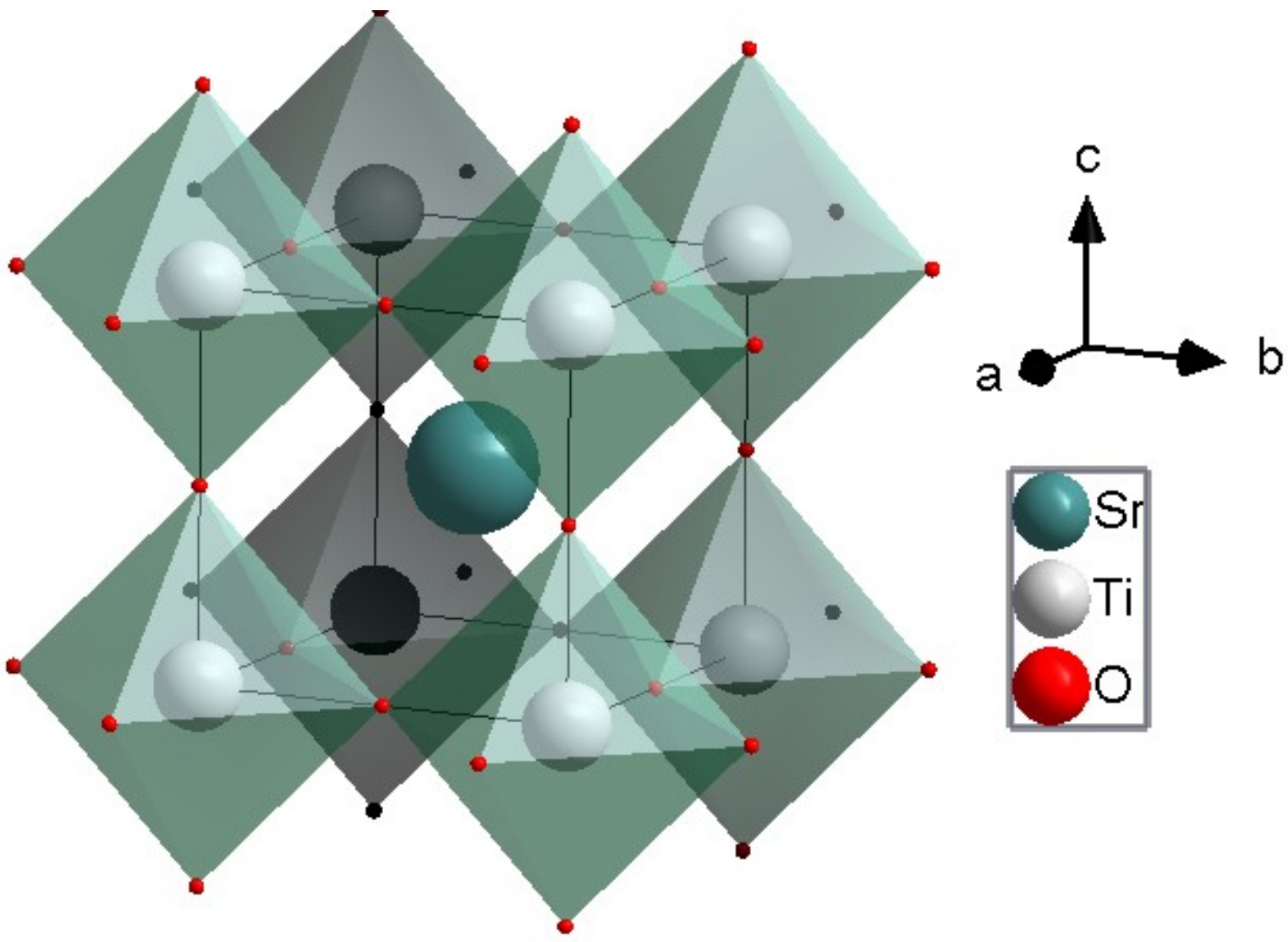

Fig. 1.1: Structure of perovskite $ABO_3$ (A: Sr and B: Ti )

Though the ideal perovskite exists in the cubic structure, there are distorted perovskite oxides in all the seven crystal systems [2-8]. The cubic perovskite is the ideal structure having Goldschmidt tolerance factor (T) as 1 (Equation 1.1) where $r_A$ and $r_B$ are the ionic radii of A and B cations [9].

$$T = \frac{r A + rO}{\sqrt{2}(rO + rB)} \qquad \text{Equation 1.1}$$

The tolerance factor helps in predicting the crystal structure of perovskites. The deviation from cubic symmetry is due to distortion caused by various possibilities like ordering of A and B-site ions, size of cations and displacement of atoms [10-11]. The octahedral tilting as mentioned by Glazer *et. al.* exists with various orientations and play a role in the stabilization of a specific crystal structure [12]. The combination of various rare earth and transition metal ions to form structures with different spin structures leads to a change in magnetic properties [13]. The structural and magnetic properties influence one another. For example in $CeCrO_3$ where magnetic interactions lead to structural changes [14]. In $RCrO_3$ (R: rare-earth) the Néel temperature decreases as the octahedral distortion increases [14-15]. Perovskite related ordered structures such as double perovskites [16-17], triple-perovskites [18-20], anion or cation-deficient perovskites having ordered structures are known, where the ordering leads to interesting magnetic properties [20-21]. For example, there are varying magnetic interactions like B – O – B' and B – O – O – B' interactions which lead to varying magnetic properties [22]. The magnetic properties get influenced by rare-earth and transition metal interactions based on factors such as spin anisotropy, spin canting, and the strength of exchange interactions [23-25].

These perovskites have various applications such as magnetocaloric, multiferroics, read and write memory devices, switches, sensors, superconducting and piezoelectric devices. The ordering of ions dictates the structure and magnetic properties [26-27]. Many of our modern technological devices rely on magnetic materials; these include electrical power generators and transformers, electric motors, telephones, computers and components of sound and video systems. Hence magnetic materials are an integral part of electronic industries today [28].

This long-range or short-range ordering arrangement of rare earth metals and transition metals at the A-site and B-site also have applications in tuning the electronic structure of the materials which as a result increases or decreases their optical band gap and photo-electrocatalytic activity [29]. We intend to obtain optimized compositions of these perovskites of interesting structural and magnetic properties.

### 1.1.1 Octahedral tilting in perovskites

The structure of perovskite is influenced by the octahedral tilting as well as by the cation displacement, taking place in the structure. For example, $NaNbO_3$ shows octahedral tilting along [001] axis and thus results in change in crystal system from cubic ($a^0a^0a^0$) to tetragonal ($a^0a^0c^+$) [17, 65]. Amongst the 23 possible tiltings depending on different degrees of tilting and various possible combinations of directions of tilting results in 23 possible tilt systems [30-35]. However most commonly only 9 tilts can be commonly found in the perovskite system (Table 1.1), with the most common octahedral tilting being $a^-b^+a^-$ or $a^-a^-a^-$ found in orthorhombic and tetragonal crystal systems respectively.

The equality of octahedral tilting being equal along [100] and [010] axis but with different degrees of tilting about [001] has been symbolized by *aac* [35-40]. The direction of octahedral tilting along an axis depending on the direction of tilting can be denoted by the symbols + (same tilt), - (opposite tilt) and 0 (no tilt) [35].

The crystal structure can be predicted by the different degrees of octahedral tilting. For example, $a^+b^+c^+$ results in body-centred crystal system while $a^-b^-c^-$ is face-centred [41]. The octahedral tilting results in a change in lattice parameters between the two octahedral centres.

When there is a different degree of octahedral tilting it results in 3 different lattice parameter changes. It is denoted by as follows [42].

a = A cosβ cosɣ
b = A cosα cosɣ
c = A cosα cosβ
(Here A denotes edge length of a cubic perovskite).

From the directions of tilt systems and their magnitude, we can also predict the orthogonality of the axes. For example, $a^+b^+c^+$ have three mutually perpendicular mirror planes. While $a^+b^-c^-$ denotes that one axis (*a)* is perpendicular to the plane containing 2 mutually inclined axis (*b* and *c*). Moreover $a^-b^-c^-$ denotes all the 3 axes are inclined to one another [40-42].

### 1.1.2 Different structures of perovskites

The ordering of A-site or B-site results in doubled unit cell axis along the ordering direction and thus results in double, triple or quintuple ordering. It has been reported earlier that ordered double perovskites ($A_2BB'O_6$) have been reported to be stable if the oxidation state difference between the A-site or B-site atoms is greater than +2 or their ionic radii difference is greater than 0.2 Å [6].

In the case of double perovskite of the form $AA'B_2O_6$ for example ($NdBaFe_2O_6$) or of the form $A_2(BB')O_6$ for example $Sr_2YbRuO_6$ results in doubling of the unit cell (Fig. 1.2) [43-46]. In the case of triple perovskite for example $Ca(Mn_{0.5}Cu_{0.5})FeReO_6$, where the degree of order is small and Ca is partially ordered at A-site and 2 metal cations may be ordered over the tetrahedral and square planer sites resulting in the tetragonal $P4_2/n$ space group [22]. Perovskite of the type $AA'BB'O_6$ : $CaMn\text{-}FeReO_6$ and $Ca(Mn_{0.5}Cu_{0.5})FeReO_6$ have columnar A/A′ and rocksalt B/B′ cation orders, as found in the recently discovered double-double perovskite [47-49]. In $MnNdMnSbO_6$, there is partial Mn/Cu order over tetrahedral and square planar A′ sites as also observed in $Ca(Mn_{0.5}Cu_{0.5})FeReO_6$ which demonstrates the triple-double cation order [50-60].

The stacking along with the ordered (axial) direction by the intermixing of double and triple perovskite ordered layers results in quintuple perovskite of the form A-A'-A-(A,A')-(A,A')-A, resulting in unit cell axis to be five times the unit cell as compared to normal perovskite [60-61].

This ordering as evident from the small half fraction reflections present at lower 2θ angles in X-ray diffraction results in strain in the crystal structure [60]. This ordering results primarily from greater size difference between A and B-site atoms which creates lattice strain in the crystal system (for example between Pr and Ba in $Pr_2Ba_3Fe_5O_{15}$) [62, 27]. If the size difference between the ordered atoms reduces then it results in weakening of the ordering (for example between Nd and Ba in $Nd_2Ba_3Fe_5O_{15}$) [27]. In $Nd_2Ba_3Fe_5O_{15}$ the quintuple structure by short-range ordering along with axial direction results in a random distribution of the cations [63,27].

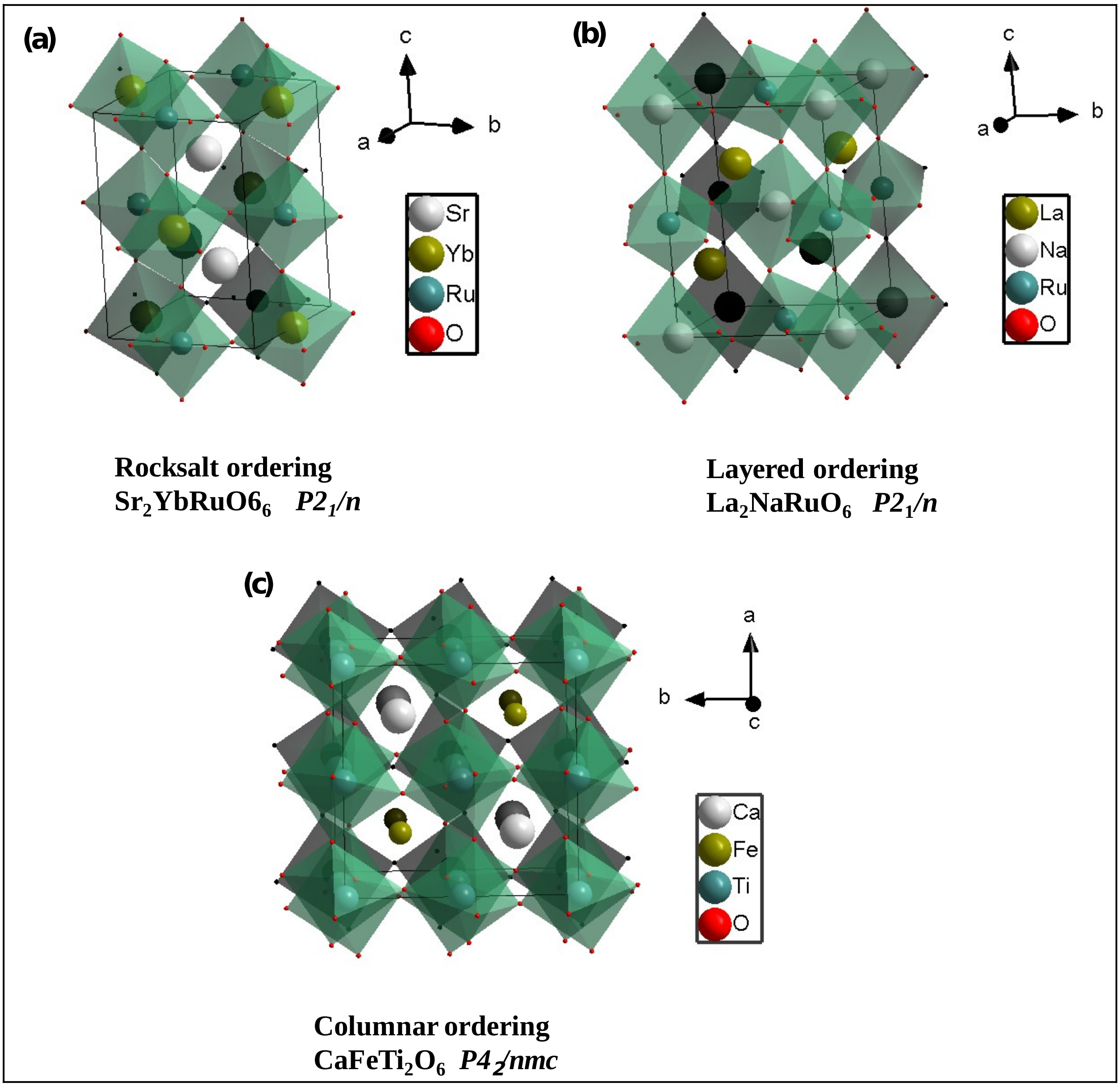

Fig. 1.2: Various ordering in perovskite structure showing a) Rocksalt ordering in $Sr_2YbRuO_6$, b) Layered ordering in $La_2NaRuO_6$ c) Columnar ordering in $CaFeTi_2O_6$ [51-53].

### 1.1.3 Factors influencing the change of space group in perovskite

A centrosymmetric space group is retained if the atoms get displaced in a non-uniform manner [70-72]. However, displacement of atoms in a particular direction result in a non-centrosymmetric space group [64-70]. It is therefore easy to suggest that the most likely space group for the structure, of A and B cations usually lie at special positions if the displacement directions are known [70-72].

When there are no tilts present, then there is a lack of necessary small weak reflections and the displacements using which we can predict the symmetry elements present in the system [72]. For example, a rhombohedral unit cell shows displacement of atoms or elongation of axes towards [111] axis: spacegroup *R3m* [73].

The Jahn-Teller distortion also manifests a tetragonal crystal system due to elongation or compression along a particular direction due to crystal field splitting. In the high-temperature phases of $NaNbO_3$, the 4 highest peaks in cubic and tetragonal crystal systems were observed [65]. Here the planes with non-integral values of Bragg peaks at low 2θ value was observed in the tetragonal phase which suggests the presence of very weak difference in reflections on pseudocubic cell i.e. (1/2, 3/2, 0), (1/2, 3/2, 1) which is equivalent to (130), (131) on the pseudocubic cell 2a*2a*cp. There are also odd-odd-even type of tilt system $a^0a^0c^+$ in *C4/mmb* [74-75, 65]. It could be possible that there are tilt reflections

(310), (130), (312) c+ tilt

(131), (151) a- tilt

The equality of *a* and *b* lattice parameters suggests that the tilt angles are equal. Two possibilities arise i.e., $a^-a^-c^+$ (*Pbnm*) and $a^-a^-c^0$ (*Imam*) [76-77]. The tilt system spacegroup is the same as that of *Pbnm* with origin halfway between Ca atoms along [001]. It is easily seen that Ti atoms are not displaced at all, as they lie in the centre of symmetry [78-80]. The cation displacement must lie along [1-10]. It is easily seen that the Ti atoms are not displaced at all, as they lie in the centre of symmetry [80].

In the composition $PbZr_{0.9}Ti_{0.1}O_3$ if the (h'h'h') peaks are split, while if the (h'00) are not split it results in ($a^-a^-a^-$) tilting with symmetry *R-3c* [64-70]. The parallel cation displacement means the spacegroup symmetry will be *R3c* [71-72]. This means tilts are lost but displacement remains to give spacegroup *R3m*. At higher temperature the symmetry becomes cubic and displacements appear to give paraelectric phase ($a^-a^-a^-$ (*R3c*) to $a^0a^0a^0$ (*R3m*) to $a^0a^0a^0$ (*Pm3m*)) [64-70].

In the case of Jahn-Teller distortion, the occupied $3d_{3x2-r2}$ and $3d_{3z2-r2}$ orbitals order in *ac* plane as exemplified by crystal structures of $LaMnO_3$ and $KCuF_3$ [80-85]. This pattern maintains a favourable coordination environment about symmetric M'-cation sites. The orbital ordering found in rocksalt perovskite $A_2MM'X_6$ is compatible with octahedral rotations about the c-axis ($a^0a^0c^-$) but appears to be compatible with $GdFeO_3$ type octahedral tilting ($a^-b^+a^-$). For example, $A_2CuMO_6$ (A: Sr/Ba, M:  W/Te) shows *I4/m* spacegroup [51-53, 85-90].

Table 1.1: Octahedral tilts in perovskite resulting in a change in space group in 15 tilt system [36].

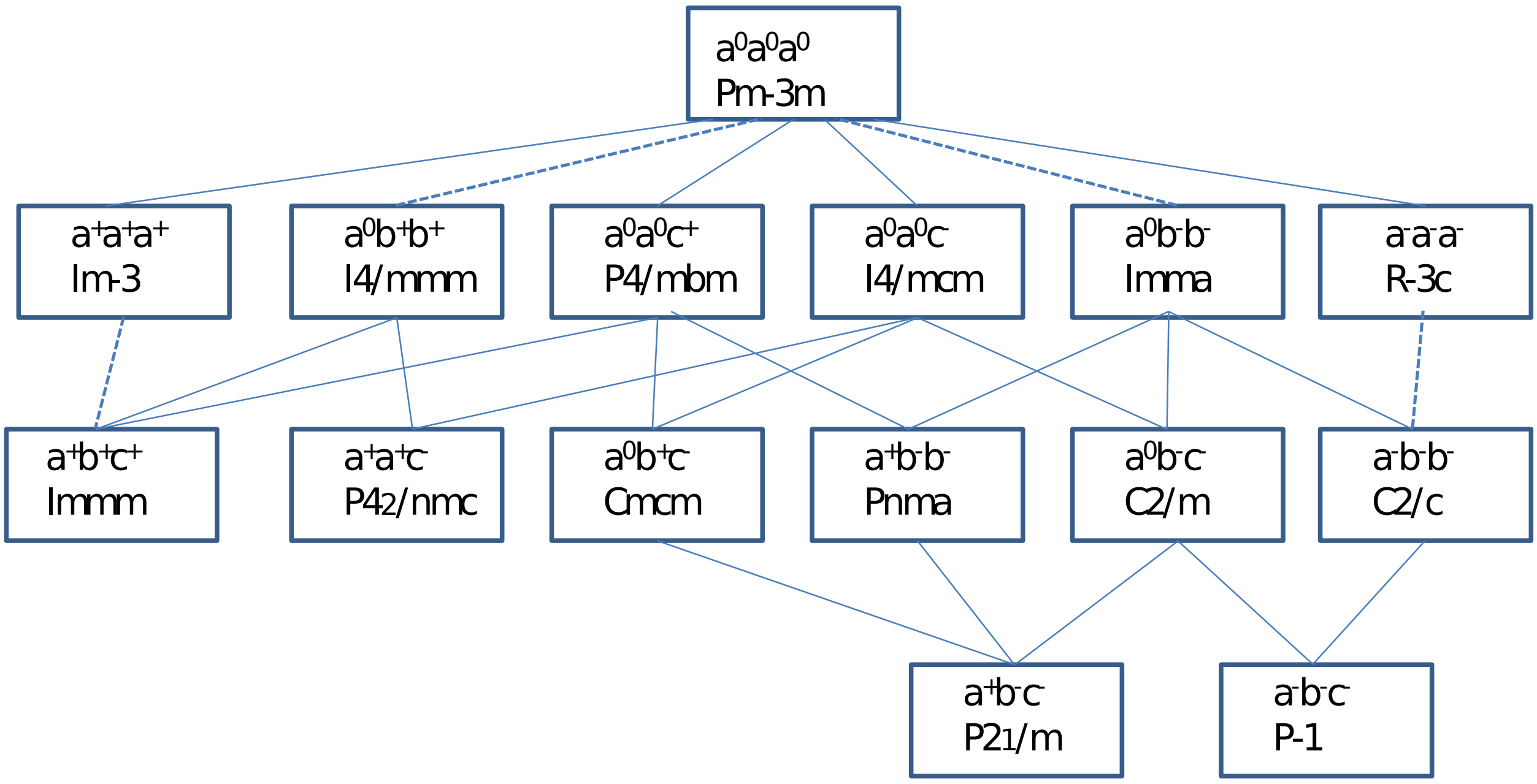

## 1.2 Magnetic properties in perovskites

### 1.2.1 Different types of the spin structure of perovskites

There are varying magnetic interactions between the B-site atoms which can be generally be segregated by two magnetic interactions that are by nearest neighbour interaction between the two magnetic interactions and secondly by the next nearest neighbour interaction (nnn) denoted by the B – O – O –B’ interactions [91-93].

The magnetic spin structure depends on the interactions between the A-site or the B-site magnetic interactions. For example, in $EuTiO_3$ the magnetic interactions between the Eu ion are denoted by $J_1$ for nearest neighbour magnetic interaction and $J_2$ denotes the next nearest neighbour magnetic interaction [94]. Depending on these magnetic interactions it demarcates the Néel temperature for the particular system. For example, on comparison of the magnetic interaction between $EuTiO_3$ and $EuZrO_3$, it results in a change in Néel temperature from 5.3 K to 4.1 K respectively [94-95]. The distance between neighbouring A-site distance is greater in $EuZrO_3$ than $EuTiO_3$ which results in a decrease in the magnetic interactions and thus a decrease in Néel temperature. Furthermore, the nearest neighbour (nn) and next-nearest neighbour (nnn) magnetic interactions results in the varying degree of antiferromagnetic spin structure (Fig 1.3).

Orbital and spin ordering can occur without charge ordering even in the case of half doped manganites, for example in $Ln_{1-x}A_xMnO_3$ (Ln: rare earth, A: Ca, Sr) where an A-type antiferromagnetic structure is observed at low temperature [96-100]. This charge ordering state leads to superstructure due to orbital ordering of $e_g$ electrons in XY plane [96-100].

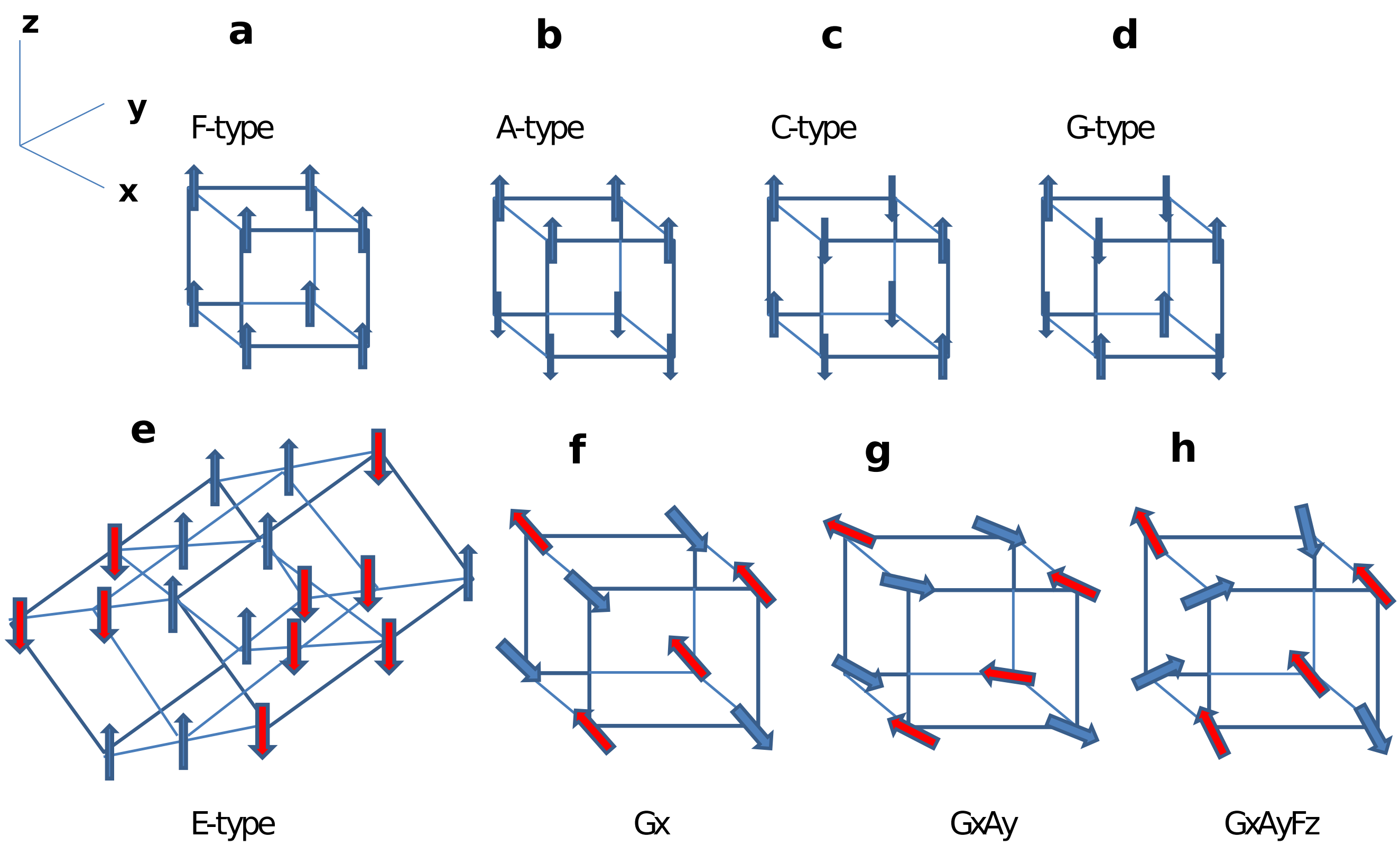

Fig. 1.3: Antiferromagnetic spin structures of (a-d) various structures of perovskites, and e-h) spin canting taking place along x, y and z directions.

### 1.2.2 Interplay of orbital ordering, octahedral tilting and cation ordering

The correlation between orbital ordering, cation ordering and octahedral tilting dictate to a great extent on the magnetic spin structure and its stabilization. For example, the Jahn-Teller distortion stabilizes the orbital ordering of $3d_{3x2-r2}$ and $3d_{3z2-r2}$ in the *ac* plane in the case of $LaMnO_3$ and $KCuF_3$ [81, 101-106] (Fig. 1.4). For example, $Sr_2MnMO_6$ (M: Sb, Ru, Ta and Nb) which crystallizes in *I4/mcm* shows octahedral tilting along the *c* axis ($a^0a^0c^-$ tilt) (Fig. 1.4) [101-106].

It has also been reported that when the long-range orbital ordering gets disrupted by the presence of a non-magnetic ion or ion which does not show orbital overlap then it results in weakening of the magnetization by blocking the magnetic exchange interaction [37]. For example, in $Sr_3CrFeMoO_9$ results in $Cr^{3+}$ - $Fe^{3+}$ which gets interrupted by the presence of $Mo^{6+}$ ion. This disruption in long-range ordering results in formation of $Cr^{3+}$ - $Cr^{3+}$ and $Fe^{3+}$ - $Fe^{3+}$ antiferromagnetic cluster interaction [107-115].

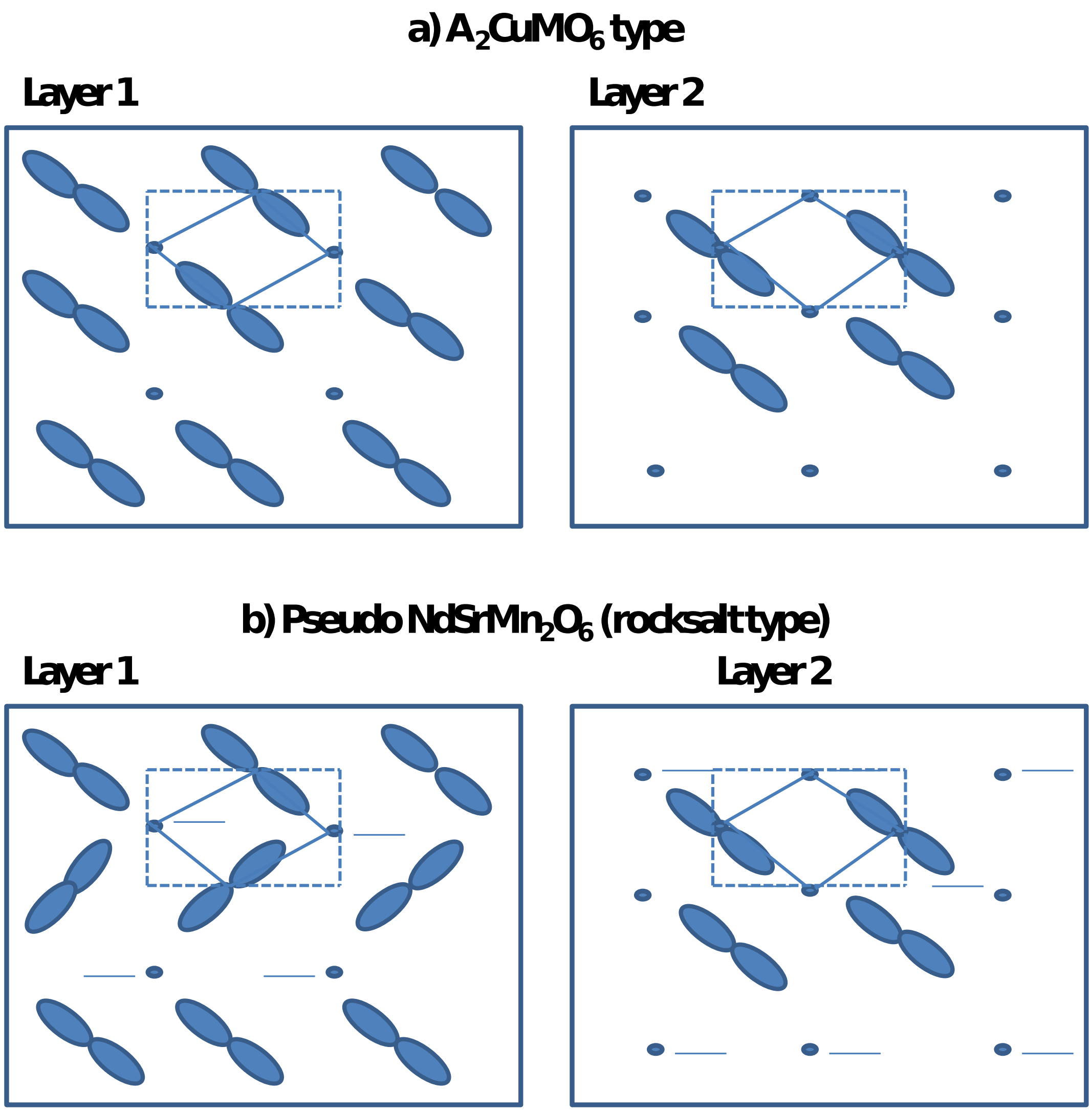

Fig. 1.4: Orbital orientation schemes of layers 1 and 2 in a)$A_2CuMO_6$ (M: Mn), b) $NdSrMn_2O_6$ . Here there is the presence of Jahn-Teller active ion at B-site, greater in (b) than (a), resulting in more symmetric octahedral coordination in (a). The solid line represents a simple cubic unit cell, while the dotted line show *a* and *c* axes of *the Pnma* unit cell [102].

### 1.2.3 Basic magnetic properties in perovskites

At a given field H, the magnetic moment tends to align in the field direction. However, at high temperatures, the thermal energy of the moments is very high, which results in paramagnetism from the randomly aligned moments. On decreasing the temperature, the randomness gets reduced and thus the moments try to interact with each other and tends to overcome the paramagnetic region [116-120]. For a paramagnetic material, the magnetic susceptibility ($\chi$) lies in the range $10^{-3} – 10^{-5}$ (>0) [116-120] (Fig. 1.5).

In transition metal cations the magnetic moment from partially filled d-shells and anions ensure spatial separation between cations. In case the magnetic moment is caused by highly localized f-electrons, which do not overlap with f-electrons on adjacent ions [40].

Langevin theory using equations (equation 1.3-1.4) aptly shows the effect of temperature on spin alignment, by taking into account the interacting moments that are placed randomly at their atomic sites (paramagnetic region) [106-124].

Here $p(\theta)$ is the probability of moments generated by individual atoms which make an angle in the range $\theta$ and $\theta+d\theta$ with H. The Curie-Weiss theory describes this interaction between the localized moments (Fig. 1.6 and equation 1.2).

Weiss assumed $H_W$ of molecular field is directly proportional to the magnetization

$H_W = \gamma M$ (Here $\gamma$ is molecular field constant)

$H_{total} = H + H_W$ (Equation 1.2 a)

Since $\chi = M/H = C/T$ (Equation 1.2 b)

$M/H+\chi M = C/T$ (Equation 1.2 c)

$M = CH / T\text{-}C\gamma$ (Equation 1.2 d)

Therefore $\chi = M/H = C/T\text{-}\theta$ (Equation 1.2 e)

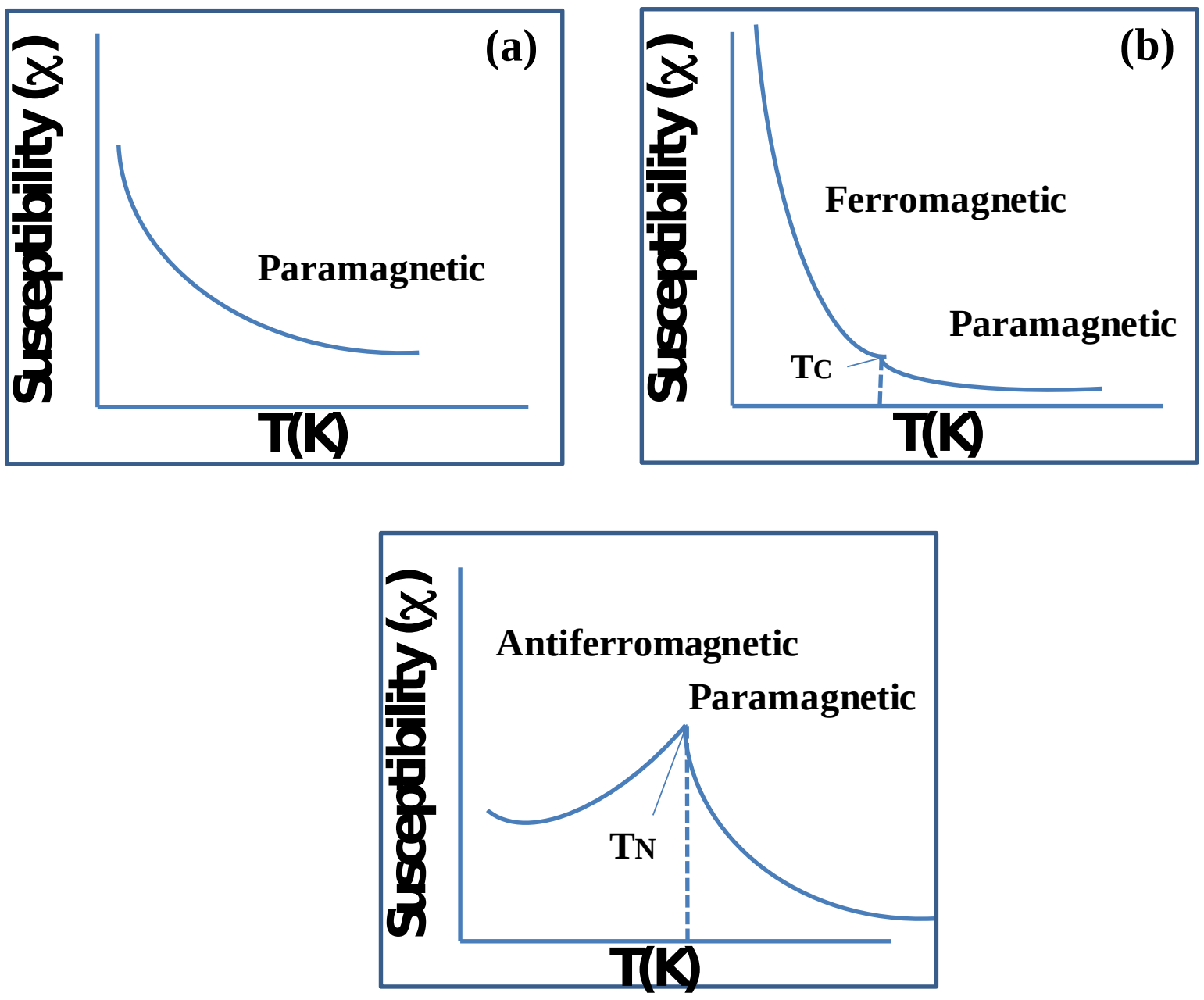

Fig. 1.5: Magnetic susceptibility vs. temperature plot of a) paramagnetism, b) ferromagnetism and c) antiferromagnetism.

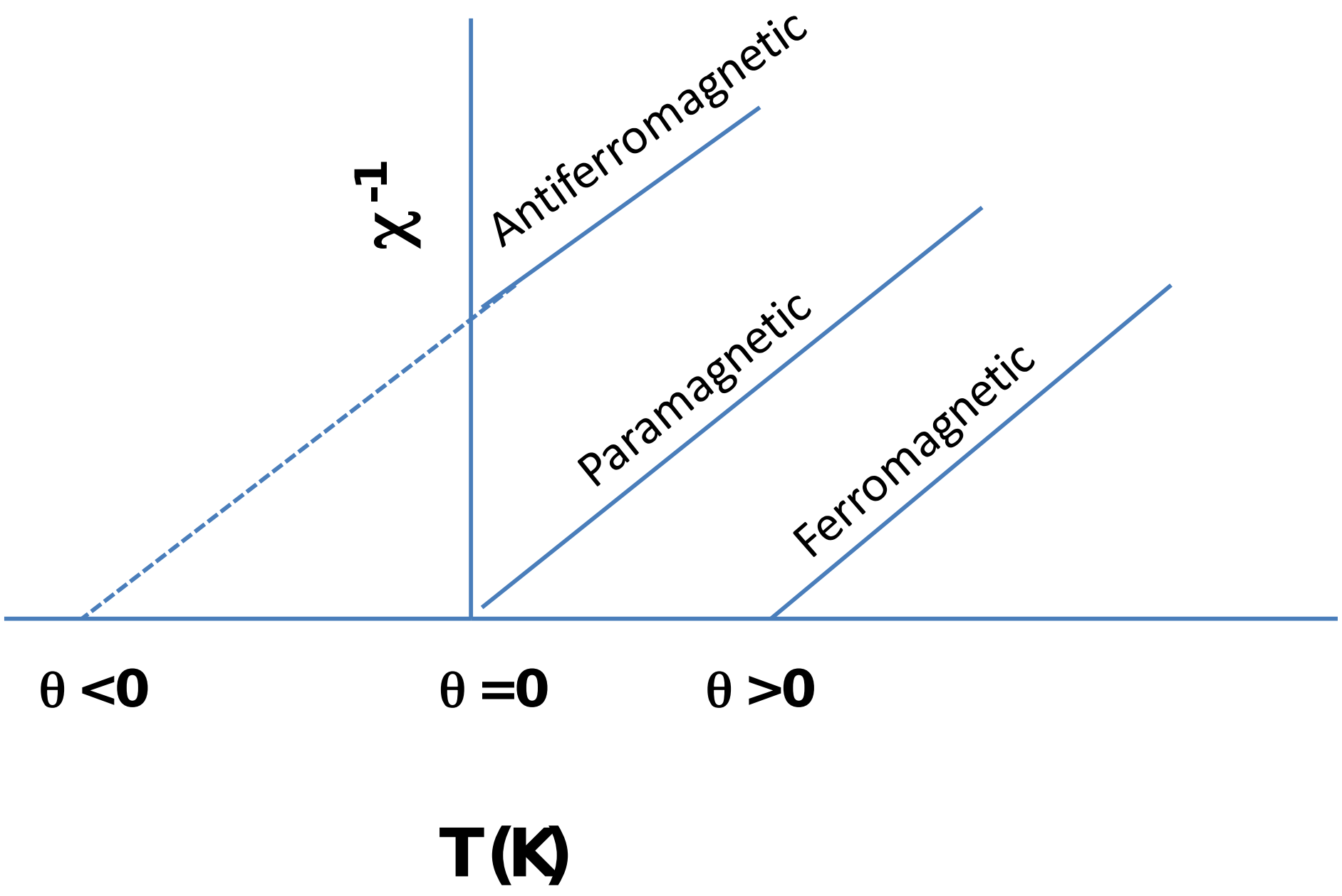

Fig 1.6: Inverse susceptibility vs. temperature plot for Curie Weiss and Curie law plot.

Langevin theory of paramagnetism (equation 1.3 and 1.4): In the paramagnetic region magnetic moments are randomly oriented, with non-interacting spins at their atomic site. As the temperature is decreased the atomic moments tend to align along the direction of the field [67]. "Energy of a magnetic moment in the direction of the field H

$E = -m.H \cos\theta$ (Equation 1.3a)

The probability of a magnetic moment making an angle θ in the direction of the field H

$e^{-E/K_BT} = e^{mH/k_BT} = e^{mH\cos\theta/k_BT}$ (Equation 1.3 b)

Probability p(θ) of an atomic moment making an angle θ and θ+dθ.

$$p(\theta) = \frac{e^{mH\cos\theta/k_BT}\sin\theta\, d\theta}{\int_0^{\pi} e^{mH\cos\theta/k_BT}\sin\theta\, d\theta} \quad \text{(Equation 1.3 c)}$$

Magnetic is the entire in presence of field H

$$M = Nm\cos\theta = Nm\int_0^{\pi}\cos\theta\, p(\theta)\, d\theta \quad \text{(Equation 1.3 d)}$$

$$= \frac{Nm\int_0^{\pi} e^{mH\cos\theta/k_BT}\cos\theta\sin\theta\, d\theta}{\int_0^{\pi} e^{mH\cos\theta/k_BT}\sin\theta\, d\theta} \quad \text{(Equation 1.3 e)}$$

$$= Nm\left[\coth\left(\frac{mH}{k_BT}\right) - \frac{k_BT}{mH}\right] = NmL(\alpha) \quad \text{(Equation 1.3 f)}$$

Where $\alpha = \frac{mH}{k_BT}$ and $L(\alpha) = \coth(\alpha) - 1/\alpha$ : Langevin function

If α is large when field is large or temperature T tends tozero, then M ≈ Nm, which means that then there will be complete alignment of spins along H.

On expanding Taylor's series

$L(\alpha) = \frac{❑}{3} - \frac{❑^3}{45} + \dots$ (Equation 1.4 a)

Thus we get $M = \frac{Nm}{3} = \frac{N m^2 H}{3 k_B T}$ (Equation 1.4 b)

$M = (\frac{N \mu_0 m^2 A}{3 k_B T} ¿ (\frac{H}{T})$

Thus $\chi = M/H = Nm^2/3k_BT = C/T$ (Equation 1.4 c)

On taking into consideration only discrete orientation due to spatial quantization

$M = NgJ\mu_B \left[ \frac{2J+1}{2J} \coth h \left( \frac{2J+1}{2J} \right) - \frac{1}{2J} \coth \left( \frac{❑}{2J} ¿ \right] = N g J \mu_B B_J(\alpha)$ (Equation 1.4 d)

The Brillouin function $B_J$ can be expanded in Taylor series

$B_J(\alpha) = \frac{J+1}{3J}\alpha - \frac{\left[ (J+1)^2 + J^2 \right] (J+1) ❑^3}{90 J^3} + \dots$ (Equation 1.4 e)

Here $\alpha = J g \mu_B H / k_BT$

$\chi = \frac{N g^2 J (J+1) \mu_B{}^2}{3 k_B T} = C/T$ (Equation 1.4 f)

Thus $C = \frac{N g^2 J (J+1) \mu_B{}^2}{3 k_B T} = \frac{N\, meff^2}{3 k_B}$ (Equation 1.4 g)

Here $m_{eff} = g\ (J(J+1))^{½}\ \mu_B$ (Equation 1.4 g)

Hence proved" [62].

## 1.3 Characterization techniques

### 1.3.1 Powder X-ray Diffraction (PXRD)

Powder X-ray diffraction is a commonly used technique to characterize our solid sample in powder form. It is used to determine the phase purity, phase crystallinity and impurity and unit cell parameters from the diffraction pattern. Every compound has its characteristic arrangement of lattices, with a specific distance between the planes which are detected by the constructive interference of the beams diffracted. Different planes diffract beams with different intensities which depends on the electron density of the elements. When there is high electron density then it results in better scattering which as a result contributes to intensity. However, for light elements, for example hydrogen, produces weak scattering, because of low electron density and as a result, its scattering gets overshadowed by heavier elements.

In the Bragg-Brentano geometry, parallel planes make an angle $\theta$ with the incident X-rays. The unreflected beam thus makes an angle $2\theta$ with the reflected beam.

Thus in the powdered sample, in the detector for a particular $\theta$, a cone of reflections are observed. These diffraction impressions would give use the d-spacing between the planes.

The interaction of X-rays which interacts with the sample and the diffracted beams constructively interferes when it fulfils the Bragg's condition and an intense peak is observed

$n\lambda = 2d\sin\theta$ Bragg's law (Equation 1.5)

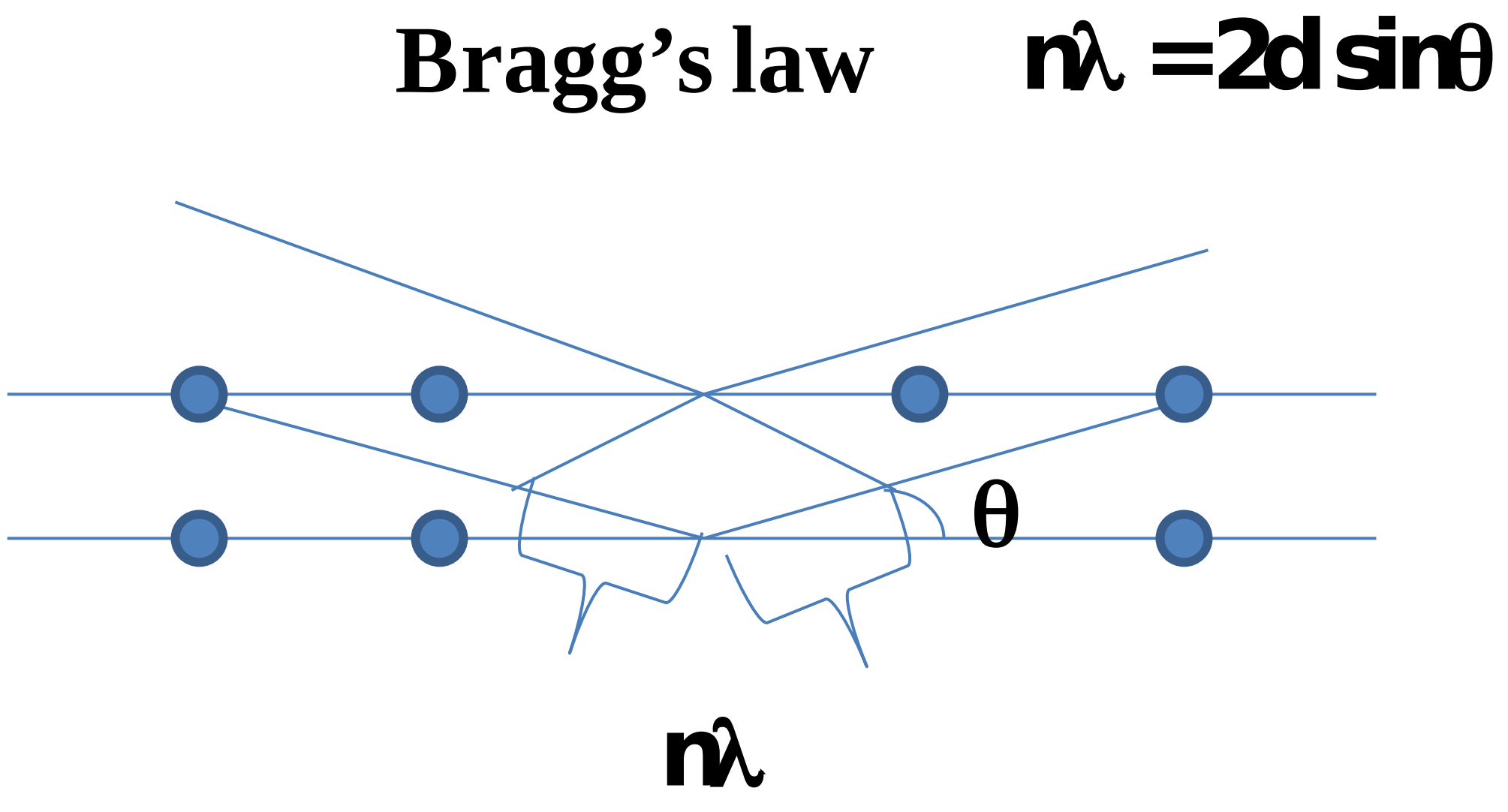

Fig. 1.7: X-ray diffracted by parallel planes obeying Bragg's law by constructive interference of incident and diffracted beam.

When there is constructive interference, then the path difference between the two rays should be $n\lambda$ (n: order of reflection) (Fig. 1.7).

In our experiment we have used Bruker D8 Advance diffractometer, operating at 30 kV and 40 mA, with Cu-Kα source ($\lambda$ = 1.54 Å), in the 2θ range 10˚-80˚. For the PXRD measurement, few milligrams (mg) of the finely grounded was taken in a quartz sample holder and was placed in the groove. The sample was smoothened with the help of a slide and was kept at affixed height and fixed place from the X-ray source. The collected diffraction data were analysed by comparing the observed data with the X-ray diffraction database (JCPDF-PDF). Xpert Highscore and PCPDF-WIN software were used for detecting phase purity and for lattice parameter calculations. The unit cell of the compounds was calculated using the least square refinement, using the TOPAZ 5.2 software. The crystal structure, spacegroup, atomic positions, occupancy, thermal parameters were calculated by simulation of data obtained and taking the known parent compound as the standard.

A slow PXRD data were collected over 2 hours were collected and the following parameters were taken into consideration for a successful refinement.

### 1.3.2 X-ray Photoelectron Spectroscopy (XPS)

XPS is the most useful surface analysis technique, measurable at a depth of 5 nm, which can be performed on a wide range of materials. The XPS technique is used for the quantitative and qualitative analysis of chemical compounds based on the binding energies of elements of electrons in their respective orbitals and oxidation states for a particular ion. XPS is performed by exciting the sample surface by a beam of monochromatic X-ray, usually, Al-K$\alpha$ or Mg-K$\alpha$ which when falls on the sample results in the emission of photoelectrons which are detected. Here to avoid impurity the sample is kept under very high or ultra-high vacuum conditions ($10^{-8}$ to $10^{-9}$ mbar). The energy analyser is hemispherical which detects the intensities and energy level and energy level splitting of the photoelectrons emitted. The peak fit software was used to deconvolute the peaks.

### 1.3.3 Energy-dispersive X-ray spectroscopy (EDAX)

EDAX technique is used for elemental compositional analysis and mapping analysis of the samples in powdered form. When an incident X-ray falls on the sample, it generates various signals, depending primarily on elastic and inelastic scattering of electrons: backscattered electrons, secondary electrons, X-rays are emitted from the sample surface. These backscattered electrons are used for compositional and elemental analysis [124].

The EDAX analysis is collected over several regions in a variety of scales (nanometer to microns) to collect the average elemental ratios of the elements and to obtain their uniformity. The FESEM EDAX measurement was performed using JEOL, JSM-7800F electron microscope. The source of the instrument was a field emission gun which was lanthanum hexaboride. The samples for SEM analysis were prepared by smearing a very small amount of sample over a carbon tape, which was fixed over an aluminium stub and was kept at a particular height from the source (field emission gun) and detector. For getting proper conductivity in a non-conductive sample, the sample was coated with a few layers of gold coating by using the gold sputtering instrument in the Argon atmosphere.

### 1.3.4 UV-Vis Diffuse Reflectance Spectroscopy (UV-Vis DRS)

UV-Vis DRS is an optical technique that is being performed on solid samples for optical measurement. We can measure the bandgap and absorbance of a sample. Shimadzu UV-2450 UV-Vis spectrophotometer with a wavelength range of 200-800 nm was used for our measurement. Here in this technique, diffuse reflectance of light is collected from all possible angles [125].

For the measurement few milligrams of the sample was mixed with barium sulphate, and was pressed uniformly to form a flat surface which was measured, by comparing the absorbance shown by a pure white material of barium sulphate as a standard. Initially, for a blank run, pure barium sulphate was used in the sample holder for the subtraction of the baseline. The sample holder was kept into an integrating sphere of a spectrophotometer to record the reflectance.

To convert the absorbance data into reflectance mode in the Kubelka-Munk equation, the following equation was used (equation 1.6).

$F(R_\infty) = (1-R_\infty)^2 / 2R_\infty$    Equation 1.6

$F(R_\infty)$ is the Kubelka-Munk function and R is the diffuse reflectance. The bandgap of the sample was obtained by extrapolating the linear portion of the absorbance coefficient vs. energy ($E_g$).

$(\alpha h\nu)^{1/n} = A(h\nu - E_g)$    Equation 1.7 (Tauc equation)

The Tauc equation (equation 1.7), A (absorbance), Eg (bandgap), h (Planck's constant), ν (frequency) and α (absorption coefficient) were used. Here the exponent value n denotes the nature of electronic transition, whether direct or indirect. For indirect transition n = 2 and for direct transition n = ½.

### 1.3.5 Photoelectrocatalytic measurement (PEC)

A photoelectrocatalytic cell uses light and current to split water into hydrogen and oxygen. In the anode electron and hole, separation takes place in presence of light. The electron flow and form $H^2$ molecule. The $OH^-$ group then flows through the electrolyte and reacts with 4 holes. The bandgap required to split water in the semiconductor is 1.23 V with the appropriate alignment of the valence band and conduction band edges. Perovskite solar cells are commercial used photoelectrochemical cells for storage to electricity.

The Mott-Schottky equation (equation 1.8) is used for finding the flat band potential, that is where the Fermi energy level of semiconductor and electrolyte surface matches. The mott-Schottky equation relates the capacitance(C) and voltage applied across the semiconductor electrolyte surface.

$$\frac{1}{C^2} = \frac{2}{\varepsilon\varepsilon o A^2 e N_d}\left( V - V_{fb} - \frac{k_B T}{e} \right) \quad \text{Equation 1.8}$$

Here ε is the dielectric constant of semiconductor and εo is the permittivity of free space. A is the area of the depletion region, Nd is the density of dopants, V is applied potential, $V_{fb}$ is flat band potential of semiconductor electrolyte interface, $k_B$ is Boltzman constant and T is temperature and e stands for the elementary charge.

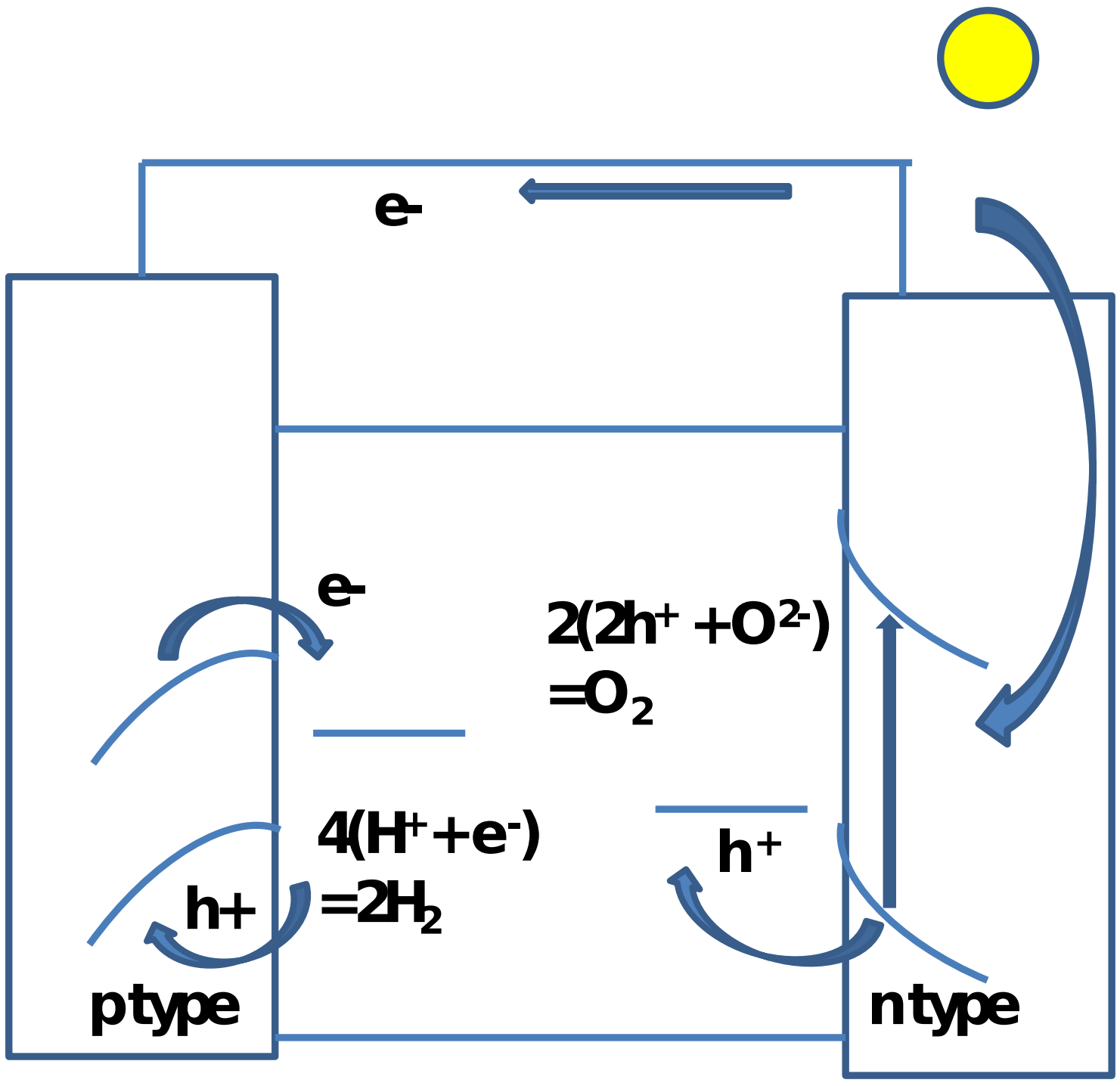

Fig. 1.8: Water splitting in a photo-electrocatalytic setup.

### 1.3.6 Magnetic Characterization

SQUID (Superconducting Quantum Interference Device) is used to measure magnetic field strength as low as microtesla and picotesla levels. Device operating at cryogenic temperatures with quantum limited sensitivity can detect field as low as $10^{-17}$ T [126-127]. In DC SQUID 2 superconducting SQUID in ring pattern are connected by 2 Josephson junction (Fig. 1.9). Josephson junction is the thin barrier in the path which separates two superconducting region surrounding the magnetic flux ϕ. Above the critical current of the Josephson junction difference in voltage between the electrodes is measured. In this magnetic flux region containing Josephson junction constant voltage and magnetic flux are maintained and the strength of the field being measured should equal to the strength of the bias magnetic field passing through the SQUID. Thus, weak magnetic field change is being detected [126-127].

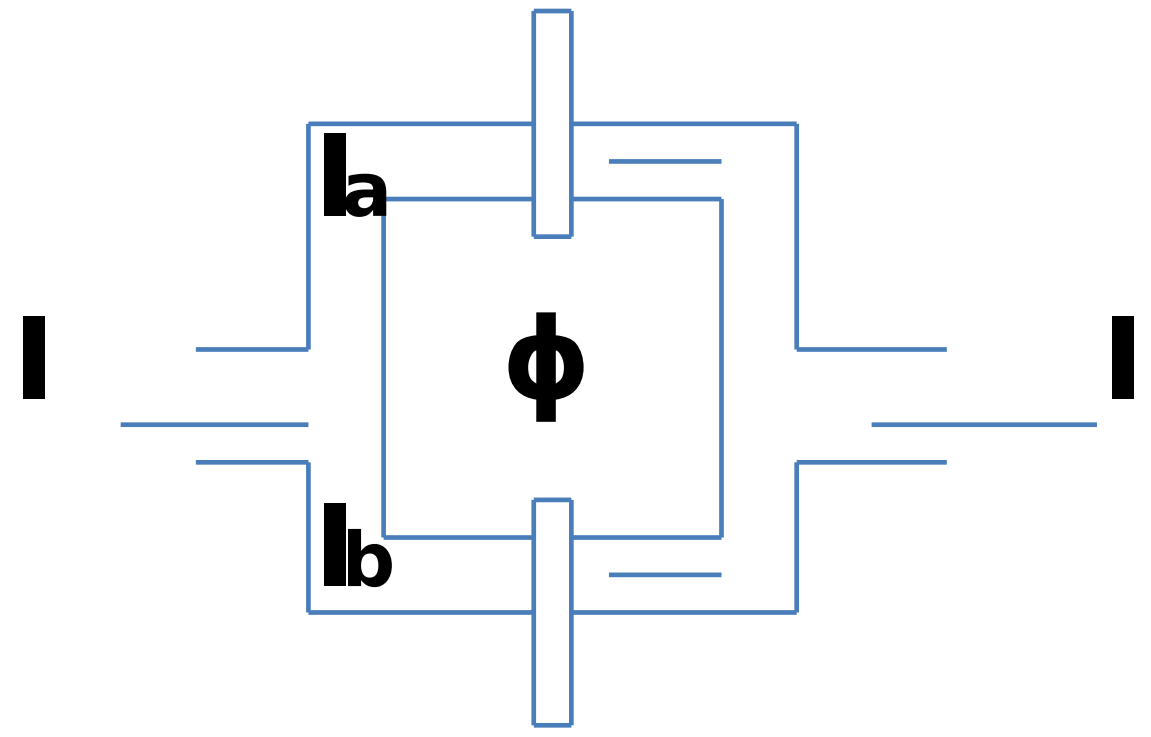

Fig. 1.9 Diagram of DC SQUID.

We have also used Physical Properties Measurement System (PPMS) to carry out some of our magnetic studies. A Vibrating Sample Magnetometer (VSM) option was used during the magnetic measurements. Temperature dependent DC susceptibility measurements were carried out following the zero field cooling (ZFC) and field cooling (FC) protocols, while the field dependent isothermal magnetization was measured only in zero field cooling protocol. Curie-Weiss law was used in the high temperature range of the inverse susceptibility data to estimate the effective magnetic moment (μeff) and the Weiss temperature ($\theta_{CW}$) of the sample.

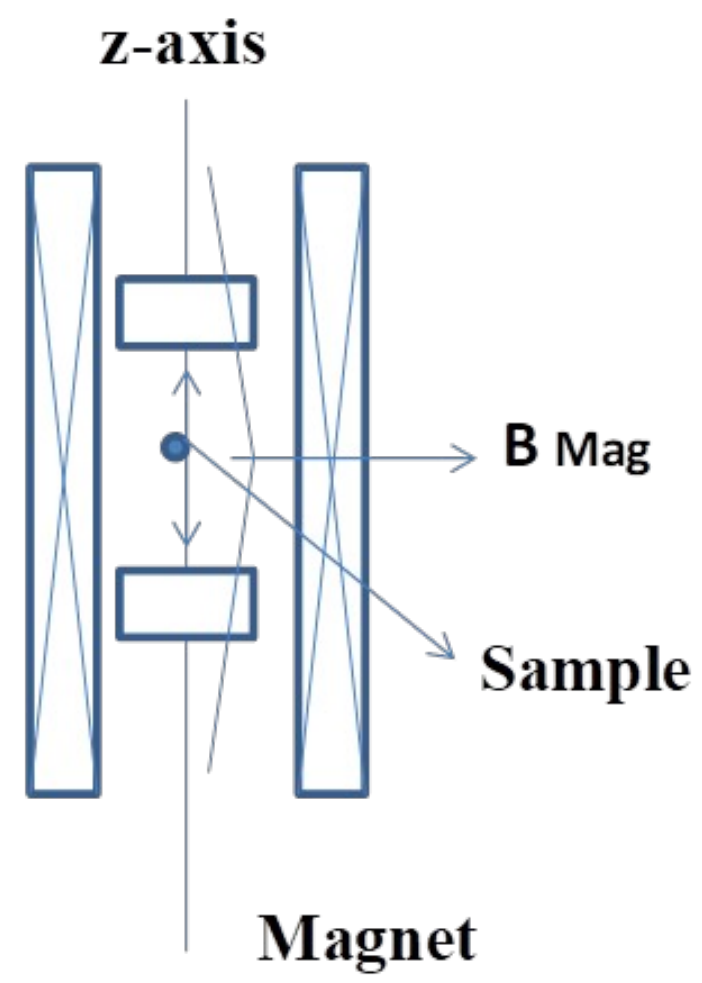

Fig 1.10: Diagram of VSM

In a Vibrating Sample Magnetometer (VSM) the AC field is being detected by coils surrounding the sample, which arises from the change in voltage, generated when the sample is being vibrated with an amplitude A, and frequency ν, along the z direction (Fig.1.10) [131]. Higher signal to noise ratio can be optimized when effective area is being covered by coil which is normal to applied magnetic field [131]. A VSM is able to detect microscopic spin structure, with domain alignment taking place in a wide range of temperature in bulk magnetic materials. Microscopic effects is being observed as a superposition of field arising from individual magnetic lattices [131-132]. It has been observed that during ferromagnetic to antiferromagnetic transition, kinetic arresting of the spins takes place creating inhomogeneous state [132]. As a result of this glass transformation, the magnetization due to field cooled cooling gets flattened out without touching zero field cooling magnetization [132]. The spin glass temperature is considered a temperature where characteristic relaxation time crosses the typical experimental time involved in measurements [132]. With the decrease in temperature the barrier height between metastable ferromagnetic to antiferromagnetic state, results in decrease in magnetization suggesting a decrease in free energy in the system, which is generally being observed in the temperature regime 25-40 K [132-133]. With decrease in temperature the metastable antiferromagnetic and ferromagnetic phases coexist in mesoscopic range, which suggests the bifurcation between the zero field cooling and field cooled cooling as a function of energy fluctuations [132-133].

# Chapter II

# Complex perovskites with Ho/Pr/La at A-site and Cr/Fe/Mn at B-site

# Chapter II

## Complex perovskites with Ho/Pr/La at A-site and Cr/Fe/Mn at B-site

### Abstract

Structure and magnetic properties were investigated with rare-earth substitution at A-site with Ho/Pr/La and transition metal ion substitutes at B-site with Fe, Cr and Mn. The following substitutions could be divided into three families as follows: a) $Ho_{1-x}Pr_x$ $(Mn_{0.5}Cr_{0.5})$ $O_3$ (x = 0, 0.3, 0.5 and 0.7), b) $Ho_{1-x}Pr_x$ $(Fe_{0.25}Mn_{0.25}Cr_{0.5})$ $O_3$ (x = 0, 0.5 and 1), and c) $La_{1-x}Pr_x$ $(Fe_{0.25}Mn_{0.25}Cr_{0.5})$ $O_3$ (x = 0, 0.5 and 1). From the Powder X-ray diffraction (PXRD) analysis all the three families were found to crystallize in the orthorhombic *Pnma* space group, and the lattice parameters were found to increase as the A-site radii increased. From the Rietveld analysis, a decrease of A-site radii resulted in increased octahedral distortion, with the difference in axial and equatorial B-O bond distance being maximum for x = 0.5. The ZFC (zero field cooled) and FC (field cooled) bifurcation in the DC magnetic susceptibility data, suggests spin-glass behaviour for all the compositions. In $Ho_{1-x}Pr_x$ $(Mn_{0.5}Cr_{0.5})$ $O_3$ (x = 0.3, 0.5 and 0.7) the ferromagnetic transition was found to increase with increase in x. From the comparison of magnetic susceptibility values at temperatures below 100 K, it was observed that $Ho_{1-x}Pr_x(Mn_{0.5}Cr_{0.5})O_3$ (x = 0.3, 0.5 and 0.7) showed an increase of magnetization value with an increase in x. However, $Ho_{1-x}Pr_x$ $(Fe_{0.25}Mn_{0.25}Cr_{0.5})$ $O_3$ (x = 0, 0.3, 0.5 and 0.7) showed decreased value of magnetization with increase in x. This suggests increased Ho-Mn magnetic interaction in the first case and competing Pr and Fe anisotropic orbital interaction in the second series. The difference between the observed and theoretical magnetic moments shows a larger deviation for $PrFe_{0.25}Mn_{0.25}Cr_{0.5}O_3$ and x = 0.5 value for all the three series, probably due to disorder-induced change of anisotropy. The maximum change in remanent magnetization was found to be maximum for $PrFe_{0.25}Mn_{0.25}Cr_{0.5}O_3$, which could be attributed reorientation of spins in the structure.

## 2.1 Introduction

Perovskite oxides of the formula $ABO_3$ (A: rare-earth element, B: transition metal), show varying structural and magnetic properties (Fig. 2.1). Here A-atom occupies the Cubo octahedral site and B-atom is octahedrally surrounded by oxygen atoms. These varied rare-earth and transition metal substitutions can have application in the field of multiferroics, magnetocaloric materials, depending on the varied interaction between the rare-earth and transition metal ions and the ionic radii [2-4]. The anisotropy of orbitals and crystal field splitting energy results in a change in transition temperature [4-6]. For example, in $HoFe_{1-x}Cr_xO_3$, on substitution of Cr in place of Fe results in the decrease of Néel temperature, $T_N$ = 647 K for x = 0 and 142 K for x = 1 [5-6], which has been reported due to the decreased Fe-Fe and increased Ho-Fe as x increased. The interaction of rare-earth and transition metal results in spin reorientation of rare-earth at low temperatures, which could be attributed to the spin-exchange interaction, Zeeman interaction and decrease of anisotropy, manifested by structural distortion [7-12]. For example, $Ho_{1-x}Pr_xFeO_3$, shows a spin reorientation temperature ($T_{SR}$) at 50 K for x = 0, and 76 K for x = 0.3. Substitution at B-site also results in a change in anisotropy and coercive field. For example, at x = 0.5 in $HoCr_{1-x}Fe_xO_3$ results in maximum anisotropy [5-7]. Furthermore, in $LaFe_{1-x}Cr_xO_3$ as the Cr concentration increased, it results in the disruption of Fe – O – Cr proportion, which leads to non-uniform distribution of ferromagnetic clusters, which was found to be maximum at x = 0.5 [13]. Thus, the interaction of partially filled 4f electron and transition metal 3d electron results in spin reorientation along with the easy axis of magnetization [14-16]. The increased value of spin reorientation transition temperature depends on the easy rotation of $Fe^{3+}$ spins from one crystal axis to another which results in complex Fe/Cr and Ho/Pr interaction [5-6]. Furthermore, the $Ho^{3+}$ ordering was found to increase with increase in x with a maximum at x = 0.5, i.e., from 8.5 K (x = 0.25) to 13 K, (x= 0.5) and 8 K (for x=1) in $HoCr_{1-x}Fe_xO_3$ [5-6]. Here we report the effect of interaction between rare-earth and transition metal ion and rare-earth interaction which may stabilize the rare-earth anisotropy. The super exchange interaction between $Pr^{3+}$ and $Mn^{3+}$ depends on the charge, crystallographic structure and relative orientation of corresponding orbitals [17-18]. The cluster glass exhibits a finite range of ferromagnetic ordering at a temperature below $T_C$ in addition to re-entrant spin glass, paramagnetic to ferromagnetic transition at a lower freezing temperature [19-20].

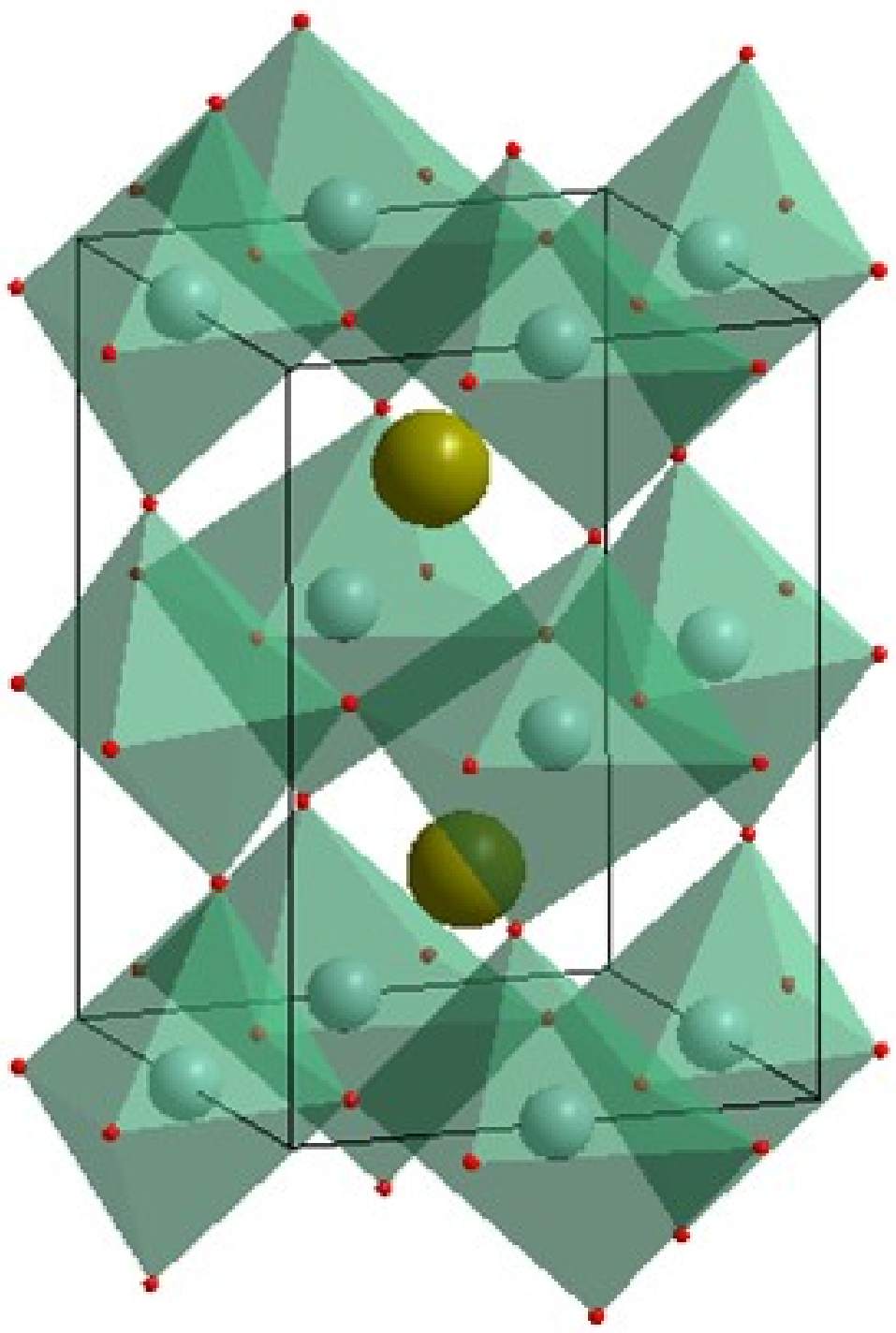

Fig. 2.1: Crystal structure of perovskite $ABO_3$ where the green dot represents A-atom and the blue dot represents B-atom which is octahedrally by oxygen atoms (red dots).

## 2.2 Experimental

### 2.2.1 Materials and synthesis

Oxides of the following three series $Ho_{1-x}Pr_x$ $(Mn_{0.5}Cr_{0.5})$ $O_3$ (x = 0, 0.3, 0.5 and 0.7),

$Ho_{1-x}Pr_x(Fe_{0.25}Mn_{0.25}Cr_{0.5})O_3$ (x = 0, 0.5 and 1) and $La_{1-x}Pr_x(Fe_{0.25}Mn_{0.25}Cr_{0.5})O_3$ (x = 0, 0.5 and 1) were synthesized in bulk form at high temperature reaction. $La_2O_3$, $Ho_2O_3$ and $Pr_6O_{11}$ were given a heat treatment at 900˚C for 12 hours, to remove moisture and carbonates. All the oxides were taken in stoichiometric amounts and a homogeneous mixture was obtained by thoroughly grinding. The following powders were calcined at 900˚C for 12 hours, and was given an intermediate grinding after cooling, and was subsequently heated at 1200˚C for 12 hours. In 0.2g of powder 20 µl of 5%, PVA solution was added, ground thoroughly and was pelletized at 10 tons of pressure and subsequently sintered at 1200˚C for 12 hours.

### 2.2.2 Characterization

Bruker D8-Advance diffractometer, with Cu-Kα radiation as the source was used to record data. The data was collected in the 2θ range from 10˚ to 80˚, having a step size of 0.02˚. TOPAZ 5.2 software was used for structural analysis of diffraction data. The background fitting, cell parameters, atomic positions and thermal parameters were obtained.

### 2.2.3 Magnetic studies

Zero field cooled (ZFC) and field cooled (FC) temperature dependent magnetization studies were performed in the temperature range of 2 K to 300 K, using a physical property measurement system (PPMS) at a magnetic field of 1000 Oe. The isothermal magnetization measurement was carried out at varying temperatures 5, 20, 70 and 300 K in the magnetic field range of ±7 T.

## 2.3 Results and discussion

### 2.3.1 Powder X-ray Diffraction studies

The PXRD of the three series of compositions were indexed with the orthorhombic *Pnma* space group (Fig. 2.2) with lattice parameters a : b : c = $\sqrt{2}a_p : \sqrt{2}a_p : 2a_p$, where $a_p$ denotes the parameter of ideal perovskite unit cell. The comparison of lattice parameters between $Ho_{1-x}Pr_x (Mn_{0.5}Cr_{0.5}) O_3$ (x = 0 and 0.5) and $Ho_{1-x}Pr_x (Fe_{0.25}Mn_{0.25}Cr_{0.5}) O_3$ (x = 0 and 0.5) suggested a decreased unit volume in the second case, which probably could be due to increased octahedral compression in this case. The peaks were shifted towards a lower 2θ value in $Ho_{1-x}Pr_x(Mn_{0.5}Cr_{0.5})O_3$ (x = 0, 0.3, 0.5 and 0.7) and $Ho_{1-x}Pr_x(Fe_{0.25}Mn_{0.25}Cr_{0.5})O_3$ (x = 0, 0.5 and 1), while in the third $La_{1-x}Pr_x(Fe_{0.25}Mn_{0.25}Cr_{0.5})O_3$ (x = 0, 0.5 and 1) peaks were shifted (Fig. 2.3) towards a higher 2θ value as the value of x increased. From the structural analysis, it was also observed that octahedral distortion increases as the A-site radii decreases, especially for x = 0.5 in all the three series, where a greater octahedral distortion was observed. Furthermore on the comparison, it was observed that greater octahedral distortion was observed in the second series $Ho_{1-x}Pr_x(Fe_{0.25}Mn_{0.25}Cr_{0.5})O_3$ (x = 0, 0.5 and 1) than in the first series. In the first and second series the *b, c* lattice parameters and volume increased with increase in x, however the *a* lattice parameter decreased with increase in x value. In the third series the *b, c* and volume decreased with increase in x and the *a* lattice parameter increases with increase in x. The octahedral B – O1 - B bond angle increased with increase in x in general suggesting greater octahedral distortion, except for intermediate values of x. In the third series however the B - O1 – B bond angle decreased with increase in x value except for x = 0.5.

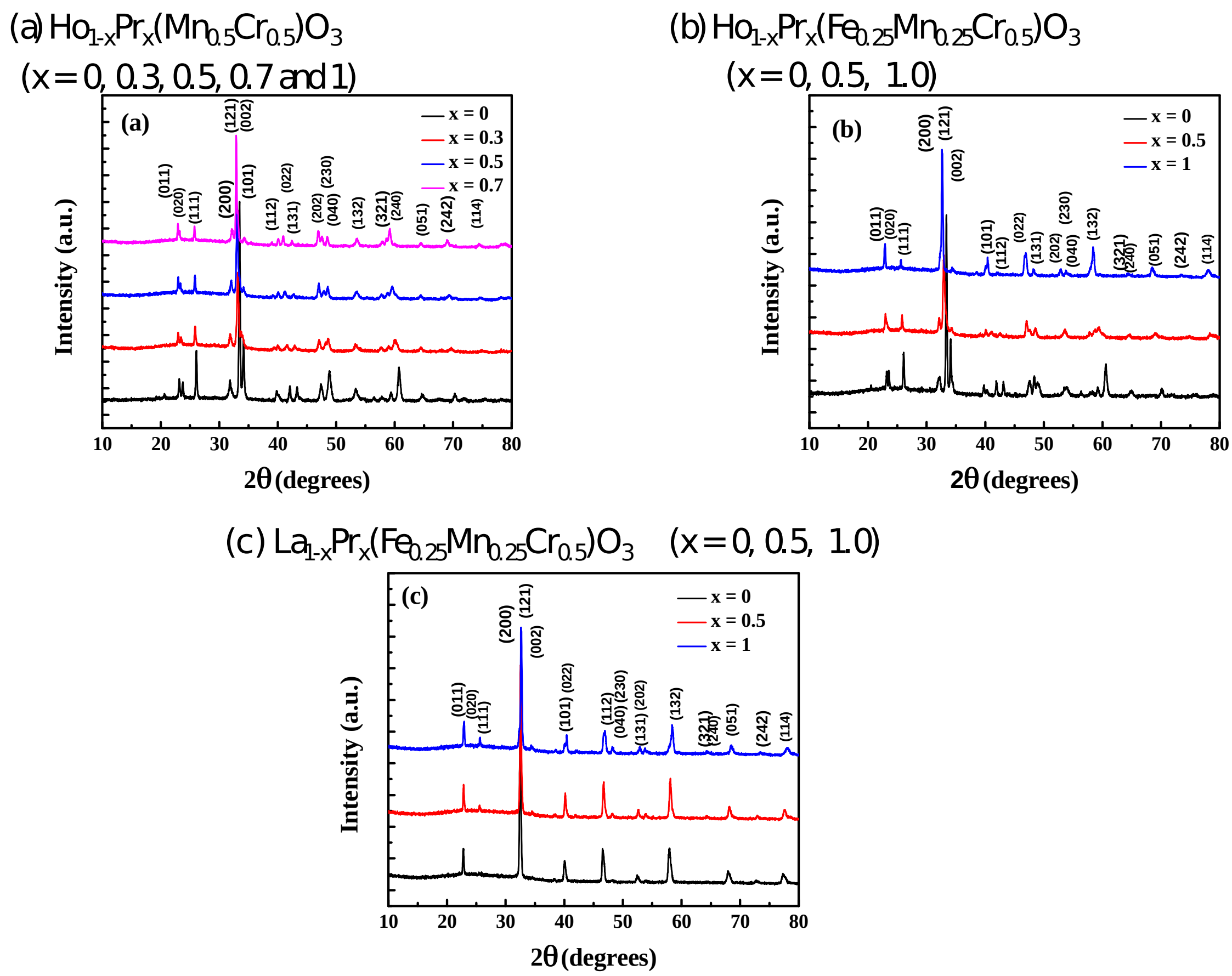

Fig 2.2: Powder X-ray diffraction data of a) $Ho_{1-x}Pr_x(Mn_{0.5}Cr_{0.5})O_3$ (x = 0, 0.3, 0.5 and 0.7), b) $Ho_{1-x}Pr_x(Fe_{0.25}Mn_{0.25}Cr_{0.5})O_3$ (x = 0, 0.5 and 1) and c) $La_{1-x}Pr_x(Fe_{0.25}Mn_{0.25}Cr_{0.5})O_3$ (x = 0, 0.5 and 1).

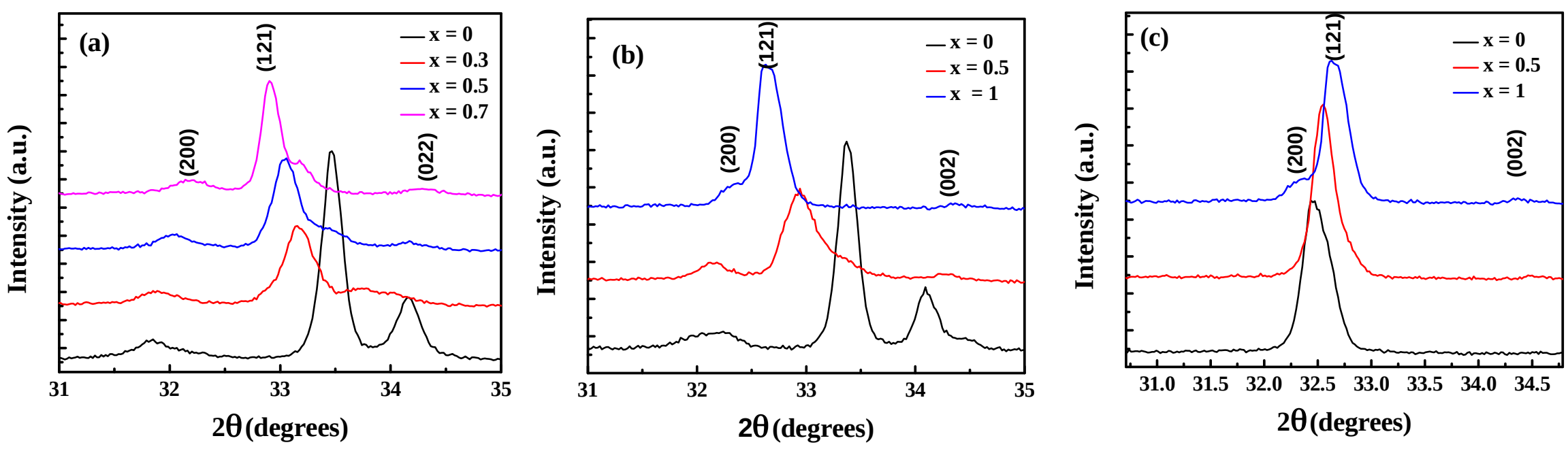

Fig 2.3: Enlarged peaks of Powder X-ray diffraction data of a) $Ho_{1-x}Pr_x(Mn_{0.5}Cr_{0.5})O_3$ (x = 0, 0.3, 0.5 and 0.7), b) $Ho_{1-x}Pr_x(Fe_{0.25}Mn_{0.25}Cr_{0.5})O_3$ (x = 0, 0.5 and 1) and c) $La_{1-x}Pr_x(Fe_{0.25}Mn_{0.25}Cr_{0.5})O_3$ (x = 0, 0.3 and 0.5).

Table 2.1: Rietveld refinement data of (a-d) $Ho_{1-x}Pr_x(Mn_{0.5}Cr_{0.5})O_3$ (x = 0, 0.3, 0.5 and 0.7), (e-g) $Ho_{1-x}Pr_x(Fe_{0.25}Mn_{0.25}Cr_{0.5})O_3$ (x = 0, 0.5 and 1) and (h-i) $La_{1-x}Pr_x(Fe_{0.25}Mn_{0.25}Cr_{0.5})O_3$ (x = 0.5 and 1).

a) $HoCr_{0.5}Mn_{0.5}O_3$ a = 5.6174(4) Å, b = 7.4735(2) Å, c = 5.2435(3) Å

| Site | x | y | z | Atom | Occ | B ($Å^2$) |
|---|---|---|---|---|---|---|
| 4c | 0.0737(3) | 0.25 | 0.0115(8) | Ho | 1 | 0.37(5) |
| 4b | 0.5 | 0 | 0 | Cr | ½ | 0.40(3) |
| 4b | 0.5 | 0 | 0 | Mn | ½ | 0.55(6) |
| 4c | -0.020(4) | 0.25 | 0.65(3) | O1 | 1 | 2.0(1) |
| 8d | 0.736(4) | -0.047(3) | 0.270(5) | O2 | 1 | 2.3(1) |

b) $Ho_{0.7}Pr_{0.3}Mn_{0.5}Cr_{0.5}O_3$ a = 5.5975(6) Å, b = 7.5347(2) Å, c = 5.3000(6) Å

| Site | x | y | z | Atom | Occ | B ($Å^2$) |
|---|---|---|---|---|---|---|
| 4c | 0.0634(1) | 0.25 | 0.012(1) | Ho | 7/10 | 0.37(5) |
| 4c | 0.0634(1) | 0.25 | 0.012 (1) | Pr | 3/10 | 0.37(5) |
| 4b | 0.5 | 0 | 0 | Mn | 1/2 | 0.55(6) |
| 4b | 0.5 | 0 | 0 | Cr | 1/2 | 0.55(6) |
| 4c | -0.061(1) | 0.25 | 0.6345(5) | O1 | 1 | 2.1(1) |
| 8d | 0.724(2) | -0.037(2) | 0.267(2) | O2 | 1 | 2.2(1) |

c) $Ho_{0.5}Pr_{0.5}Mn_{0.5}Cr_{0.5}O_3$ a = 5.5854(6) Å, b = 7.5979(8) Å, c = 5.3539(6) Å

| | | | | | | |
|---|---|---|---|---|---|---|
| 4c | 0.0601(2) | 0.25 | 0.0095(1) | Ho | ½ | 0.9(1) |
| 4c | 0.0601(2) | 0.25 | 0.0095(1) | Pr | ½ | 0.9(1) |
| 4b | 0.5 | 0 | 0 | Mn | ½ | 0.7(3) |
| 4b | 0.5 | 0 | 0 | Cr | ½ | 0.7(3) |
| 4c | -0.058(2) | 0.25 | 0.624(2) | O1 | 1 | 2.0(5) |
| 8d | 0.720(3) | -0.047(1) | 0.269(3) | O2 | 1 | 2.37(5) |

d) $Ho_{0.3}Pr_{0.7}Mn_{0.5}Cr_{0.5}O_3$ a = 5.5589(6) Å, b = 7.6439(7) Å and c = 5.3938(2) Å

| | x | y | z | Atom | Occ | B ($Å^2$) |
|---|---|---|---|---|---|---|
| 4c | 0.053(1) | 0.25 | 0.009(2) | Ho | 3/10 | 1.1(1) |
| 4b | 0.053(1) | 0.25 | 0.009(2) | Pr | 7/10 | 1.1(1) |
| 4c | 0.5 | 0 | 0 | Mn | ½ | 1.5(1) |
| 4c | 0.5 | 0 | 0 | Cr | ½ | 1.5(1) |
| 4c | -0.052(1) | 0.25 | 0.6004(1) | O | 1 | 1.9(2) |
| 8d | 0.734(3) | -0.042(3) | 0.287(3) | O | 1 | 1.9(2) |

e) $HoFe_{0.25}Mn_{0.25}Cr_{0.5}O_3$ a = 5.5603(4) Å, b = 7.5181(5) Å, c = 5.2475(3) Å

| | x | y | z | Atom | Occ | B ($Å^2$) |
|---|---|---|---|---|---|---|
| 4c | 0.0612(4) | 0.25 | 0.0245(8) | Ho | 1 | 0.4(2) |
| 4b | 0.5 | 0 | 0 | Fe | ¼ | 0.9(1) |
| 4b | 0.5 | 0 | 0 | Mn | ¼ | 0.9(1) |
| 4b | 0.5 | 0 | 0 | Cr | ½ | 0.9(1) |
| 4c | -0.051(1) | 0.25 | 0.516(2) | O1 | 1 | 2.0(6) |
| 8d | 0.84(1) | 0.012(2) | 0.165(1) | O2 | 1 | 2.9(5) |

f) $Ho_{0.5}Pr_{0.5}Fe_{0.25}Mn_{0.25}Cr_{0.5}O_3$ a = 5.5550(6) Å, b = 7.6317(6) Å, c = 5.3620(2) Å

| | x | y | z | Atoms | Occ | B (Å$^2$) |
|---|---|---|---|---|---|---|
| 4c | 0.0565(2) | 0.25 | 0.014(2) | Ho | ½ | 1.7(1) |
| 4c | 0.0565(2) | 0.25 | 0.014(2) | Pr | ½ | 1.7(1) |
| 4b | 0.5 | 0 | 0 | Fe | ¼ | 1.9(1) |
| 4b | 0.5 | 0 | 0 | Mn | ¼ | 1.9(1) |
| 4b | 0.5 | 0 | 0 | Cr | ½ | 1.9(1) |
| 4c | -0.053(1) | 0.25 | 0.64(2) | O | 1 | 1.3(1) |
| 8d | 0.749(2) | 0.0136(2) | 0.236(4) | O | 1 | 1.6(4) |

g) $PrFe_{0.25}Mn_{0.25}Cr_{0.5}O_3$ a = 5.5854(6) Å, b = 7.5979(8) Å, c = 5.3539(6) Å

| Site | x | y | z | Atom | Occ | B (Å$^2$) |
|---|---|---|---|---|---|---|
| 4c | 0.0418(5) | 0.476(5) | 0.262(3) | Pr | 1 | 0.55(2) |
| 4b | 0.5 | 0 | 0 | Fe | ¼ | 0.9(2) |
| 4b | 0.5 | 0 | 0 | Mn | ¼ | 0.9(2) |
| 4b | 0.5 | 0 | 0 | Cr | ½ | 0.9(2) |
| 4c | 0.476(5) | 0.25 | 0.065(4) | O | 1 | 1.7(2) |
| 8d | 0.262(4) | 0.044(2) | -0.305(4) | O | 1 | 1.8(1) |

h) $LaFe_{0.25}Mn_{0.25}Cr_{0.5}O_3$ a = 5.5087(3) Å, b = 7.7917(6) Å, c = 5.4652(8) Å

| | x | y | z | Atom | Occ | B (Å$^2$) |
|---|---|---|---|---|---|---|
| 4c | 0.0175(2) | 0.25 | 0.0106(2) | La | 1 | 0.40(3) |
| 4b | 0.5 | 0 | 0 | Fe | ¼ | 0.55(6) |
| 4b | 0.5 | 0 | 0 | Mn | ¼ | 0.55(6) |
| 4b | 0.5 | 0 | 0 | Cr | ½ | 0.76(1) |
| 4c | -0.017(1) | 0.25 | 0.506(2) | O1 | 1 | 2.0(1) |
| 8d | 0.759(3) | -0.030 | 0.324(2) | O2 | 1 | 2.5(2) |

i) $La_{0.5}Pr_{0.5}Fe_{0.25}Mn_{0.25}Cr_{0.5}O_3$ a = 5.5005 (2) Å, b = 7.7779 (2) Å and c = 5.4711 (2) Å

| | x | y | z | Atom | Occ | B (Å$^2$) |
|---|---|---|---|---|---|---|
| 4c | 0.028(2) | 0.25 | 0.0095(3) | La | ½ | 1.2(1) |
| 4c | 0.028(2) | 0.25 | 0.0095(3) | Pr | ½ | 1.2(1) |
| 4b | 0.5 | 0 | 0 | Fe | ¼ | 1.8(3) |
| 4b | 0.5 | 0 | 0 | Mn | ¼ | 1.8(3) |
| 4b | 0.5 | 0 | 0 | Cr | ½ | 1.8(3) |
| 4c | -0.027(3) | 0.25 | 0.561(4) | O | 1 | 2.0(3) |
| 8d | 0.752(3) | -0.0034(5) | 0.289(2) | O | 1 | 1.8(1) |

| Composition | a (Å) | b (Å) | c (Å) | B - $O_1$ B - $O_2$ (Å) | A - $O_1$ A - $O_2$ (Å) | B -$O_1$ - B B–$O_2$– B (Å) | $O_2$-B-$O_1$ $O_1$ -A - $O_2$ (Å) | A - B B - B A – A (Å) | Volume ($Å^3$) |
|---|---|---|---|---|---|---|---|---|---|
| $HoCr_{0.5}Mn_{0.5}O_3$ | 5.6174 (4) | 7.4735 (5) | 5.2435 (3) | 2.03 (1) 1.98 (1) | 2.66 (3) 2.43 (1) | 133.61 (1) 157.93 (1) | 118.81 (1) 80.05 (2) | 3.73 (3) 3.84 (2) 3.83 (1) | 220.13 (6) |
| $Ho_{0.3}Pr_{0.7}Mn_{0.5}Cr_{0.5}O_3$ | 5.5589 (6) | 7.6439 (7) | 5.3938 (2) | 2.01 (2) 2.05 (2) | 3.24 (2) 2.68 (2) | 144.49(1) 157.71 (2) | 116.37(1) 70.38 (2) | 3.62 (2) 3.87 (1) 3.87 (1) | 229.19 (3) |
| $Ho_{0.5}Pr_{0.5}Mn_{0.5}Cr_{0.5}O_3$ | 5.5854 (6) | 7.5979 (8) | 5.3539 (6) | 2.04 (1) 2.03 (3) | 3.53 (2) 2.70 (1) | 137.50 (1) 156.38 (2) | 121.17 (1) 68.51 (1) | 3.66 (1) 3.87 (2) 3.86(1) | 227.21 (8) |
| $Ho_{0.7}Pr_{0.3}Mn_{0.5}Cr_{0.5}O_3$ | 5.5975 (6) | 7.5347 (1) | 5.3000 (6) | 2.04 (1) 1.91 (1) | 3.38 (1) 2.62 (1) | 134.35 (1) 160.52 (1) | 120.44 (1) 68.52(1) | 3.67 (1) 3.85 (1) 3.83 (1) | 223.53 (1) |

Table 2.2a: Structural analysis of $Ho_{1-x}Pr_xCr_{0.5}Mn_{0.5}O_3$ for x = 0, 0.3, 0.5 and 0.7.

Table 2.2b: Structural analysis of $Ho_{1-x}Pr_x$ $(Fe_{0.25}Mn_{0.25}Cr_{0.5})$ $O_3$ (x = 0, 0.5 and 1).

| **Composition** | **a (Å)** | **b (Å)** | **c (Å)** | **B - $O_1$ B - $O_2$ (Å)** | **A - $O_1$ A - $O_2$ (Å)** | **B -$O_1$ - B B–$O_2$– B (Å)** | **$O_2$-B-$O_1$ $O_1$ -A -$O_2$ (Å)** |
|---|---|---|---|---|---|---|---|
| **$HoFe_{0.25}Mn_{0.25}Cr_{0.5}O_3$** | 5.5603(4) | 7.5181 (5) | 5.2475 (3) | 1.90(1) 1.97(1) | 2.74(3) 2.20(1) | 162.33(1) 141.46(1) | 98.50(1) 67.40(2) |
| **$Ho_{0.5}Pr_{0.5}Fe_{0.25}Mn_{0.25}Cr_{0.5}O_3$** | 5.5550(6) | 7.6317 (6) | 5.3620 (2) | 2.07(2) 1.99(2) | 2.09(2) 2.65(2) | 134.18(1) 173.16(2) | 107.34(1) 68.53(2) |
| **$PrFe_{0.25}Mn_{0.25}Cr_{0.5}O_3$** | 5.5854(6) | 7.5979 (8) | 5.3539 (6) | 2.04(1) 2.03(3) | 3.53(2) 2.70(1) | 137.50(1) 156.38(2) | 121.17(1) 68.51(1) |

Table 2.2c: Structural analysis of $La_{1-x}Pr_x(Fe_{0.25}Mn_{0.25}Cr_{0.5})O_3$ (x = 0, 0.5 and 1).

| **Composition** | **a (Å)** | **b (Å)** | **c (Å)** | **B - $O_1$ B - $O_2$ (Å)** | **A - $O_1$ A - $O_2$ (Å)** | **B -$O_1$ - B B–$O_2$– B (Å)** | **$O_2$-B-$O_1$ $O_1$ -A - $O_2$ (Å)** |
|---|---|---|---|---|---|---|---|
| **$LaFe_{0.25}Mn_{0.25}Cr_{0.5}O_3$** | 5.5087(3) | 7.7917(6) | 5.4652(8) | 1.95(1) 2.29(1) | 2.74(1) 2.71(1) | 172.78(1) 159.37(1) | 99.15(1) 72.81(1) |
| **$La_{0.5}Pr_{0.5}Fe_{0.25}Mn_{0.25}Cr_{0.5}O_3$** | 5.5005(1) | 7.7779(1) | 5.4711(1) | 1.98(1) 2.10(1) | 3.08(1) 2.74(1) | 158.59(1) 171.37(1) | 100.90(1) 67.73(1) |
| **$PrFe_{0.25}Mn_{0.25}Cr_{0.5}O_3$** | 5.5854(6) | 7.5979(8) | 5.3539(6) | 2.04(1) | 3.53(2) | 137.50(1) | 121.17(1) |

| | | | |
|---|---|---|---|
| 2.03(3) | 2.70(1) | 156.38(2) | 68.51(1) |

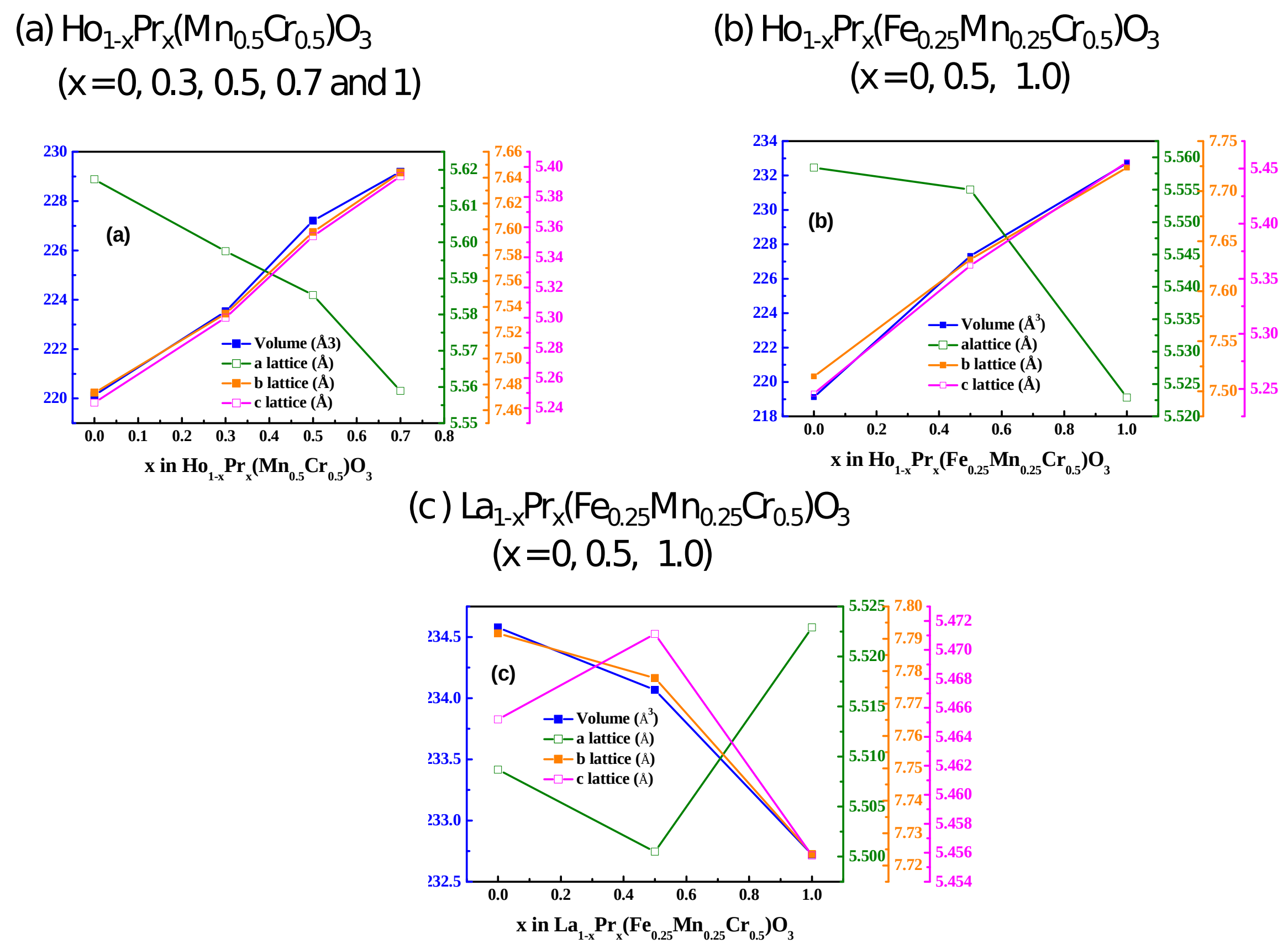

Fig. 2.4: Graphical representation of lattice parameters and volume for a) $Ho_{1-x}Pr_x(Mn_{0.5}Cr_{0.5})O_3$ (x = 0, 0.3, 0.5 and 0.7), b) $Ho_{1-x}Pr_x(Fe_{0.25}Mn_{0.25}Cr_{0.5})O_3$ (x = 0, 0.5 and 1) and c) $La_{1-x}Pr_x(Fe_{0.25}Mn_{0.25}Cr_{0.5})O_3$ (x = 0, 0.5 and 1).

### 2.3.2 Magnetization studies

The temperature dependent DC magnetic susceptibility ~~performed~~ in ZFC (zero field cooled) and FC (field cooled) conditions was measured at 1000 Oe as shown in (Fig. 2.5). We can observe that magnetic susceptibility gets increased with decreasing temperature for all the samples a) $Ho_{1-x}Pr_x(Mn_{0.5}Cr_{0.5})O_3$ (x = 0, 0.3, 0.5 and 0.7), b) $Ho_{1-x}Pr_x(Fe_{0.25}Mn_{0.25}Cr_{0.5})O_3$ (x = 0, 0.5 and 1) and c) $La_{1-x}Pr_x(Fe_{0.25}Mn_{0.25}Cr_{0.5})O_3$ (x = 0, 0.5 and 1). As we decrease the temperature, the susceptibility undergoes a ferromagnetic like transition below a certain temperature $T_C$. Further decrease in temperature results in bifurcation of the ZFC and FC data below a certain temperature $T_{SR}$, suggests absence of long-range ordering in the system. In case of $Ho_{1-x}Pr_x(Mn_{0.5}Cr_{0.5})O_3$ (x = 0, 0.3, 0.5 and 0.7), as we increase the Pr concentration in the system a systematic increase in the $T_C$ and $T_{SR}$ is observed. The value of magnetic moment below $T_C$ is also found to increase with increasing Pr concentration. Similarly, for $Ho_{1-x}Pr_x(Fe_{0.25}Mn_{0.25}Cr_{0.5})O_3$ (x = 0.5 and 1) and $La_{1-x}Pr_x(Fe_{0.25}Mn_{0.25}Cr_{0.5})O_3$ (x = 0, 0.5 and 1) the magnetic moments are found to get increased with increasing Pr concentration. Only an exception is found in case of $Ho_{1-x}Pr_x(Fe_{0.25}Mn_{0.25}Cr_{0.5})O_3$ with x = 0, where the low temperature magnetic susceptibility is quite large than the other two concentrations. Therefore, in this case we can say that Pr doping at Ho site at first decrease the magnetic moment but then follows the trend like other systems as we keep on increasing the Pr concentration in the system. Change in the $T_C$ and $T_{SR}$ follows the similar trend for all the systems studied.

In the first series, decreased value of magnetization with increasing Ho concentration could be attributed to the increased Ho-Mn interaction [10-12].

From the literature, it is known that an increased number of unpaired electrons, half-filled resulting in a higher value of the azimuthal quantum number, which results in greater anisotropy of orbitals [13], hence it suggests greater Ho anisotropy in our case in the first series. In the second series however due to incorporation of Fe at Mn site, results in system to show greater anisotropy, which, as reported in the literature [14] competes with the Pr orbital [15], and thus decreases its magnetic moment. The antiferromagnetic alignment of the spins of rare-earth concerning the net moment generated by increased canted antiferromagnetic transition metal ions, in the case of $PrFe_{0.25}Mn_{0.25}Cr_{0.5}O_3$ was observed from the decreased value of magnetic susceptibility (Fig. 2.5b) [16]. Furthermore, as the Pr concentration increases, in the first case $Ho_{1-x}Pr_x$ $(Mn_{0.5}Cr_{0.5})$ $O_3$ (x = 0.5 0.7 and 1), a uniform increase in Curie temperature was observed, suggesting a gradual change in anisotropy in the system of the rare-earth as x increases. Also, the bifurcation temperature between ZFC and FC increased as the Pr concentration increased.

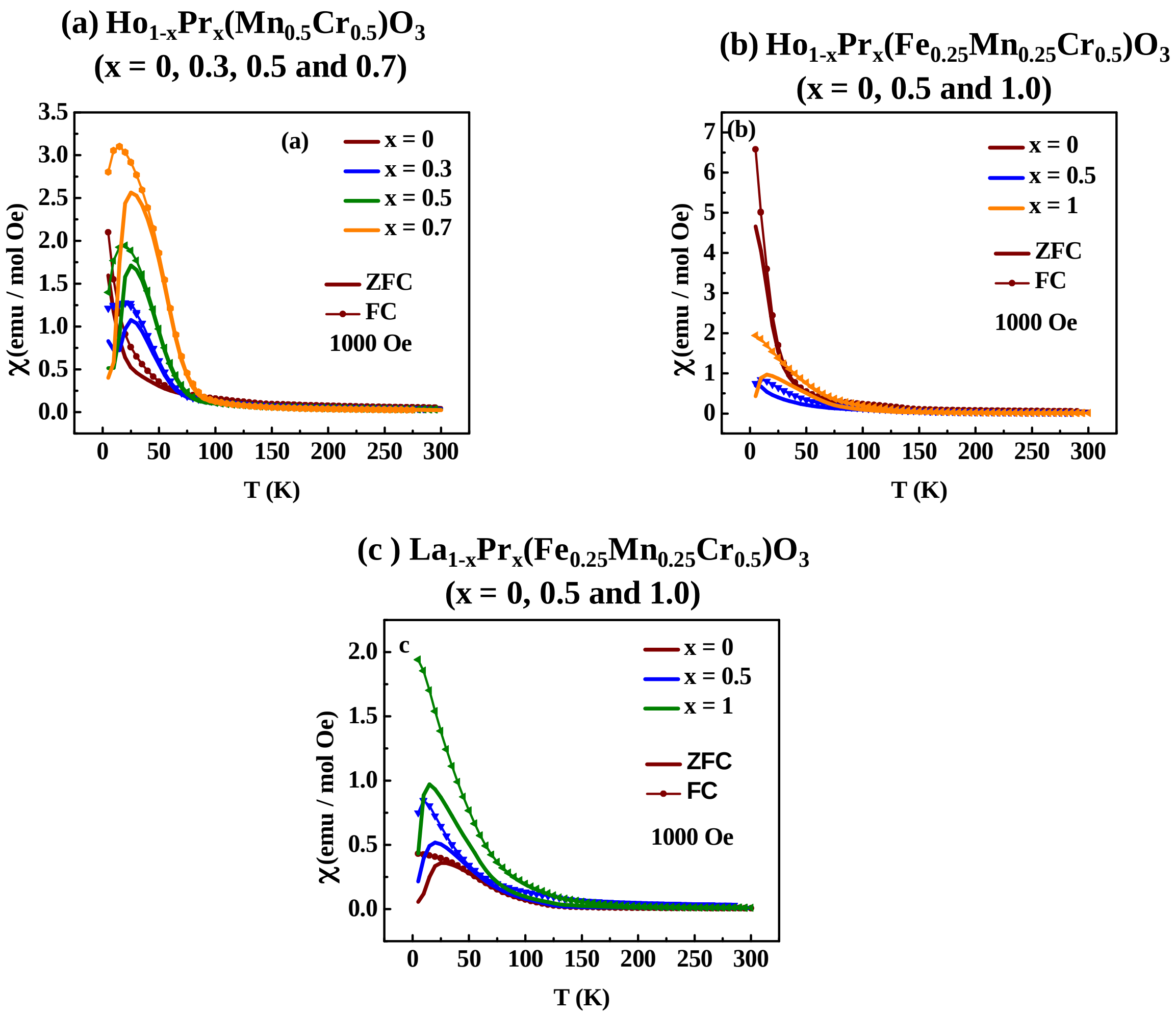

Fig 2.5: DC magnetic susceptibility studies of a) $Ho_{1-x}Pr_x(Mn_{0.5}Cr_{0.5})O_3$ (x = 0, 0.3, 0.5 and 0.7), b) $Ho_{1-x}Pr_x(Fe_{0.25}Mn_{0.25}Cr_{0.5})O_3$ (x = 0, 0.5 and 1) and c) $La_{1-x}Pr_x(Fe_{0.25}Mn_{0.25}Cr_{0.5})O_3$ (x = 0, 0.5 and 1) measured in temperature range 5-300 K.

From the increase in magnetic susceptibility plot amongst all the compositions, a maximum Curie Weiss temperature was obtained in $PrFe_{0.25}Mn_{0.25}Cr_{0.5}O_3$ ($\theta_{CW}$ = -58 K), which suggests a stable spin structure in this case [17]. The Curie-Weiss fit was taken in the linear region at high temperatures in the range between 250-300 K (Fig. 2.7 and 2.8). The difference between the calculated and theoretical magnetic moment was found to be similar and maximum change (Fig 2.7) was observed when x was 0.5, due to increased disorder in the system, resulting in competing for anisotropy of rare-earth orbitals. The magnetic moments (theoretical and calculated) decreased with an increase in x in the first two cases and opposite in the third case, as the rare moment changes (Fig. 2.7 a, b and c).

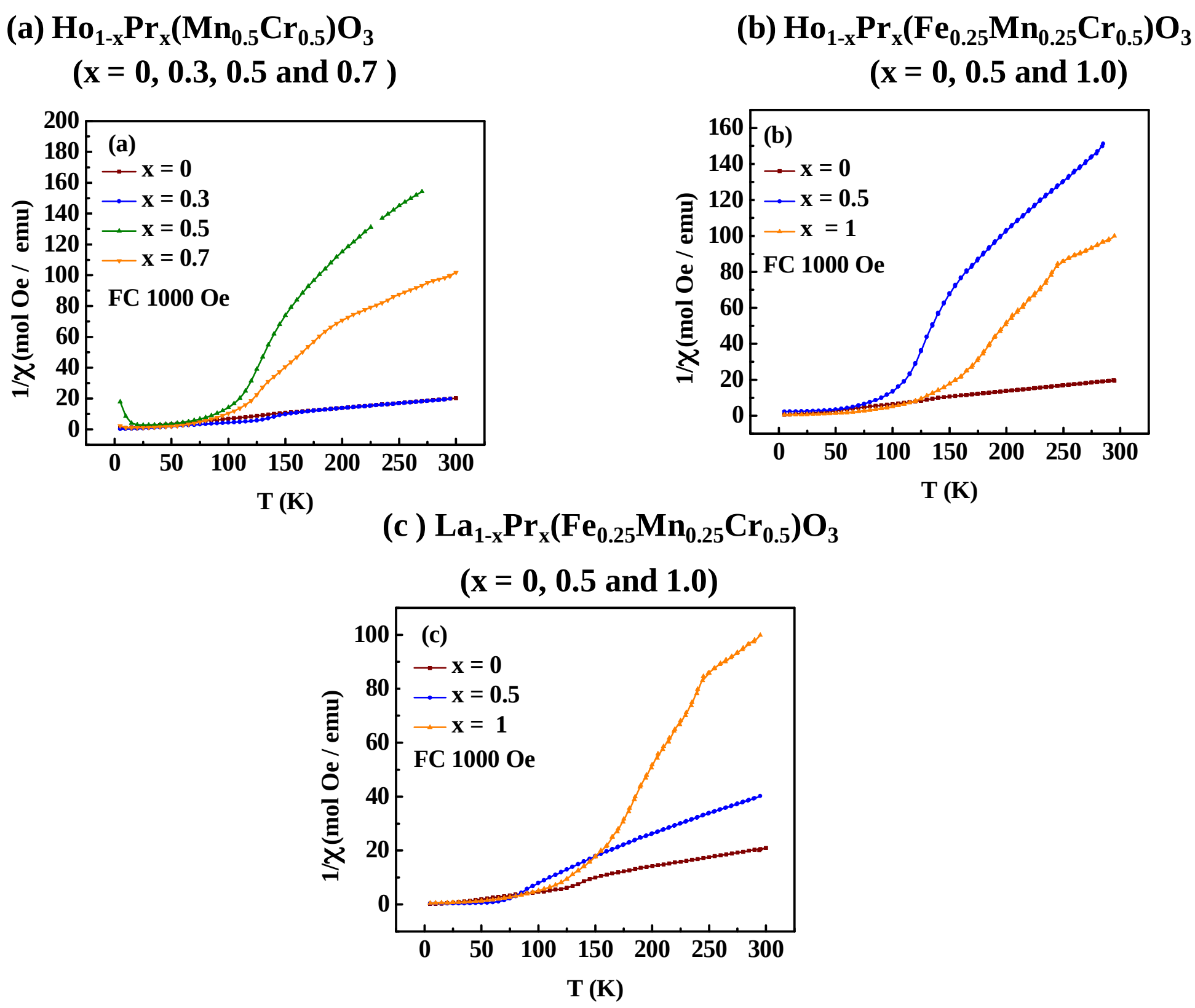

Fig 2.6: Inverse magnetic susceptibility plot of a) $Ho_{1-x}Pr_x$ $(Mn_{0.5}Cr_{0.5})$ $O_3$ (x = 0, 0.3, 0.5 and 0.7), $Ho_{1-x}Pr_x$ $(Fe_{0.25}Mn_{0.25}Cr_{0.5})$ $O_3$ (x = 0, 0.5 and 1) and c) $La_{1-x}Pr_x(Fe_{0.25}Mn_{0.25}Cr_{0.5})O_3$ (x = 0, 0.3 and 0.5) measured in temperature range 5-300 K.

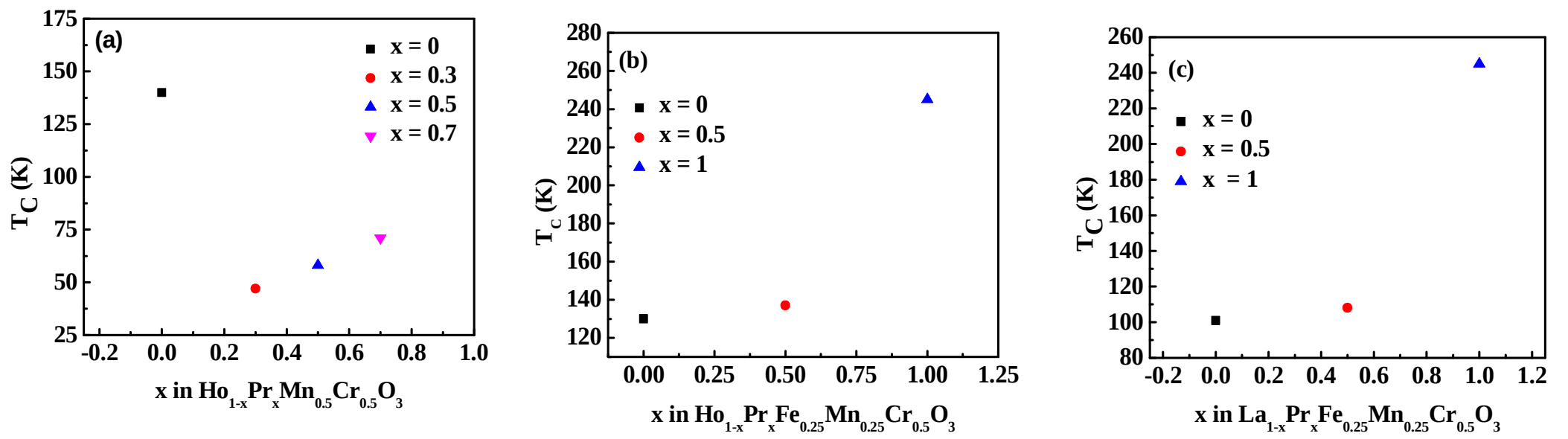

Fig 2.7: Curie temperature plot measured at 1000 Oe for a) $Ho_{1-x}Pr_x$ $(Mn_{0.5}Cr_{0.5})$ $O_3$ (x = 0, 0.3, 0.5 and 0.7), b) $Ho_{1-x}Pr_x$ $(Fe_{0.25}Mn_{0.25}Cr_{0.5})$ $O_3$ (x = 0, 0.5 and 1) and c) $La_{1-x}Pr_x$ $(Fe_{0.25}Mn_{0.25}Cr_{0.5})$ $O_3$ (x = 0, 0.3 and 0.5).

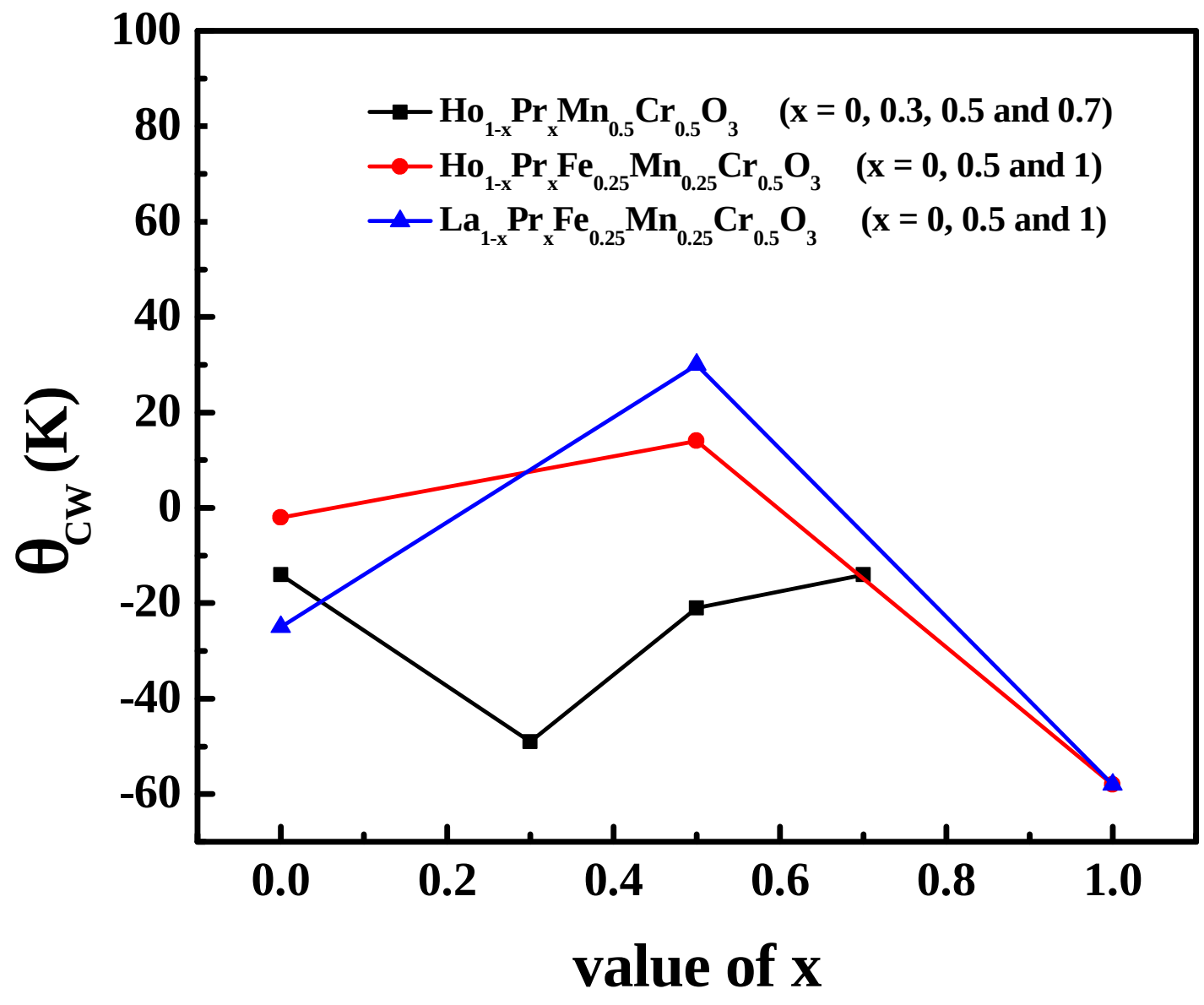

Fig 2.8: Curie Weiss temperature plot for a) $Ho_{1-x}Pr_x$ $(Mn_{0.5}Cr_{0.5})$ $O_3$ (x = 0, 0.3, 0.5 and 0.7), b) $Ho_{1-x}Pr_x(Fe_{0.25}Mn_{0.25}Cr_{0.5})O_3$ (x = 0, 0.5 and 1) and c) $La_{1-x}Pr_x(Fe_{0.25}Mn_{0.25}Cr_{0.5})O_3$ (x = 0, 0.5 and 1) measured from inverse magnetic susceptibility.

The isothermal magnetization data was performed at 300 K, (Fig. 2.10) which suggests a paramagnetic behaviour for all the compositions. A weak hysteresis obtained at 5 K and 20 K data suggest weak ferromagnetism, attributed to the canted spin structure. The maximum magnetization obtained for a higher value of x, at 5 K in 7 T field range was observed for a higher value of x for the first two cases and not in the third case. The coercive field in the first case increased with x while in the second series it was found to decrease with an increase in x (Fig. 2.11). While in the third series coercive field increased as the Pr concentration increased with an increase in x.

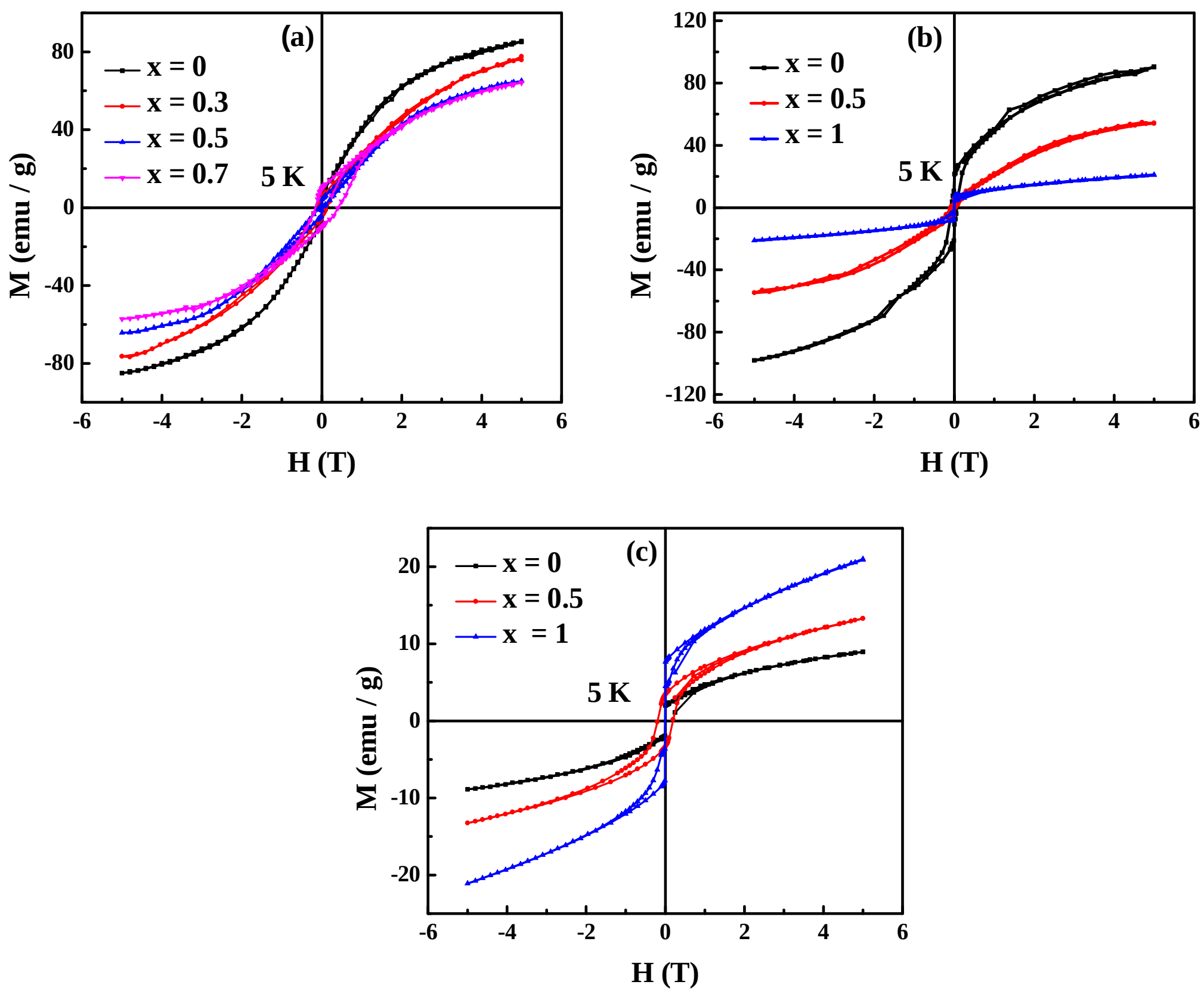

Fig 2.9: Isothermal magnetisation plot at 5 K for a) $Ho_{1-x}Pr_x(Mn_{0.5}Cr_{0.5})O_3$ (x = 0, 0.3, 0.5 and 0.7), b) $Ho_{1-x}Pr_x(Fe_{0.25}Mn_{0.25}Cr_{0.5})O_3$ (x = 0, 0.5 and 1) and c) $La_{1-x}Pr_x(Fe_{0.25}Mn_{0.25}Cr_{0.5})O_3$ (x = 0, 0.5 and 1).

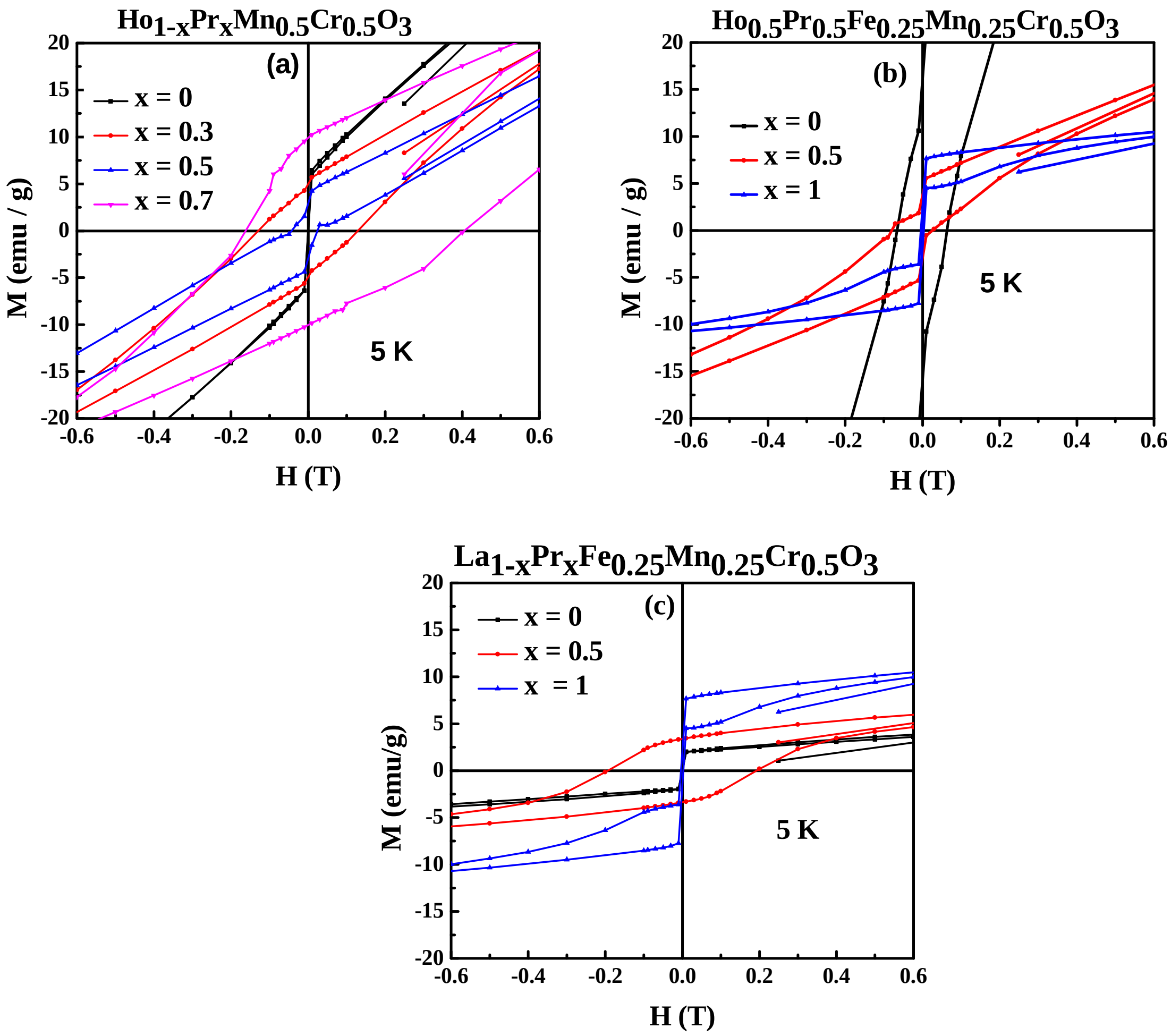

Fig 2.10: Enlarged isothermal magnetization plot at 5 K for a) $Ho_{1-x}Pr_x$ $(Mn_{0.5}Cr_{0.5})$ $O_3$ (x = 0, 0.3, 0.5 and 0.7), b) $Ho_{1-x}Pr_x$ $(Fe_{0.25}Mn_{0.25}Cr_{0.5})$ $O_3$ (x = 0, 0.5 and 1) and c) $La_{1-x}Pr_x$ $(Fe_{0.25}Mn_{0.25}Cr_{0.5})$ $O_3$ (x = 0, 0.5 and 1).

Remanent magnetization at 5 K increased as x increased in the first and third case and was found to decrease in the second case, suggesting easy alignment of rare-earth ion in the former series (Fig.2.10) [20]. The remanent magnetization at 7 T increases in the first series, while it decreases in the second series that is attributed to rare-earth anisotropy competing with transition metal ion in the first case (Fig. 2.11) [22].

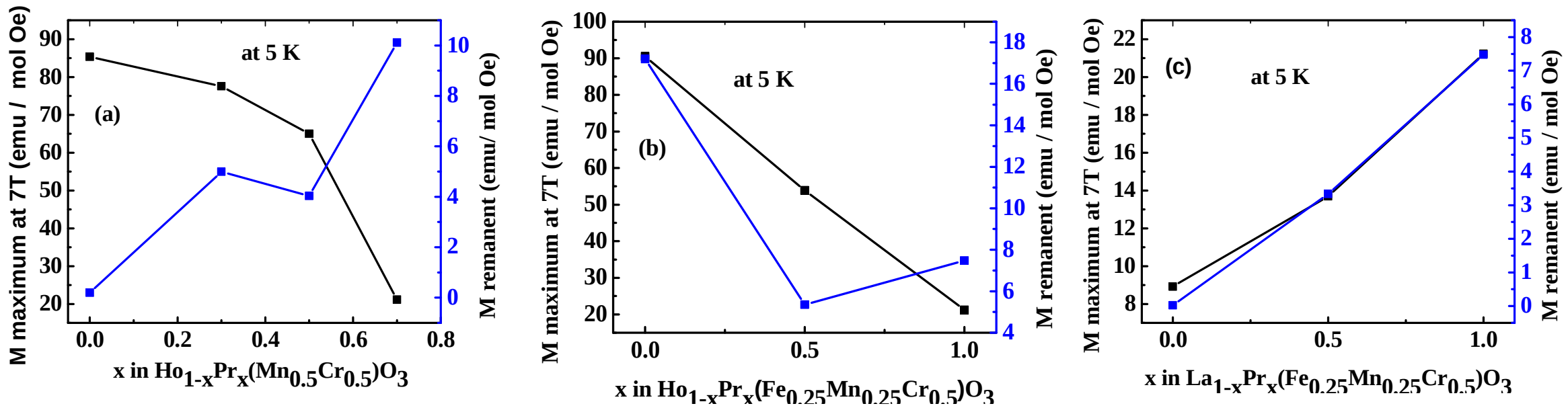

Fig 2.11: Maximum magnetization obtained at 7 T field and comparison with remanent magnetization data at 5 K for a) $Ho_{1-x}Pr_x (Mn_{0.5}Cr_{0.5}) O_3$ (x = 0, 0.3, 0.5 and 0.7),

b) $Ho_{1-x}Pr_x (Fe_{0.25}Mn_{0.25}Cr_{0.5}) O_3$ (x = 0, 0.5 and 1) and c) $La_{1-x}Pr_x (Fe_{0.25}Mn_{0.25}Cr_{0.5}) O_3$ (x = 0, 0.5 and 1).

## 2.4 Conclusion

Structural investigation of three series and were refined in orthorhombic *Pnma* space group which showed octahedral distortion increases and lattice volume decreases as the A-site radii decrease. The DC magnetic susceptibility bifurcation between ZFC and FC suggest spin glass-like behaviour. Increased magnetic susceptibility in $Ho_{1-x}Pr_x (Mn_{0.5}Cr_{0.5}) O_3$ (x = 0.3, 0.5 and 0.7) and decrease in $Ho_{1-x}Pr_x (Fe_{0.25}Mn_{0.25}Cr_{0.5}) O_3$ (x = 0, 0.5 and 1) as x increases which could be attributed to Ho-Mn interaction and competing anisotropy of orbitals of Pr and Fe atoms. In $PrFe_{0.25}Mn_{0.25}Cr_{0.5}O_3$ maximum Curie Weiss temperature suggests the stable spin structure in this composition.

# Chapter III

# Investigation of structural and magnetic properties of a Ru-deficient perovskite $PrMn_{0.5}Ru_{0.35}O_3$

# Chapter III

## Investigation of structural and magnetic properties of a Ru-deficient perovskite $PrMn_{0.5}Ru_{0.35}O_3$

### Abstract

Structure and magnetic properties on substitution of Ru at Mn site have been investigated. The Pr-Mn-Ru-O phase was synthesized by a high-temperature solid-state route in their stoichiometric ratios. The phase was found to show Ru-deficiency from X-ray Photoelectron Spectroscopy (XPS) and Energy Dispersive X-ray Analysis (EDAX), with Ru deficiency of 0.15(2) in stoichiometry. This phase of $PrMn_{0.5}Ru_{0.35}O_3$ was found to crystallize in *Pnma* spacegroup, which showed a decreased *b* lattice parameter in comparison with the reported compositions $PrFe_{0.5}Mn_{0.5}O_3$ and $LaMn_{0.5}Ru_{0.5}O_3$. From the zero-field cooled (ZFC) and field cooled (FC) DC magnetic susceptibility data, a bifurcation between ZFC and FC was observed at temperatures below 31 K, and the ferromagnetic transition was obtained at temperatures below 90 K. Furthermore, the magnetic anomalies at 18 K and 4.5 K could be attributed to the disorder-induced magnetic spin ordering of Mn and Ru spins and 'Pr' ordering. From the X-ray Photoelectron Spectroscopy (XPS) it was observed that Mn and Ru ions showed majorly +4 and +5 oxidation states of the ions.

## 3.1 Introduction

Perovskite oxides of the formula $ABO_3$ (A: rare-earth, B: transition metal) showed varying structural and magnetic properties [1]. These structure-dependent magnetic properties have their applications in the field of multiferroics, magnetoelectric and colossal magnetoresistance etc, which can be manipulated with varying B-site substitution [1-2]. The parent compound $PrMnO_3$ shows double-exchange ferromagnetic interactions at temperatures below 100 K [3]. Also the compound $SrRu_{1-x}O_3$ (x<0.09) shows varied range of ferromagnetic transitions, depending on the concentration of Ru vacancy 163 K (x = 0) to 45 K ( x= 0.09) [4]. Furthermore, another Ru deficient perovskite compound $(La_{0.52}Ba_{0.48})(Mn_{0.51}Ru_{0.49})O_3$ shows Ru and Mn spin ordering at 80 K and 195 K respectively, [5] depending on their ordering temperatures. Attempts to synthesise Ru based perovskite under high pressure have also resulted in Ru vacancy. For example, $PrRuO_3$ was synthesized under high pressure (15 GPa) and high temperatures (1200˚C) which has resulted in Ru deficiency of 0.08 in stoichiometry [8].

Thus these antiferromagnetic and ferromagnetic transition temperatures depend to a great extent on ionic radii and oxidation states of these B-site ions [7-8]. There are various kinds of magnetic exchange interactions at the Fe and Mn containing B-sites: Fe-O-Fe, Fe-O-Mn, Mn-O-Mn which results in spin glass transition temperature for $PrMn_{0.5}Fe_{0.5}O_3$ at 200 K, while $LaMn_{0.5}Fe_{0.5}O_3$ shows ferromagnetic transition at 85 K and spin glass transition at 220 K [9-11]. Various factors such as Dzyaloshinskii-Moriya interaction, super-exchange interaction and magnetic anisotropy result in varied magnetic properties [10-11].

On substitution of Ru at B-site results in perovskite compounds at high temperatures which often results in Ru vacancy, due to the formation of $RuO_4$ volatile complex, when heated at temperatures greater than 1000˚C [1,4, 12-18]. Also, the greater structural instability also manifests Ru vacancy. For example, it has been reported in $SrRu_{1-x}O_3$ (x<0.09), as inferred from X-ray and neutron diffraction data that $Sr^{2+}$ tends to occupy Ru deficient B-site [4,18], particularly along the axial *b* direction.

Furthermore, perovskite compounds of the type $ARuO_3$ (A: Sr, La, Ca, Ru) show that as the A-site ionic radii decreases, it results in greater Ru-O-Ru orbital overlap [19]. Structure dependent changes in magnetic properties are also related to $CaMn_{0.5}Ru_{0.5}O_3$ which shows G-type antiferromagnetic transition at 95 K and ferromagnetic transition at 230 K [14, 20]. This greater orbital overlap between Mn and Ru ions favours an A-type antiferromagnetic structure [20-22]. Here in this study, we report the effect of substitution of Ru at the Mn site in $PrMnO_3$ and its structure and magnetic properties.

## 3.2 Experimental

### 3.2.1 Materials and synthesis

$PrMn_{0.5}Ru_{0.5}O_3$ was prepared by the solid-state route. $Pr_6O_{11}$ (Sigma Aldrich), $Mn_2O_3$ (Sigma Aldrich) and $RuO_2$ (Sigma Aldrich) were taken in stoichiometric ratios of Pr:Mn:Ru as 1:0.5:0.5. $Pr_6O_{11}$ was given a heat treatment at 900˚C for 12 hours so to remove moisture and carbonate content. The final reactants were weighed (0.2g) and ground and was heated in powder form at 900˚C for 12 hours, and with intermediate grinding after cooling, was then subsequently heated at 1200˚C for 12 hours.

### 3.2.2 Powder X-ray diffraction study (PXRD)

Bruker D8-Advance diffractometer, which has a source of Cu-K$\alpha$ was used for the collection of diffraction data. The data was collected from 10˚ to 70˚ as a 2$\Theta$ angle. The Rietveld refinement on the PXRD data was performed using TOPAZ 5.2 software. The lattice parameters, background subtraction, site occupancy and thermal parameters were refined using this software.

### 3.2.3 Energy dispersive X-ray analysis (EDAX)

The FESEM EDAX measurement was performed using JEOL, JSM-7800F electron microscope. The source of the instrument was a field emission gun which was lanthanum hexaboride for generating electrons. The sample was mounted on a C-tape and the elements Pr, Mn and Ru were detected successfully and quantitatively analysed. Here vacuum was applied before switching on the source.

### 3.2.4 X-ray photoelectron spectroscopy (XPS)

XPS studies were performed on the powdered sample on a Shimadzu ESCA-3400 spectrophotometer with Mg K$\alpha$ source (1253.6 eV) operating at 10 kV and 10 mA. The sample was used in the powder and an ultra-high vacuum of $10^{-8}$ torr pressure was maintained before applying the source.

### 3.2.5 Magnetic studies

Zero field cooled (ZFC) and field cooled (FC) temperature dependent magnetic studies of the compound were carried out in the temperature range of 2 K to 390 K using a physical property measurement system (PPMS) at a magnetic field of 1000 Oe and isothermal magnetization measurement was carried out at a magnetic field of ±7 T at varying temperatures of 2 K, 50 K and 300 K.

## 3.3 Results and discussion

### 3.3.1 X-ray Diffraction studies (XRD)

The structure of $PrMn_{0.5}Ru_{0.35}O_3$ was investigated by powder X-ray diffraction pattern. The PXRD pattern could be indexed successfully with orthorhombic *Pnma* spacegroup with lattice parameters a = *5.6777(4)* Å, b = *7.7026(3)* Å and c = *5.4415(3)* Å. In the comparison of the *b* lattice parameters with other compositions such as $LaMn_{0.5}Ru_{0.5}O_3$, $PrFe_{0.5}Mn_{0.5}O_3$ and $LaFe_{0.5}Mn_{0.5}O_3$ [9-10] (Table 3.2), it was observed that $PrMn_{0.5}Ru_{0.35}O_3$ shows decreased *b* lattice parameter. On further structural analysis, it was observed that $PrMn_{0.5}Ru_{0.35}O_3$ shows greater octahedral tilting along the *b* axis, with Mn/Ru – $O_{1(axial)}$ – Mn/Ru bond angle being *157.15(1)*°, while equatorial bond angle was *168.88(1)*° (Table 3.3). This difference in equatorial and axial bond angles was found to be greater in $PrMn_{0.5}Ru_{0.35}O_3$, in comparison with other related perovskites (Table 3.2). This bond angle anisotropy and compression of *b* lattice parameter along the axial direction suggests B-site Ru vacancy, particularly along the axial direction [4, 20, 24].

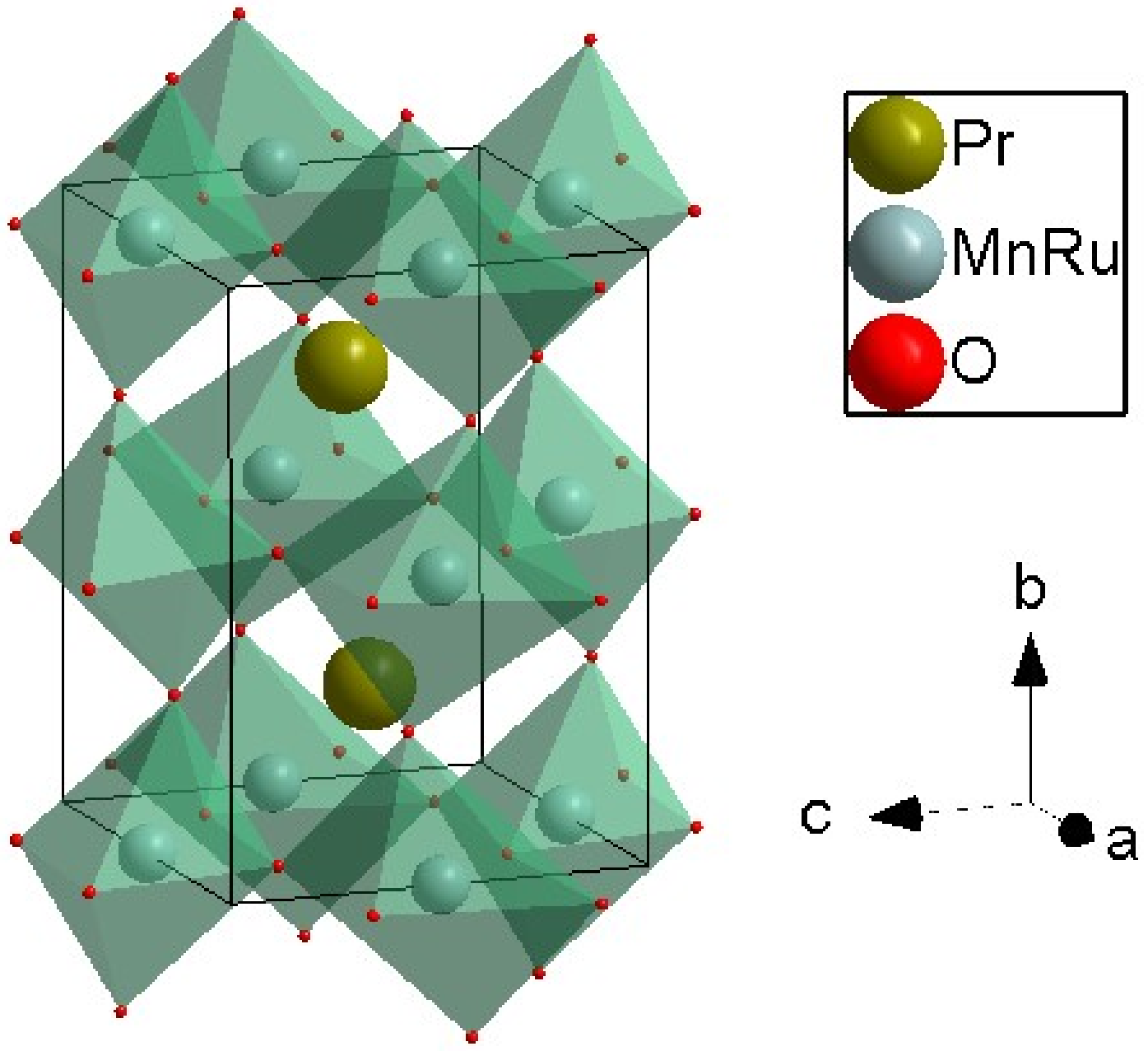

Fig 3.1: Crystal structure of $PrMn_{0.5}Ru_{0.5}O_3$ having orthorhombic symmetry (space group *Pnma)*, corresponding to $GdFeO_3$ structure with octahedral tilting $a^-b^+a^-$.

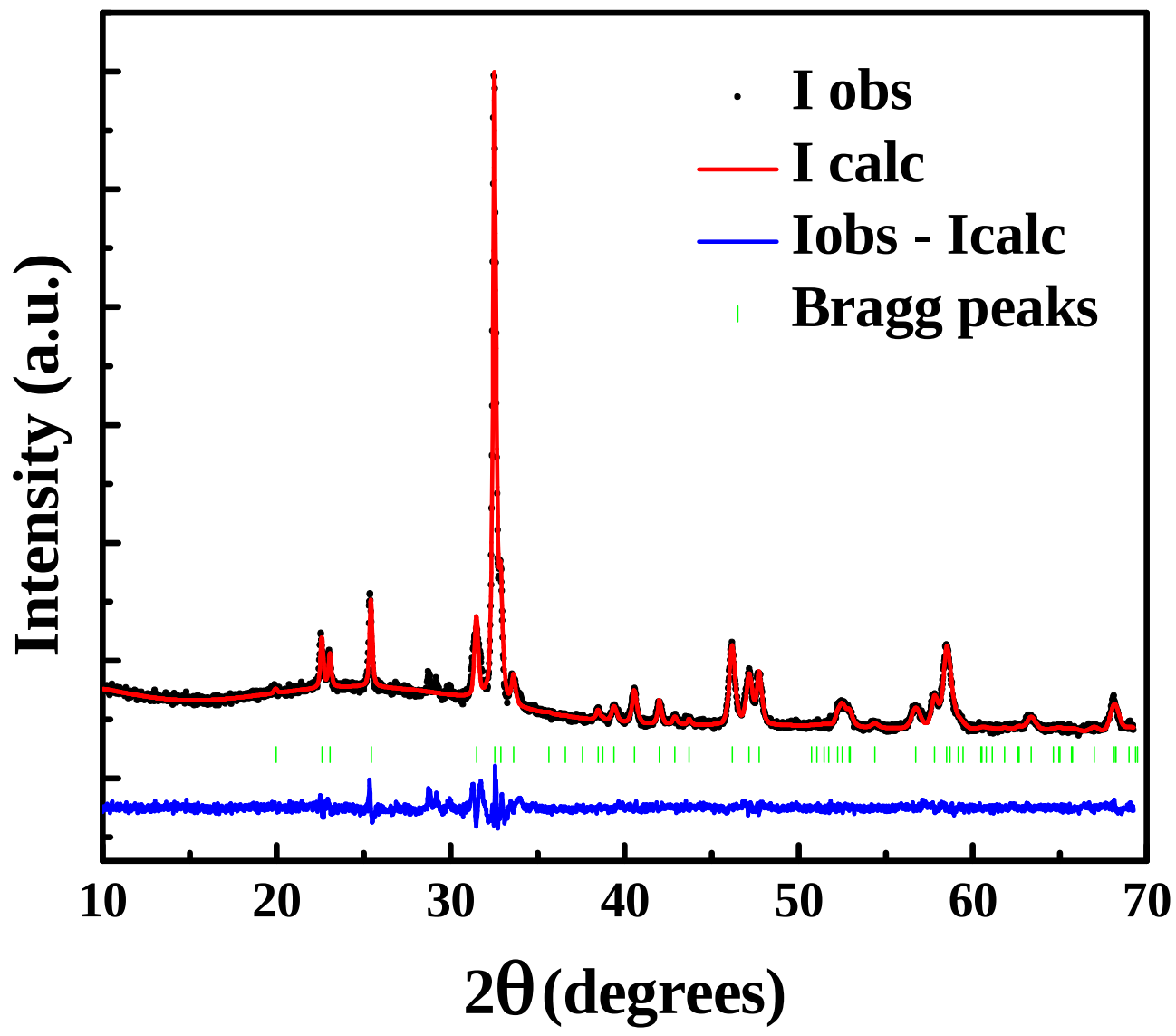

Fig 3.2.: The room temperature powder X-ray diffraction data (black) of $PrMn_{0.5}Ru_{0.5}O_3$. The Rietveld refinement profiles, Bragg peak positions (green vertical lines), and the residuals are shown as in blue lines.

Table 3.1: Details of the parameters obtained after Rietveld refinement of the Powder XRD data of $PrMn_{0.5}Ru_{0.5}O_3$ *a = 5.6777(4) Å, b = 7.7026(3) Å, c = 5.4415(3) Å*, Spacegroup: *Pnma. Rexp : 2.79 Rwp : 4.42 Rp : 3.27 GOF : 1.58*

| *Site* | *x* | *y* | *z* | *Atom* | *Occ* | $B_{iso}$ |
|---|---|---|---|---|---|---|
| *4c* | *0.0486(5)* | *0.25* | *0.012(1)* | *Pr* | *1* | *1.3(1)* |
| *4b* | *0* | *0* | *0.5* | *Mn/Ru* | *0.5* | *1.7(1)* |
| *4c* | *0.486(6)* | *0.25* | *-0.070(5)* | *O* | *1* | *4.2(1)* |
| *8d* | *0.310(3)* | *0.496(5)* | *0.237(6)* | *O* | *1* | *1.3(2)* |

Table 3.2: Comparison of the bond angle and bond distances of the system $PrMn_{0.5}Ru_{0.35}O_3$ with respect to $RMn_{0.5}A_{0.5}O_3$ (R: Pr, La and A: Ru, Fe): as reported in the literature.

| **Composition** | **a (Å)** | **b (Å)** | **c (Å)** | **B - O1**<br>**B - O2**<br>**Min (Å)** | **A - O1**<br>**A - O2**<br>**Min (Å)** | **B -O1 - B**<br>**B–O2– B**<br>**Min(˚)** | **O2-B-O1**<br>**O1 -A -O2**<br>**Max(˚)** | **A - B**<br>**B - B**<br>**A – A**<br>**Min (Å)** | **Volume (Å³)** | **Ref.** |
|---|---|---|---|---|---|---|---|---|---|---|
| $PrMn_{0.5}Ru_{0.35}O_3$ | 5.6777 (4) | 7.7026 (3) | 5.4415 (3) | 1.97(1)<br>1.68(1) | 2.43(1)<br>2.59 (1) | 157.15(1)<br>168.88(1) | 98.07(1)<br>67.27 (1) | 3.21(1)<br>3.93(1)<br>3.89(1) | 237.97 (6) | This work |
| $PrFe_{0.5}Mn_{0.5}O_3$ | 5.4728 (2) | 7.7496 (3) | 5.5284 (2) | 1.98(2)<br>1.88(2) | 2.28 (2)<br>2.48 (2) | 155.23(1)<br>158.34(2) | 107.86(1)<br>168.52 (2) | 3.21(2)<br>3.89(1)<br>3.90(1) | 234.47 (3) | [10] |
| $LaMn_{0.5}Ru_{0.5}O_3$ | 5.5593 ((6) | 7.8309 (1) | 5.5484 (8) | 2.02(1)<br>1.92(1) | 2.32 (2)<br>2.56 (1) | 151.18(1)<br>161.79(2) | 96.89(1)<br>62.74 (1) | 3.23(1)<br>3.93(2)<br>3.93(1) | 241.54 (8) | [16] |
| $LaFe_{0.5}Mn_{0.5}O_3$ | 5.5351 (3) | 7.7938 (1) | 5.4990 (1) | 1.96(1)<br>2.01(2) | 2.55 (1)<br>2.41 (2) | 168.25(3)<br>156.41(4) | 98.17 (1)<br>68.20 (2) | 3.33(2)<br>3.90(2)<br>3.90(2) | 237.22 (3) | [9] |

The decreased *b* lattice parameter and increased distortion along the axial direction further suggest increased vacancies along the axis. The axial *b*-lattice parameter of $LaFe_{0.5}Mn_{0.5}O_3$ and $LaMn_{0.5}Ru_{0.5}O_3$ was found to increase from 7.7938(1) Å and 7.8309(1) Å respectively (Table 3.2). However, the *b*-lattice parameter increased in $PrMn_{0.5}Ru_{0.35}O_3$ in comparison with $PrFe_{0.5}Mn_{0.5}O_3$ from 7.7026(3) Å to 7.7496(3) Å, despite the octahedral distortion in the axial direction being greater in $PrFe_{0.5}Mn_{0.5}O_3$ (155.23(1)˚) in comparison with $PrMn_{0.5}Ru_{0.35}O_3$ (157.15(1)˚). We predict here greater Ru-vacancy along the *b* axial direction in $PrMn_{0.5}Ru_{0.35}O_3$.

### 3.3.2 Energy dispersive X-ray analysis (EDAX) studies

EDAX analysis showed the ratio of Pr:Mn: Ru to be *1:0.5:0.35* (Fig. 3.3). From the quantitative analysis, no peaks of all the detected elements were found to overlap and thus their standard deviation was found to be within the instrument detection limit.

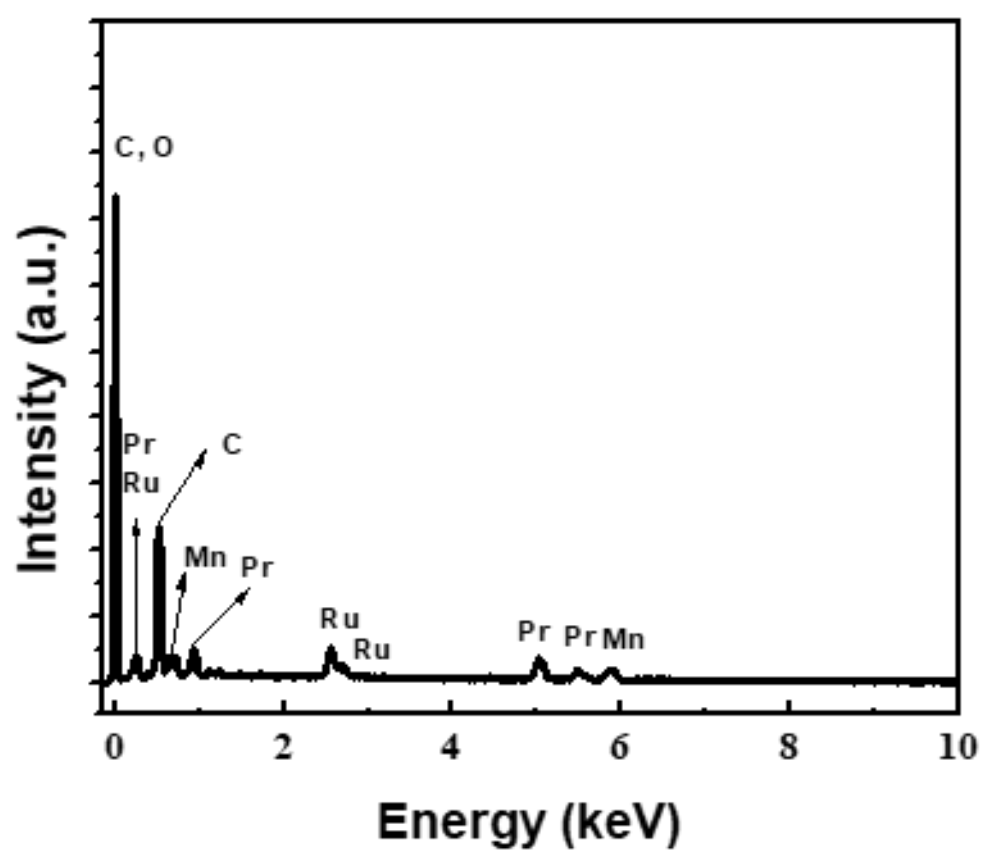

Fig 3.3: EDAX analysis of $PrMn_{0.35}Ru_{0.5}O_3$ showing Pr : Mn : Ru ratio to be 1: 0.5 : 0.35.

### 3.3.3 X-ray photoelectron spectroscopy (XPS) studies

The XPS studies were performed on powdered samples and the spectral lines were fit using the Peakfit software. The collected data was analyzed by taking into account background subtraction and spectral normalization. The peaks were deconvoluted with the corresponding binding energy of elements and taking the peak width, peak shape and the FWHM as flexible parameters. The survey spectrum was also standardized to C-1s peak at 284.26 eV. The varied oxidation states of Mn were obtained in +2, +3 and +4 oxidation states and Ru showed +4, +5 and 0 oxidation states (Table 3.5). The stoichiometric ratios of Pr: Mn: Ru was quantitatively analyzed by taking into account the inelastic mean free path and photoionization cross-section of the elements [29-30].

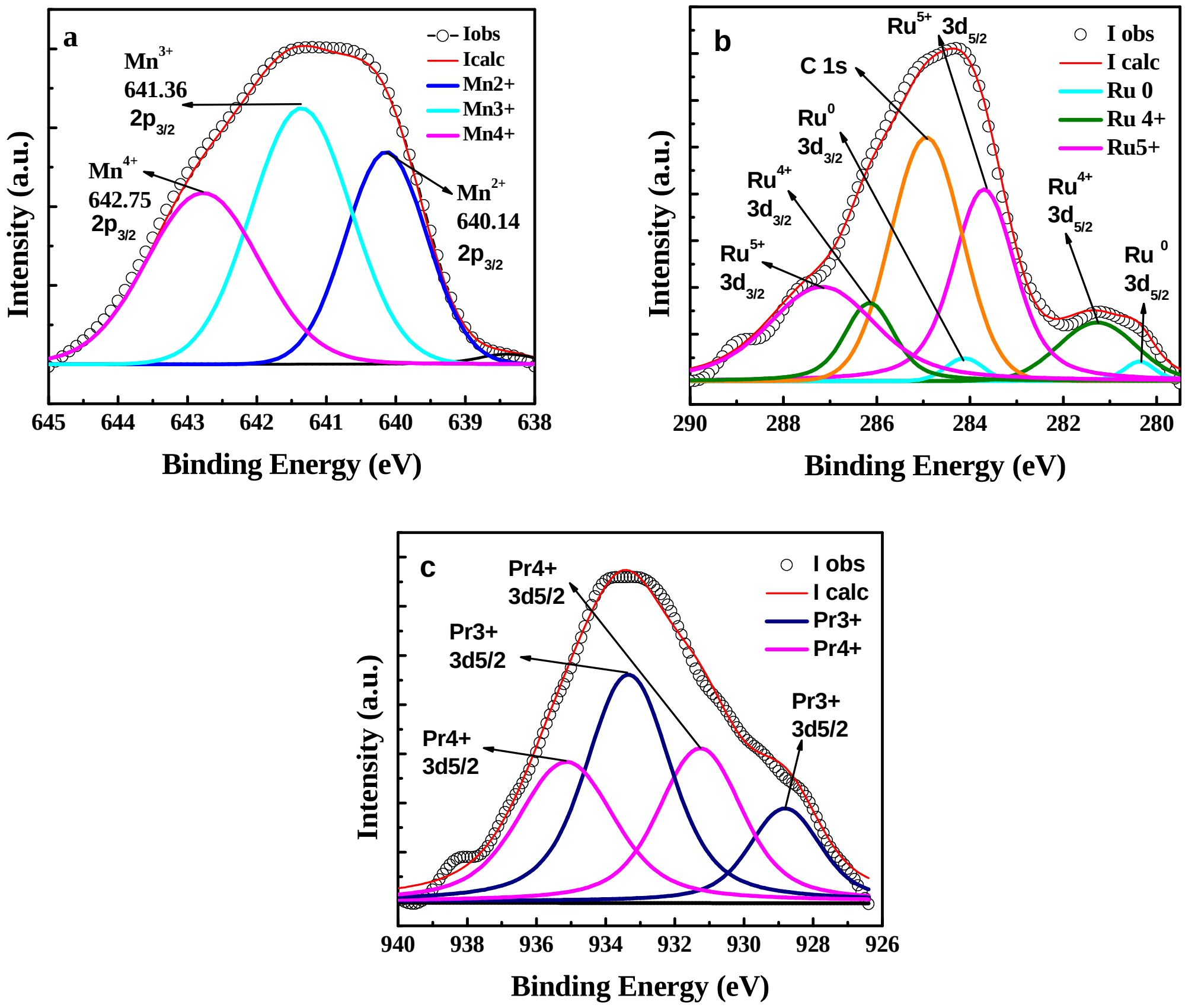

Fig. 3.4: XPS analysis showing the region of binding energies for a) Mn b) Ru and c) Pr.

Table 3.3: XPS analysis of $PrMn_{0.5}Ru_{0.35}O_3$ showing the region of binding energies for a) Mn and b) Ru and c) Pr in $PrMn_{0.5}Ru_{0.35}O_3$.

| **S.No** | **Centroid** | **FWHM** | **%Area** | **ΣArea** |
|---|---|---|---|---|
| $Pr^{3+}$ | 929.14 | 2.43 | 49.651 | 21221.27 |
| $3d_{5/2}$ | 933.33 | 2.87 | | |
| $Pr^{4+}$ | 931.40 | 2.85 | 50.349 | 21520.35 |
| $3d_{5/2}$ | 934.97 | 3.27 | | |

ΣArea = 42741.62, On dividing by: photoionization cross-section = 32.85 barns and inelastic mean free path = 122 nm, we get =**10.665**

| | | | | |
|---|---|---|---|---|
| $Mn^{2+}$ $2p_{3/2}$ | 640.14 | 1.38 | 27.646 | 788.22 |
| $Mn^{3+}$ $2p_{3/2}$ | 641.36 | 1.69 | 40.944 | 1167.33 |
| $Mn^{4+}$ $2p_{3/2}$ | 642.75 | 1.92 | 31.410 | 895.52 |

ΣArea = 2851.07, On dividing by: photoionization cross-section = 4.63 barns and inelastic mean free path = 106 nm, we get = **5.809**

| | | | | |
|---|---|---|---|---|
| $Ru^0$ $3d_{5/2}$ | 280.37 | 0.78 | 4.373 | 175.873 |
| $3d_{3/2}$ | 284.12 | 0.95 | | |
| $Ru^{4+}3d_{5/2}$ | 281.27 | 1.96 | 25.583 | 1028.715 |
| $3d_{3/2}$ | 286.15 | 1.26 | | |
| $Ru^{5+}3d_{5/2}$ | 283.69 | 1.57 | 70.044 | 2816.491 |
| $3d_{3/2}$ | 287.17 | 2.72 | | |

ΣArea (3d3/2) = 1870.69, On dividing by: photoionization cross-section = 5.17 barns and inelastic mean free path = 90 nm, we get =**4.020**

Here ratio of stoichiometry of Mn : Ru = 0.5(4.02/5.809) = **0.5: 0.35**

Ratio of Pr: Mn: Ru: **1: 0.54: 0.37** from ratio of 10.665: 5.809: 4.02

The EDAX and XPS measurement suggest Ru deficiency at B-site in 0.15(2) in stoichiometry.

### 3.3.4 Magnetization studies

DC magnetic susceptibility studies in both ZFC and FC were performed at 1000 Oe (Fig. 3.5a). Temperature-dependent magnetic susceptibility showed a spontaneous ferromagnetic like behaviour at temperatures below 90 K (Fig 3.5), followed by a bifurcation at 31 K which may be attributed to the spin-glass like nature of the sample. It has also been observed that there is a rapid increase in ZFC and FC moments at temperatures below 18 K, which could be attributed to the alignment of the ‘Pr’ moments as has been reported for $PrMn_{0.5}Fe_{0.5}O_3$ and $Nd_{0.75}Sr_{1.25}CoO_4$ [10,31]. From the differential magnetic susceptibility plot, a rapid increase in magnetization in the ZFC plot at 4.5 K could be attributed to the ordering of both Mn and Ru spins as has also been reported in the composition $(La_{0.52}Ba_{0.48})(Mn_{0.51}Ru_{0.49})O_3$ [5-6]. The antiferromagnetic ($Mn^{4+}$ - O – $Ru^{5+}$) and ferromagnetic ($Mn^{3+}$ - O – $Ru^{4+}$) lead to such behavior [5, 15]. It is worth mentioning that in $Ba_3MRu_2O_9$ (M : Y, In, La, Sm, Eu and Lu) a magnetic anomaly due to antiferromagnetic coupling between $Ru_2O_9$ dimers is seen with a decreased effective magnetic moment in the temperature range 4.5 – 12.5 K [37]. This is not expected in the case of $PrMn_{0.5}Ru_{0.35}O_3$, where the magnetic moment increases at a temperature below 4.5 K. Thus, the magnetic interaction due to Ru – O –Ru cluster interaction is insignificant, signifying the interactions of Mn/Ru below 4.5 K in $PrMn_{0.5}Ru_{0.35}O_3$. A similar anomaly has been reported for $PrBa_2Cu_3O_7$, where the 'Pr' moment starts ordering at a temperature below 4.5 K, evident from the increase of magnetic moment with a probable antiferromagnetic transition at 1.6 K [38]. Similarly, we conclude that the sharp increase in the magnetic moment at temperatures below 4.5 K in $PrMn_{0.5}Ru_{0.35}O_3$, may signify anti-ferromagnetic interactions due to 'Pr' moments.

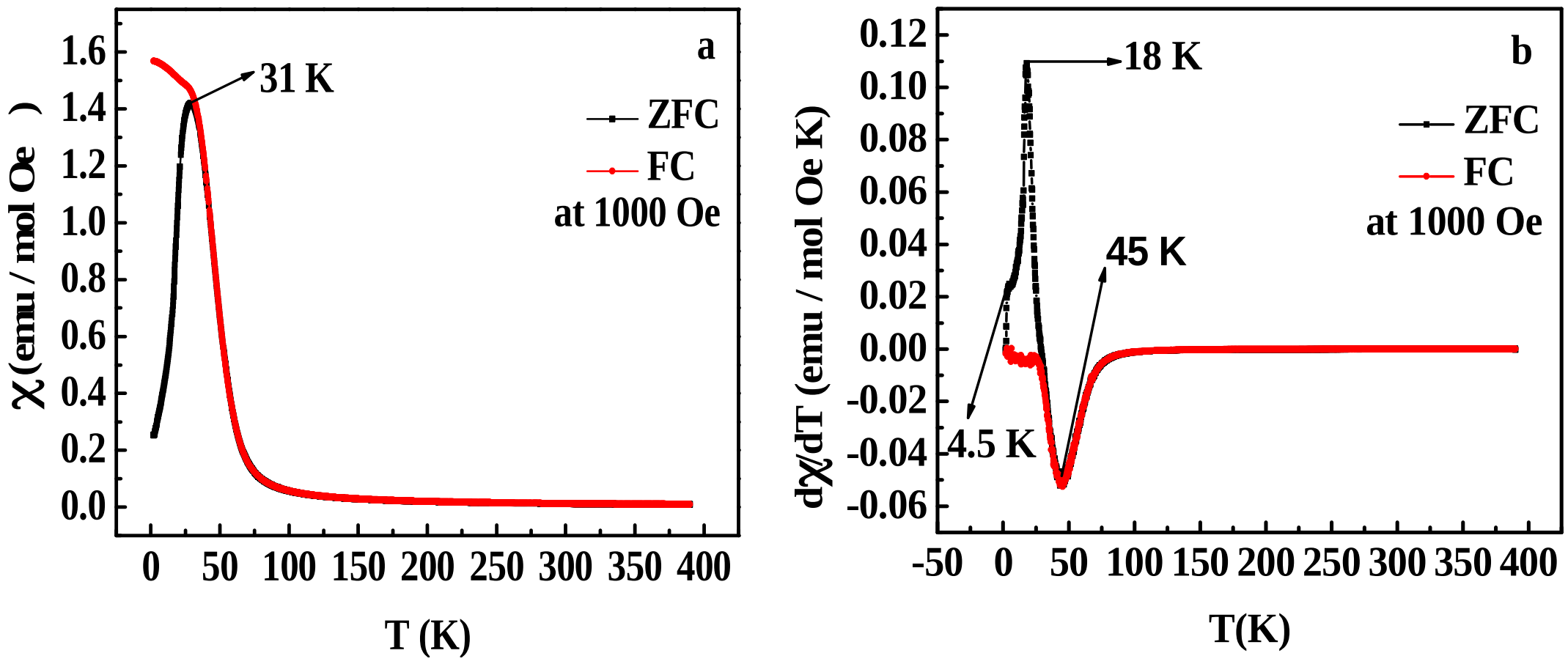

Fig 3.5: a) DC magnetic susceptibility plot of $PrMn_{0.5}Ru_{0.35}O_3$ in the temperature range 2 K-390 K. and b) dχ/dT with respect to temperature at 1000 Oe field.

From the inverse magnetic susceptibility plot (Fig. 3.5), the Curie-Weiss fitting of the 1/χ vs. temperature data, it was found that the Weiss temperature $\theta_{CW}$ = 21 K (equation 3.1) and the effective magnetic moment $\mu_{effective}$ = 5.104 $\mu_B$/f.u., which is comparable to the calculated value of 5.248 $\mu_B$/f.u. The paramagnetic behaviour was fitted in the linear region in the temperature range 200-390 K. This deviation from linearity at temperatures below 182 K, showing weak ferromagnetic behaviour probably arises from the canted antiferromagnetic spins of Mn ions. For example, $La_{1-x}Ca_xMnO_3$ shows double exchange ferromagnetic interaction between $Mn^{3+}$ and $Mn^{2+}$ at a temperature below 150 K [15, 40].

$1/\chi = C/T-\theta$ Curie Weiss law Equation 3.1

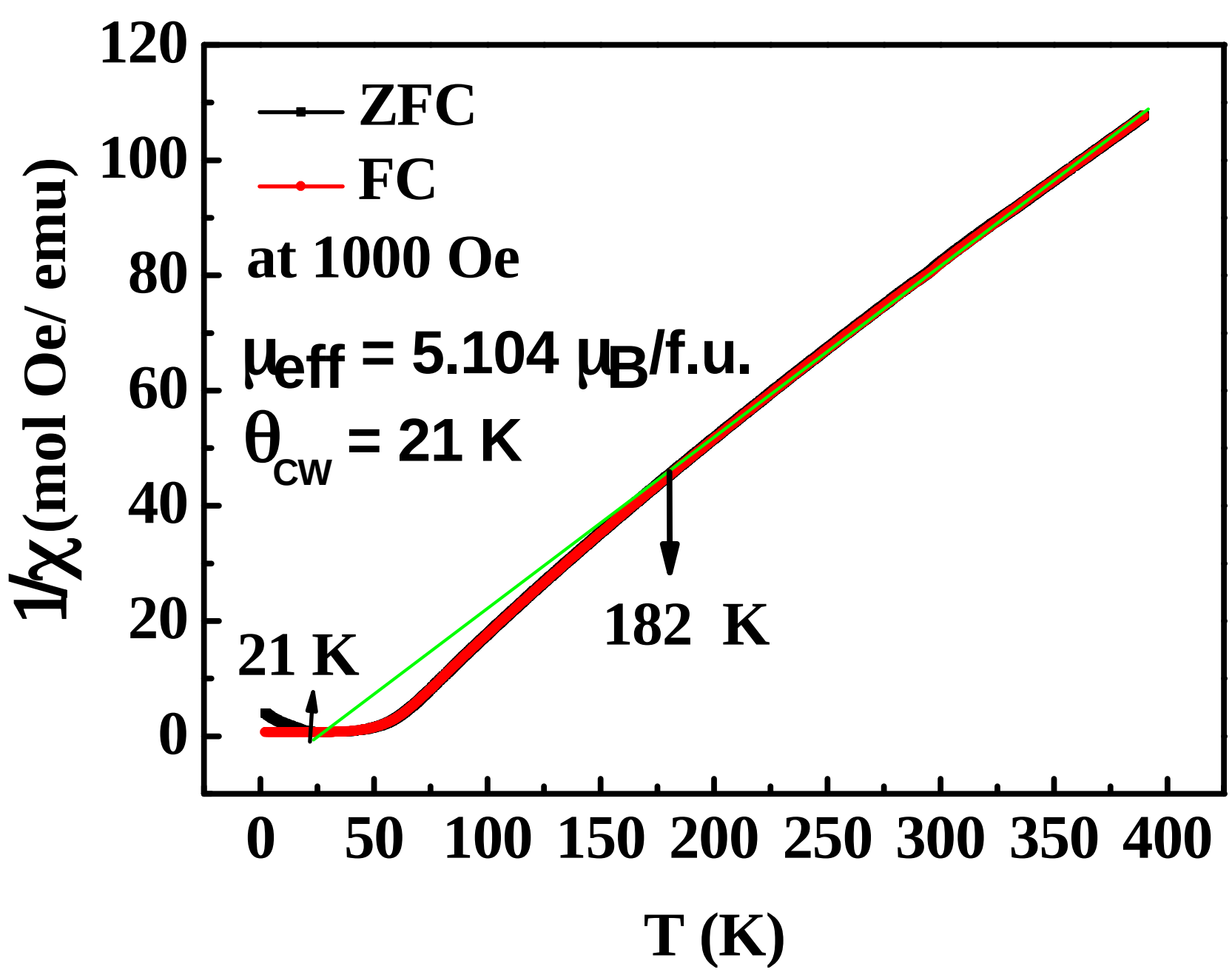

Fig 3.6: Inverse magnetic susceptibility plot at 1000 Oe for $PrMn_{0.5}Ru_{0.35}O_3$.

Isothermal field dependent magnetization with applied magnetic field of ±7 T measured at 300 K, 50 K and 2 K are shown in Fig. 3.7. The field dependent magnetization at 300 K shows paramagnetic nature. The isotherms start deviating from the linearity and show ferromagnetic nature by further decreasing the temperature to 50 K and 2 K. At 2 K, the field dependent magnetization data demonstrate a significant coercive field of ~ 600 Oe, while the signal does not get saturated by applying the magnetic field as high as 7 T. Presence of significant magnetic coercivity without a saturation in the magnetization at high magnetic field indicate. The presence of antiferromagnetic as well as ferromagnetic interactions or absence of any long-range ferromagnetic ordering. The saturation magnetization was calculated on taking into account various quantitative magnetic moments of transition metal ions: $Mn^{2+}$ (5.9 $\mu_B$/f.u.), $Mn^{3+}$ (3.8 $\mu_B$/f.u.) and $Mn^{4+}$ (4.9 $\mu_B$/f.u.) and $Ru^{4+}$ (low spin 2.83 $\mu_B$/f.u.) and $Ru^{5+}$ (low spin 1.73 $\mu_B$/f.u.) [10, 20, 38-41].

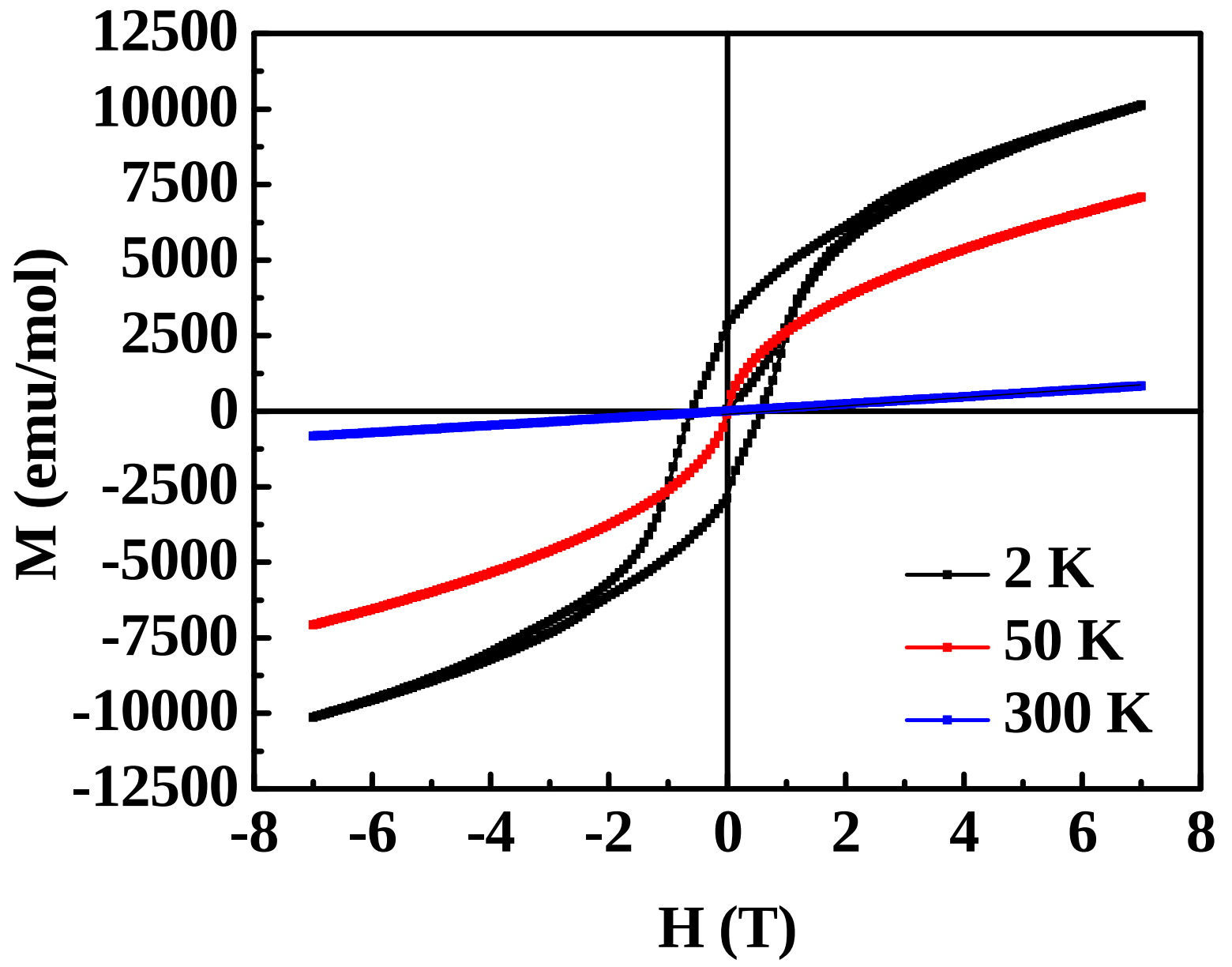

Fig 3.7: Isothermal magnetization of $PrMn_{0.5}Ru_{0.35}O_3$ at the indicated temperatures in the magnetic field range of -7 T to +7 T.

On correlating the structural properties with the magnetic properties, it can be concluded that anisotropic orbital interaction between the B- atoms, as a result, manifests increased octahedral distortion and B-site Ru deficiency along the *b* axis which results in increased bifurcation between the ZFC and FC due to short-range magnetic interaction.

### 3.4 Conclusions

The structure and magnetic properties of $PrMn_{0.5}Ru_{0.35}O_3$ was analyzed and was refined in the orthorhombic *Pnma* space group. The EDAX and XPS analysis showed a Ru vacancy of 0.15(2) in stoichiometry per unit cell. The increased octahedral distortion and compression of the *b* lattice parameter further suggests the increased structural distortion in the structure. From the DC magnetic susceptibility data, it was found to show ferromagnetic transition at temperatures below 90 K and a magnetic anomaly obtained at 18 K could be attributed to disorder-induced ferromagnetic ordering. Furthermore, the increased magnetic moment in ZFC at temperatures below could be attributed to the magnetic ordering of the 'Pr' spins and Mn – O – Ru ordering. From the isothermal magnetization data performed at 2 K and 50 K the non-saturation nature of the magnetization at high temperatures could be attributed to competing ferromagnetic as well as antiferromagnetic interaction or absence of any long-range ordering in the system. The core-level XPS studies show that majority of the Ru ions are in +3(40.94%) and +5 (49.39%) oxidation states.

# Chapter IV

# Investigation of structural and magnetic properties of $La_{0.5-x}Sm_xPr_{0.5}CrO_3$ (x = 0, 0.1, 0.2, 0.3, 0.4 and 0.5)

# Chapter IV

# Investigation of structural and magnetic properties of $La_{0.5-x}Sm_xPr_{0.5}CrO_3$ (x = 0, 0.1, 0.2, 0.3, 0.4 and 0.5)

## Abstract

The magnetic and structural properties in $La_{0.5-x}Sm_xPr_{0.5}CrO_3$ (x = 0, 0.1, 0.2, 0.3, 0.4 and 0.5) has been investigated. The powder X-ray diffraction was satisfactorily refined in the orthorhombic *Pnma* space group. The structural anomaly showed that octahedral distortion increased with an increase in x value. DC magnetic susceptibility at 100 Oe in zero field cooled (ZFC), showed negative magnetization in negative magnitude for x = 0.4 and 0.5 and decreasing magnetization value for x = 0.3. The field cooled cooling (FCC) and field cooled warming (FCW) magnetization measured at 1000 Oe showed that the Néel temperature decreased from 275 K (x = 0) to 246 K (x = 0.5). Furthermore, FCC and FCW magnetization for x = 0 and 0.1 in the temperature range 160-25 K suggest the presence of a magnetic phase transition. From the isothermal magnetization data measured at 5 K, the magnetization value at 7 T was found to increase with the increase in value of x. In $La_{0.1}Sm_{0.4}Pr_{0.5}CrO_3$ the negative exchange bias temperature was obtained at temperatures below 230 K. The influence of canted antiferromagnetism with the change in rare earth moment has been investigated.

## 4.1 Introduction

Complex perovskites of the formula $ABO_3$ (A: rare earth, B: transition metal) have their applications in the field of colossal magnetoresistance, high-temperature superconductors, magneto-caloric materials and exchange bias memory devices etc. [1]. In the magnetic studies of several perovskites, the phenomena of negative magnetization have been observed due to the alignment of rare earth spins antiferromagnetically with respect to the canting of the B –site spin alignment [1-8]. The change in spin orientation of rare earth with respect to transition metal ions results in non-overlapping of field cooled cooling (FCC) and field cooled warming (FCW) curves as has been observed in $CeCrO_3$, $Lu_2FeO_4$ etc. [9-10]. It has been reported for $La_{1-x}Sm_xCrO_3$ that the spin reorientation transition temperature ($T_{SR}$) depends on Sm-Cr interaction which decreases as the Sm concentration increases [13]. The canted antiferromagnetically aligned spins also result in applications in transient random-access memory (TRAM) based on its positive and negative exchange bias effect [13]. The decreased anisotropy plays a major role in the increased degree of spin canting and thus as a result leads to the easy orientation of canted spins which matches with the spin orientation of rare earth spins using the equation 4.1.

$$M_{total} = M_{Cr} + C\,(H_I + H) / (T - \theta) \qquad \text{Equation 4.1}$$

Thus, when $M_{total}$ becomes zero, which is known as compensation temperature which is proportional to the internal field ($H_I$), depends on the anisotropy of the system and the moment generated from the canted antiferromagnetically aligned $M_{Cr}$ and other factors such as Néel temperature ($T_N$) and Weiss temperature ($\theta_{CW}$) [11-12]. The observation of a more distorted structure results in a more magnetic moment from Cr ions ($M_{Cr}$), which as a result increases the net magnetic field ($H_I$) [13]. For example, the compound $La_{0.15}Pr_{0.85}CrO_3$ shows $H_I$ and $M_{Cr}$ at 2800-3000 Oe and 15-40 emu/mol respectively, while in $GdCrO_3$ the values are 5500-6400 Oe and 400-500 emu/mol respectively (equation 4.1) [7, 8].

The factors which play a major role in the overall value of negative magnetization are specific to magnetic structure and f-d orbital exchange interactions. The Néel temperature also gets decreased as the structural distortion increases, depending primarily on the B - O - B' bond angle and due to the decreased exchange interactions [12, 15-18].

The correlation between the Néel temperature and compensation temperature depends on factors such as unidirectional anisotropy, structural distortion and Dzyaloshinskii-Moriya interaction [18-20]. For example, $La_{1-x}Sr_xMnO_3$ (x = 0.52 - 0.56) shows magnetic hysteresis loop at temperatures close to the Néel temperature, due to the interaction between 2D antiferromagnetism ($dx^2$-$y^2$ orbitals) and 3D ferromagnetism [11]. This canted antiferromagnetic structure of spins particularly close to the Néel temperature was also observed from the isothermal exchange bias shift. For example, the parent compound $La_{1-x}Pr_xCrO_3$ shows positive exchange bias at 140 Oe at low magnetic fields 3-15 kOe [8, 17, 25-26]. However, at higher field regions the antiferromagnetic anisotropy increases, which increases the antiferromagnetic coupling. The negative exchange bias shift has been attributed due to strengthened antiferromagnetic coupling from the unidirectional anisotropy [20]. While the positive exchange bias results from the increased alignment of ferromagnetic spins from antiferromagnetic domains, aligned parallel to the magnetic field [20]. The negative exchange bias has been reported for $Ni_{50}Mn_{37}In_{13}$ to be the highest in ZFC at 1300 Oe [28]. Studying the magnetic property of various rare-earth substituted perovskites, having different spin structures resulting in varied magnetic exchange interactions has been investigated.

Table 4.1: Compositions earlier reported with negative magnetization with their Néel temperature ($T_N$) and temperature at which magnetization becomes zero ($T_{comp}$).

| S.No | Composition | $T_N$ (K) | $T_{comp}$ (K) | Reference |
|---|---|---|---|---|
| 1. | $LaCrO_3$ | 288 | - | [35] |
| 2. | $PrCrO_3$ | 240 | - | [36] |
| 3. | $LaCr_{0.8}Mn_{0.2}CrO_3$ | 282 | 80 | [37] |
| 4. | $La_{0.9}Pr_{0.1}CrO_3$ | 280 | 17 | [15] |
| 5. | $La_{0.2}Cr_{0.8}CrO_3$ | 264 | 45 | [38] |
| 6. | $CeCrO_3$ | 260 | 133 | [22] |
| 7. | $La_{0.15}Pr_{0.85}CrO_3$ | 245 | 230 | [38] |
| 8. | $GdCrO_3$ | 160 | 130 | [7] |
| 9. | $SmCrO_3$ | 150 | 110 | [39] |
| 10. | $DyCrO_3$ | 144 | 143 | [38] |
| 11. | $EuCrO_3$ | 130 | 50 | [40] |
| 12. | $YbCrO_3$ | 118 | 18 | [41] |

## 4.2 Experimental

### 4.2.1 Materials and synthesis

$La_2O_3$ (Sigma Aldrich), $Sm_2O_3$ (Sigma Aldrich), $Pr_6O_{11}$ (Sigma Aldrich) and $Cr_2O_3$ (Sigma Aldrich) were weighed in their respective stoichiometric ratios. However, $La_2O_3$, $Sm_2O_3$ and $Pr_6O_{11}$ were given a heat treatment at 900˚C for 12 hours before weighing. All the oxides were ground to form a homogeneous powder mixture. $La_{0.5-x}Sm_xPr_{0.5}CrO_3$ 9 (x = 0, 0.1, 0.2, 0.3, 0.4 and 0.5) were synthesized by solid-state reaction method. The compound was calcined at 900˚C for 12 hours and then heated at 1200˚C for 12 hours. The obtained powder (0.2 g) was further pelletized at 10 tons after adding 20 µl of 5% PVA solution and then heated at 1200˚C for 12 hours. The heating and cooling rates were fixed at 150˚C/hour.

### 4.2.2 Powder X-ray diffraction (PXRD)

Bruker D8-Advance diffractometer, the instrument was used for the recording of diffraction data. Here Cu-Kα was used as a source. The analysis of powder X-ray diffraction data was collected in the 2q range of 10˚ to 80˚ at a rate of 1˚ per minute. TOPAZ 5.2 software was used for analysing the cell parameters, background fitting, atomic positions and isotropic thermal parameters which were refined using this software.

### 4.2.3 Magnetization studies

Zero field cooled (ZFC) and field cooled (FC) Temperature dependent magnetic studies were performed using a physical property measurement system (PPMS) in the temperature range of 5-300 K at a magnetic field of 100 Oe at Zero field cooling (ZFC) mode. Temperature dependent magnetization studies in Field cooled cooling (FCC) and Field cooled warming (FCW) conditions were performed at 10,000 Oe. The isothermal field dependent magnetization measurements were carried out at a magnetic field of ±5 T for all the compositions. For $La_{0.1}Sm_{0.4}Pr_{0.5}CrO_3$, the isothermal magnetization measurements were performed in the temperature range of 250-300 K in both ZFC and ZFW modes at ± 5 T fields.

## 4.3 Results and discussion

### 4.3.1 X-ray diffraction studies

The structure of $La_{0.5-x}Sm_xPr_{0.5}CrO_3$ (x = 0, 0.1, 0.2, 0.3, 0.4 and 0.5) were investigated using powder X-ray diffraction data. The *c*-lattice parameter and B-$O_1$-B' bond angle was found to decrease as the value of x increased (Table). The PXRD simulation study was performed using the parent compound $La_{0.5}Pr_{0.5}CrO_3$ [17]. The decrease in *c* lattice parameter as the rare-earth A-site radii decreased and increased octahedral distortion thus creating more distorted structure as x increases.

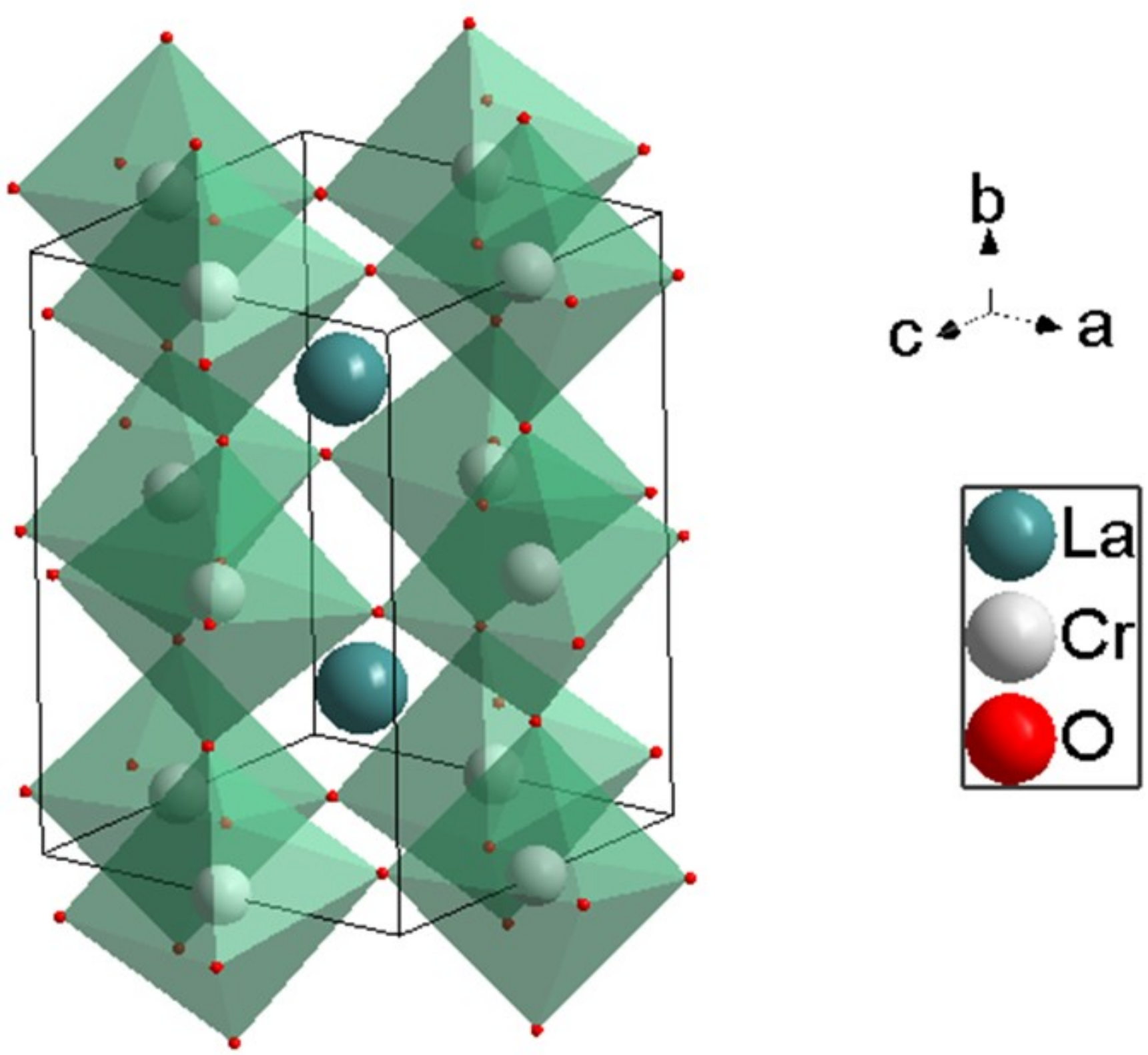

Fig. 4.1: Structure of $La_{0.5}Pr_{0.5}CrO_3$ perovskite with orthorhombic *Pnma* structure.

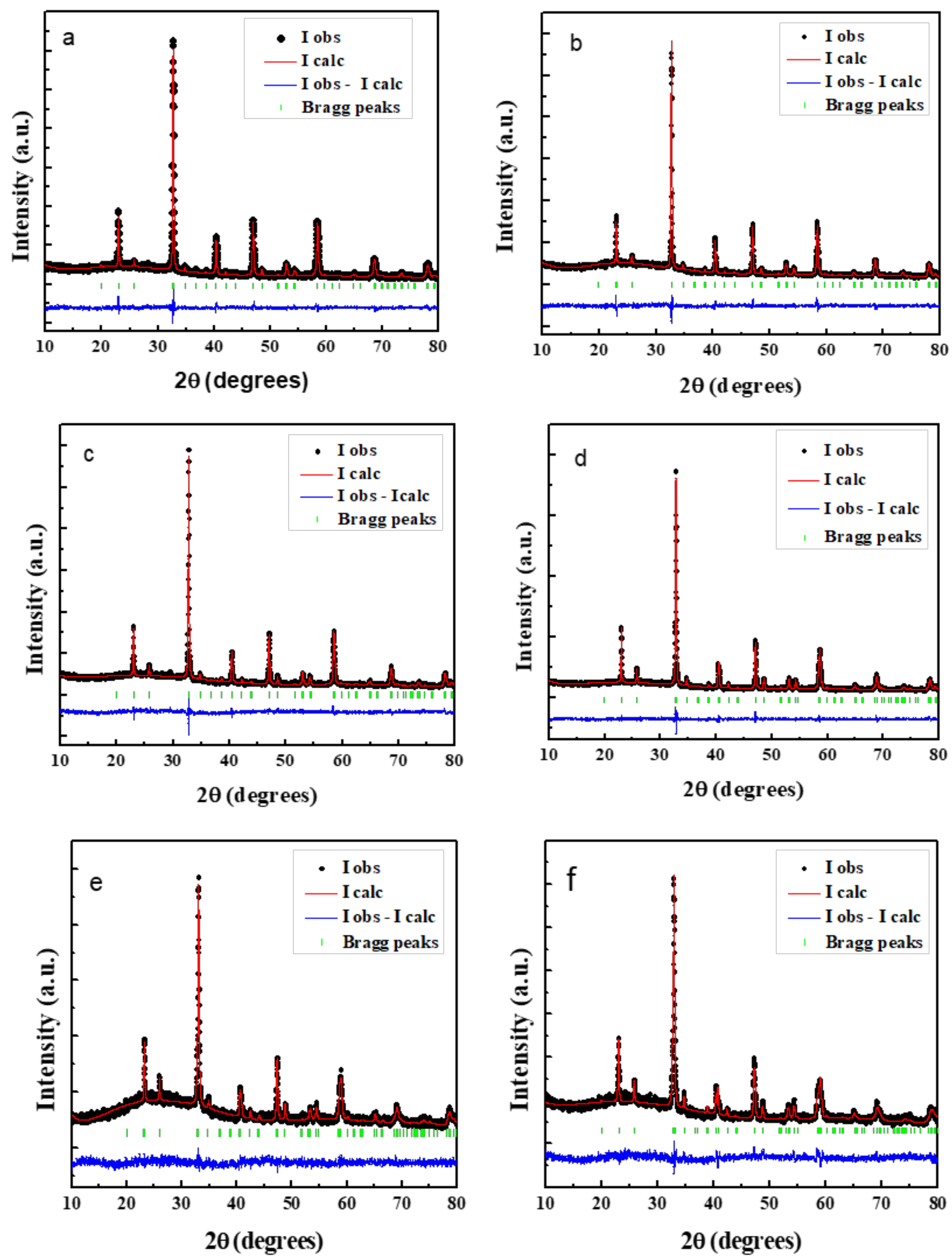

Fig. 4.2: Rietveld refinement of the power X-ray diffraction data of $La_{0.5-x}Sm_xPr_{0.5}CrO_3$ for a) x = 0, b) x = 0.1, c) x = 0.2, d) x = 0.3, e) x = 0.4 and f) x = 0.5 refined with orthorhombic *Pnma* spacegroup. Here the black points represent the observed diffraction data, and the red plot

represents the calculated data and the green vertical lines signifies Bragg peaks and the blue line represents the difference between observed and calculated intensities.

Table 4.2a: Rietveld refinement of $La_{0.5}Pr_{0.5}CrO_3$ with orthorhombic *Pnma* spacegroup. a = *5.4633(1)* Å, b = *7.7465(2)* Å and c = *5.4739(1)* Å

| Atom | Site | x | y | z | Occ | Beq (Å$^2$) |
|---|---|---|---|---|---|---|
| La/Pr | 4c | 0.0284(4) | 0.25 | -0.0100(7) | 0.5/0.5 | 1.34(6) |
| Cr | 4b | 0 | 0 | 0.5 | 1 | 1.16(6) |
| O1 | 4c | 0.489(3) | 0.25 | 0.024(4) | 1 | 1.14(4) |
| O2 | 8d | 0.293(3) | 0.034(2) | -0.290(3) | 1 | 1.57(2) |

Table 4.2b: Rietveld refinement of $La_{0.4}Sm_{0.1}Pr_{0.5}CrO_3$ with orthorhombic *Pnma* spacegroup.

a = *5.4699(2)* Å, b = *7.7392(1)* Å and c = *5.45564(2)* Å

| Atom | Site | x | y | z | Occ | Beq (Å$^2$) |
|---|---|---|---|---|---|---|
| La/Sm/Pr | 4c | 0.0312(3) | 0.25 | -0.0107(6) | 0.4/0.1/0.5 | 1.30(6) |
| Cr | 4b | 0 | 0 | 0.5 | 1 | 1.30(5) |
| O1 | 4c | 0.482(3) | 0.25 | 0.039(4) | 1 | 2.3(3) |
| O2 | 8d | 0.309(3) | 0.028(2) | -0.29(2) | 1 | 1.3(1) |

Table 4.2c: Rietveld refinement of $La_{0.3}Sm_{0.2}Pr_{0.5}CrO_3$ with orthorhombic *Pnma* spacegroup.

a = *5.4731(2)* Å, b = *7.7147(2)* Å and c = *5.4527(2)* Å

| **Atom** | **Site** | **x** | **y** | **z** | **Occ** | **Beq (Å$^2$)** |
|---|---|---|---|---|---|---|
| La/Sm/Pr | 4c | 0.0334(4) | 0.25 | -0.0104(6) | 0.3/0.2/0.5 | 1.22(4) |
| Cr | 4b | 0 | 0 | 0.5 | 1 | 1.22(4) |
| O1 | 4c | 0.465(4) | 0.25 | 0.073(4) | 1 | 1.7(5) |
| O2 | 8d | 0.303(3) | 0.024(2) | -0.293(3) | 1 | 1.1(1) |

Table 4.2d : Rietveld refinement of $La_{0.2}Sm_{0.3}Pr_{0.5}CrO_3$ with orthorhombic *Pnma* spacegroup.

a = *5.4727(2)* Å, b = *7.7001(2)* Å and c = *5.4368(1)* Å

| **Atom** | **Site** | **x** | **y** | **z** | **Occ** | **Beq (Å$^2$)** |
|---|---|---|---|---|---|---|
| La/Sm/Pr | 4c | 0.0324(5) | 0.25 | 0.003(1) | 0.2/0.3/0.5 | 2.05(6) |
| Cr | 4b | 0 | 0 | 0.5 | 1 | 1.06(5) |
| O1 | 4c | 0.48(1) | 0.25 | 0.088(3) | 1 | 1.4(4) |
| O2 | 8d | 0.258(4) | 0.013(4) | -0.26(1) | 1 | 1.6(3) |

Table 4.2e: Rietveld refinement of $La_{0.1}Sm_{0.4}Pr_{0.5}CrO_3$ with orthorhombic *Pnma* spacegroup.

a = *5.4739(1)* Å, b = *7.6848(2)* Å and c = *5.4226(2)* Å

| **Atom** | **Site** | **x** | **y** | **z** | **Occ** | **Beq (Å²)** |
|---|---|---|---|---|---|---|
| La/Sm/Pr | 4c | 0.060(7) | 0.25 | -0.03(1) | 0.1/0.4/0.5 | 2.0(1) |
| Cr | 4b | 0 | 0 | 0.5 | 1 | 2.0(2) |
| O1 | 4c | 0.46(5) | 0.25 | 0.021(1) | 1 | 1.0(1) |
| O2 | 8d | 0.30(4) | 0.06(1) | -0.28(2) | 1 | 1.0(2) |

Table 4.2f: Rietveld refinement of $Sm_{0.5}Pr_{0.5}CrO_3$ with orthorhombic *Pnma* spacegroup.

a = *5.4748(1)* Å, b = *7.6698(2)* Å and c = *5.4048(1)* Å

| Atom | Site | x | y | z | Occ | Beq (Å²) |
|---|---|---|---|---|---|---|
| Sm/Pr | 4c | 0.0431(5) | 0.25 | -0.014(1) | 0.5/0.5 | 1.5(1) |
| Cr | 4b | 0 | 0 | 0.5 | 1 | 1.4(1) |
| O1 | 4c | 0.440(4) | 0.25 | 0.081(4) | 1 | 1.9(1) |
| O2 | 8d | 0.289(4) | 0.040(3) | -0.297(4) | 1 | 1.0(1) |

Table 4.3: Comparison of lattice parameters of $La_{0.5-x}Sm_xPr_{0.5}CrO_3$ for x = 0, 0.1, 0.2, 0.3, 0.4 and 0.5.

| Composition | a (Å) | b (Å) | c (Å) | Fe/Mn - $O_1$<br>Fe/Mn - $O_2$<br>(Å) | R/Sr - $O_1$<br>R/Sr - $O_2$<br>(Å) | B - $O_1$ - B<br>B - $O_2$ - B<br>(˚) | $O_2$ - B - $O_1$<br>$O_1$ - A - $O_2$<br>(˚) |
|---|---|---|---|---|---|---|---|
| $La_{0.5}Pr_{0.5}CrO_3$ | 5.4633(4) | 7.7465(5) | 5.4759(4) | 1.942(1) | 2.523(1) | 171.46(1) | 86.10(1) |
| | | | | 1.967(1) | 2.691(1) | 155.76(1) | 60.41(1) |
| $La_{0.4}Sm_{0.1}Pr_{0.5}CrO_3$ | 5.4699(3) | 7.7392(5) | 5.4564(3) | 1.949(1) | 2.895(1) | 166.18(1) | 90.15(10 |
| | | | | 2.050(1) | 2.757(1) | 154.14(1) | 67.51(1) |
| $La_{0.3}Sm_{0.2}Pr_{0.5}CrO_3$ | 5.4731(2) | 7.7147(4) | 5.4527(3) | 1.978(1) | 2.414(1) | 154.20(1) | 91.08(1) |
| | | | | 2.014(1) | 2.755(1) | 155.77(1) | 65.16(1) |
| $La_{0.2}Sm_{0.3}Pr_{0.5}CrO_3$ | 5.4727(2) | 7.7001(3) | 5.4368(2) | 1.987(1) | 3.058(1) | 151.39(1) | 95.04(1) |
| | | | | 1.939(1) | 2.707(1) | 172.77(1) | 68.84(1) |
| $La_{0.1}Sm_{0.4}Pr_{0.5}CrO_3$ | 5.4739(4) | 7.6848(7) | 5.4226(6) | 1.937(1) | 2.206(1) | 165.50(1) | 104.91(1) |
| | | | | 2.082(1) | 2.846(1) | 147.83(1) | 70.99(1) |
| $Sm_{0.5}Pr_{0.5}CrO_3$ | 5.4748(5) | 7.6698(6) | 5.4048(4) | 1.994(1) | 3.116(1) | 148.14(1) | 93.81(1) |
| | | | | 2.001(1) | 2.677(1) | 153.59(1) | 68.35(1) |

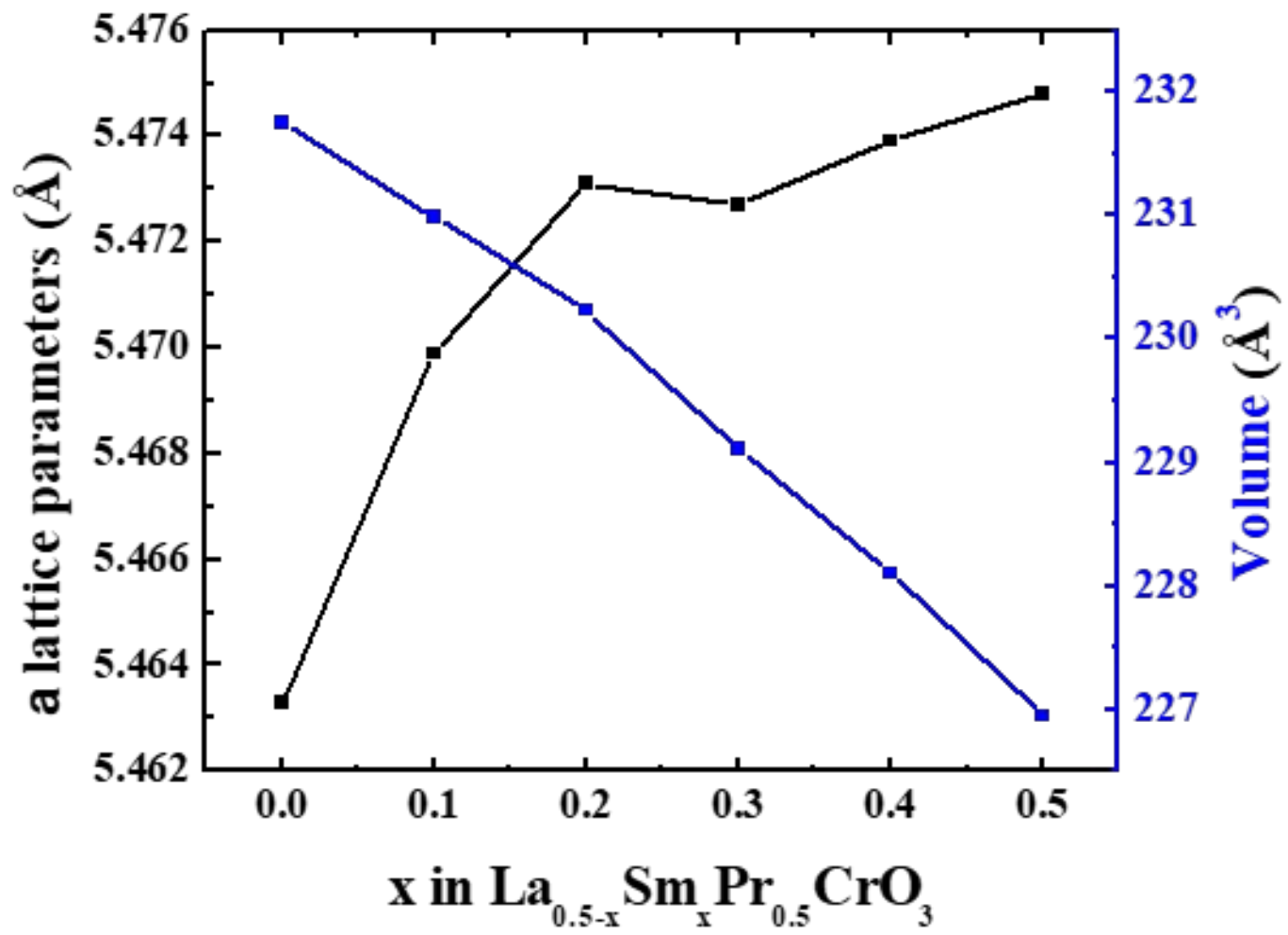

Fig. 4.3: Lattice volume and a lattice parameter in $La_{0.5-x}Sm_xPr_{0.5}CrO_3$.

Here the lattice volume and *b* and *c* lattice parameter decreased with an increase in value of x, as the Sm concentration increased. The '*a*' lattice parameter however increased as the value of x increased which could be attributed to the decreased octahedral distortion of B - O – B bond angle along *a*-axis with an increase in x value.

### 4.3.2 Magnetization studies

Temperature dependent DC magnetic susceptibility in ZFC mode was performed for $La_{0.5-x}Sm_xPr_{0.5}CrO_3$ (x= 0, 0.1, 0.2, 0.3, 0.4 and 0.5). The positive value of magnetization was observed for x = 0, 0.1 and 0.2, which shows an increase in magnetization as the value of x increased in the low-temperature region (below 180 K). Furthermore, it was also observed that at temperatures below 25 K, magnetization was found to increase rapidly due to the ordering of the rare-earth moments. It has been reported in the literature that the compositions $LaCrO_3$ and $Sm_{1-x}La_xCrO_3$ have attributed this reorientation of spins from *c* to *a*n axis at temperatures 25 K and 160 K [25].

On comparison of the magnetization value for x = 0.3 and 0.2, it was observed that the positive value of magnetization was less for x = 0.3 than x = 0.2. We predict that for x = 0.3, 0.4 and 0.5 increased octahedral distortion facilitates an increased degree of canting and favours rare earth ordering [35]. Furthermore, it was also observed that the compensation temperature for x = 0.4 was found to be greater than x = 0.5 (Fig.4.4b), probably due to the increased rare-earth and transition metal interaction for x = 0.4. As the magnetization begins to decrease from x = 0.3, due to decreased anisotropy, resulting in increased opposite alignment of spins to the direction of magnetic field (Fig.4.4a) [12].

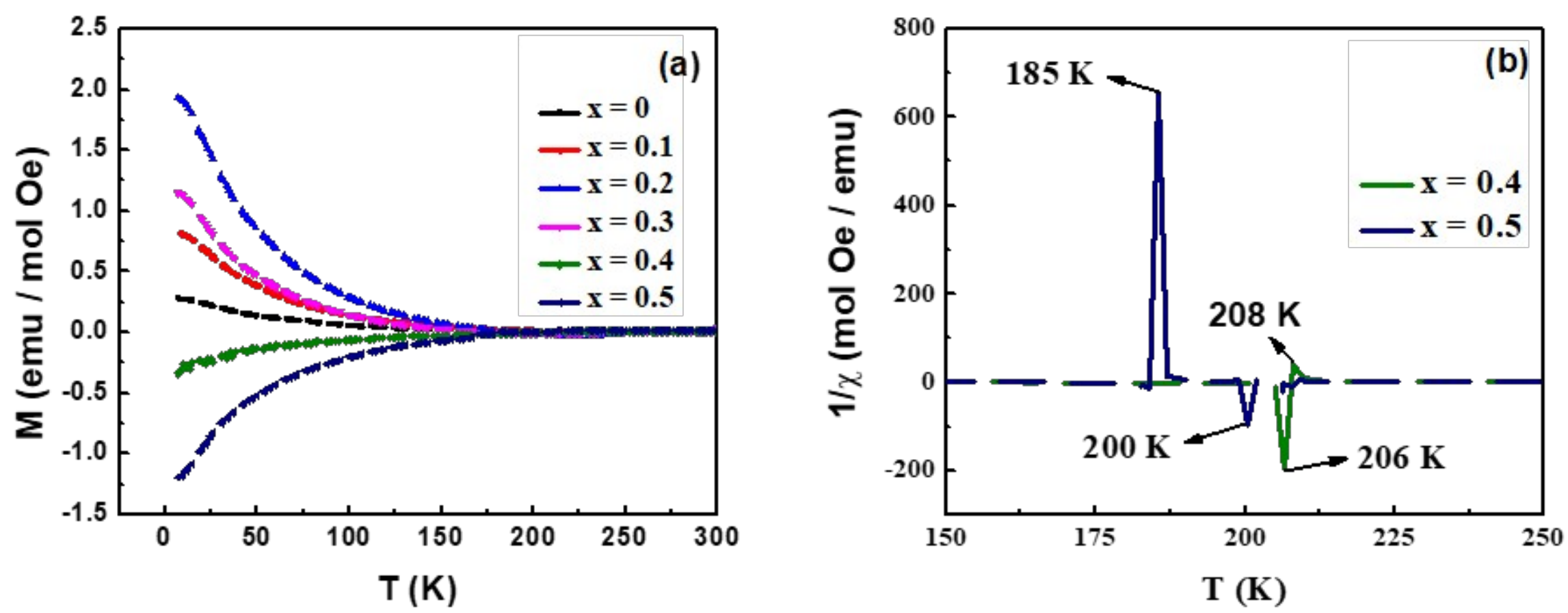

Fig.4.4: Magnetic susceptivity plot of zero field cooled (ZFC) at 100 Oe for $La_{0.5-x}Sm_xPr_{0.5}CrO_3$ for x = 0, 0.1, 0.2, 0.3, 0.4 and 0.5 and b) Inverse magnetic susceptibility plot of $La_{0.5-x}Sm_xPr_{0.5}CrO_3$.

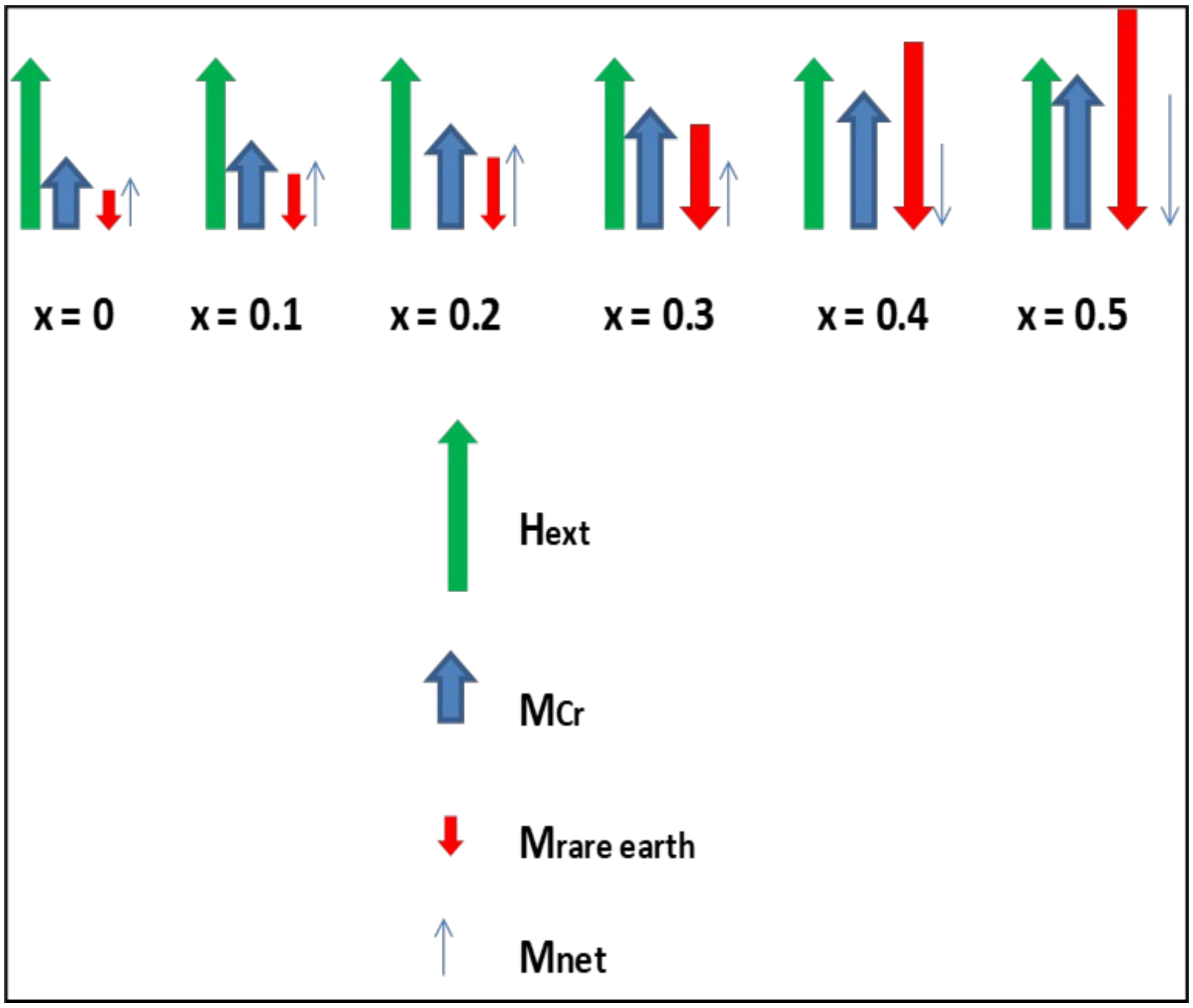

Fig. 4.5: Magnetic spin structure below the temperature where magnetization crosses zero ($T_{comp}$) in zero field cooled warming data at 100 Oe for $La_{0.5-x}Sm_xPr_{0.5}CrO_3$ for x = 0, 0.1, 0.2, 0.3, 0.4 and 0.5.

The temperature dependent FCC and FCW DC magnetic susceptibility data at 10,000 Oe are shown in Fig. 4.6. The studies show paramagnetic behavior in at high temperature range 250-300 K for all the samples. Below this temperature, where a deviation from Curie Weiss law due to weak ferromagnetism was generated by canted antiferromagnetic alignment of the moments [12]. At temperatures below 50 K, this rapid increase in magnetization could be attributed to rare-earth ordering [12].

From the inverse magnetic susceptibility plot, the Néel temperature was found to decrease as the value of x increases from 275 K (x = 0) to 246 K (x = 0.5) (Fig. 4.6b and Table 4.3). The Weiss temperature suggests antiferromagnetic behaviour which increased in magnitude as the value of x increases, except for x = 0, which shows a weak ferromagnetic behaviour. A rapid increase in the moment at temperatures below 50 K could be attributed to rare earth ordering (Fig.4.8 and 4.9) [21-22]. It was also observed that for x = 0 and 0.1, the bifurcation between FCC and FCW in the temperature range 160-25 K, further suggests magnetic phase transition taking place. This magnetic spin reorientation of rare earth, at temperatures close to 25 K, resulting in first-order phase transition has been reported in the literature [21-22].

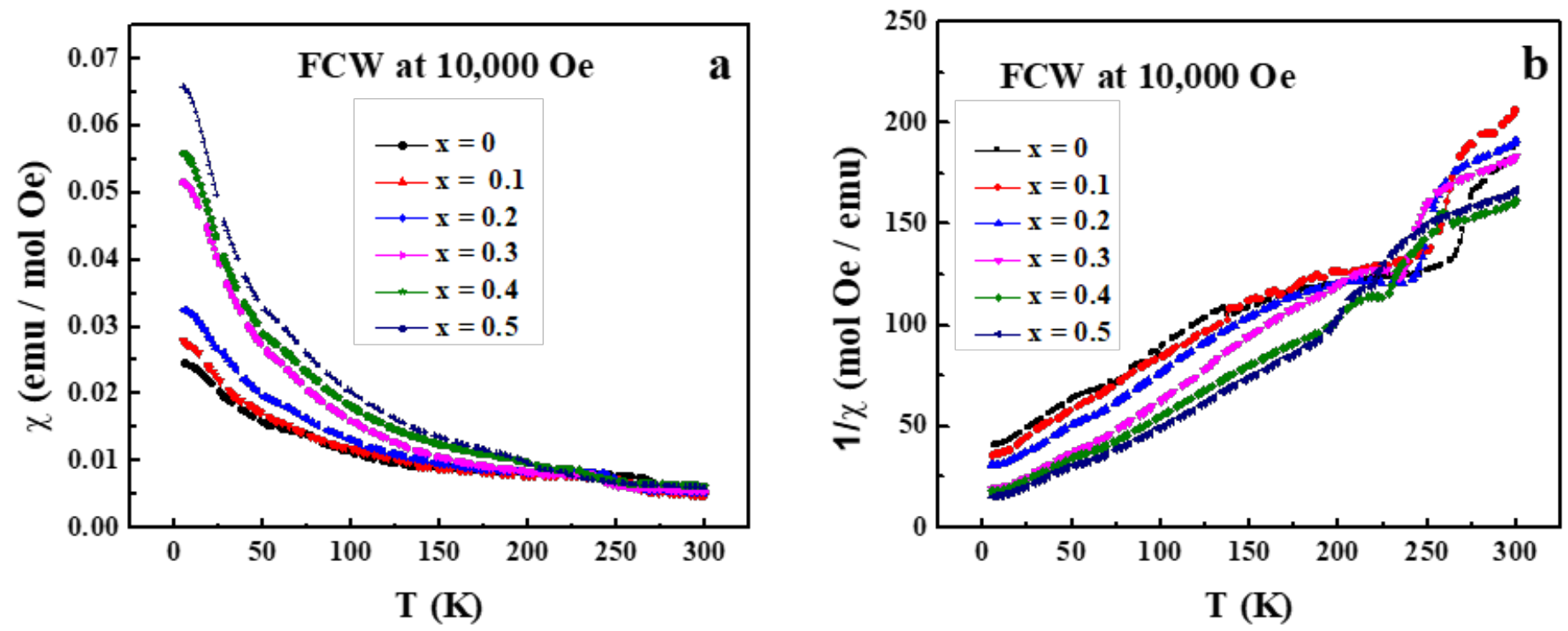

Fig. 4.6 a) Magnetic susceptibility for field cooled warming (FCW) plot at 10,000 Oe for $La_{0.5-x}Sm_xPr_{0.5}CrO_3$ for x = 0, 0.1, 0.2, 0.3, 0.4 and 0.5, b) Inverse magnetic susceptibility plot for FCW at 10,000 Oe for $La_{0.5-x}Sm_xPr_{0.5}CrO_3$ for x = 0, 0.1, 0.2, 0.3, 0.4 and 0.5.

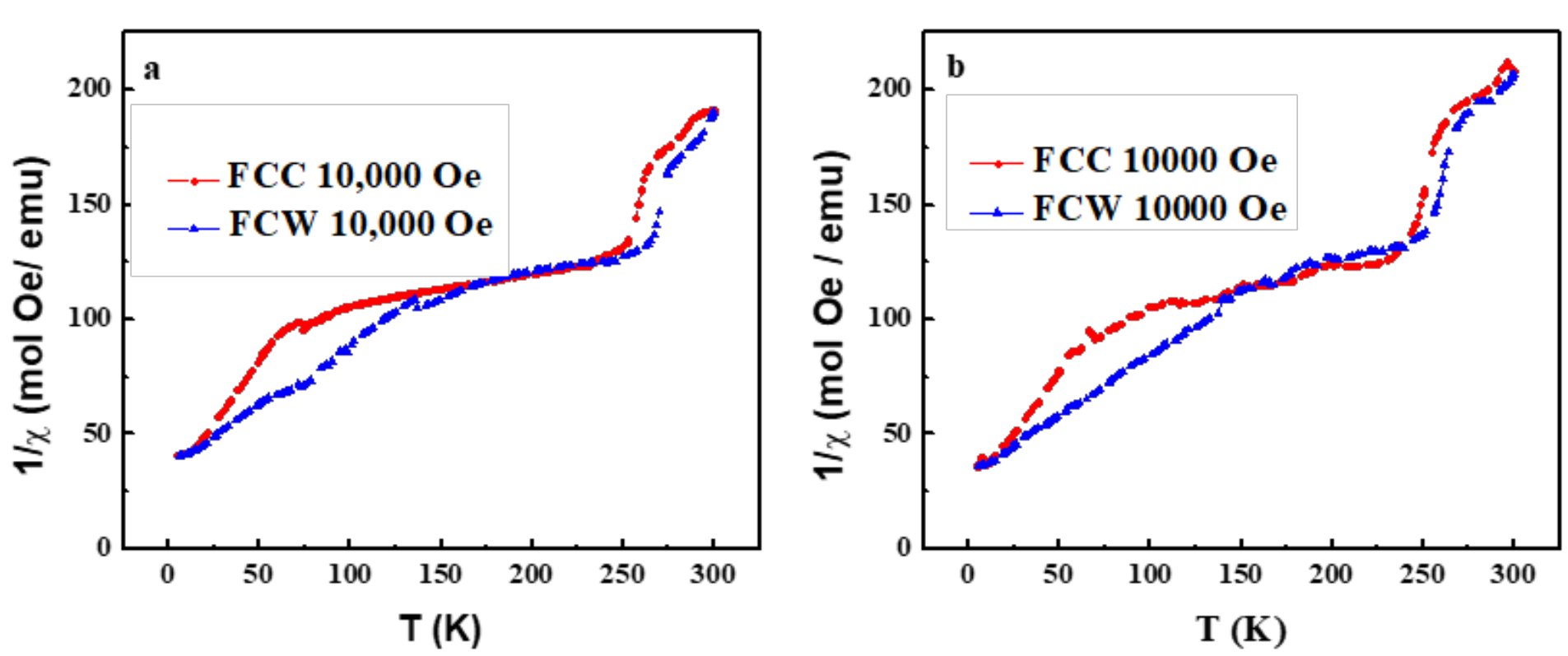

Fig. 4.7: Inverse magnetic susceptibility for field cooled cooling (FCC) and field cooled warming (FCW) at 10,000 Oe for a) x = 0 and b) x = 0.1.

The decrease in spin anisotropy plays a major role in the increased degree of spin canting and thus as a result leads to the easy orientation of canted spins which matches with the spin orientation of the rare-earth spins. A rapid increase in the moment at temperatures below 50 K was observed which could be attributed to rare-earth ordering and a bifurcation in FCC and FCW for x = 0 and 0.1 in the temperature range 25-160 K which further suggests magnetic phase transition.

The isothermal magnetization curve at 5 K, in the magnetic field range of ±5 T, shows a weak hysteresis loop, which was found to increase as the value of x increased (Fig. 4.8). The remanent magnetization and coercive field were found to increase as the value of x increased (Fig. 4.9 and Table 4.3). From the virgin curves (Fig. 4.9a) it was observed that for x = 0, 0.1 and 0.2, at zero magnetic fields was found to show positive value of magnetization, however for x = 0.4 and 0.5, it was found to show the negative value of magnetization, which could be attributed to the presence of ferromagnetic and antiferromagnetic characteristics respectively.

The isothermal magnetization for zero-field cooling (ZFC) and zero field warming (ZFW), performed in $La_{0.1}Sm_{0.4}Pr_{0.5}CrO_3$ showed that in the temperature below 230 K, the loops were found to shift towards a negative field direction, thus showing a negative exchange bias effect. These exchanges bias effect could be attributed to the canted antiferromagnetic spin structure along with rare-earth alignment at temperatures close to its Néel temperature [32]. The exchange bias shift has been plotted (Fig. 4.11) and could be denoted by $H_{EB} = (H_1-H_2)/2$ and the coercive field by $H_C = (H_1+H_2)/2$ (Fig 4.12) [33].

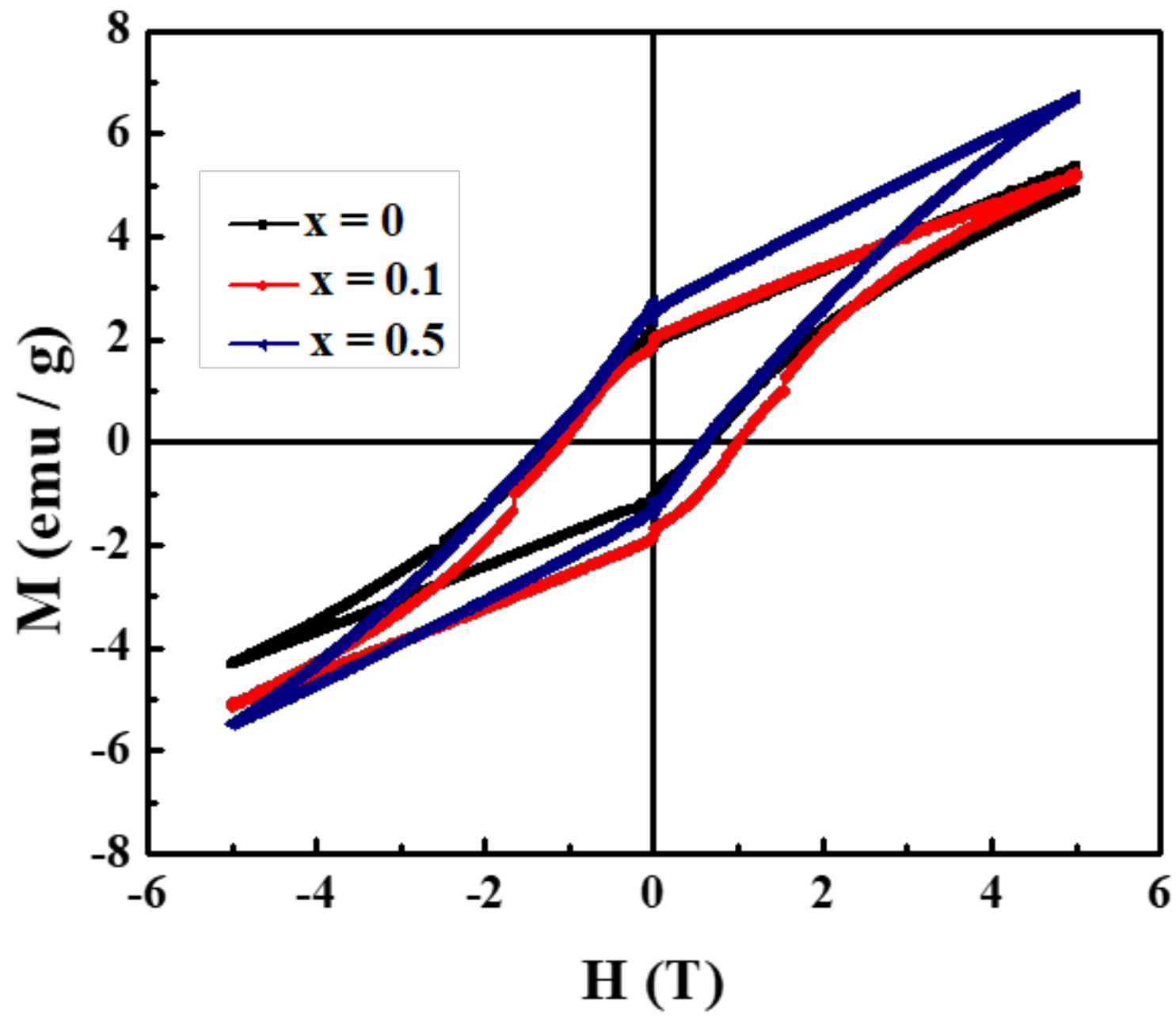

Fig. 4.8: Isothermal magnetization plot at 5 K in $La_{0.5-x}Sm_xPr_{0.5}CrO_3$ for x = 0, 0.1 and 0.5 with magnetic field applied ± 5 T.

Here it was observed that the coercive field and remanent magnetization increased as the value of x increased at 5 K temperature which could be attributed to increased structural distortion and increased canted spin moments as the Sm concentration increased.

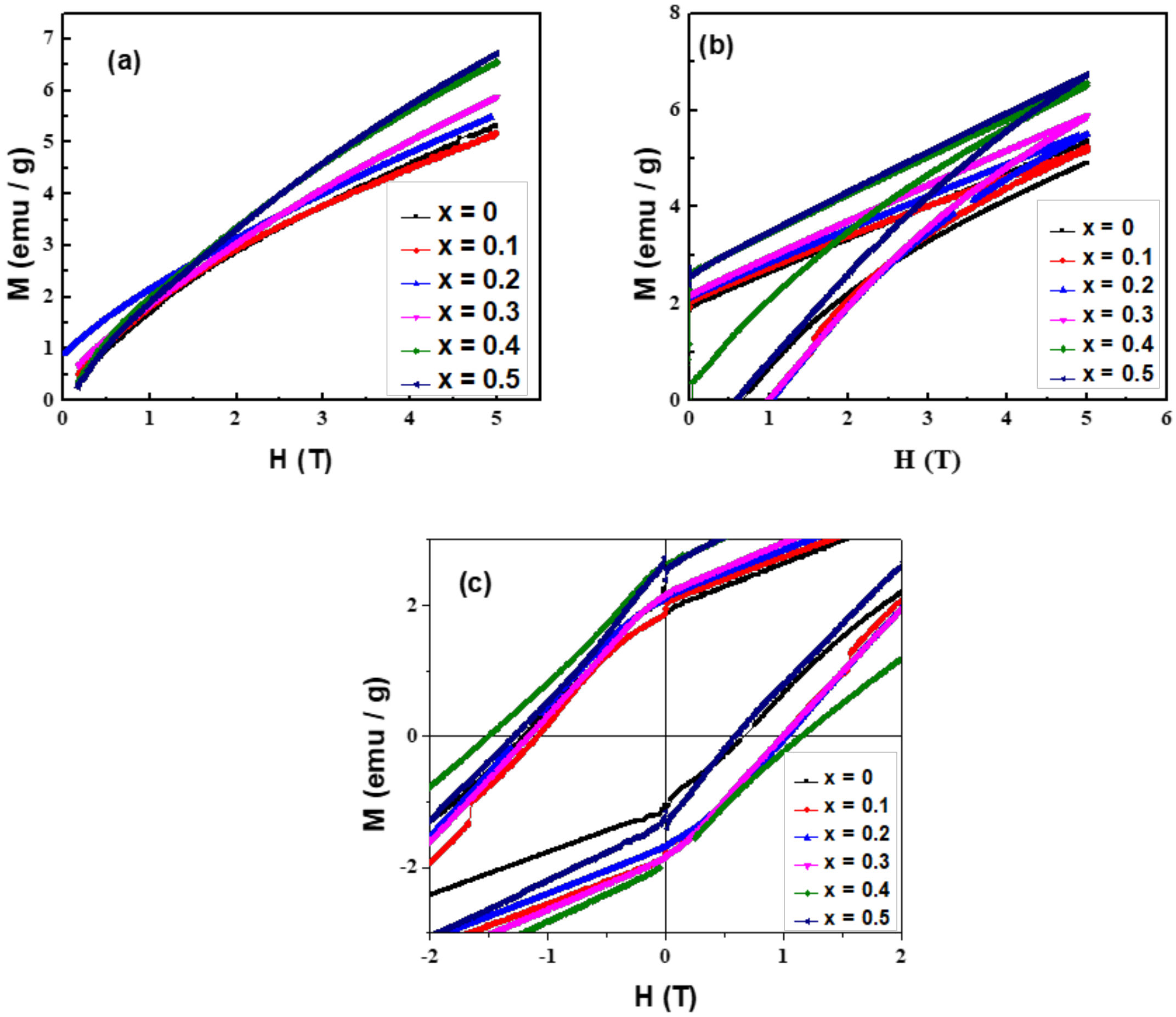

Fig. 4.9: Isothermal magnetization plot of $La_{0.5-x}Sm_xPr_{0.5}CrO_3$ for x = 0, 0.1, 0.2, 0.3, 0.4 and 0.5 where at 5 K for a) Virgin magnetic plot by increasing the magnetic field initially from 0 to 5 T.

Here the virgin curve (Fig. 4.9a) measured at 5 K temperature shows that the highest value of magnetization was achieved at a higher value of x in the 5 T magnetic field range. In the isothermal magnetization plot at 5 K (Fig. 4.9b and c), it was also observed that the coercive field and remanent magnetization value was found to increase with the increase in value of x. These factors could be attributed to increased spin reorientation, as a result of reduced anisotropy manifested by ordered antiferromagnetic alignment of rare earth with respect to transition metal ions as the value of x increases.

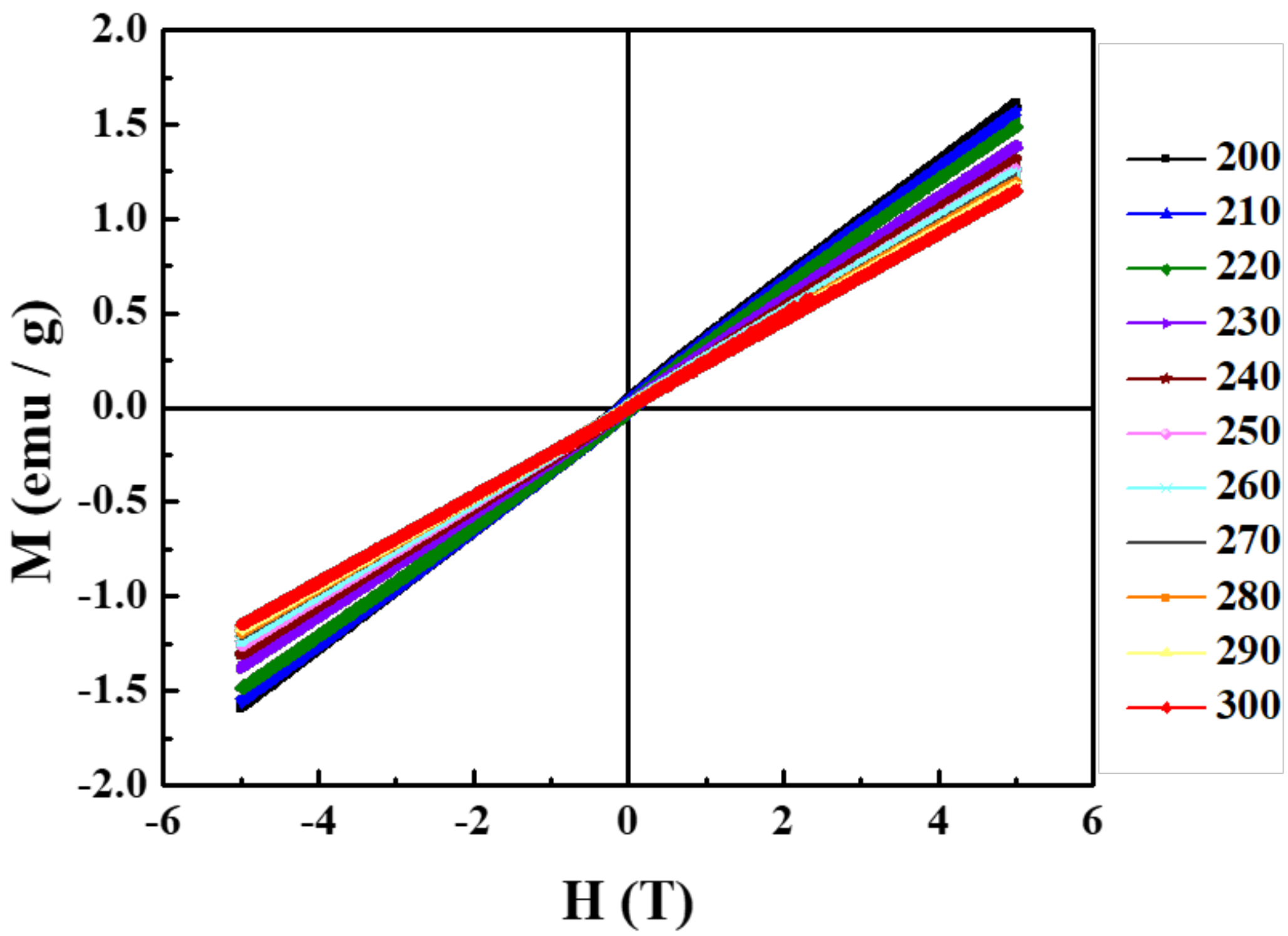

Fig. 4.10: Isothermal magnetization measured for $La_{0.1}Sm_{0.4}Pr_{0.5}CrO_3$ in field range ± 5 T for temperatures 200-300 K.

It was earlier observed that in the composition $La_{0.1}Sm_{0.4}Pr_{0.5}CrO_3$ negative magnetization was observed at temperatures below 208 K in ZFC measurement at 100 Oe field (Fig. 4.4b). Here measuring isothermal magnetization (Fig.4.10) for the composition $La_{0.1}Sm_{0.4}Pr_{0.5}CrO_3$ in the temperature range 200-300 K (at a temperature interval of 10 K) resulted in a non-overlapping magnetization curve. The highest value of magnetization measured at the 5 T field decreased as the temperature was increased which could be attributed to the presence of canted spin moments.

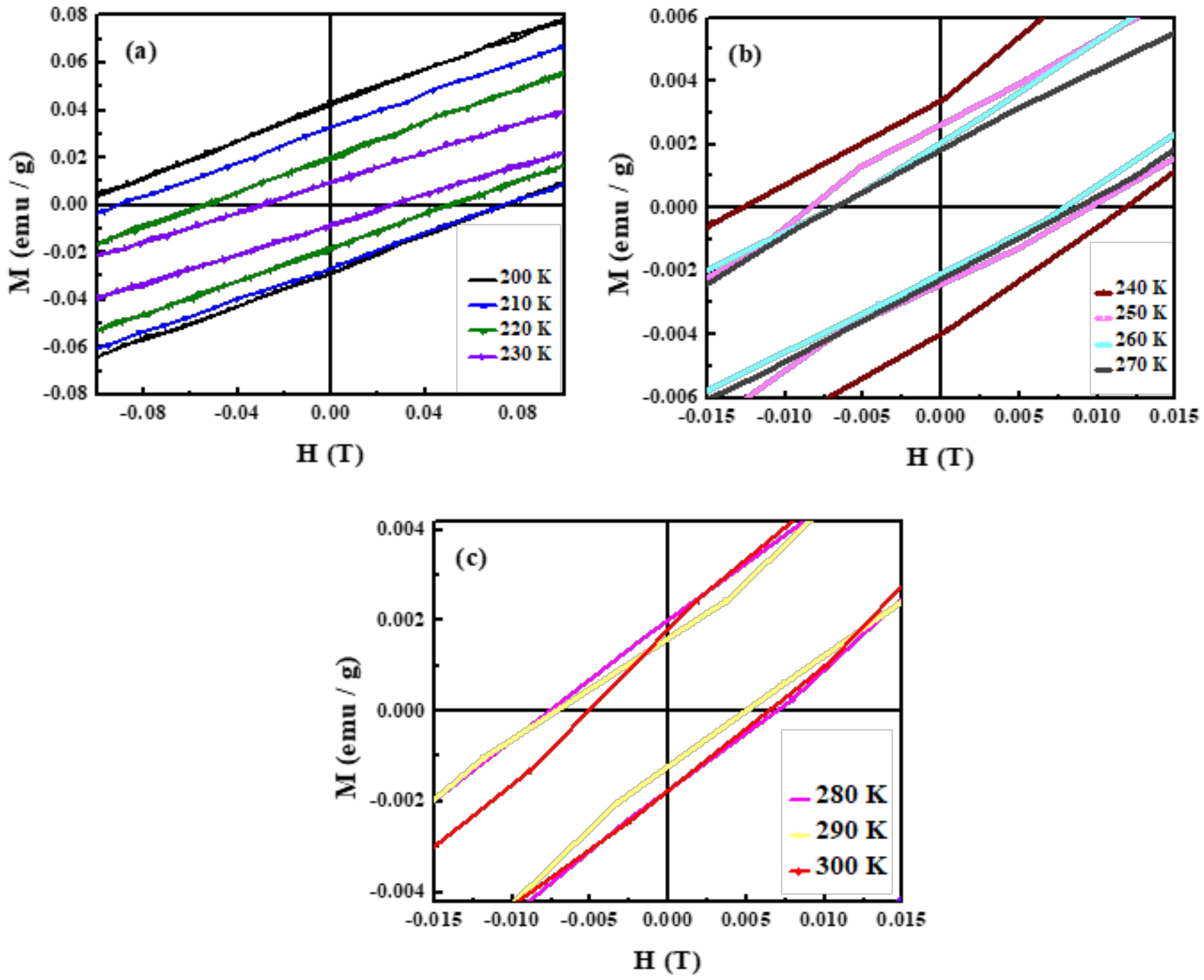

Fig. 4.11: Enlarged view of the Isothermal magnetization measured for $La_{0.1}Sm_{0.4}Pr_{0.5}CrO_3$ in field range ± 5 T for temperatures a) 200-230 K and b) 240-270 K and c) 280-300 K.

These canted spin moments as observed in Fig. 4.10 could be attributed to this negative exchange bias present in $La_{0.1}Sm_{0.4}Pr_{0.5}CrO_{3.}$

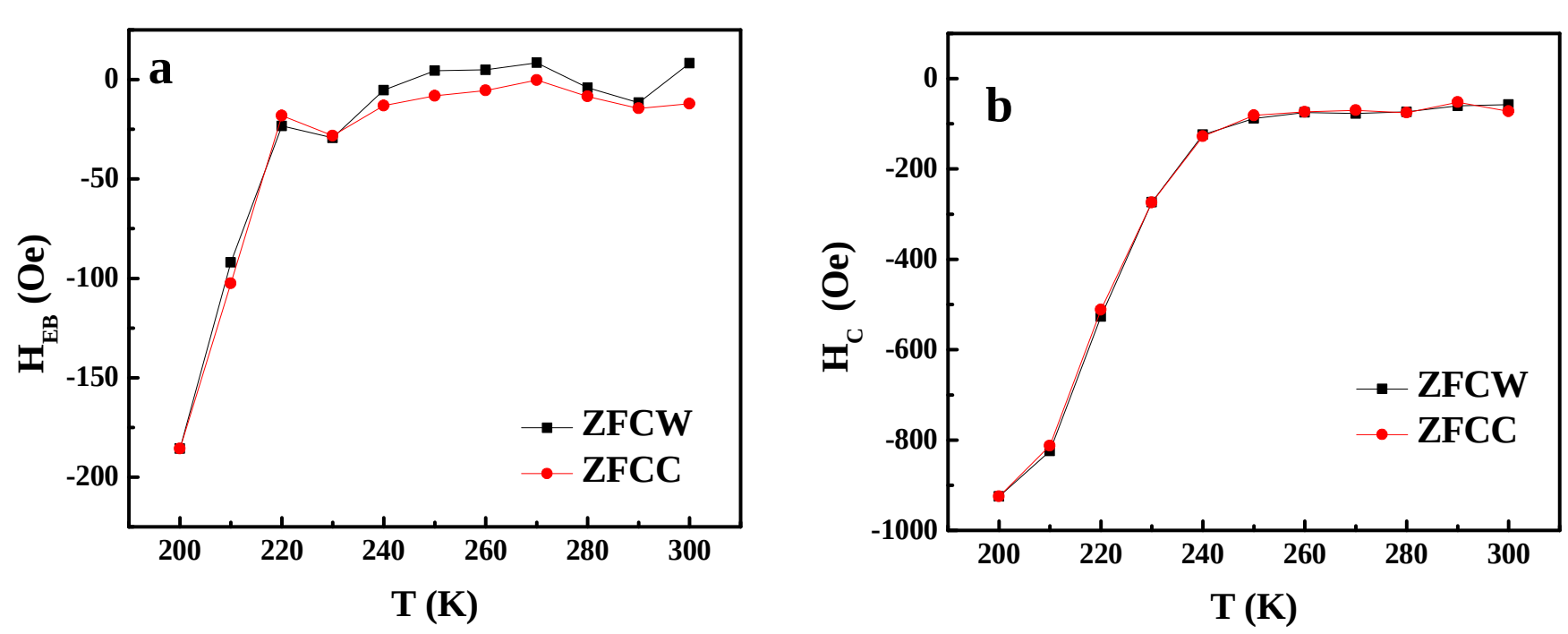

Fig. 4.12: Plot of a) exchange bias in ZFCW and ZFCC isothermal magnetization for Isothermal magnetization measured for $La_{0.1}Sm_{0.4}Pr_{0.5}CrO_3$ in the field range ± 5 T for temperatures 200-300

K. and b) Coercive field (Hc) for ZFCW and ZFCC for $La_{0.1}Sm_{0.4}Pr_{0.5}CrO_3$ in the field range ± 5 T for temperatures 200-300 K.

This degree of anisotropy present in the crystal system depends on the octahedral distortion which leads to a correlation between the Neel temperature and spin reorientation temperature based on unidirectional anisotropy, structural distortion and Dzyaloshinskii Moriya interaction.

Table 4.4: Néel temperature and effective magnetic moment obtained for $La_{0.5-x}Sm_xPr_{0.5}CrO_3$ with x= 0, 0.1, 0.2, 0.3, 0.4 and 0.5.

| S.No | Composition | $T_N$ (K) | $\theta_{CW}$ (K) | $\mu_{observed}$ ($\mu_B$/f.u.) | $\mu_{theoretical}$ ($\mu_B$/f.u.) | $M_R$ (emu/mol Oe) |
|---|---|---|---|---|---|---|
| 1. | $La_{0.5}Pr_{0.5}CrO_3$ | 246 | 60 | 2.857 | 4.624 | 1.915 |
| 2. | $La_{0.4}Sm_{0.1}Pr_{0.5}CrO_3$ | 252 | -27 | 2.957 | 4.630 | 2.176 |
| 3. | $La_{0.2}Sm_{0.3}Pr_{0.5}CrO_3$ | 253 | -163 | 3.427 | 4.640 | 2.243 |
| 4. | $La_{0.3}Sm_{0.2}Pr_{0.5}CrO_3$ | 263 | -166 | 3.100 | 4.647 | 2.248 |
| 5. | $La_{0.4}Sm_{0.1}Pr_{0.5}CrO_3$ | 267 | -167 | 3.087 | 4.655 | 2.577 |
| 6. | $Sm_{0.5}Pr_{0.5}CrO_3$ | 275 | -171 | 4.864 | 4.662 | 2.744 |

Tuning the rare-earth composition based on its size and degree of anisotropy could be optimized to get the appropriate magnetic spin structure and the transition temperatures

## 4.4 Conclusions

New phase of $La_{0.5-x}Sm_xPr_{0.5}CrO_3$ (x = 0, 0.1, 0.2, 0.3,0.4 and 0.5) was analyzed and its canted spin structure containing the rare earth in perovskite structure was investigated. The Neel temperature was found to decrease with an increase in x, due to the increased structural distortion. Increased spin reorientation, as a result of reduced anisotropy manifests ordered antiferromagnetic alignment of rare earth with respect to transition metal ions resulting in negative magnetization for x = 0.4 and 0.5 values. The FCC and FCW were found to be non-overlapping for x = 0 and 0.1, which suggests the spin reorientation transition and magnetic transition as well. The ZFCC and ZFCW show a negative exchange bias for $La_{0.1}Sm_{0.4}Pr_{0.5}CrO_3$, at temperatures below 230 K, at temperatures closer to the Néel temperature and closer to the compensation temperature. With the decrease in temperature, there is a decrease in the degree of antiferromagnetic ordering competing with the ferromagnetic moment, which ultimately leads to a decrease in magnetic moment.

We tried obtaining the value of internal magnetization ($M_R$) and internal field ($H_I$) generated from canted spins of Cr ions. As we know that as the octahedral distortion increases it results in greater degree of canted spins moments from $Cr^{3+}$ ions, which generates greater values of net magnetic moments. We tried comparing the $H_I$ and $M_R$ values of $La_{0.3}Sm_{0.2}Pr_{0.5}CrO_3$ and $La_{0.1}Sm_{0.4}Pr_{0.5}CrO_3$, where the $H_I$ and $M_R$ values increased as the values of x increases. The $M_R$ and $H_I$ obtained in case of $La_{0.1}Sm_{0.4}Pr_{0.5}CrO_3$ was obtained from extrapolating the isothermal magnetization curve at various temperatures and finding their common cross-sectional points, which was found to be 720 emu/mol and 12574 Oe. The $M_R$ and $H_I$ for case of $La_{0.3}Sm_{0.2}Pr_{0.5}CrO_3$ was found to decrease and was obtained to be 474 emu/mol and 2278 Oe.

# Chapter V

# Structural and magnetic properties of $R_{0.5}Sr_{0.5}Fe_{0.5}Mn_{0.5}O_3$ (R = Gd, Nd or Pr) perovskites

# Chapter V

# Structural and magnetic properties of $R_{0.5}Sr_{0.5}Fe_{0.5}Mn_{0.5}O_3$ (R = Gd, Nd or Pr) perovskites

## Abstract

Perovskite compounds of the formula $R_{0.5}Sr_{0.5}Fe_{0.5}Mn_{0.5}O_3$ (R = Gd, Nd and Pr) were synthesized successfully by the high-temperature solid-state route. The Rietveld refinement analysis of the powder X-ray diffraction data shows that $R_{0.5}Sr_{0.5}Fe_{0.5}Mn_{0.5}O_3$ (R = Gd, Nd and Pr) were successfully refined in the orthorhombic space group *Pnma*. On substitution of rare-earths with size increasing from Gd to Nd to Pr shows that the octahedral distortion decreases. The octahedral distortion arises primarily due to octahedral tilts and Jahn-Teller distortion arising from the $Mn^{3+}$ and $Fe^{4+}$ ions. $Gd_{0.5}Sr_{0.5}Fe_{0.5}Mn_{0.5}O_3$ shows canted antiferromagnetic due to $Mn^{3+}$ spins, observed at temperatures close to 49 K and low temperatures below 3 K, $Gd^{3+}$ ordering takes place. The compositions $Nd_{0.5}Sr_{0.5}Fe_{0.5}Mn_{0.5}O_3$ and $Pr_{0.5}Sr_{0.5}Fe_{0.5}Mn_{0.5}O_3$ show spin-glass like nature at temperatures 47 K and 58 K respectively. The isothermal magnetization further confirmed the existence of canted antiferromagnetic in Gd and spin-glass like behaviour in $Nd_{0.5}Sr_{0.5}Fe_{0.5}Mn_{0.5}O_3$ and $Pr_{0.5}Sr_{0.5}Fe_{0.5}Mn_{0.5}O_3$. From the X-ray photoelectron spectroscopy measurement (XPS) a significant amount of Fe and Mn ions were found in +3 and +4 oxidation states.

## 5.1 Introduction

Perovskite maganites and ferrites hold a matter of importance in the condensed matter as they show the great correlation between magnetic and electronic properties for their possible applications in the field of multiferroics, colossal magnetoresistance and complex magnetic phases [1-4]. The oxidation states of the B-site ions and structural distortion play a major role in their nature of magnetic interactions: double exchange interactions and super-exchange interactions [5-6].

The structural distortions are governed to some extent by the tolerance factor depending on the A and B-site ionic radii and Jahn-Teller distortion. The compounds which have Jahn-Teller active ions at B-site such as $Mn^{3+}$ and $Fe^{4+}$ induces a greater degree of anisotropy to the system [6-7].

On substitution at both A and B-site positions depending on the arrangement of A and B-site atoms based on charge and orbital ordering plays a major role in short- or long-range magnetic exchange interaction and their transitions. For example, $R_{0.5}Ca_{0.5}MnO_3$ (R = La/Pr/Nd/Sm) has an orthorhombic Pnma structure while $R_{0.5}Sr_{0.5}MnO_3$ (R = Gd/Nd/Pr) shows the transition from *Pnma* to *Imma* to *I4/mcm* as the size of the rare earth increases from Gd to Nd to Pr [4, 8]. As the size difference between the A-site substituted atoms decreases, that is as the rare-earth size decreases from La to Pr to Gd in $R_{0.5}Sr_{0.5}MnO_3$ (R = La, Nd and Pr) it results in the short-range ordering of the A-site cations which possibly further enhances the Jahn-Teller distortion [4, 8].

The effect of substitution of Mn by Fe in the composition $RFe_{1-x}Mn_xO_3$ (R= Gd, Pr and Dy) was reported to increase the magnetocrystalline anisotropy and super exchange interaction resulted in a decrease in Curie temperature. The orbital ordering plays a major role in magnetic behaviour based on the double exchange interaction between $Fe^{3+}$ ($t_2g^3$ $eg^2$, high spin) and $Mn^{3+}$ ($t_2g^3$ $eg^1$, high spin) [8-13]. For example, the composition $LaFe_{0.5}Mn_{0.5}O_3$ which has $Mn^{3+}$ and $Fe^{3+}$ ions show a weak ferromagnetic transition at a temperature close to 60 K, arising from the disorder-induced short-range ferromagnetic ordering [14]. Also, it has been reported that $YFe_{1-x}Mn_xO_3$ antiferromagnetic spin-glass behaviour in the temperature range 150-350 K, due to canted moments arising from Fe – O – Fe and Mn – O – Mn and Fe – O – Mn interactions [15-18]. Similarly, manganites of the formula $RMnO_3$ (R = Ho, Er, Tm, Yb or Lu) shows the antiferromagnetic transition in the temperature range 70-130 K and the ferrites series of the series $RFeO_3$ (R = Lu, Er or Tb) was reported to weak ferromagnetic transition in the temperature range 120 – 140 K [19-20]. Here we intend to study the variation of A-site substitution in $R_{0.5}Sr_{0.5}Fe_{0.5}Mn_{0.5}O_3$ (R = Gd, Nd or Pr) on their structural and magnetic properties [21-22].

## 5.2 Experimental

### 5.2.1 Materials and synthesis

Oxides of the type $R_{0.5}Sr_{0.5}Fe_{0.5}Mn_{0.5}O_3$ (R = Gd, Nd and Pr) were synthesized by the conventional solid-state route. $Gd_2O_3$ (Sigma Aldrich), $Nd_2O_3$ (Sigma Aldrich), $Pr_6O_{11}$ (Alfa Aesar), $SrCO_3$ (Merck) and $Fe_2O_3$ (Sigma Alrich) were preheated for 12 hours at 900˚C. All the oxides taken in stoichiometric amounts were ground thoroughly in an agate mortar and pestle. The above mixture was heated in an alumina crucible at 900˚C for 12 hours and was ground after cooling and then subsequently heated at 1200˚C for 12 hours. The subsequent powder (0.2g) was mixed with 20 µL of 5% PVA solution and was pelletized at 10 tons pressure and was further sintered at 1200˚C for 12 hours.

### 5.2.2 Characterization

The phases were investigated by powder X-ray diffraction using Bruker D8 Advance diffractometer (Cu K$\alpha$ radiation) which was equipped within the filter. The K$\alpha_2$ radiation was stripped off using the software. The background was modelled using Chebyshev polynomial and the profile was stimulated with pseudo-Voigt function. The scanning factor, cell parameters, background fitting and isotropic thermal parameters for all the atoms were refined.

### 5.2.3 Magnetic studies

Zero field cooled (ZFC) and field cooled (FC) temperature dependent magnetization studies were performed using Magnetic Property Measurement System (MPMS – SQUID), Quantum Design, USA, in the temperature range 2 K to 390 K, at a magnetic field of 1000 Oe. The isothermal field dependent magnetization studies for all the three compositions were performed at varying temperatures in the magnetic field range of +7 T to – 7 T.

### 5.2.4 X-ray Photoelectron Spectroscopy (XPS)

XPS (X-ray Photoelectron Spectroscopy) studies were performed using Shimadzu ESCA-3400 spectrophotometer (Kyoto, Japan) on powdered samples. Here Mg-K$\alpha$ (1253.6 eV) was taken as the source being operated at 10 kV and 10 mA.

## 5.3 Results and Discussion

### 5.3.1 X-ray diffraction studies

The X-ray patterns of the calcined compounds reveal orthorhombic unit cell parameters a:b:c = $\sqrt{2}$ap:$\sqrt{2}$ap: 2ap, where ap = 3.9 Å (the parameter of ideal cubic perovskite unit cell). The refinement gives satisfactorily agreement factors Rwp (weighted residual factor), Rp (Bragg factor) and $\chi^2$. The refined occupancies, cell parameters, thermal parameters and positional parameters are provided in Table 5.1a, 5.1b and 5.1c. Fig. 5.2 shows the Rietveld refinement from powder X-ray studies. The initial structure model was taken from reference [19, 31]. Earlier reports showed their parent compounds with no B-site substitution in $R_{0.5}Sr_{0.5}MnO_3$ (R = Gd, Nd and Pr) were also reported in the orthorhombic *Pnma* spacegroup. The cell parameters and unit cell volume are plotted in Fig. 5.3a. It was also observed that the peak splitting was found to be more in Pr, followed by Nd and Gd. For $R_{0.5}Sr_{0.5}Fe_{0.5}Mn_{0.5}O_3$ (R = Gd, Nd and Pr) the difference between the *b* and lattice parameters were found to be *0.006(1)* Å, *0.026(1)* Å and *0.041(2)* Å respectively. The degree of peak splitting of (200) and (020) peaks further suggests this anomaly.

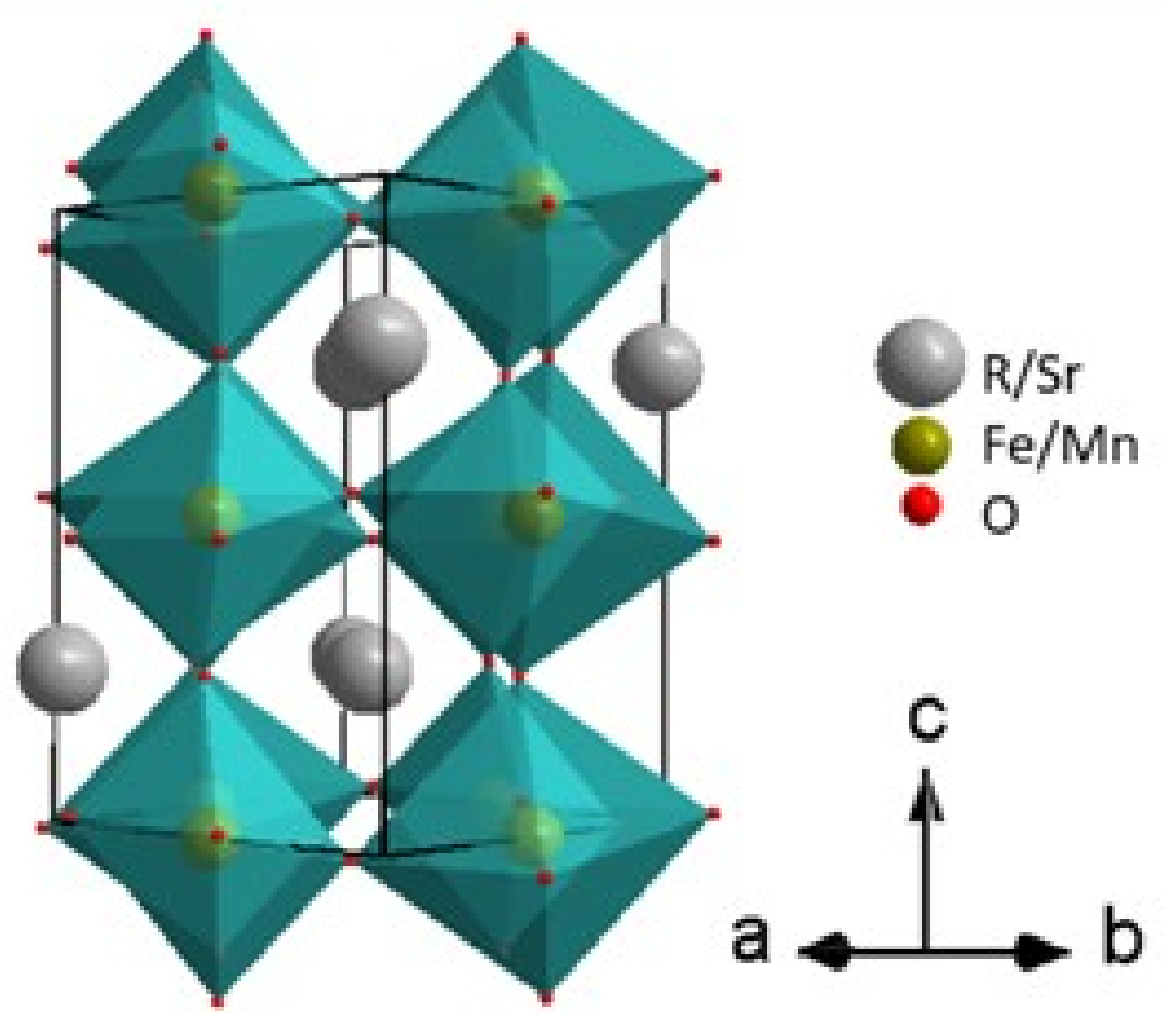

Fig 5.1 (a) Polyhedral representation of the structure of $R_{0.5}Sr_{0.5}Fe_{0.5}Mn_{0.5}O_3$ (R: Gd, Nd and Pr).

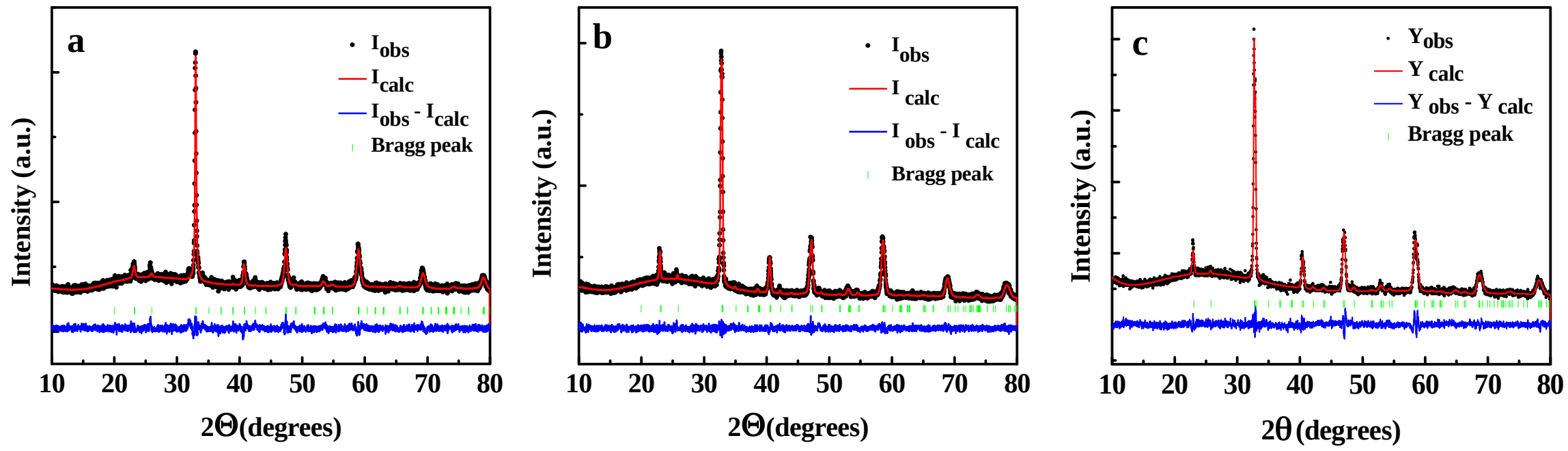

Fig. 5.2: The room temperature powder X-ray diffraction data (black) of (a) $Gd_{0.5}Sr_{0.5}Fe_{0.5}Mn_{0.5}O_3$ (b) $Nd_{0.5}Sr_{0.5}Fe_{0.5}Mn_{0.5}O_3$ and (c) $Pr_{0.5}Sr_{0.5}Fe_{0.5}Mn_{0.5}O_3$, refined in orthorhombic *Pbnm* space group. The calculated profile from Rietveld refinement, the fitted Bragg peak positions and the residual are shown in red lines, green (short vertical bars) and blue lines in the panel.

Here, the *b* lattice parameter showed decreased value than expected for $Gd_{0.5}Sr_{0.5}Fe_{0.5}Mn_{0.5}O_3$ however the lattice volume was found to be higher in comparison to $Nd_{0.5}Sr_{0.5}Fe_{0.5}Mn_{0.5}O_3$ and $Pr_{0.5}Sr_{0.5}Fe_{0.5}Mn_{0.5}O_3$, suggesting the presence of Jahn-Teller distortion in $Gd_{0.5}Sr_{0.5}Fe_{0.5}Mn_{0.5}O_3$ (Fig. 5.3a).

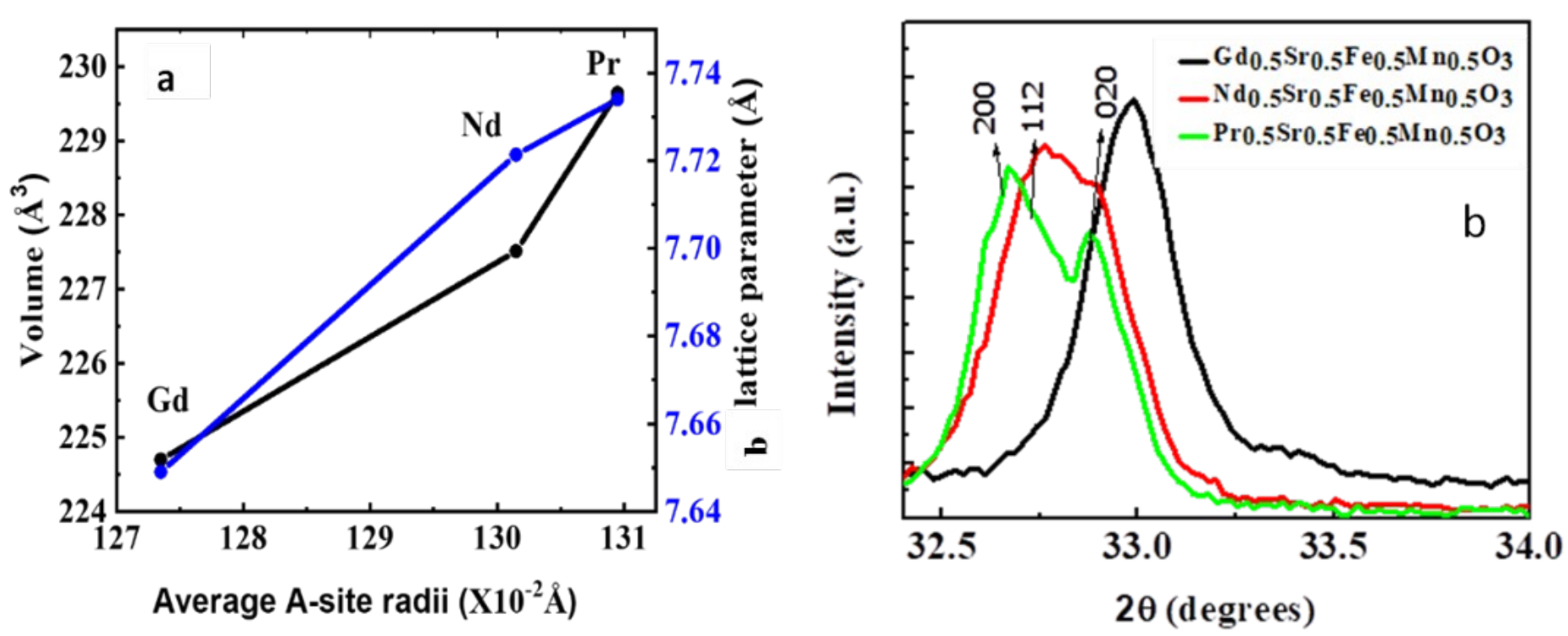

Fig 5.3: (a) Variation of lattice parameter *c* and cell volume. (b) Peak splitting and shift of the (200) and (020) PXRD peaks, with variation in A-site cation.

It was observed that as the size of the rare-earth increased, the Jahn-Teller distortion decreased, as inferred from the *b* lattice parameter and the difference in the bond lengths of Fe/Mn – $O_1$ (axial) and Fe/Mn – $O_2$ (equatorial). The difference in the Mn/Fe – $O_1$ and Mn/Fe – $O_2$ bond lengths (*0.250(6)* Å) was found to be maximum in the case of $Gd_{0.5}Sr_{0.5}Fe_{0.5}Mn_{0.5}O_3$ as compared to $Nd_{0.5}Sr_{0.5}Fe_{0.5}Mn_{0.5}O_3$ and $Pr_{0.5}Sr_{0.5}Fe_{0.5}Mn_{0.5}O_3$ where this difference was *0.044(3)* Å and *0.076(6)* Å respectively. Due to the size mismatch between the rare-earth and $Sr^{2+}$ ion on the A-site, a short-range ordering of the charges takes place. This could be postulated that the higher octahedral distortion, primarily from Jahn-Teller distortion and short-range ordering of charges, could result in a greater degree of charge ordering in $Gd_{0.5}Sr_{0.5}Fe_{0.5}Mn_{0.5}O_3$ .

Table 5.1a: Rietveld refinement of Gd0.5Sr0.5Fe0.5Mn0.5O3, obtained from powder X-ray diffraction data. Space group = Pbnm (# 62), a = 5.423(1) Å, b= 5.417(1) Å, c = 7.649(1) Å.

| **Atom** | **Site** | **x** | **y** | **z** | **Occ** | **Beq (Å$^2$)** |
|---|---|---|---|---|---|---|
| Gd/Sr | 4c | 0.997(5) | 0.020(1) | 0.25 | 0.5/0.5 | 1.4(1) |
| Fe/Mn | 4b | 0.5 | 0 | 0 | 0.5/0.5 | 1.5(1) |
| O1 | 4c | 0.99(3) | 0.467(7) | 0.25 | 1 | 3.4(6) |
| O2 | 8d | 0.725(6) | 0.209(5) | 0.964(4) | 1 | 3.7(4) |

*$R_p$ (%) = 3.47, $R_{wp}$ (%) = 4.38, GOF = 1.21, Vol. = 224.70(1) Å$^3$*

Table 5.1b: Rietveld refinement of $Nd_{0.5}Sr_{0.5}Fe_{0.5}Mn_{0.5}O_3$, obtained from powder X-ray diffraction. *Space group = Pbnm (# 62), a = 5.4412(2) Å, b = 5.4151(1) Å, c = 7.7214(1) Å*

| Atom | Site | x | y | z | Occ | Beq (Å$^2$) |
|---|---|---|---|---|---|---|
| Nd/Sr | 4c | 0.992(1) | 0.0210(9) | 0.25 | 0.5/0.5 | 1.6(1) |
| Fe/Mn | 4b | 0.5 | 0 | 0 | 0.5/0.5 | 1.7(1) |
| O1 | 4c | 0.961(6) | 0.496(5) | 0.25 | 1 | 3.0(4) |
| O2 | 8d | 0.783(7) | 0.205(5) | 0.973(3) | 1 | 4.2(4) |

*$R_p$ (%) = 3.59, $R_{wp}$ (%) = 4.51, GOF = 1.08, Vol. = 227.51(1) Å$^3$*

Table 5.1c: Rietveld refinement of $Pr_{0.5}Sr_{0.5}Fe_{0.5}Mn_{0.5}O_3$, obtained from powder X-ray diffraction. *Space group = Pbnm (# 62), a = 5.4695(1) Å, b=5.4287(1) Å, c = 7.7340(1) Å*

| Atom | Site | x | y | z | Occ | Beq (Å$^2$) |
|---|---|---|---|---|---|---|
| Pr/Sr | 4c | 0.994(3) | 0.010(9) | 0.25 | 0.5/0.5 | 1.5(1) |
| Fe/Mn | 4b | 0.5 | 0 | 0 | 0.5/0.5 | 1.6(1) |
| O1 | 4c | 0.98(1) | 0.521(9) | 0.25 | 1 | 3.2(4) |
| O2 | 8d | 0.75(2) | 0.23(1) | 0.99(1) | 1 | 4.5(4) |

*$R_p$ (%) = 4.35, $R_{wp}$ (%) = 5.26, GOF = 1.21, Vol. = 229.64(1) Å$^3$*

Table 5.2: The bond angle and bond distances of the system $R_{0.5}Sr_{0.5}Fe_{0.5}Mn_{0.5}O_3$ (R:Gd, Nd, Pr).

| Composition | a (Å) | b (Å) | c (Å) | Fe/Mn - $O_1$<br>Fe/Mn - $O_2$ | R/Sr - $O_1$<br>R/Sr - $O_2$ | B - $O_1$ - B<br>B - $O_2$ - B | $O_2$ - B - $O_1$<br>$O_1$ - A - $O_2$ |
|---|---|---|---|---|---|---|---|
| $Gd_{0.5}Sr_{0.5}Fe_{0.5}Mn_{0.5}O_3$ | 5.423<br>(1) | 5.417<br>(1) | 7.649<br>(1) | 1.928(4)<br>2.187(3) | 2.411(1)<br>2.828(2) | 168.698(1)<br>162.964(2) | 92.34(2)<br>68.54(1) |
| $Nd_{0.5}Sr_{0.5}Fe_{0.5}Mn_{0.5}O_3$ | 5.4412<br>(1) | 5.4151<br>(1) | 7.7214<br>(1) | 1.942(1)<br>1.997(2) | 2.578(1)<br>2.619(2) | 167.388(1)<br>158.463(1) | 91.73(2)<br>65.93(4) |
| $Pr_{0.5}Sr_{0.5}Fe_{0.5}Mn_{0.5}O_3$ | 5.4695<br>(6) | 5.4287<br>(1) | 7.7340<br>(1) | 1.940(1)<br>2.006(1) | 2.593(2)<br>2.693(1) | 170.657(1)<br>173.503(1) | 87.73(4)<br>59.62(5) |

### 5.3.2 X-ray Photoelectron Spectroscopy studies

XPS studies on Gd, Nd and Pr samples were performed in the energy range up to 1200 eV. The survey spectrum showed peaks of Gd-3d, Fe-2p, Nd-3d, Pr-3d and O-1s of XPS peaks along with observation of broad Auger peaks. On substitution of rare-earth (Gd, Nd and Pr) in the A-site results in varying oxidation states of the Mn and Fe ions. The Fe-$2p_{3/2}$ was deconvoluted into 3 peaks which are as follows, $Fe^{4+}$ (712.6 eV), $Fe^{3+}$ (710.2 eV) and $Fe^{2+}$ (708.5 eV) [23-24]. The Mn-$2p_{3/2}$ could also be deconvoluted into three peaks as $Mn^{4+}$ (643.2 eV), $Mn^{3+}$ (641.3 eV) and $Mn^{2+}$ (640 eV) [25-27] as given the Table 5.3. Furthermore, O-1s could also be deconvoluted into lattice oxygen and O-non stoichiometry caused to balance the lattice charge as peaks at binding energies 530 and *528* eV respectively [25-28].

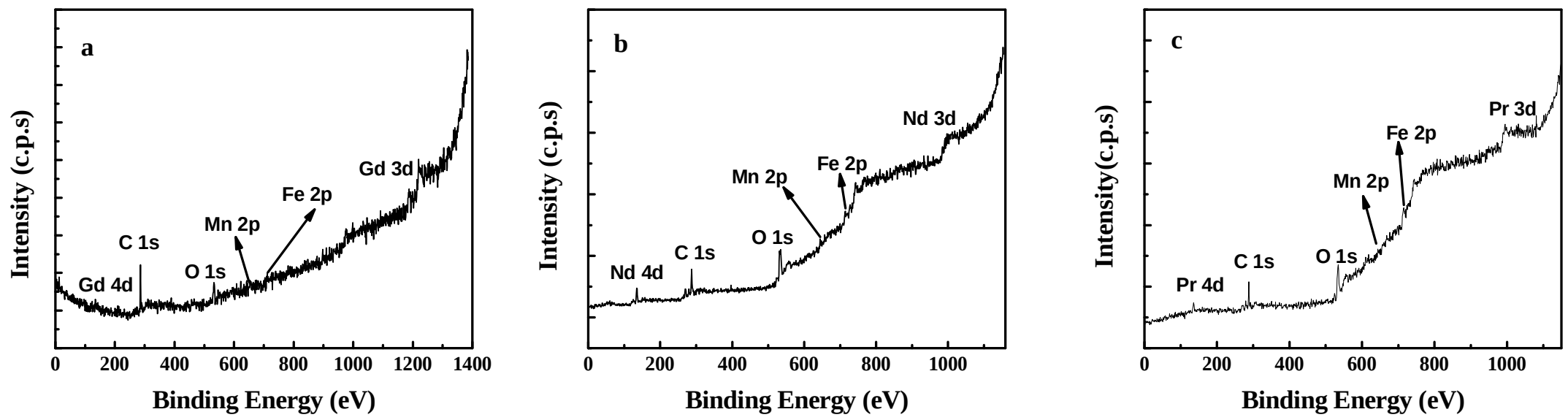

Fig. 5.4: Survey scan of a) $Gd_{0.5}Sr_{0.5}Fe_{0.5}Mn_{0.5}O_3$ b) $Nd_{0.5}Sr_{0.5}Fe_{0.5}Mn_{0.5}O_3$ and c) $Pr_{0.5}Sr_{0.5}Fe_{0.5}Mn_{0.5}O_3$

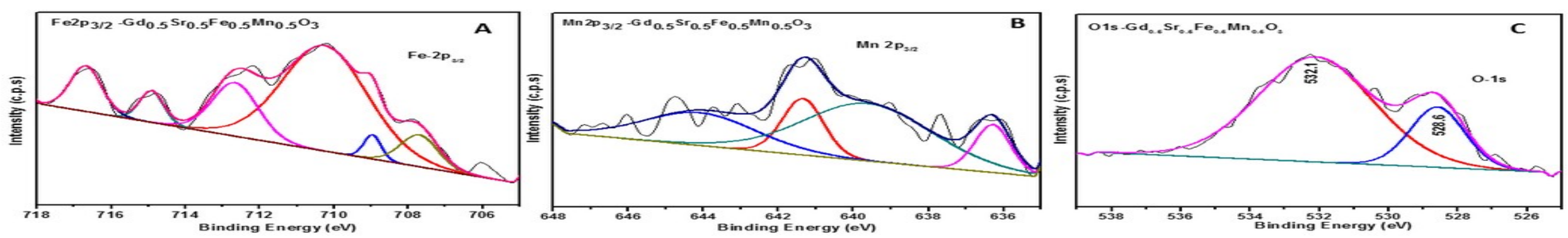

Fig 5.5: High resolution X-Ray Photoelectron Spectra (XPS) for A) Fe($2p_{3/2}$) B) Mn($2p_{3/2}$) and C) O(1s) ions in $Gd_{0.5}Sr_{0.5}Fe_{0.5}Mn_{0.5}O_3$.

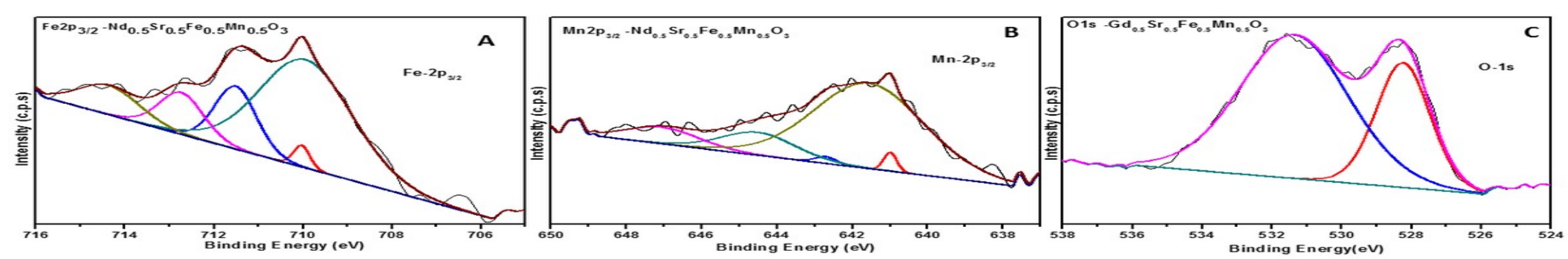

Fig 5.6: High resolution X-Ray Photoelectron Spectra (XPS) for A) Fe($2p_{3/2}$) B) Mn($2p_{3/2}$) and C) O(1s) ions in $Nd_{0.5}Sr_{0.5}Fe_{0.5}Mn_{0.5}O_3$.

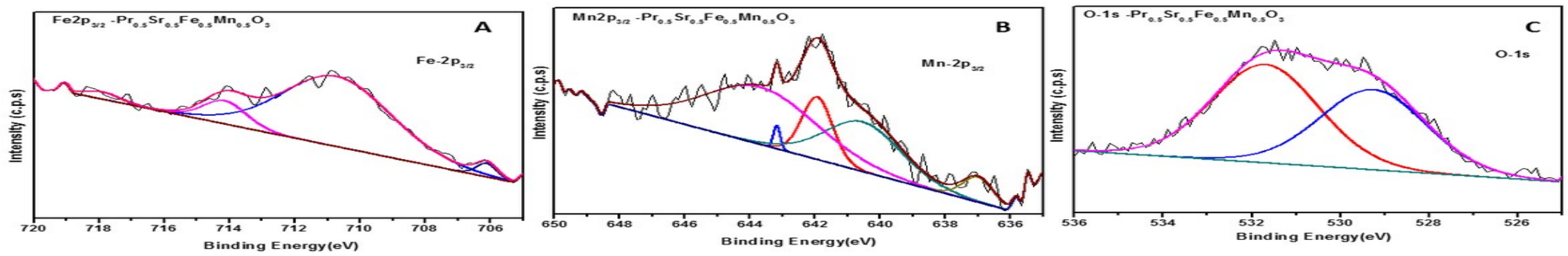

Fig 5.7: High resolution X-Ray Photoelectron Spectra (XPS) for A) Fe($2p_{3/2}$) B) Mn($2p_{3/2}$) and C) O(1s) ions in $Pr_{0.5}Sr_{0.5}Fe_{0.5}Mn_{0.5}O_3$

Table 5.3: XPS analysis showing percentage ratios of Mn and Fe ions in a)$Gd_{0.5}Sr_{0.5}Fe_{0.5}Mn_{0.5}O_3$, b) $Nd_{0.5}Sr_{0.5}Fe_{0.5}Mn_{0.5}O_3$ and c) $Pr_{0.5}Sr_{0.5}Fe_{0.5}Mn_{0.5}O_3$.

| **Sr. No.** | **Samples** | **$Fe^{4+}$ (Area %)** | **$Fe^{3+}$ (Area %)** | **$Fe^{2+}$ (Area %)** | **$Mn^{4+}$ (Area %)** | **$Mn^{3+}$ (Area %)** | **$Mn^{2+}$ (Area %)** |
|---|---|---|---|---|---|---|---|
| 1 | $Gd_{0.5}Sr_{0.5}Fe_{0.5}Mn_{0.5}O_3$ | 16.12 | 58 | 2.7 | 27 | 15 | 48 |
| 2 | $Nd_{0.5}Sr_{0.5}Fe_{0.5}Mn_{0.5}O_3$ | 15.8 | 20.11 | 65 | 13 | 74 | 11.9 |
| 3 | $Pr_{0.5}Sr_{0.5}Fe_{0.5}Mn_{0.5}O_3$ | 10.9 | 83 | 6.1 | 69 | 14 | 27 |

### 5.3.3 Magnetization studies

Zero field cooled (ZFC) and field cooled (FC) DC magnetic susceptibility studies performed on Gd, Nd and Pr samples, performed with an external magnetic field of *1000* Oe in the temperature range 2 – 390 K (Fig. 5.8). For $Gd_{0.5}Sr_{0.5}Fe_{0.5}Mn_{0.5}O_3$ it was observed that ZFC and FC magnetization data was found to coincide with each other. The ZFC and FC for $Gd_{0.5}Sr_{0.5}Fe_{0.5}Mn_{0.5}O_3$ (Fig.5.8a) were found to overlap with each other almost in the entire temperature range and shows no tendency of prominent peak or saturation in the entire temperature range. Such behaviour suggests the major contribution of paramagnetic behaviour due to the paramagnetic contribution originating from the $Gd^{3+}$ ions. However certain minor deviation between ZFC and FC at temperatures close to *49* K could be due to the canted antiferromagnetic ordering of B-site ions containing $d^4$ configuration which arrange antiferromagnetically around this temperature range. From the dχ/dT vs. T plot (Fig. 5.9a), a sharp transition at temperatures below 3 K could be attributed to the incommensurate antiferromagnetic ordering of the $Gd^{3+}$ moments [32] and could be associated with the net alignment of the spins of Gd antiparallel to the net moment generated from the canted alignment of the B-site spins. Furthermore, the inverse magnetic susceptibility plot shows that in the high-temperature region linear portion of the data could be successfully fit (green line) to the Curie-Weiss law.

$$\chi = C / T\text{-}\theta \qquad \text{(Equation 5.1, Curie-Weiss law)}$$

Here C is the Curie constant and θ is the Weiss temperature. Here the Weiss temperature was found to be very close to origin which further indicates the dominance of paramagnetic behaviour. However, it was also observed that the deviation from linearity was observed at temperatures below 49 K, with Weiss temperature observed at low-temperature θ = -3 K, (Equation 5.1) suggesting antiferromagnetic behaviour.

From the inverse magnetic susceptibility plot of $Gd_{0.5}Sr_{0.5}Fe_{0.5}Mn_{0.5}O_3$ the effective magnetic moment ($\mu_{observed}$) was found to be 7.22 $\mu_B$/f.u. which was found to coincide with the theoretically calculated magnetic moment (calculated) to be 7.35 $\mu_B$/f.u. Here the various spin moments of the transition metal ions were taken into consideration for the effective moment calculation.

Temperature dependence of DC magnetic susceptibility for $Nd_{0.5}Sr_{0.5}Fe_{0.5}Mn_{0.5}O_3$ and $Pr_{0.5}Sr_{0.5}Fe_{0.5}Mn_{0.5}O_3$ (Fig. 5.8 b and c) shows the spin-glass transition temperature to be *47* K and *58* K respectively. On further decreasing the temperature shows that the FC magnetization increases, while the ZFC magnetization reaches maxima at *38* and *43* K respectively for Nd and Pr and on further decreasing temperature results in a downward feature. This bifurcation between ZFC and FC shows a spin-glass like nature for $Nd_{0.5}Sr_{0.5}Fe_{0.5}Mn_{0.5}O_3$ and $Pr_{0.5}Sr_{0.5}Fe_{0.5}Mn_{0.5}O_3$ compositions.

In the inverse magnetic susceptibility plot, the effective magnetic moment calculated for $Nd_{0.5}Sr_{0.5}Fe_{0.5}Mn_{0.5}O_3$ and $Pr_{0.5}Sr_{0.5}Fe_{0.5}Mn_{0.5}O_3$ compositions ($\mu_{observed}$) were found to be *5.02* $\mu_B$/f.u and *5.12* $\mu_B$/f.u respectively. Both these values were observed to correspond to the theoretical magnetic moment calculated by taking into account the XPS concentration of the respective ions.

The isothermal magnetization data performed at 300 K for all three compositions, was found to show paramagnetic behaviour. The "S-shape like behaviour" for $Gd_{0.5}Sr_{0.5}Fe_{0.5}Mn_{0.5}O_3$ at 20 K, which showed a slim hysteresis and at high fields it was found to increase linearly, without showing any indication of saturation up to the magnetic fields as high as 7 T. This small hysteresis could be originating from the canted antiferromagnetically aligned $Mn^{3+}$ spins [29-32]. In the case of $Nd_{0.5}Sr_{0.5}Fe_{0.5}Mn_{0.5}O_3$ on further decreasing the temperature from 70 K to 20 K to 2 K, it was observed that the coercive field and remanent magnetization was found to increase. The non-saturation nature of hysteresis at fields as high as 7 T could be attributed to the existence of both ferromagnetic and antiferromagnetic short-range ordering of the spins. For $Nd_{0.5}Sr_{0.5}Fe_{0.5}Mn_{0.5}O_3$ and $Pr_{0.5}Sr_{0.5}Fe_{0.5}Mn_{0.5}O_3$ the isothermal magnetization was measured at 2 K and 20 K shows weak non-saturating hysteresis which suggests short-range ordering or spin glass nature in the compositions [29-32].

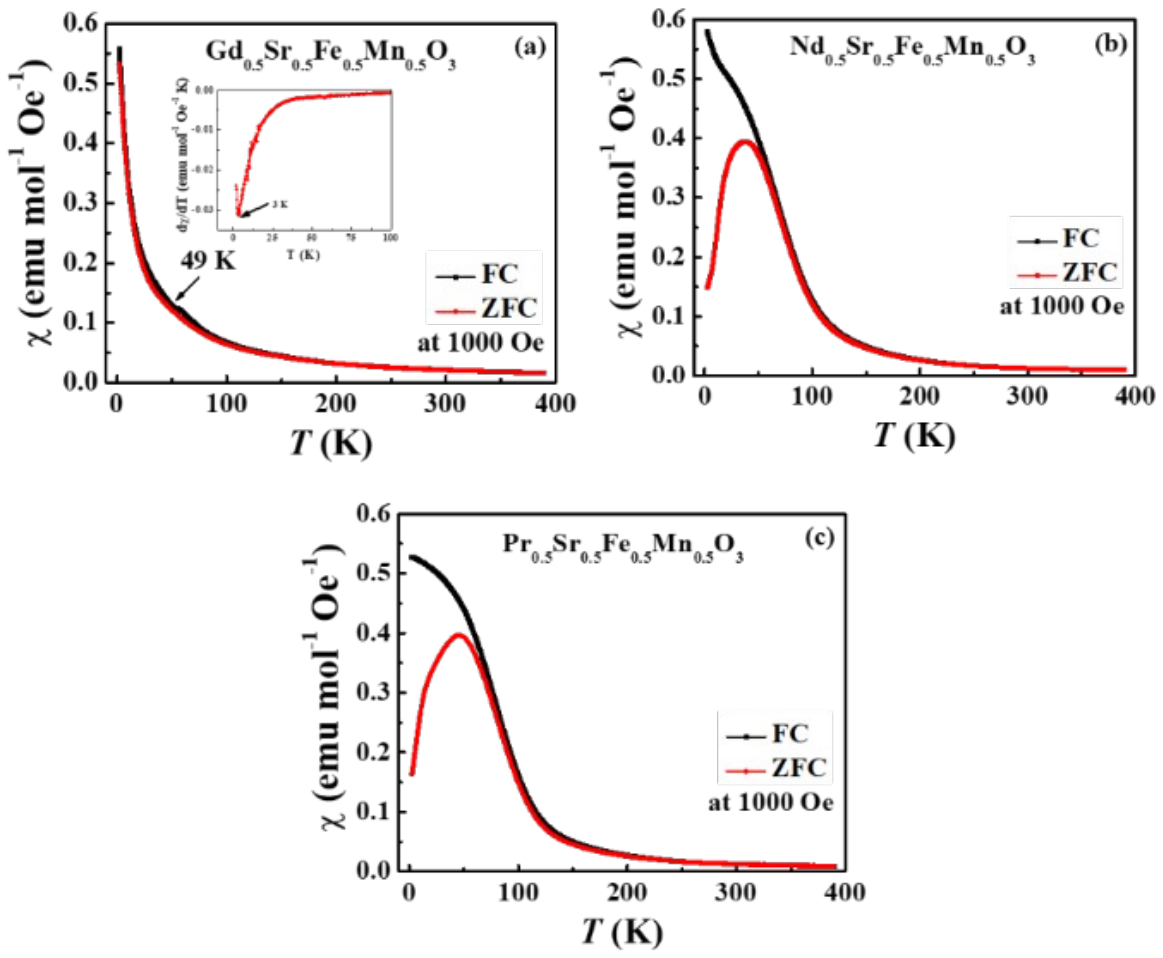

Fig 5.8: (a) Temperature dependent magnetic susceptibility (ZFC and FC) plot of (a) $Gd_{0.5}Sr_{0.5}Fe_{0.5}Mn_{0.5}O_3$ (Inset shows dχ/dT plot with respect to temperature). (b) $Nd_{0.5}Sr_{0.5}Fe_{0.5}Mn_{0.5}O_3$ and (c) $Nd_{0.5}Sr_{0.5}Fe_{0.5}Mn_{0.5}O_3$.

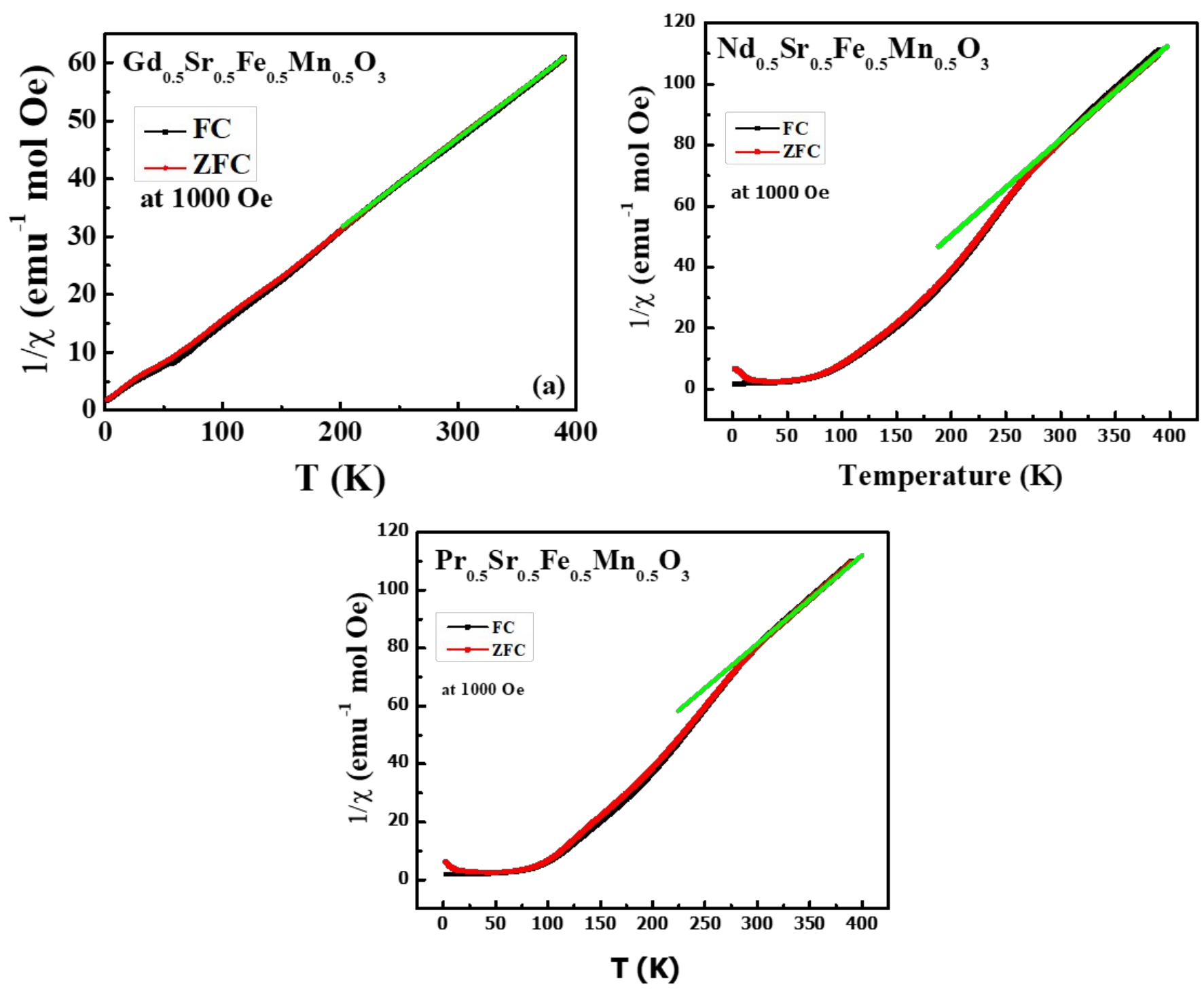

Fig. 5.9: Temperature dependent inverse magnetic susceptibility plot of (a) $Gd_{0.5}Sr_{0.5}Fe_{0.5}Mn_{0.5}O_3$, (b) $Nd_{0.5}Sr_{0.5}Fe_{0.5}Mn_{0.5}O_3$ and (c) $Pr_{0.5}Sr_{0.5}Fe_{0.5}Mn_{0.5}O_3$ in the temperature range of 2 K –390 K.

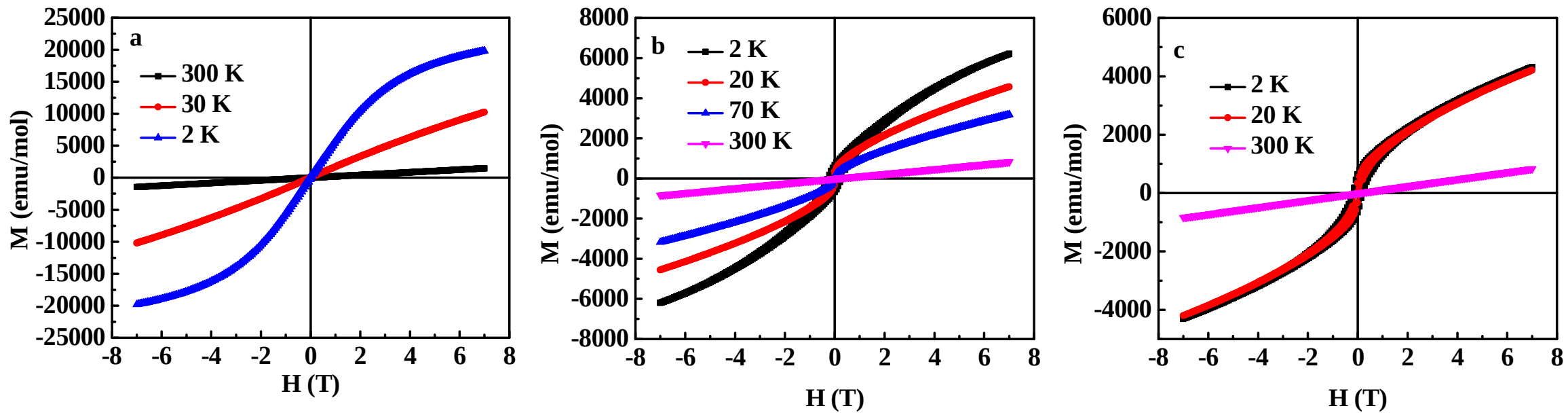

Fig. 5.10: Magnetic field dependent isothermal magnetization of (a) $Gd_{0.5}Sr_{0.5}Fe_{0.5}Mn_{0.5}O_3$ (b) $Nd_{0.5}Sr_{0.5}Fe_{0.5}Mn_{0.5}O_3$ and (c) $Pr_{0.5}Sr_{0.5}Fe_{0.5}Mn_{0.5}O_3$ in the magnetic field range of -7 T to +7 T.

Table 5.4: Experimentally obtained and calculated magnetic moment parameters.

| Composition | $\mu_{observed}$ $\mu_B$/f.u. | $\mu_{theoretical}$ $\mu_B$/f.u. |
|---|---|---|
| $Gd_{0.5}Sr_{0.5}Fe_{0.5}Mn_{0.5}O_3$ | 7.223 | 7.355 |
| $Nd_{0.5}Sr_{0.5}Fe_{0.5}Mn_{0.5}O_3$ | 5.022 | 5.478 |
| $Pr_{0.5}Sr_{0.5}Fe_{0.5}Mn_{0.5}O_3$ | 5.112 | 5.699 |

5.3.4: Magnetic moment calculation a) observed b) theoretical

a) Magnetic moment calculation of observed magnetic moment (μobserved): Fig. 5.9a, b and c

1. For $Gd_{0.5}Sr_{0.5}Fe_{0.5}Mn_{0.5}O_3$

$x_1 = 388.41$   $x_2 = 325.49$

$y_1 = 60.70$   $y_2 = 50.97$

$$C = \frac{x2-x1}{y2-y1} = \frac{325.49-388.41}{60.70-50.97} = \frac{62.92}{9.73} = 6.47$$

$\sqrt{C} = 2.54$

$\mu$ effective = $2.84\sqrt{C} = 7.223\ \mu_B$ / f.u.

2. For $Nd_{0.5}Sr_{0.5}Fe_{0.5}Mn_{0.5}O_3$

   $x_1 = 389.29$ $\quad x_2 = 308.63$

   $y_1 = 110.70$ $\quad y_2 = 84.82$

   $$C = \frac{x2-x1}{y2-y1} = \frac{389.29-308.63}{110.70-84.82} = \frac{80.67}{25.80} = 3.13$$

   $\sqrt{C} = 1.768$

   $\mu$ effective = $2.84\sqrt{C} = 5.022\ \mu_B$ / f.u.

3. For $Pr_{0.5}Sr_{0.5}Fe_{0.5}Mn_{0.5}O_3$

   $x_1 = 389.47$ $\quad x_2 = 300.41$

   $y_1 = 109.28$ $\quad y_2 = 81.77$

   $$C = \frac{x2-x1}{y2-y1} = \frac{389.47-300.41}{109.28-81.77} = \frac{89.07}{27.51} = 3.24$$

   $\sqrt{C} = 1.80$

   $\mu$ effective = $2.84\sqrt{C} = 5.112\ \mu_B$ / f.u.

   b) Theoretical magnetic moment calculation of $R_{0.5}Sr_{0.5}Fe_{0.5}Mn_{0.5}O_3$ (R: Gd, Nd and Pr)

   $Fe^{4+}$ (2.21), $Fe^{3+}$ (5.80), $Fe^{2+}$ (4.90)

   $Mn^{4+}$ (3.8), $Mn^{3+}$ (4.9), $Mn^{2+}$ (5.9)

   1. For $Gd_{0.5}Sr_{0.5}Fe_{0.5}Mn_{0.5}O_3$
      From XPS
      $Fe^{4+}: Fe^{3+}: Fe^{2+}$:: 16.12% : 58% : 2.7%
      $Mn^{4+}: Mn^{3+}: Mn^{2+}$:: 27% : 15% : 48%
      $$\mu \text{ effective} = [\ 0.5(7.94)^2 + 0.5\{0.161(2.21)^2 + 0.58(5.8)^2 + 0.027(4.9)^2\} + 0.5\{0.27(3.8)^2 + 0.15(4.9)^2 + 0.48(5.9)^2\}]^{1/2}$$
      $$= (54.09)^{1/2} = 7.355\ \mu_B \text{ / f.u.}$$

2. For $Nd_{0.5}Sr_{0.5}Fe_{0.5}Mn_{0.5}O_3$

   From XPS

   $Fe^{4+}$: $Fe^{3+}$: $Fe^{2+}$:: 15.80% : 20.11% : 65%

   $Mn^{4+}$: $Mn^{3+}$: $Mn^{2+}$:: 13% : 74% : 11.90%

   $$\mu \text{ effective} = [\,0.5(3.62)^2 + 0.5\{0.158(2.21)^2 + 0.201(5.8)^2 + 0.65(4.9)^2\} + 0.5\{0.13(3.8)^2 + 0.74(4.9)^2 + 0.12(5.9)^2\}]^{1/2}$$

   $$= (30.13)^{1/2} = 5.478\ \mu_B \text{ / f.u.}$$

3. For $Pr_{0.5}Sr_{0.5}Fe_{0.5}Mn_{0.5}O_3$

   From XPS

   $Fe^{4+}$: $Fe^{3+}$: $Fe^{2+}$:: 10.90% : 83% : 6.1%

   $Mn^{4+}$: $Mn^{3+}$: $Mn^{2+}$:: 69% : 14% : 27%

   $$\mu \text{ effective} = [\,0.5(3.58)^2 + 0.5\{0.109(2.21)^2 + 0.83(5.8)^2 + 0.061(4.9)^2\} + 0.5\{0.69(3.8)^2 + 0.14(4.9)^2 + 0.27(5.9)^2\}]^{1/2}$$

   $$= (32.49)^{1/2} = 5.699\ \mu_B \text{ / f.u.}$$

## 5.4 Conclusions

New series of perovskite oxides $R_{0.5}Sr_{0.5}Fe_{0.5}Mn_{0.5}O_3$ (R: Gd, Nd, and Pr) has been synthesized as pure phases and characterized. The powder X-ray diffraction studies confirmed the orthorhombic structure (*Pnma*) for all the above phases. The structural analysis from Rietveld refinement shows that the structures were found to show octahedral distortion primarily from octahedral tilting and Jahn-Teller distortion which was found to be maximum for $Gd_{0.5}Sr_{0.5}Fe_{0.5}Mn_{0.5}O_3$ followed by $Nd_{0.5}Sr_{0.5}Fe_{0.5}Mn_{0.5}O_3$ and $Pr_{0.5}Sr_{0.5}Fe_{0.5}Mn_{0.5}O_3$ compositions. The elongation of the *b* lattice parameter and the increased difference between the $B\text{-}O_{axial}$ and $B\text{-}O_{equatorial}$ confirmed maximum Jahn-Teller distortion in $Gd_{0.5}Sr_{0.5}Fe_{0.5}Mn_{0.5}O_3$. A canted antiferromagnetic nature observed for Gd at temperatures below 49 K could be attributed to the canted alignment of $Mn^{3+}$ spins along with the antiferromagnetic alignment of $Gd^{3+}$ spins at low temperatures. A spin glass-like behaviour for $Nd_{0.5}Sr_{0.5}Fe_{0.5}Mn_{0.5}O_3$ and $Pr_{0.5}Sr_{0.5}Fe_{0.5}Mn_{0.5}O_3$ at temperatures 47 K and 58 K was observed. The non-saturating nature of the hysteresis loop at 2 K for $Nd_{0.5}Sr_{0.5}Fe_{0.5}Mn_{0.5}O_3$ and $Pr_{0.5}Sr_{0.5}Fe_{0.5}Mn_{0.5}O_3$ suggests the existence of antiferromagnetic and ferromagnetism or short-range ordering of the spins. The XPS analysis confirmed the presence of varied oxidation states of $Mn^{2+}$, $Mn^{3+}$, $Mn^{4+}$ and $Fe^{2+}$, $Fe^{3+}$ and $Fe^{4+}$ showing major oxidation states in +3 and +4 oxidation states for all the three compositions. The spins moments of the ions were taken into account for the calculation of the theoretical magnetic moment which was found to coincide with the observed magnetic moment.

# Chapter VI

# Investigation of photoelectrocatalytic and magnetic properties of $Sr_2YbRu_{1-x}Ta_xO_6$ (x = 0, 0.25, 0.5, 0.75 and 1)

# Chapter VI

## Investigation of photoelectrocatalytic and magnetic properties of $Sr_2YbRu_{1-x}Ta_xO_6$ (x = 0, 0.25, 0.5, 0.75 and 1)

### Abstract

The photoelectrocatalytic and magnetic properties on the effect of substitution of Ta at Ru site in $Sr_2YbRuO_6$ has been investigated. DC magnetic susceptibility studies indicate that $Sr_2YbRuO_6$ shows a strong Yb – O – Ru long range antiferromagnetic ordering ($T_N$) as evident from the highest Curie Weiss temperature ($\theta_{CW}$ = -148 K), amongst $Sr_2YbRu_{1-x}Ta_xO_6$ (x = 0, 0.25, 0.5, 0.75 and 1), which are as follows -125 K (x = 0.25), -118 K (x = 0.5), -102 K (x = 0.75) respectively. Increased Ta concentration results in band gap to increase linearly in $Sr_2YbRu_{1-x}Ta_xO_6$, 1.78(1) eV (x =0) to 2.08(1) eV (x = 0.75). In photoelectrocatalytic measurement maximum OER evolution current density is 17 μA/cm$^2$ (light) at 0.8 V vs. Ag/AgCl and maximum photo response current was observed for $Sr_2YbRuO_6$. From the XPS analysis Ru and Ta were found to be majorly in +5/+4 oxidation states.

## 6.1 Introduction

Perovskite compounds of the formula $ABO_3$, shows varying magnetic, electronic properties, depending on their structural properties [1]. In the double perovskite of the type $A_2BB'O_6$ due to B-site ordering results in doubling of a unit cell and results in stable double perovskite structure if the difference in oxidation state is found to be greater than +2 or if their ionic radii difference is greater than 0.2 Å [2]. In this study, the parent compositions $Sr_2YbRuO_6$ and $Sr_2YbTaO_6$ have both been reported in monoclinic *$P2_1/n$* spacegroup because of the ordered arrangement of B atoms of Yb and Ru/Ta atoms, because of the greater difference in their charge and ionic radii [3-4]. These ordered arrangements result in varied magnetic interactions between the B-atoms and thus result in varied spin structures, which are B – O – B' and relatively weaker B – O – O – B interaction [5]. For example, compositions of the type $A_2RRuO_6$ (R: rare-earth, A: Ca, Sr and Ba) show majorly three types of the magnetic structure, depending on strength of B - O – O – B interactions. In type I structure Ru - O – O –Ru was found to be negligible, with relatively stronger Ru - O – Yb interactions, for example, $Sr_2YbRuO_6$ (Fig. 6.2) [4]. In Type II there is Ru - O – Yb interaction and an increased Ru – O – O Ru interaction, compared to type I, for example, $Sr_2TmRuO_6$ [7]. While in Type III magnetic structure shows a relatively stronger Ru - O – O Ru interaction, for example, $Ba_2LaRuO_6$ [5-7, 16]. In the case of rare-earths having a low magnetic moment, it results in increased Ru–O-O-Ru interaction, for example, $Ba_2LaRuO_6$ [5-7, 16]. It has been reported in the literature that a more stable antiferromagnetic structure results in higher Weiss temperature, for example, $Ba_2TmRuO_6$ (type II), with μeffective= 7.92 $\mu_B$/f.u. and $\theta_{CW}$ = -34 K [7-8]. While in type I spin structure $Ba_2YbRuO_6$ shows a higher $\theta_{CW}$ = -181 K and $\mu_{effective}$ = 4.2 $\mu_B$/f.u [7-8]. In the parent composition $Sr_2YbRuO_6$ it has been mentioned that ferromagnetically coupled spins along the a-b plane between B-atoms are arranged antiferromagnetically along with the *c* – axis ($T_N$ = 48 K), resulting in type I, A-type antiferromagnetic spin structure (Fig. 6.2) [7]. While $Sr_2YbTmO_6$ has its spins parallel along the c-axis $T_N$ = 42 K. It can be concluded that $Sr_2YbRuO_6$ shows higher $T_N$ amongst other $Sr_2YbRO_6$ (R: rare-earth) types, because of its type I antiferromagnetic structure with increased Yb – O – Ru magnetic order [7-9, 15].

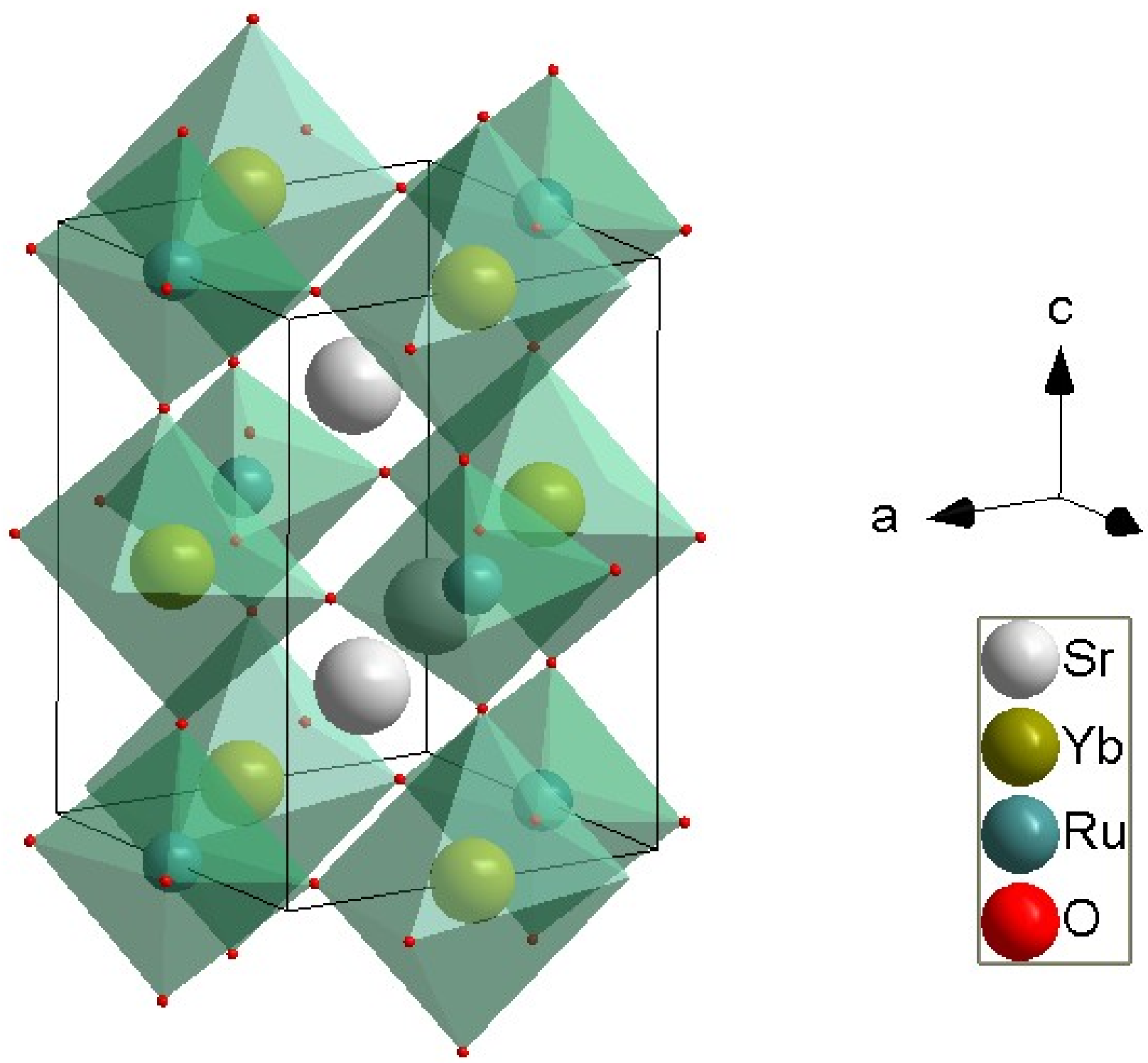

Fig 6.1: Polyhedral representation of the structure of $Sr_2YbRuO_6$.

In literature, ruthenates show electronic properties depending on their structural properties. In magnetic semiconductors, the optical spectrum particularly the valence band and conduction band edge positions get influenced by magnetic exchange interactions and thus magnetic ordering [10]. This exchange interaction mediated coupled localized magnetic moment, either long-range or short-range results in delocalization of conduction electron. Modelling studies have shown contributions of localized carriers to spin moment [11-12].

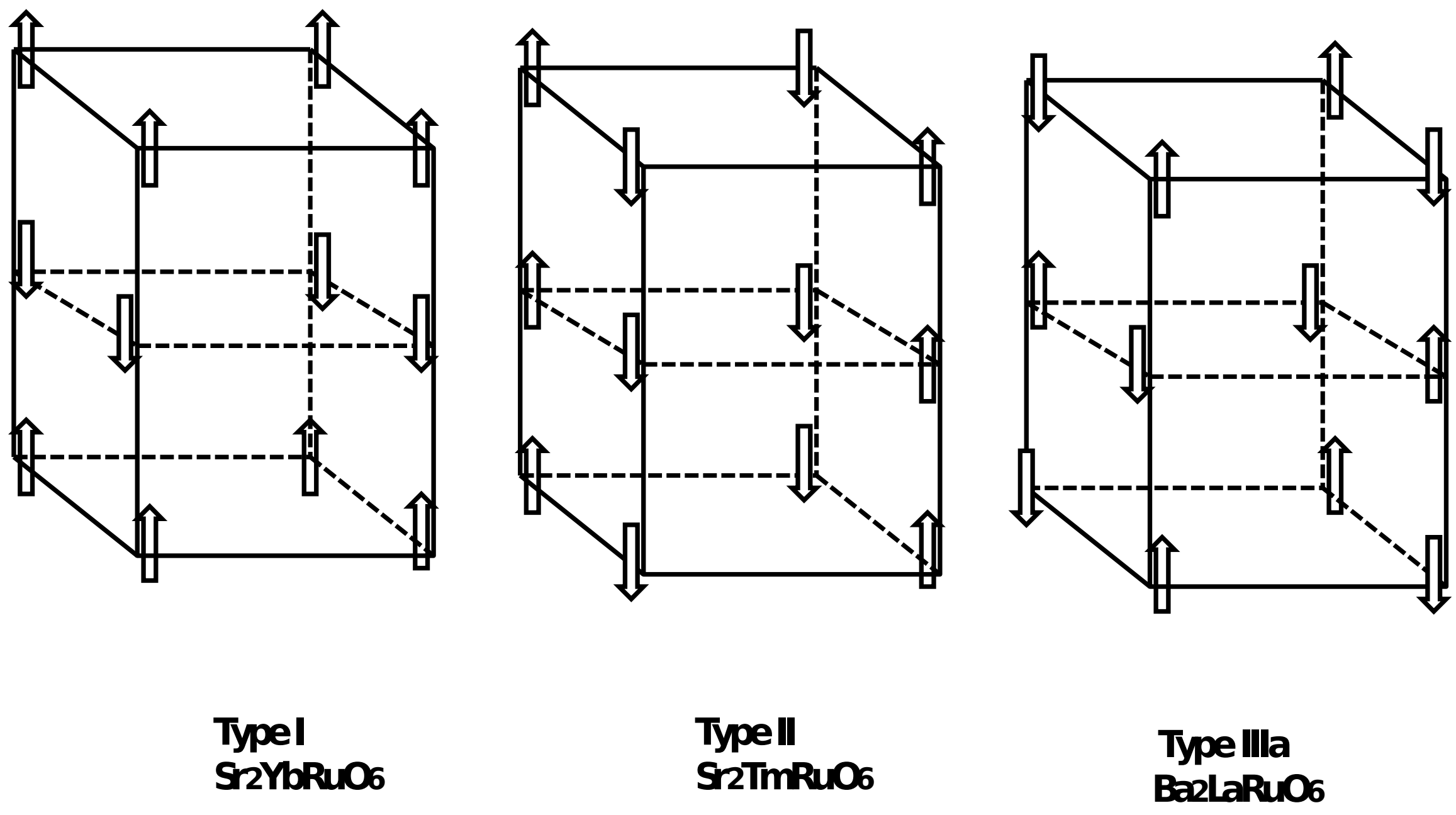

Fig. 6.2: Magnetic structure of Type I, II and III antiferromagnetic structures.

In the literature on perovskite-type oxides, it has been reported that the valence band energy level is determined primarily by O-2p orbitals, while the conduction band energy level is determined by the d-orbitals of the metal cations [13]. Thus, the bandgap of the semiconductor perovskite is determined by octahedral distortion and orbital ordering [15-17]. For example, doping of Ni at B-site in $Ba_2InTaO_6$ results in a decrease of bandgap from 3.82 eV to 3.25 eV in $Ba_2In_{0.9}Ni_{0.1}TaO_6$, because of disruption in its long-range d-d overlap [14]. Thus, the long-range/short-range orbital overlap is the key for tuning the bandgap positions and photo electrocatalysts for various reactions. For example, Na in $SrRuO_3$ results in the increase of the oxidation state of Ru and enhances its OER activity [14]. The structural properties also result in a change in electronic properties and thus photo electrocatalytic activity. For example, $SrRu_{1-x}Cr_xO_3$ (x<0.3) [18] shows metallic ferromagnetic properties with $T_C$ in the range of 160-190 K, while for x < 0.7, it exhibits a cubic phase, showing insulating and paramagnetic behaviour. The indirect metal-oxygen-metal interaction changes the d orbital bandwidth, and thus depends on the connectivity between the neighbouring octahedra [19]. For example, severe octahedral distortion in $Ca_2In_{0.9}Ni_{0.1}TaO_6$ and decreased d-d charge transfer from Ni to In/Ta results in insulating behaviour [20].

The charge mobility from semiconductor to electrolyte holds importance for photo electrocatalytic charge carrier density [21]. For example, in $CaLaScRuO_6$ higher filling of $e_g$ orbital of Ru, enhances the –OH adsorption [22]. Generally, alkaline medium is preferred for oxygen evolution activity (OER) activity in Ru containing compositions for advanced kinetics and optimum Ru-OH binding strength [18, 21-23].

Here we intend to study the photo electrocatalytic, magnetic and structural properties and their correlations.

## 6.2 Experimental

### 6.2.1 Materials and synthesis

Oxides of the type $Sr_2YbRu_{1-x}Ta_xO_6$ (x = 0, 0.25, 0.5, 0.75 and 1) in air at high temperature. Stoichiometric quantities of $SrCO_3$ (Sigma Aldrich), $Yb_2O_3$ (Sigma Aldrich), $RuO_2$ (Sigma Aldrich) and $Ta_2O_5$ (Sigma Aldrich) was sintered at 1150˚C for 4 hours after thoroughly grinding. The as-obtained powder (0.2 g) was mixed with 20 μl of 5% PVA solution, pelletised at 10 tons pressure and then sintered at 1200˚ C for 4hours. For the case of $Sr_2YbTaO_6$, the pellet was ground again and subsequently pelletized and was given an additional heat treatment at 1400˚C for 4 hours. Phase purity was confirmed from powder X-ray diffraction and lattice parameters were obtained by Rietveld refinement of the powder diffraction data collected on a Bruker D8 Advance diffractometer equipped with a CuKα source with Ni filter. The $K\alpha_2$ reflections were stripped off using the software.

### 6.2.2 Magnetic studies

Magnetization studies were performed in the temperature range of 5 K to 300 K using a Quantum design, Physical Property Measuring System Vibrating Sample Magnetometer, USA, at a magnetic field of 100 Oe for zero-field cooled (ZFC) and at 1000 Oe for field cooled warming (FCW), and isothermal magnetization measurement was performed in the magnetic field range of +5 T to -5 T at varying temperatures.

### 6.2.3 Diffuse reflectance spectroscopy measurement (DRS)

Diffuse reflectance spectroscopy (DRS) was performed on a Shimadzu UV-2450 UV-visible spectrophotometer, using dried $BaSO_4$ as a standard, in the wavelength range 400 nm to 800 nm.

### 6.2.4 Photoelectrocatalytic measurement (PEC)

Photoelectrochemical experiments were performed on a Metrohm Autolab (PGSTAT M204) instrument, equipped with a Xe arc lamp (400 W, 200-1000 nm range). The working electrode was prepared by drop-casting from a suspension onto an ITO glass electrode, having an area approximately $1cm^2$. The suspension was prepared as follows: 4 mg of the powdered catalyst was dispersed in 1.4 ml of isopropyl alcohol and 0.6 ml of water, followed by the addition of the binder as Nafion (20 μl). A three-electrode photoelectrochemical cell was used to obtain the current-voltage (I-V) characteristics in a 0.1M aqueous KOH solution.

a) **Mott-Schottky plots:** Mott-Schottky studies were performed by using the as-prepared ITO electrode as a working electrode, platinum wire as counter electrode and Ag/AgCl as a reference electrode. Mott-Schottky analysis was carried out at a frequency of 1 kHz.

b) **Oxygen evolution current density in light and dark:** LSV (linear sweep voltammetry) scan was performed in light and dark in the potential range 0 to 0.8 V versus Ag/AgCl reference electrode with a scan rate of 10 mV/s.

### 6.2.5 X-ray photoelectron spectroscopy (XPS)

XPS studies were performed on the powdered sample on a Shimadzu ESCA-3400 spectrophotometer with Mg Kα source (1253.6 eV) operating at 10 kV and 10 mA. The sample was used in powdered form for measurement. The chamber was maintained an ultra-high vacuum of $10^{-8}$ torr during measurement.

## 6.3 Results and discussion

### 6.3.1 Powder X-ray diffraction studies

The PXRD pattern of the calcined compounds $Sr_2YbRu_{1-x}Ta_xO_6$ was indexed in the monoclinic $P2_1/n$ space group (Fig. 6.3). The *c* lattice parameter was found to increase with the increase in x. The shift of the PXRD 2θ peak towards a lower 2θ value with an increase in Ta concentration (Fig. 6.4), suggests lattice parameter expansion as the larger Ta replaces the Ru site. The octahedral distortion along the axial direction increased with the increase in Ta concentration. Also, the change in Sr – O bond distances was found to be maximum for a higher value of x, suggesting a greater degree of distortion. Thus, Ta substitution results in increased octahedral distortion [24].

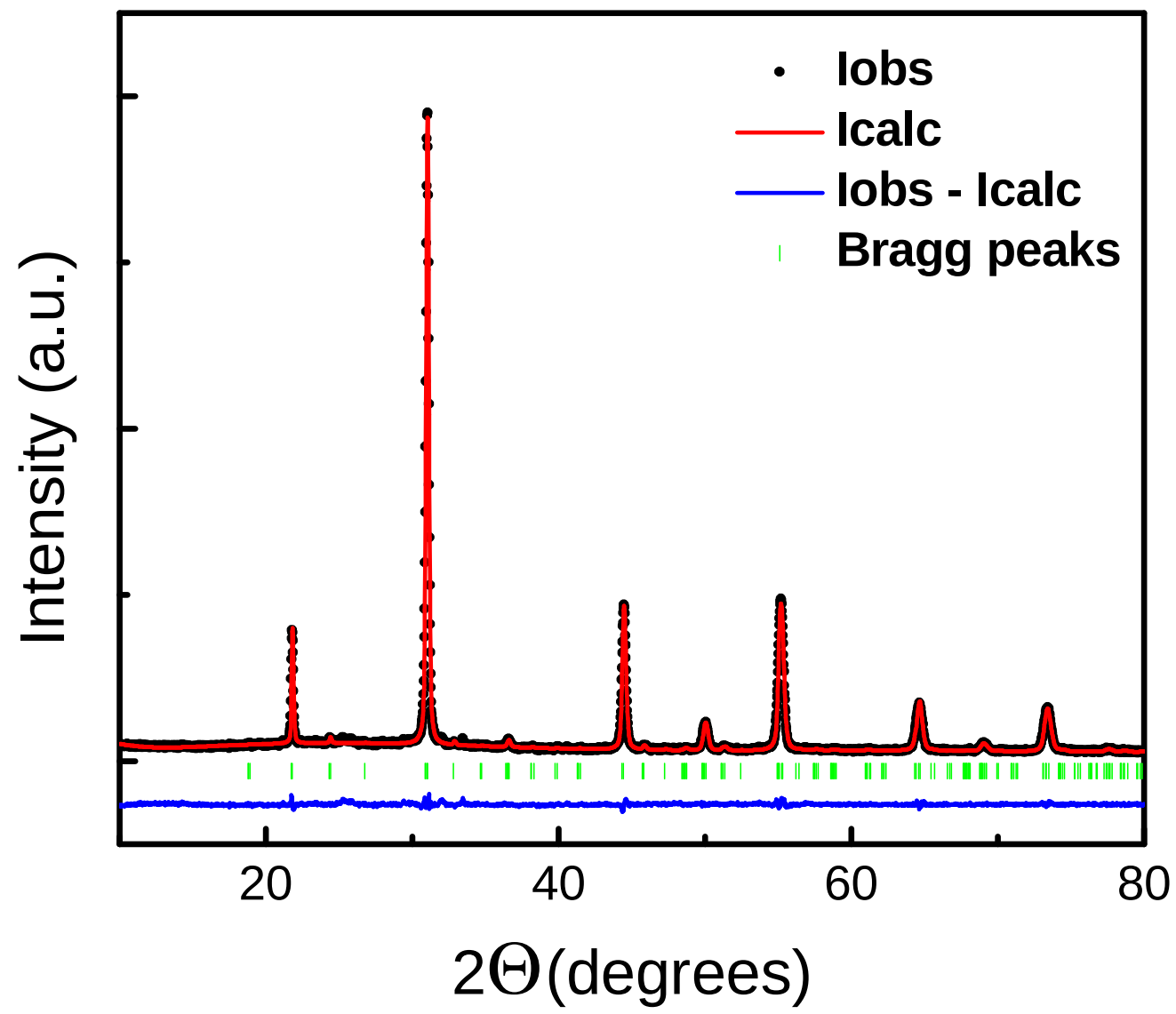

Fig 6.3: The room temperature powder X – ray diffraction data (black) of $Sr_2YbRu_{0.5}Ta_{0.5}O_6$

Refined in monoclinic *P2₁/n* space group. The profile from Rietveld refinement, the fitted Bragg peak positions and the residual are shown in green red lines (short vertical bars) and blue lines in the panel.

Table 6.1: Rietveld refinement of a) $Sr_2YbRuO_6$, b) $Sr_2YbRu_{0.75}Ta_{0.25}O_6$ c) $Sr_2YbRu_{0.5}Ta_{0.5}O_6$ d) $Sr_2YbRu_{0.25}Ta_{0.75}O_6$ and e) $Sr_2YbTaO_6$ obtained from powder X-ray diffraction data.

| a) **$Sr_2YbRuO_6$ a = 5.7926(1) Å, b = 5.7326(1) Å, c = 8.1138(1) Å, β = 90.24(1)°** | | | | | | |
|---|---|---|---|---|---|---|
| Site | x | y | z | Atom | Occ | $B_{iso}$ |
| 4e | 0.009(3) | 0.028(1) | 0.248(5) | Sr | 1 | 2.37(5) |
| 2d | 0.5 | 0 | 0 | Yb | 1 | 0.40(3) |
| 2c | 0 | 0.5 | 0 | Ru | 1 | 0.55(6) |
| 4e | 0.27(1) | 0.222(8) | 0.041(8) | O1 | 1 | 6.0(1) |
| 4e | 0.33(1) | 0.29(1) | 0.46(1) | O2 | 1 | 6.0(1) |
| 4e | 0.98(1) | 0.452(9) | 0.25(2) | O3 | 1 | 8.0(2) |

| b) | $Sr_2YbRu_{0.75}Ta_{0.25}O_6$ a = 5.7929(4) Å, b = 5.7472(3) Å, c = 8.1321(3) Å,β=90.157(7)° | | | | | |
|---|---|---|---|---|---|---|
| 4e | 0.007(2) | 0.018(4) | 0.26(1) | Sr | 1 | 2.2(1) |
| 2d | 0.5 | 0 | 0 | Yb | 1 | 2.4(2) |
| 2c | 0 | 0.5 | 0 | Ru | 3/4 | 0.9(1) |
| 2c | 0 | 0.5 | 0 | Ta | 1/4 | 0.9(1) |
| 4e | 0.25(3) | 0.26(3) | 0.025(6) | O1 | 1 | 6.0(5) |
| 4e | 0.41(1) | 0.30(2) | 0.52(1) | O2 | 1 | 6.0(6) |
| 4e | 0.98(1) | 0.53(1) | 0.27(1) | O3 | 1 | 8.0(5) |

| c) | $Sr_2YbRu_{0.5}Ta_{0.5}O_6$ a = 5.7963(4) Å, b = 5.7648(4) Å, c = 8.1539(5) Å, β = 90.213(1)° | | | | | |
|---|---|---|---|---|---|---|
| 4e | 0.002(1) | 0.010(2) | 0.24(1) | Sr | 1 | 2.37(5) |
| 2d | 0.5 | 0 | 0 | Yb | 1 | 0.40(3) |
| 2c | 0 | 0.5 | 0 | Ru | 1/2 | 0.55(6) |
| 2c | 0 | 0.5 | 0 | Ta | 1/2 | 0.55(6) |
| 4e | 0.25(4) | 0.30(1) | 0.03(1) | O1 | 1 | 6.0(1) |
| 4e | 0.345(5) | 0.240(8) | 0.45(1) | O2 | 1 | 6.0(1) |
| 4e | 0.915(8) | 0.51(2) | 0.200(5) | O3 | 1 | 8.0(2) |

| d) | $Sr_2YbRu_{0.25}Ta_{0.75}O_6$ a = 5.8001(1) Å, b = 5.7727(2) Å, c = 8.1701(2) Å, β= 90.136(3)° | | | | | |
|---|---|---|---|---|---|---|
| 4e | 0.005(2) | 0.012(1) | 0.256(1) | Sr | 1 | 1.33(4) |
| 2d | 0.5 | 0 | 0 | Yb | 1 | 2.09(7) |
| 2c | 0 | 0.5 | 0 | Ru | 1/4 | 0.96(5) |
| 2c | 0 | 0.5 | 0 | Ta | 3/4 | 0.96(5) |
| 4e | 0.27(1) | 0.27(2) | 0.020(6) | O1 | 1 | 1.6(5) |
| 4e | 0.30(1) | 0.26(1) | 0.475(6) | O2 | 1 | 3.3(6) |
| 4e | 0.99(2) | 0.433(6) | 0.26(1) | O3 | 1 | 0.9(1) |

| e) | $Sr_2YbTaO_6$ a = 5.8149(3) Å, b = 5.7839(3) Å, c = 8.1921(3) Å, β = 89.905(5)° | | | | | |
|---|---|---|---|---|---|---|
| 4e | 0.009(3) | 0.028(1) | 0.248(5) | Sr | 1 | 1.2(2) |
| 2d | 0.5 | 0 | 0 | Yb | 1 | 1.3(1) |
| 2c | 0 | 0.5 | 0 | Ta | 1 | 1.3(1) |
| 4 | 0.27(1 | 0.22(1 | 0.041( | O | 1 | 2.4( |
| e | ) | ) | 8) | 1 | | 2) |
| 4e | 0.33(1) | 0.28(1) | 0.46(1) | O2 | 1 | 2.3(2) |
| 4e | 0.98(1) | 0.452(9) | 0.25(2) | O3 | 1 | 1.2(1) |

Here it was observed that the *a*, *b* and *c* lattice parameters and volume increased with increase in x vlaue in $Sr_2YbRu_{1-x}Ta_xO_6$ (x = 0, 0.25, 0.5, 0.75 and 1).

**Table 6.2:** The bond angle and bond distances of the system $Sr_2YbRu_{1-x}Ta_xO_6$ (x= 0, 0.25, 0.5, 0.75 and 1).

| | **$Sr_2YbRuO_6$ (x = 0)** | **$Sr_2YbRu_{0.75}Ta_{0.25}O_6$ (x = 0.25)** | **$Sr_2YbRu_{0.5}Ta_{0.5}O_6$ (x = 0.5)** | **$Sr_2YbRu_{0.25}Ta_{0.75}O_6$ (x = 0.75)** | **$Sr_2YbTaO_6$ (x = 1)** |
|---|---|---|---|---|---|
| Sr-O1 (Å) | 2.522(1) | 2.655(1) | 2.614(1) | 2.609(1) | 2.531(1) |
| Sr-O2 | 2.242(1) | 2.328(1) | 2.307(1) | 2.587(1) | 2.284(1) |
| Sr-O3 | 2.436(1) | 2.729(1) | 2.446(1) | 2.432(1) | 2.458(1) |
| Yb-O1 | 1.871(1) | 2.078(1) | 2.269(1) | 2.057(1) | 1.877(1) |
| Yb-O2 | 2.282(1) | 2.629(1) | 2.531(1) | 2.258(1) | 2.326(1) |
| Yb-O3 | 2.049(1) | 1.915(1) | 2.496(1) | 2.007(1) | 2.071(1) |
| Ru-O1 | 2.259(1) | 2.023(1) | 1.868(1) | 2.059(1) | 2.280(1) |
| Ru-O2 | 1.960(1) | 1.832(1) | 2.531(1) | 1.885(1) | 1.925(1) |
| Ru-O3 | 2.051(1) | 2.173(1) | 2.496(1) | 2.153(1) | 2.070(1) |
| Yb-O1-Ru | 161.19(1) | 168.51(1) | 162.12(1) | 167.43(1) | 161.08(1) |
| Yb-O2-Ru | 147.63(1) | 131.51(1) | 149.38(1) | 161.75(1) | 149.31(1) |
| Yb-O3-Ru | 163.25(1) | 168.27(1) | 151.62(1) | 158.29(1) | 163.28(1) |

As the value of x increases in $Sr_2YbRu_{1-x}Ta_xO_6$ (x = 0, 0.25, 0.5, 0.75 and 1) the Sr-O1 bond distance was found to decrease with increase in x value for x = 0.25, 0.5, 0.75 and 1. The octahedral distortion (Yb – O – Ru) was found to greater at intermediate values of x particularly for x = 0.25.

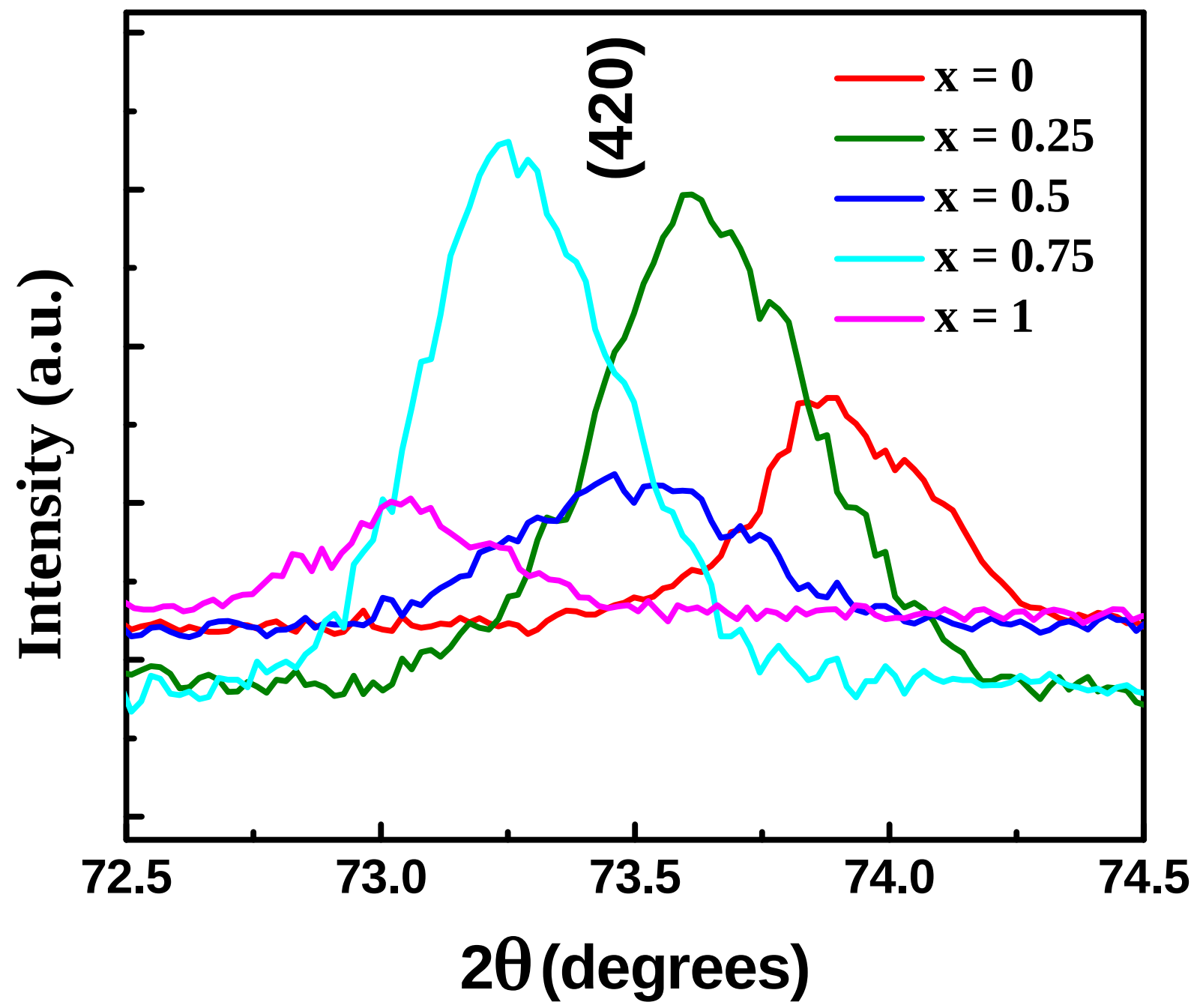

Fig 6.4: Peak splitting of the (332) peak, obtained from powder XRD studies.

Here, as the value of x increases that is as the Ta concentration increases in $Sr_2YbRu_{1-x}Ta_xO_6$ (x = 0, 0.25, 0.5, 0.75 and 1) peaks were found to shift uniformly towards a lower 2θ value suggesting lattice expansion with an increase in Ta concentration.

### 6.3.2 Magnetic studies

The field cooled DC magnetic susceptibility studies, at 1000 Oe, (Fig. 6.5), shows antiferromagnetic transition and Curie-Weiss ($\theta_{CW}$) temperature in the range 5 – 49 K and 148 to 102 K respectively, which also decreases as the Yb – O – Ru ordering decreases, with increased concentration of Ta at B-site (Fig. 6.7).

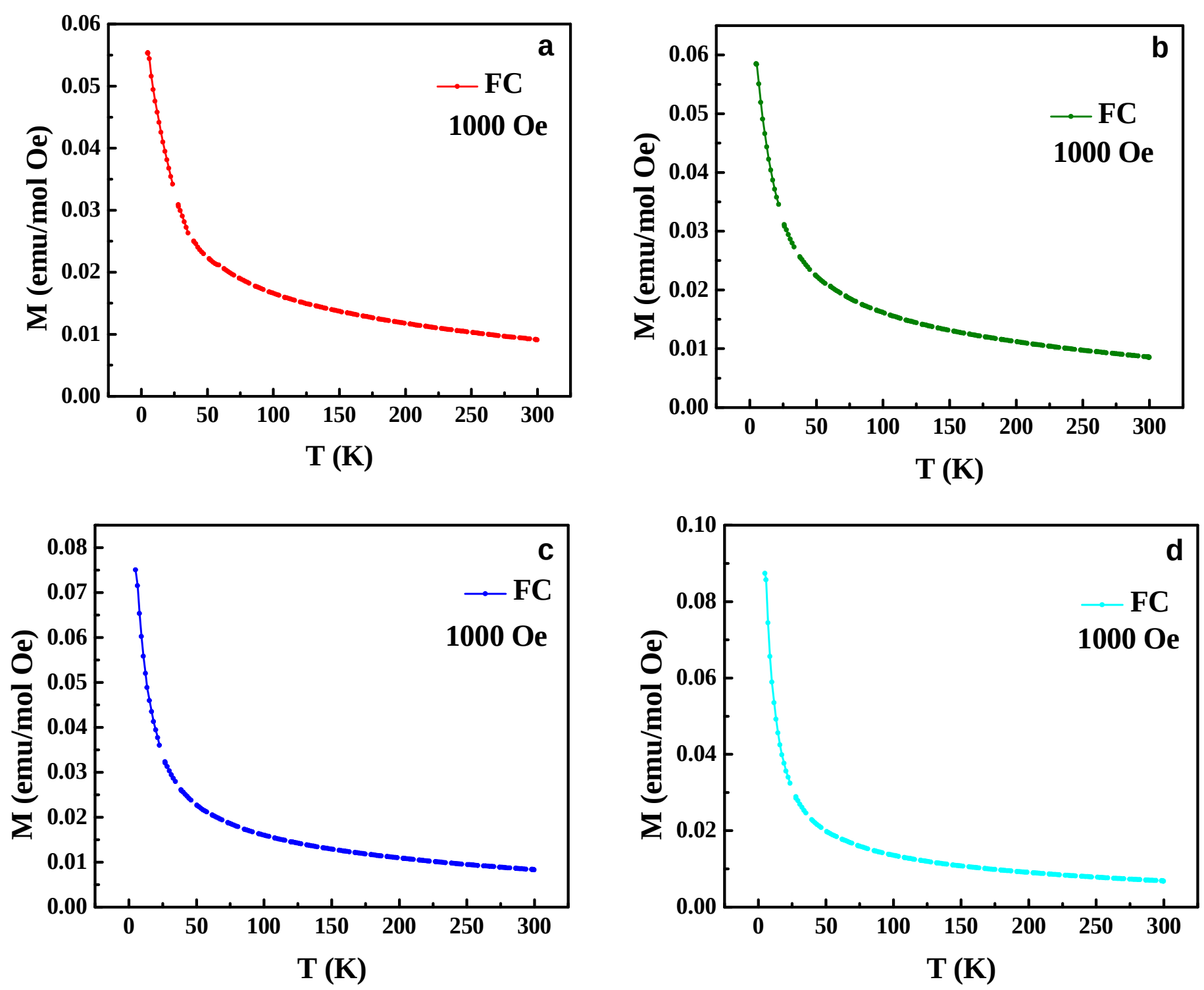

Fig. 6.5: Temperature dependent magnetic susceptibility plot of $Sr_2YbRu_{1-x}Ta_xO_6$ $0 \leq x \leq 1$.

The $\theta_{CW}$ are listed in Fig. 6.7, Table 6.3. The field cooled magnetization data for all the compositions showed a similar trend with no sharp transitions but majorly antiferromagnetic behaviour as inferred from the negative $\theta_{CW}$. The $Sr_2YbRuO_6$ has perfect antiferromagnetic Type I structure as mentioned earlier, which probably results in increased $\theta_{CW}$ as compared to Ta-doped $Sr_2YbRu_{1-x}Ta_xO_6$ (x = 0.25, 0.5 and 0.75). Also, from the $\chi$ v/s T plot (Fig. 6.6) the magnetization at 5 K was maximum for $Sr_2YbRu_{0.25}Ta_{0.75}O_6$ and decreased with the as the x value decreased, probably due to the increased ordering of Yb moments or due to increased Ru – O – O – Ru interactions, resulting in enhanced frustration of antiferromagnetic spins [25-26].

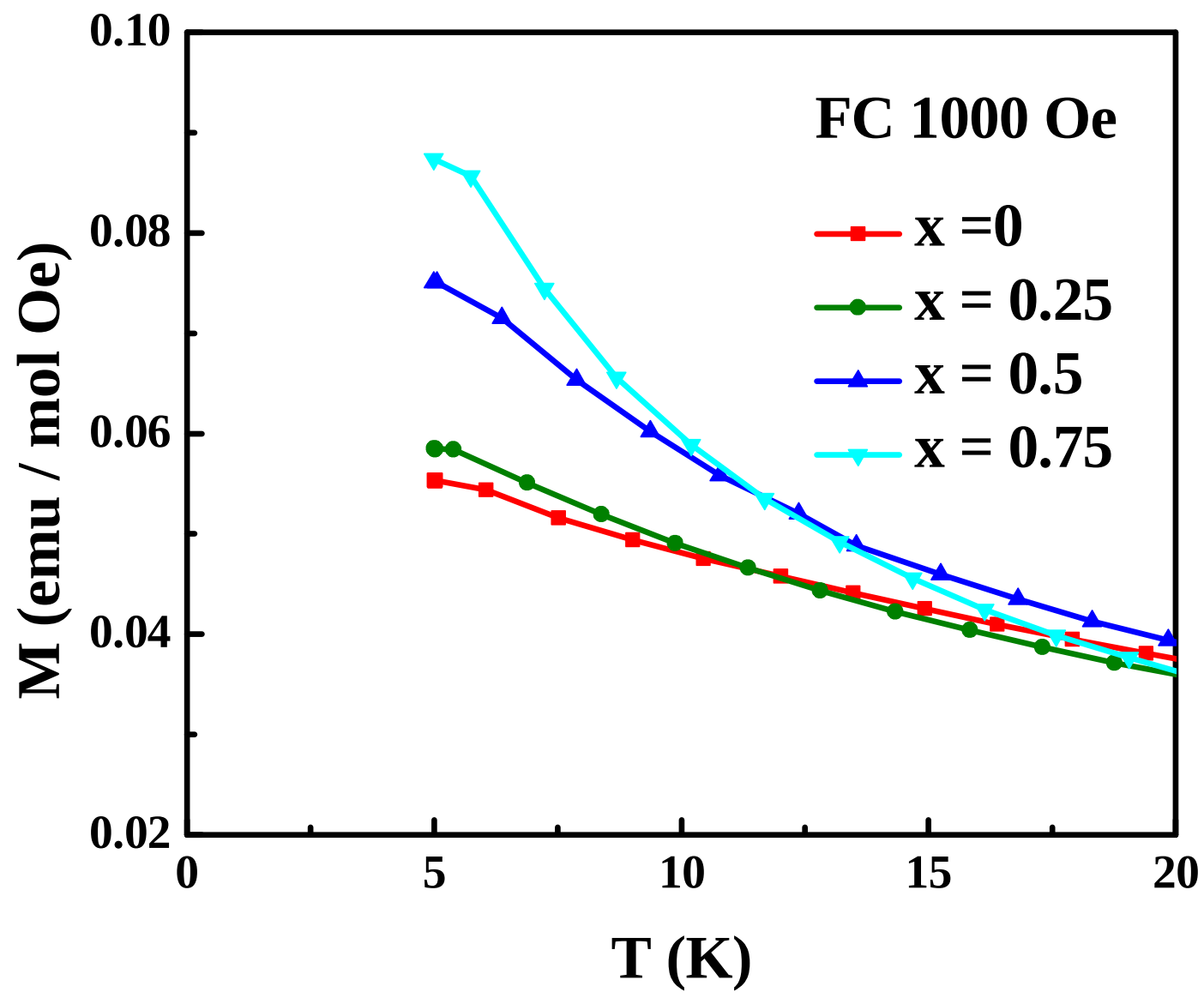

Fig. 6.6: Enlarged plot at the low temperature of temperature-dependent magnetic susceptibility plot of $Sr_2YbRu_{1-x}Ta_xO_6$ $0 \leq x \leq 1$.

It was observed that as the value of x increases in $Sr_2YbRu_{1-x}Ta_xO_6$ (x = 0, 0.25, 0.5, 0.75 and 1) the magnetization value increases at 5 K temperature (Fig. 6.6). This could be attributed to the net decrease in magnetization moment from the opposite alignment of the net moment generated from the canted Ru–O–Ru moments opposite to the Yb moment, which is found to be maximum at lower values of x.

From the isothermal magnetization measured at 5 K, it was observed that the magnetization increased with an increase in Ta doping concentration. No saturation magnetization was observed even at the 7 T magnetic field, due to the absence of long-range ordering. This further proves the fact that the long-range order Yb-O-Ru decreases on Ta doping, due to the increased canting of spins. It was also observed that substituting Ta, plays a major role in the disruption of Yb –O –Ru interactions and plays a lesser role in the disruption of the Ru-O-O-Ru interactions.

Taking into account the moments of $Yb^{3+}$(4.54 $\mu_B$/f.u.) and $Ru^{5+}$(3.87 $\mu_B$/f.u.) the 'm' $_{effective}$ was found to be approximately equal to 'm' $_{calculated}$ (Table 6.3) [25-26]. A linear trend of increase in $\mu_{effective}$ was observed as the concentration of Ru increases (as expected). However, it was observed that the $\mu_{observed}$ was slightly less as compared to its respective $\mu_{calculated}$ values for all the members of the series, which may arise due to huge degree of crystal field splitting of $Yb^{3+}$ moments and also due to covalent effect of $Yb^{3+}$ moments and canting of spins [27-29].

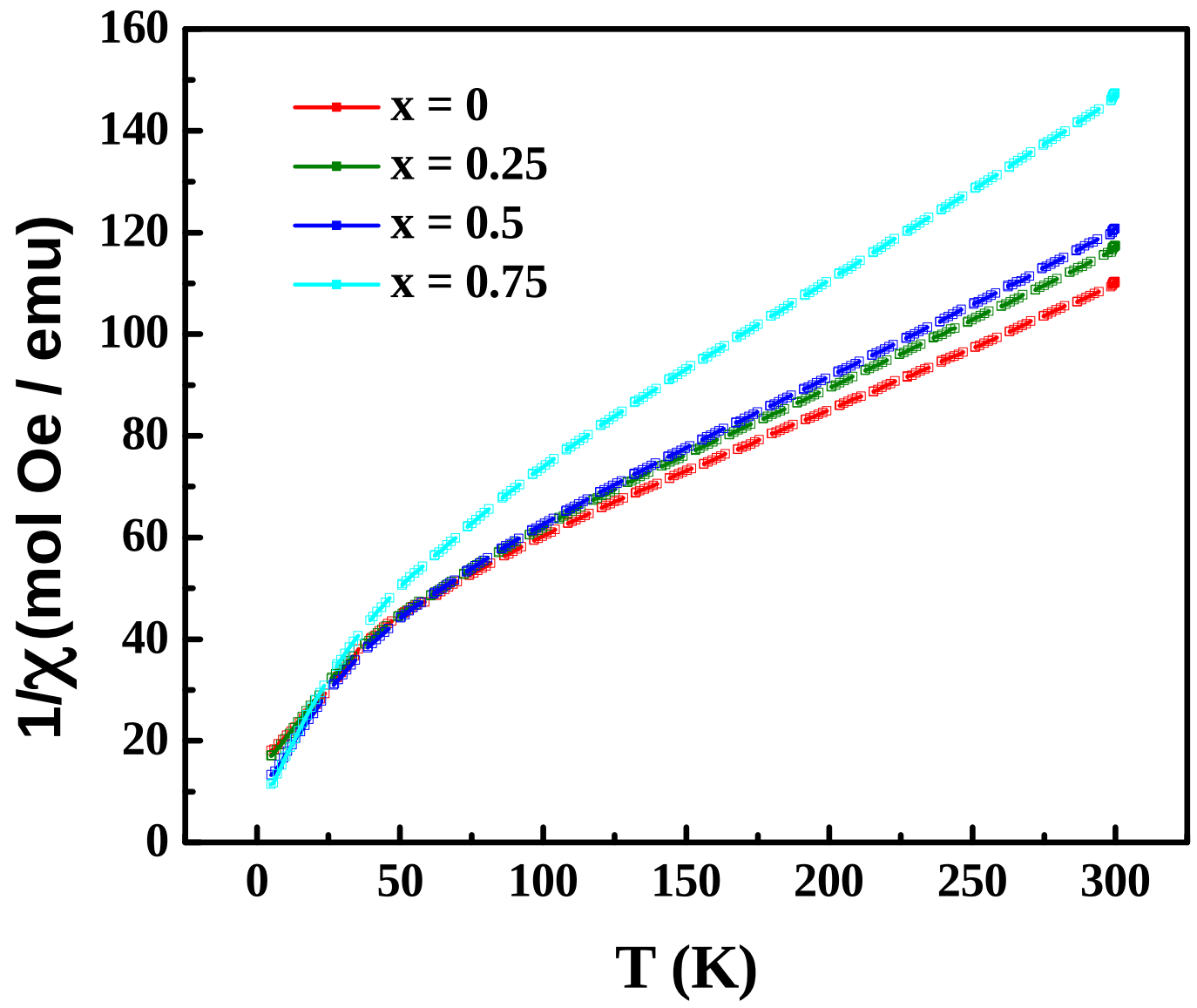

Fig.6.7: Inverse magnetic susceptibility plots of $Sr_2YbRu_{1-x}Ta_xO_6$

Here it was observed that the effective magnetic moment increases with the increase in the value of x in $Sr_2YbRu_{1-x}Ta_xO_6$ (x = 0, 0.25, 0.5,0.75 and 1).

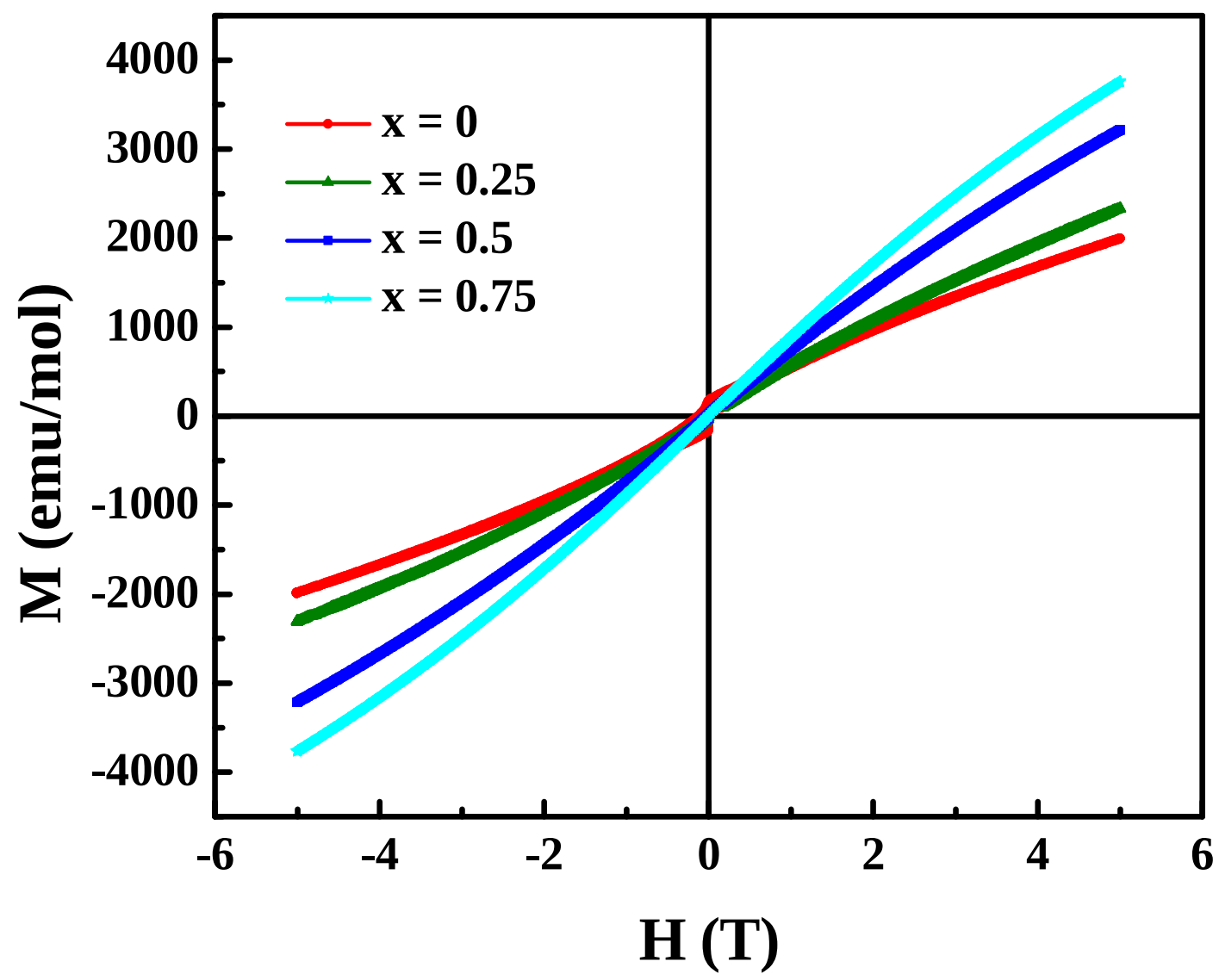

Fig 6.8: Isothermal magnetization at 5 K for $Sr_2YbRu_{1-x}Ta_xO_6$.

Here the magnetization value at the 5 T field increased with the increase in the value of x which could be attributed to the alignment of net moment generated from the canted antiferromagnetic alignment of Ru spin opposite to the Yb moments.

Table 6.3: Experimentally obtained and calculated magnetic moment parameters.

| Composition | μ observed $\mu_B$/f.u. | μ theoretical $\mu_B$/f.u. | Néel temperature (K) | Curie-Weiss temperature (K) |
|---|---|---|---|---|
| $Sr_2YbRuO_6$ | 5.966 | 5.594 | 51 | -148 |
| $Sr_2YbRu_{0.75}Ta_{0.25}O_6$ | 5.643 | 5.469 | 40 | -125 |
| $Sr_2YbRu_{0.5}Ta_{0.5}O_6$ | 5.301 | 5.380 | 49 | -118 |
| $Sr_2YbRu_{0.25}Ta_{0.75}O_6$ | 4.935 | 4.721 | 43 | -102 |

### 6.3.3 Diffuse reflectance studies

Visually, the as-obtained powders were coloured (Fig. 6.9). Thus, diffuse reflectance spectroscopy – DRS (Fig. 6.9) was carried out in the visible region to determine the bandgaps of the oxides Fig. 6.10). It was seen that the bandgap increases from 1.78(1) eV (x = 0) to 2.08(1) eV (x = 0.75) following a linear trend on increasing the Ta concentration in $Sr_2YbRu_{1-x}Ta_xO_6$ (x = 0, 0.25, 0.5 and 0.75) (Fig. 6.11). However, for the case of $Sr_2YbRuO_6$ the bandgap (1.78(1) eV) is higher than predicted by this trend. This higher bandgap in the parent compound $Sr_2YbRuO_6$ (Fig. 6.11), might be correlated with the increased magnetic ordering [30].

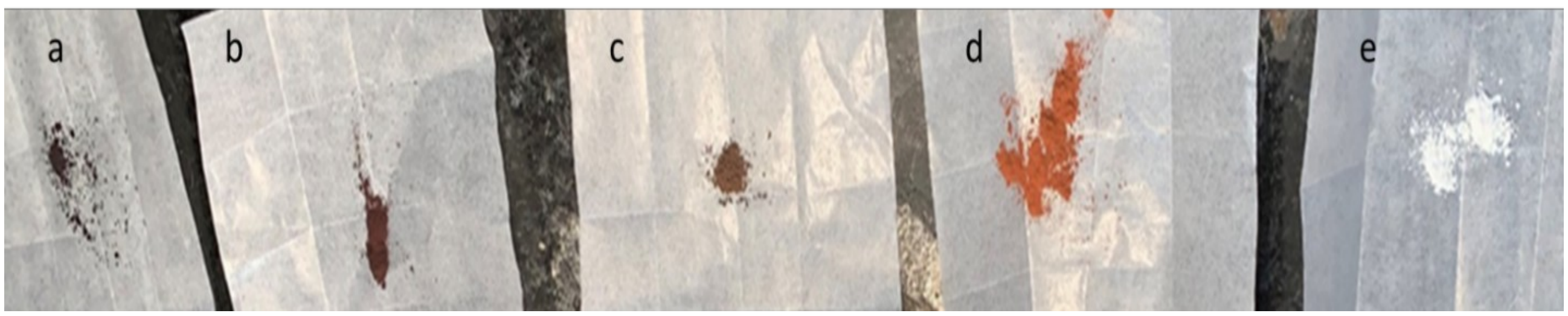

Fig 6.9: Colour of the powder for a) $Sr_2YbRuO_6$, b) $Sr_2YbRu_{0.75}Ta_{0.25}O_6$, c) $Sr_2YbRu_{0.5}Ta_{0.5}O_6$, d) $Sr_2YbRu_{0.25}Ta_{0.75}O_6$, e) $Sr_2YbTaO_6$

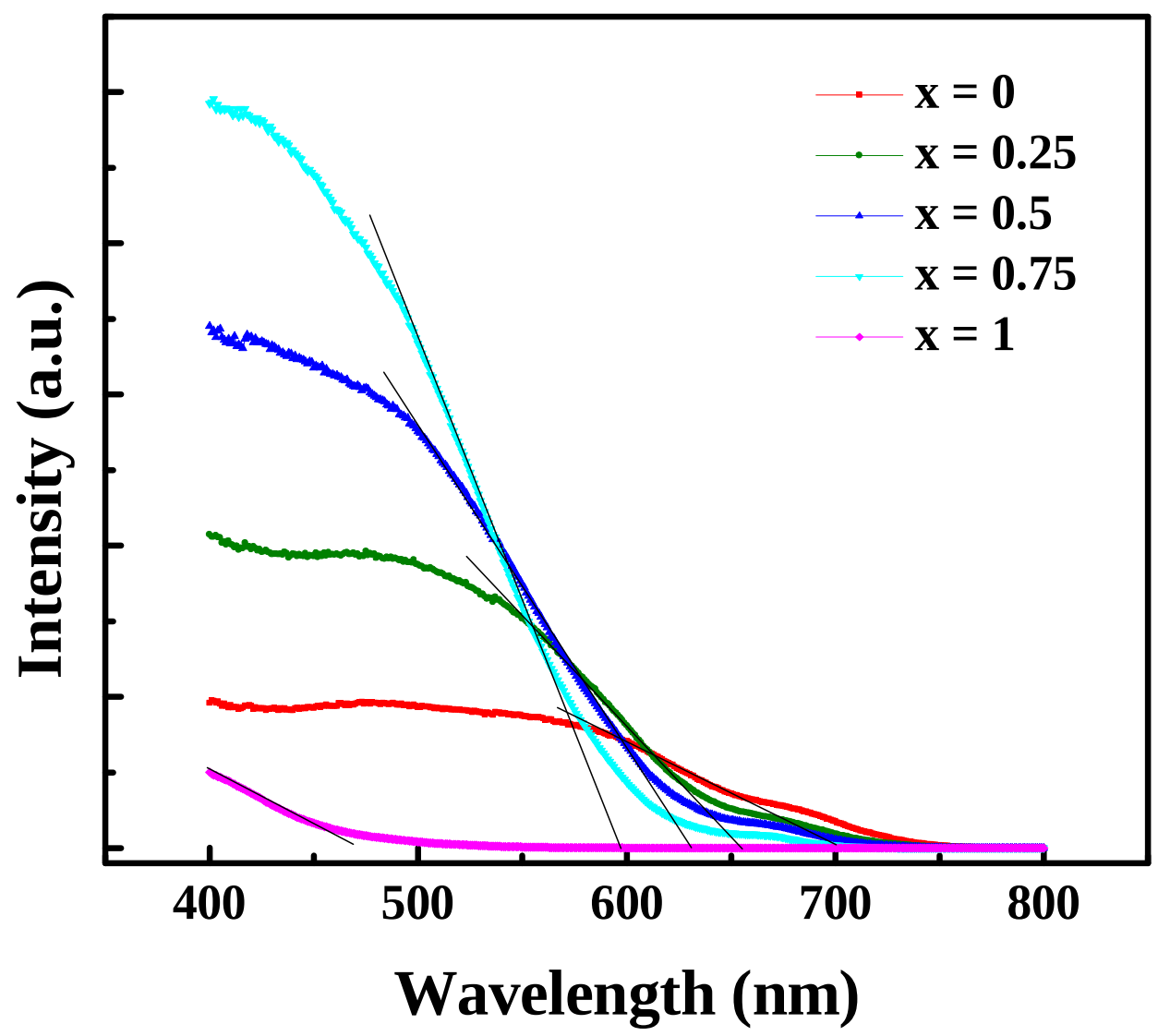

Fig 6.10: Diffuse reflectance spectroscopy (DRS) of the system $Sr_2YbRu_{1-x}Ta_xO_6$ (x = 0, 0.25,0.5, 0.75 and1).

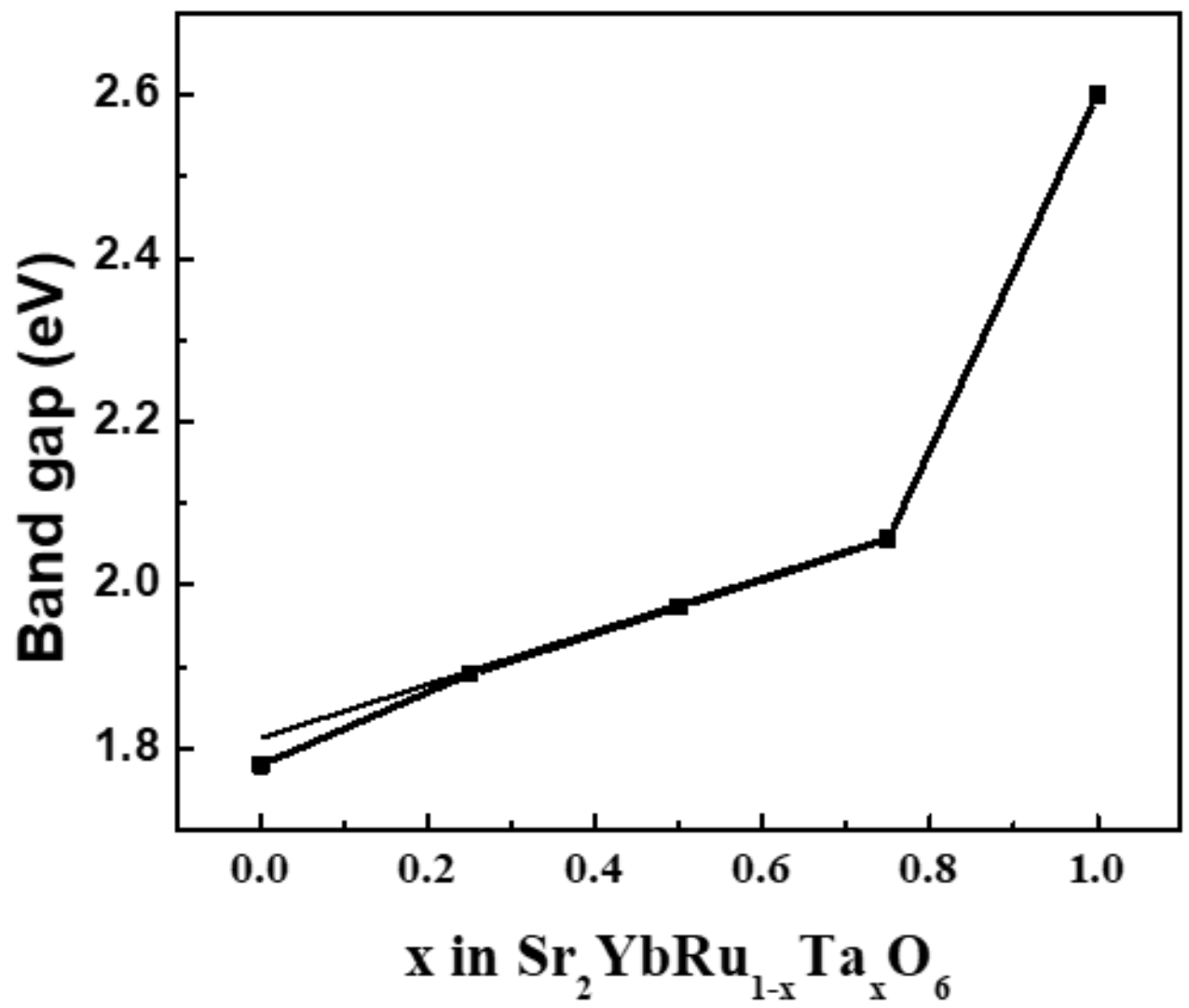

Fig 6.11: Comparison of change in band gap for a) $Sr_2YbRuO_6$ b) $Sr_2YbRu_{0.75}Ta_{0.25}O_6$ c) $Sr_2YbRu_{0.5}Ta_{0.5}O_6$ d) $Sr_2YbRu_{0.25}Ta_{0.75}O_6$ and e) $Sr_2YbTaO_6$.

Table 6.4: Band gap for $Sr_2YbRu_{1-x}Ta_xO_6$ (x = 0, 0.25, 0.5, 0.75 and 1).

| **Compound** | **Band Gap (eV)** |
|---|---|
| $Sr_2YbRuO_6$ | 1.78(1) |
| $Sr_2YbRu_{0.75}Ta_{0.25}O_6$ | 1.90(1) |
| $Sr_2YbRu_{0.5}Ta_{0.5}O_6$ | 1.97(1) |
| $Sr_2YbRu_{0.25}Ta_{0.75}O_6$ | 2.08(1) |
| $Sr_2YbTaO_6$ | 2.65(1) |

### 6.3.4 Photoelectrocatalytic studies

#### a) Mott Schottky measurement

These semiconducting oxide materials have applications in light-harvesting where we can tailor the bandgap of materials for better charge separation and obtaining optimal photoelectrochemical performance [30-35]. Mott-Schottky analysis was performed to determine the nature of these semiconductors as well as obtain the flatband potentials (Fig 6.12, Table 6.4). It was observed from the slope of the Mott-Schottky plots that all the species of $Sr_2YbRu_{1-x}Ta_xO_6$ (x = 0, 0.25, 0.5, 0.75 and 1) exhibited n-type behavior [35-39]. For n-type semiconductors, the flatband potential corresponds to the position of the conduction band, hence we could obtain the conduction band positions for all the compositions. Using the positions of the conduction bands obtained from the Mott-Schottky analysis and the bandgaps obtained from the DRS analysis, the positions of the valence bands were calculated as depicted in Fig. 6.13.

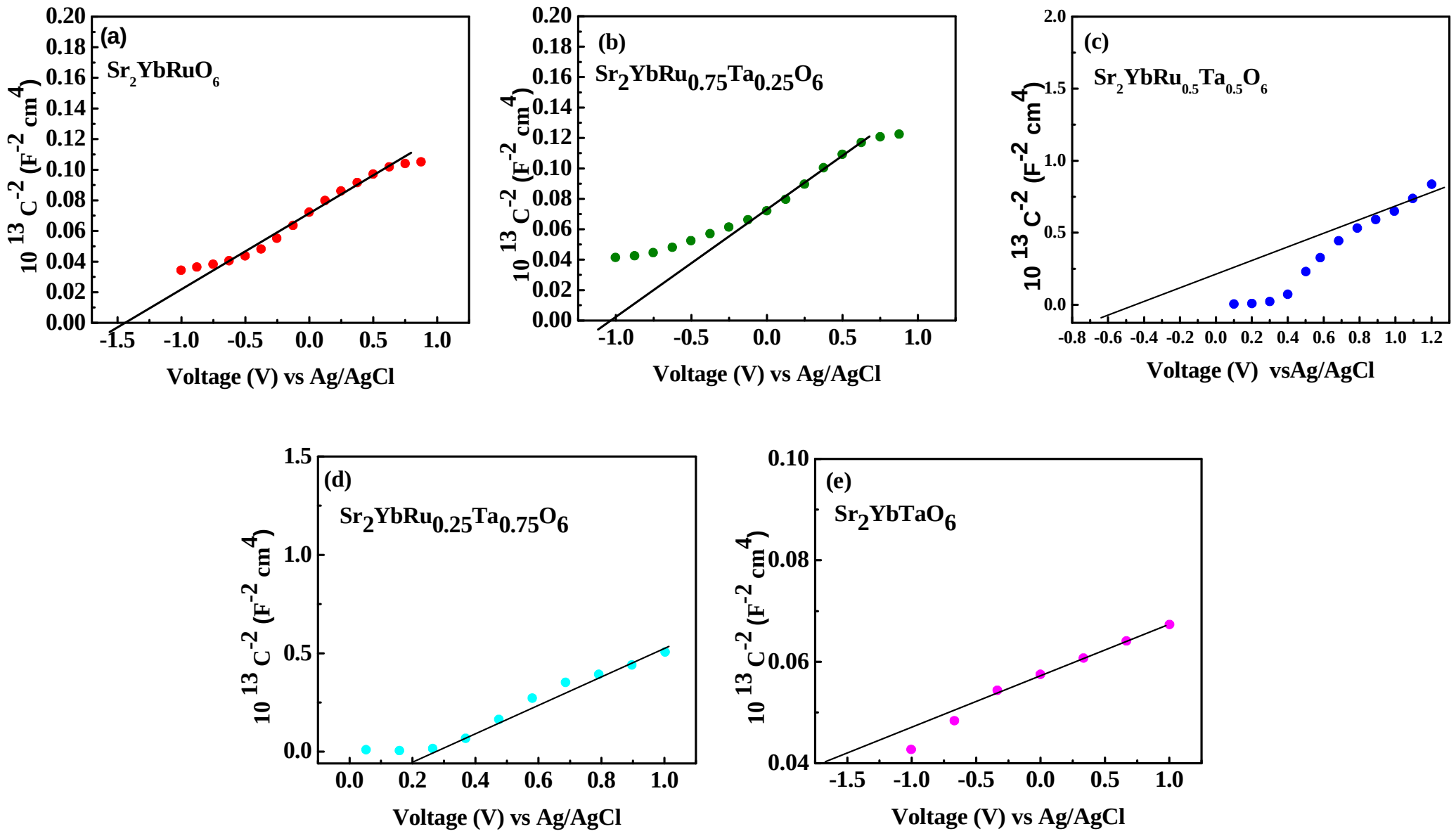

Fig 6.12: Mott Schottky plot for a) $Sr_2YbRuO_6$, b) $Sr_2YbRu_{0.75}Ta_{0.25}O_6$ c) $Sr_2YbRu_{0.5}Ta_{0.5}O_6$ d) $Sr_2YbRu_{0.25}Ta_{0.75}O_6$ and e) $Sr_2YbTaO_6$

Table 6.5: Analysis of Mott Schottky plot for flat band potential with respect to Ag/AgCl.

| Sr.No | Composition | Flat band potential (V) |
|---|---|---|
| 1. | $Sr_2YbRuO_6$ | -1.149 |
| 2. | $Sr_2YbRu_{0.75}Ta_{0.25}O_6$ | -1.050 |
| 3. | $Sr_2YbRu_{0.5}Ta_{0.5}O_6$ | -0.795 |
| 4. | $Sr_2YbRu_{0.25}Ta_{0.75}O_6$ | 0.212 |
| 5. | $Sr_2YbTaO_6$ | -1.630 |

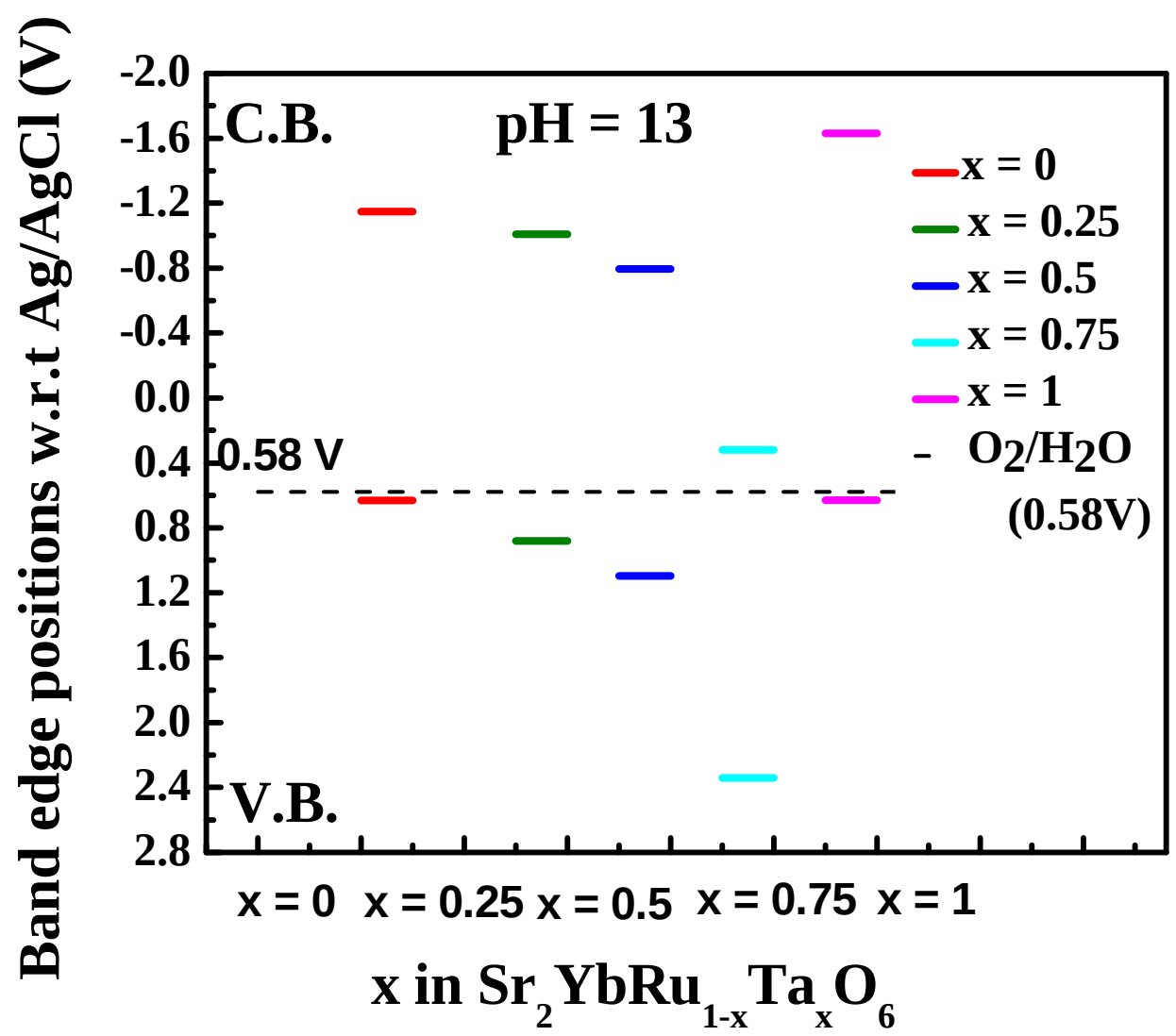

Fig 6.13: Band edge positions for a) $Sr_2YbRuO_6$ b) $Sr_2YbRu_{0.75}Ta_{0.25}O_6$ c) $Sr_2YbRu_{0.5}Ta_{0.5}O_6$, d) $Sr_2YbRu_{0.25}Ta_{0.75}O_6$ e) $Sr_2YbTaO_6$.

**b) Oxygen evolution current density in light and dark**

It was observed (Fig. 6.13) that the position of the conduction and valence bands was suitable to split water electrochemically and produce $O_2$ (OER reaction). On absorbing light radiation, a photoelectron moves from the valence band to the conduction band, producing a hole in the valence band. This hole aids the oxidation of water to produce molecular oxygen. Thus, a photoelectrochemical OER reaction was carried out in an alkaline medium using these perovskites as catalysts. It was observed that the photocurrent density was maximum in the case of $Sr_2YbRuO_6$, and decreases with Ta concentration (Fig. 6.14 and 6.15). $Sr_2YbRuO_6$ exhibited the maximum OER current density (17 µA/cm$^2$) and maximum dark current (12 µA/cm$^2$) at 0.8 V with respect to Ag/AgCl. Fig. 6.15 shows plots of the photocurrent density (current obtained in light- current obtained in dark) for all the species. The onset potential were found to be 0.21, 0.41, 0.43, 0.57, 0.59 and 0.62 V (w.r.t Ag/AgCl), respectively for x = 0, 0.25, 0.5, 0.75 and 1 in $Sr_2YbRu_{1-x}Ta_xO_6$ which are comparable to the state-of-art OER electrocatalyst $RuO_2$ [35]. It was seen that both the dark current (the electrochemical component) and the light current decrease as the amount of Ru decreases in the series. In the literature, it has been reported that perovskite oxides containing Ru are known to have benchmark current densities and lower onset potentials for OER. Thus, the trends observed for the dark current may be easily explained based on this current directly being correlated to the percentage of Ru in the composition. The trends in the photocurrent (light current - dark current) density may be explained by the position of the valence band w.r.t the potential for OER (Fig. 6.13). It is seen that energetically the transfer of holes from the valence band of the perovskite to the oxidation potential level of water is most feasible for x=0 in-phase ($Sr_2YbRuO_6$) in $Sr_2YbRu_{1-x}Ta_xO_6$ followed by the x= 0.25 phase. It is energetically not very favourable for compositions with x = 0.5 – 1.0 to carry out OER, hence almost negligible photocurrent density is observed for higher x values.

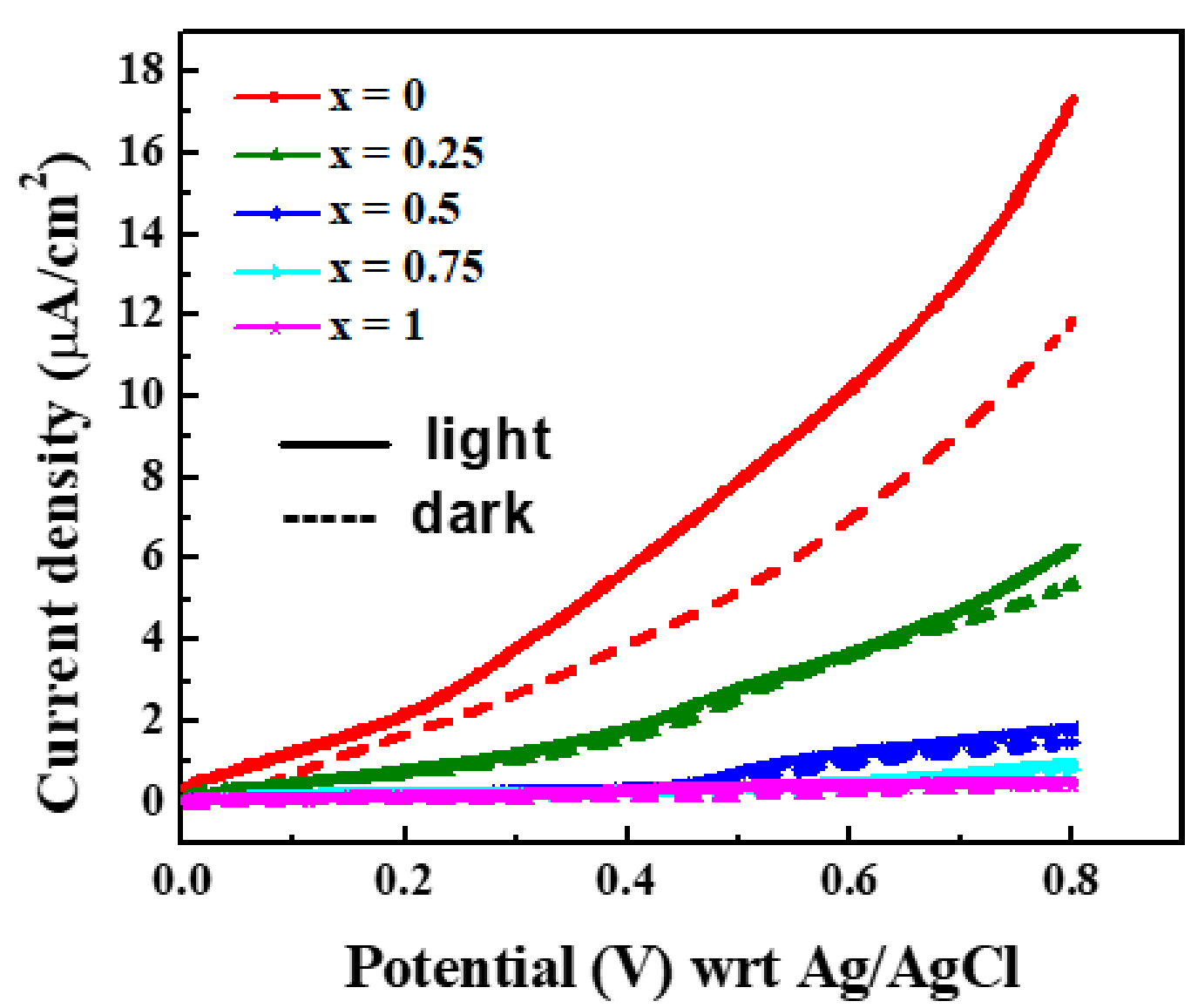

Fig. 6.14: Photoelectrocatalytic current density measurement in presence of light bold line) and in dark (dotted line), with respect to Ag/AgCl, using Pt as a counter electrode.

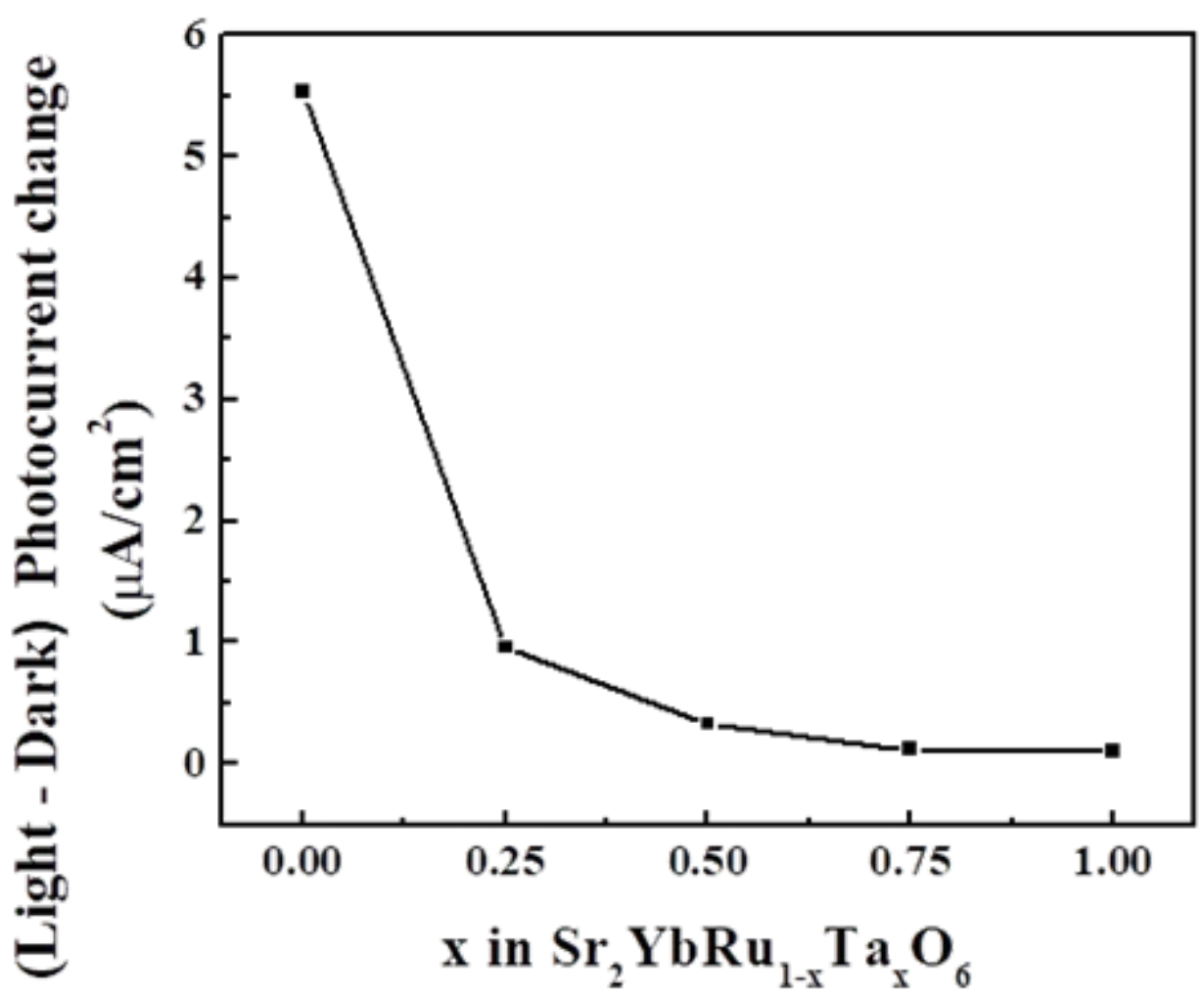

Figure 6.15: Photocurrent in $Sr_2YbRu_{1-x}Ta_xO_6$ as a function of composition.

**6.3.5 X-ray photoelectron spectroscopy analysis of a) Ta and b) Ru**

To investigate the valance states of Ta and Ru, the room temperature X-ray photoelectron spectroscopy (XPS) studies were performed on our powdered samples under an ultravacuum condition. The core level spectra of Ta and Ru ions have been shown in Fig. 6.16 and Fig. 6.17, respectively, and the spectral lines were fit using the Peakfit software. The collected data was analyzed by taking into account background subtraction and spectral normalization. The peaks were deconvoluted with the corresponding binding energy of elements and taking the peak width, peak shape and the FWHM as flexible parameters. The survey spectrum was also standardized to C-1s peak at 284.26 eV. Table 6.6 represents the standard binding energy values of the Ta ions used for the XPS spectra analysis and their corresponding intensity contributions in the core level spectra. Table 6.7 enlisted the standard binding energy values of Ru ions and their corresponding contributions that are used to analyse the XPS spectra. The varied oxidation states of Ta were obtained in +5, +4 and +1 oxidation states and Ru showed +4, +5 and 0 (negligible) oxidation states (Table 6.8). The stoichiometric ratios of Ru:Ta was quantitatively analyzed by taking into account the inelastic mean free path and photoionization cross-section of the elements [29-30] and listed in Table 6.8. The Ta and Ru ratio for all the samples are in acceptable agreement with that of the nominal compositions we have considered.

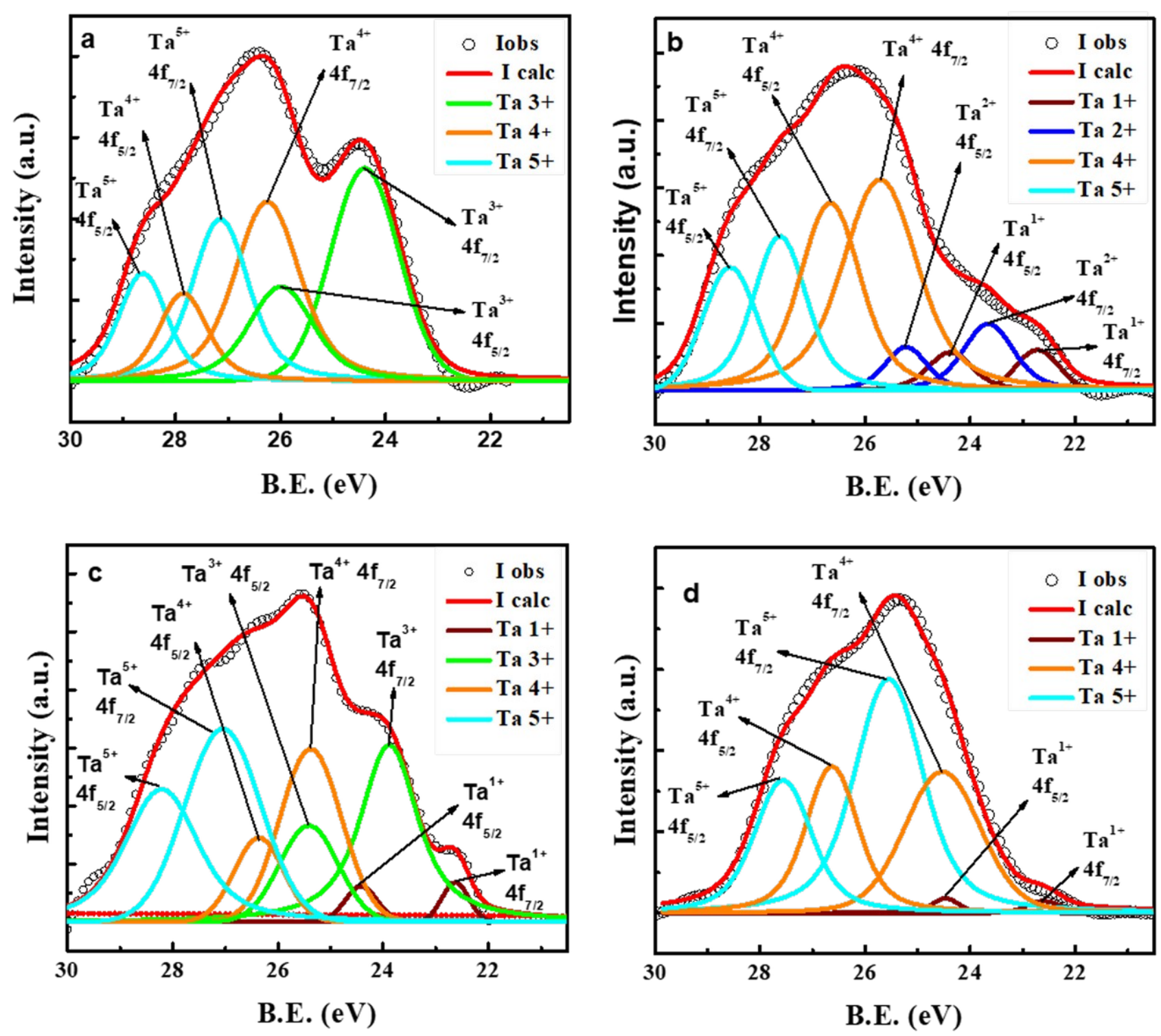

Fig 6.16: XPS analysis of Ta ion for a) $Sr_2YbRu_{0.75}Ta_{0.25}O_6$, b) $Sr_2YbRu_{0.5}Ta_{0.5}O_6$, c)$Sr_2YbRu_{0.75}Ta_{0.25}O_6$ and d) $Sr_2YbTaO_6$.

Table 6.6: XPS analysis to obtain percentages of Ta species.

| | | $Ta^{1+}$ | $Ta^{2+}$ | $Ta^{3+}$ | $Ta^{4+}$ | $Ta^{5+}$ |
|---|---|---|---|---|---|---|
| 1. | **$Sr_2YbRu_{0.75}Ta_{0.25}O_6$** | | | 24.365 [1.535] | 25.735[1.452 ] | 26.893 [1.271] |
| | | | | 26.097 | 27.692[1.106 | 28.341 |
| | | | | [1.468] | ] | [1.089] |
| | **Intensity (%)** | | | **32.282** | **25.124** | **42.594** |
| **2.** | **$Sr_2YbRu_{0.5}Ta_{0.5}O_6$** | 22.723 [0.951] | 23.717 [1.139] | | 25.691[1.683] | 26.600 [1.369] |
| | | 24.410 [1.007] | 25.224 [0.950] | | 27.518[1.205] | 28.568 [1.205] |
| | **Intensity (%)** | **6.553** | **10.634** | | **48.205** | **34.608** |
| **3.** | **$Sr_2YbRu_{0.25}Ta_{0.75}O_6$** | 22.632 [1.089] | | 23.886 [1.386] | 25.379[1.353] | 27.046 [1.205] |
| | | 24.422 [0.950] | | 25.413 [1.436] | 27.361[1.304] | 28.209 [1.370] |
| | **Intensity (%)** | **4.945** | | **24.277** | **49.665** | **21.113** |
| **4.** | **$Sr_2YbTaO_6$** | 22.599 [0.743] | | | 24.545[1.700] | 25.537 [1.535] |
| | | 24.487 [0.594] | | | 26.578[1.172] | 27.454 [1.287] |
| | **Intensity (%)** | **1.770** | | | **42.861** | **56.363** |

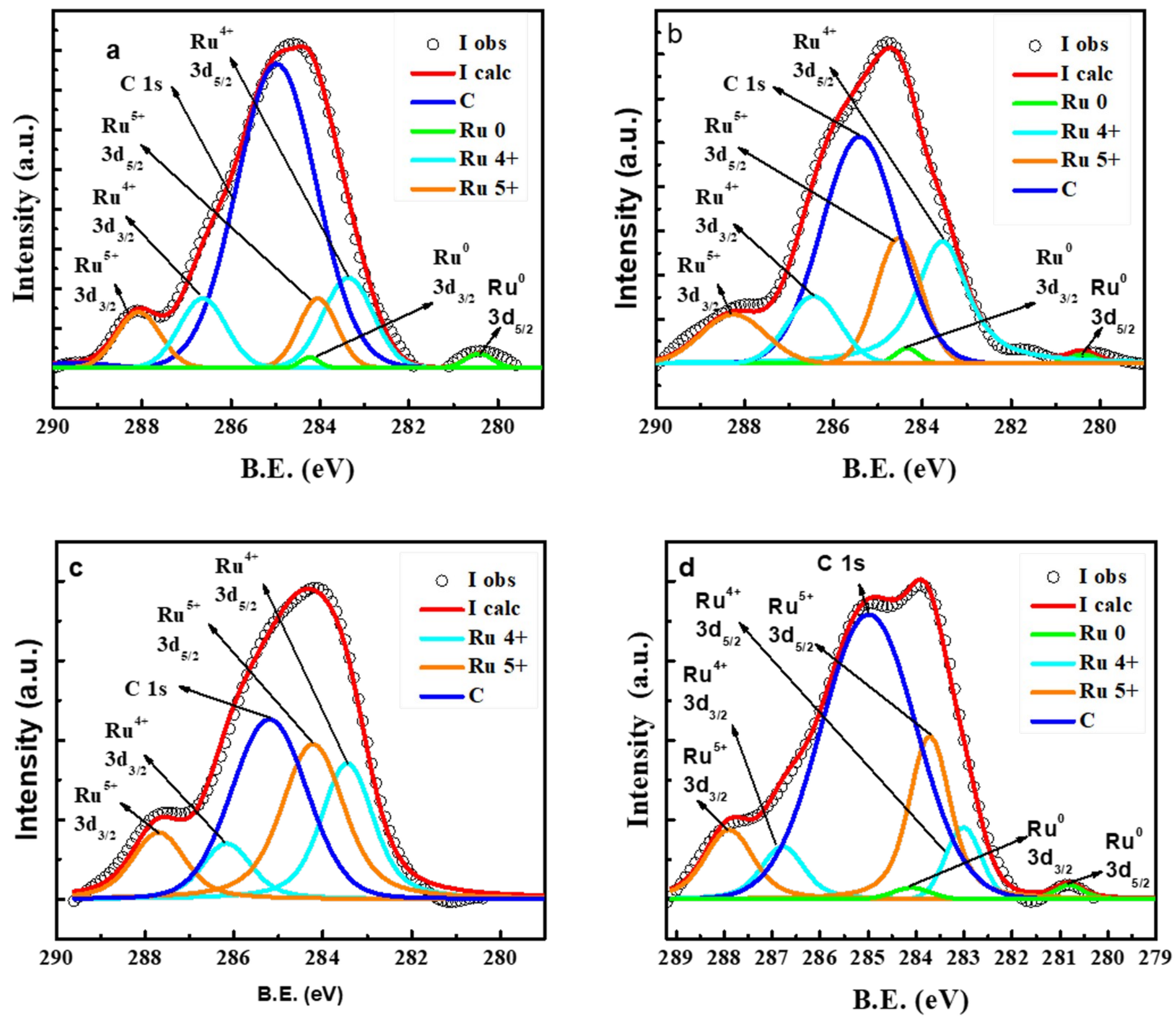

Fig 6.17: XPS analysis of Ru ion in a) $Sr_2YbRuO_6$ b) $Sr_2YbRu_{0.75}Ta_{0.25}O_6$ c) $Sr_2YbRu_{0.5}Ta_{0.5}O_6$ and d) $Sr_2YbRu_{0.25}Ta_{0.75}O_6$.

Table 6.7: XPS analysis to obtain percentages of Ru species.

| | Composition | $Ru^{5+}$ | $Ru^{4+}$ | $Ru^{0}$ | C |
|---|---|---|---|---|---|
| **1.** | **$Sr_2YbRuO_6$** | 284.050[1.012] | 283.370[1.320] | 280.429[0.722] | 284.976 |
| | | 288.082[1.156] | 286.636[1.239] | 284.235[0.495] | |
| | **Intensity (%)** | **37.881** | **57.676** | **4.443** | |
| **2.** | **$Sr_2YbRu_{0.75}Ta_{0.25}O_6$** | 284.553[1.099] | 283.839[1.382] | 280.611[1.099] | 285.368 |
| | | 288.087[1.299] | 286.409[1.099] | 285.240[0.804] | |
| | **Intensity (%)** | **44.512** | **52.147** | **3.341** | |
| **3.** | **$Sr_2YbRu_{0.5}Ta_{0.5}O_6$** | 284.228[1.650] | 283.486[1.382] | | 285.192 |
| | | 287.523[1.403] | 286.087[1.320] | | |
| | **Intensity (%)** | **54.088** | **43.073** | | |
| **4.** | **$Sr_2YbRu_{0.25}Ta_{0.75}O_6$** | 283.731[1.011] | 283.018[0.867] | 280.894[0.701] | 284.984 |
| | | 287.701[1.196] | 286.686[1.052] | 284.101[0.928] | |
| | **Intensity (%)** | **65.864** | **29.040** | **5.097** | |

Table 6.8: XPS analysis to obtain percentages of Ru and Ta species.

| | Composition | Ru 3d $_{3/2}$ | Ru 3d $_{5/2}$ | Ta 4f $_{5/2}$ | Ta 4f $_{7/2}$ |
|---|---|---|---|---|---|
| **1** | **$Sr_2YbRuO_6$ (area)** | 838.694 | 1079.844 | | |
| **2** | **$Sr_2YbRu_{0.75}Ta_{0.25}O_6$ (area)** | 773.314 | 1834.478 | 342.036 | 630.994 |
| | Total stoichiometry of Ru/Ta (3d $_{3/2}$ and 4f $_{5/2}$) | 1834.478/8.48/134 = 1.615 | | 342.036/3.99/138 = 0.619 | |
| | Ratio of stoichiometry of Ru/Ta | 1.620/0.619 = 2.701 | | | |
| **3** | **$Sr_2YbRu_{0.5}Ta_{0.5}O_6$ (area)** | 1710.54 | 3155.818 | 1434.646 | 2612.493 |
| | Total stoichiometry of Ru/Ta (3d $_{3/2}$ and 4f $_{5/2}$) | 3155.818/8.48/134 = 2.777 | | 1434.646/3.99/138 = 2.606 | |
| | Ratio of stoichiometry of Ru/Ta | 2.777/2.606 = 1.065 | | | |
| **4** | **$Sr_2YbRu_{0.25}Ta_{0.75}O_6$ (area)** | 1119.521 | 1791.515 | 1792.510 | 3223.503 |
| | Total stoichiometry of Ru/Ta (3d $_{3/2}$ and 4f $_{5/2}$) | 1791.515/8.48/134 = 1.577 | | 1792.51/3.99/138 = 3.255 | |
| | Ratio of stoichiometry of Ru/Ta | 1.577/3.255 = 0.485 | | | |
| **5** | **$Sr_2YbTaO_6$ (area)** | | | 5292.988 | 1778.384 |

1. For Ru
   The summation of area under peak was divided by inelastic mean free path = 134 nm

   For $3d_{3/2}$ it was divided by photoionization cross section = 5.84

   For $3d_{5/2}$ it was divided by photoionization cross section = 8.48

2. For Ta
   The summation of area under peak was divided by inelastic mean free path = 138 nm

   For $4f_{5/2}$ it was divided by photoionization cross section = 3.99

   For $4f_{7/2}$ it was divided by photoionization cross section = 5.08

Further we have calculated the effective magnetic moment of the samples using the magnetic contributions of the Ru and Ta ions in their respective oxidation states that are obtained from the XPS analysis. The magnetic moments for $Sr_2YbRu_{1-x}Ta_xO_6$ were obtained to be 5.594 $\mu_B$/f.u. (x = 0), 5.469 $\mu_B$/f.u. (x = 0.25), 5.380 $\mu_B$/f.u. (x = 0.5), 5.220 $\mu_B$/f.u. (x = 0.75) and 4.721 $\mu_B$ /f.u. (x = 1). Details of the calculations are given below. The values of the magnetic moments obtained from the calculations are in well agreement with that obtained from the magnetic susceptibility measurements as given in Table 6.3. We can see that the trend of decrease in effective magnetic moment with the decrease in Ru concentration in $Sr_2YbRu_{1-x}Ta_xO_6$ also follows for the effective magnetic moment values obtained using the contributions from the XPS data.

Magnetic moments of the Ru, Yb and Ta ions were taken as follows:

$Ru^{5+}$ (4 $\mu_B$/f.u.), $Ru^{4+}$ (2.83 $\mu_B$/f.u.), $Yb^{3+}$ (4.54 $\mu_B$/f.u.)

$Ta^{4+}$ (1.7 $\mu_B$/f.u.), $Ta^{3+}$ (2.83 $\mu_B$/f.u.), $Ta^{2+}$ (3.87 $\mu_B$/f.u.), $Ta^{1+}$ (4.9 $\mu_B$/f.u.)

Calculations for magnetic moments using the contributions of ions from XPS analysis:

1. For $Sr_2YbRuO_6$

   From XPS

   $Ru^{5+}$: $Ru^{4+}$: $Ru^{0}$:: 37.88% : 57.68% : 4.44%

   $$\mu_{effective} = [0.379(4)^2 + 0.577\,(2.83)^2 + (4.54)^2]^{1/2}$$
   $$= (31.297)^{1/2} = 5.594\ \mu_B/\text{f.u.}$$

2. For $Sr_2YbRu_{0.75}Ta_{0.25}O_6$

   From XPS

   $Ru^{5+}$: $Ru^{4+}$: $Ru^{0}$:: 44.51% : 52.15% : 3.34%

   $Ta^{5+}$: $Ta^{4+}$: $Ta^{3+}$:: 42.59% : 25.12% : 32.28%

   $$\mu_{effective} = [0.75\{0.445(4)^2 + 0.522(2.83)^2\} + 0.25\{0.251(1.7)^2 + 0.323(2.83)^2\} + (4.54)^2]^{1/2}$$
   $$= (29.91)^{1/2} = 5.469\ \mu_B/\text{f.u.}$$

3. For $Sr_2YbRu_{0.5}Ta_{0.5}O_6$

   From XPS

   $Ru^{5+}$: $Ru^{4+}$:: 54.09% : 43.07%

   $Ta^{5+}$: $Ta^{4+}$: $Ta^{2+}$: $Ta^{1+}$: 34.61%: 48.51% : 10.63%: 6.55%

   $$\mu_{effective} = [0.5\{0.541(4)^2 + 0.431(2.83)^2\} + 0.5\{0.485(1.7)^2 + 0.106(3.87)^2 + 0.066(4.9)^2\} + (4.54)^2]^{1/2}$$
   $$= (28.948)^{1/2} = 5.380\ \mu_B/\text{f.u.}$$

4. For $Sr_2YbRu_{0.25}Ta_{0.75}O_6$

   From XPS

   $Ru^{5+}$: $Ru^{4+}$: $Ru^{0}$:: 65.86% : 20.04% : 5.10%

   $Ta^{5+}$: $Ta^{4+}$: $Ta^{3+}$: $Ta^{1+}$: 21.11%: 49.67%: 24.28% : 4.95%

   $$\mu_{effective} = [0.25\{0.659(4)^2 + 0.20(2.83)^2\} + 0.75\{0.497(1.7)^2 + 0.243(2.83)^2 + 0.049(4.9)^2\} + (4.54)^2]^{1/2}$$

   $$= (27.25)^{1/2} = 5.220\ \mu_B/\text{f.u.}$$

5. For $Sr_2YbTaO_6$

   From XPS

   $Ta^{5+}$: $Ta^{4+}$: $Ta^{1+}$: 56.36%: 42.86%: 1.77%

   $$\mu_{effective} = [\{0.429(1.7)^2 + 0.018(4.9)^2\} + (4.54)^2]^{1/2}$$

   $$= (22.28)^{1/2} = 4.721\ \mu_B/\text{f.u.}$$

There have been several reports in the literature which show catalytic reactions for oxygen evolution in alkaline medium on Ru based materials [40-44]. It has been reported that the onset potential value at 10 $A/cm^2$ indicates occurrence of a high OER for Ru based composite in 1M KOH with an overpotential of 12 mV, comparable to that of Ru nanosheet in 0.5 M $H_2SO_4$ at an overpotential of 20 mV [42-43]. Low index faceted growth of branched Ru nanoparticles has been reported to show improved stability for OER performance, as compared to polycrystalline Ru nanoparticles, with a current density of 10 $mA/cm^2$ at an overpotential of 180 mV [42-44].

On analyzing the intercept of Mott Schottky plots (Fig. 6.12) the flat band potential ($V_{fb}$) was calculated and were found to be -1.149 (x = 0), -1.050 (x = 0.25), -0.795 (x = 0.5), 0.212 (x = 0.75) and -1.630 (x = 1) respectively. It was observed that the obtained values are in the range with the values reported in the literature for similar catalysts [45-46]. These data present a balance between charge carrier density ($N_d$): a kinetic parameter and flat band potential ($V_{fb}$): a thermodynamic parameter which governs the photocurrent [47]. In $Sr_2YbRu_{1-x}Ta_xO_6$ it was observed that the photoelectrocatalytic current decrease with increase in x which could be attributed to decrease in charge carriers, greater optical band gap, as well the position of the conduction band edges obtained away from the fermi energy level, in case of $Sr_2YbTaO_6$. The greater energy gap between the valence band and conduction band as the value of x increases could also be attributed to decrease in current density (fig. 6.13). Further similar trend in light and dark for all values of x suggests that it is thermodynamically driven where larger density of states are present closer to the Fermi energy level of redox potential, at low values of x, for higher current density (fig. 6.14 and 6.15). In similar reports in n type semiconductor, it has been observed that the $V_{fb}$ is directly proportional to the energy of the conduction band edge, thus it is very close to conduction band edge than that for a more negative $V_{fb}$, thermodynamically. Here we have assumed the Fermi energy level of electrolyte to be conduction band edge for the following n type materials [48-49]. The slopes of $1/C^2$ vs V (Mott Schottky plot) for all compositions were found to be all positive indicating n type behaviour in $Sr_2YbRu_{1-x}Ta_xO_6$ (x = 0, 0.25, 0.5, 0.75 and 1) as evident from the XPS measurements, also showing excess negative charge per unit cell originating from Ru and Ta ions which were obtained to be 0.79 (x = 0), 0.74 (x = 0.25), 1.68 (x = 0.5), 1.11 (x = 0.75) and 0.45 (x = 1) respectively. The charge carrier density in $Sr_2YbRu_{1-x}Ta_xO_6$ (x = 0, 0.25, 0.5, 0.75 and 1) was found to be $2.64*10^{16}$(x = 0), $1.959*10^{16}$ (x = 0.25), $2.989*10^{15}$ (x = 0.5), $1.922*10^{15}$ (x = 0.75) and $1.284*10^{17}$ (x = 1). It was found that as the excess negative charge found from XPS in the system decreases from 0.797 for x = 0 to 0.745 for x = 0.25, it results in decrease in charge carrier density also. However the decrease in charge carrier density for x = 0.5, despite showing greater negative charge in the system, as compared to x = 0.25, could be attributed to band positions valence band and conduction band with respect to Fermi energy level [50] and, increased octahedral distortion which further results in charge trapping. The charge carrier density for x = 0.75 despite showing less distorted structure as compared to x = 0.5 possess lower value, of charge carrier density as compared to x = 0.5 due to low value of negative charge present in system. The position of the valence band maxima and conduction band maxima must lie lower and higher in energy than the oxidation and reduction reaction free energy in order for these reactions to be thermodynamically favourable [50]. It has been reported in literature that sodium doping increases the oxidation state of ruthenium thereby displacing positively O-p band and Ru-d band centres, weakening Ru adsorbate bonds [51]. $SrRuO_3$ contains $Ru^{4+}(4d^4)$ with relatively extended d orbitals, $Ru^{4+}$ which adopts a low spin configuration ($t_2g^3$ $t_2g^1$) favouring Jahn-Teller distortion, that split the three fold $t_2g$ into low dxz orbitals and two higher dyz and dxy [59]. The structure sensitive ordered ($Ru^{4+}$) to disordered ($Ru^{5+}$) transition and the development of active sites for OER is determined by synergy between electronic effects (surface energy) and morphological effects (surface defects) [52]. Activity is controlled by surface density of defects for example in the stability of Ru cations rather than binding energy between substrate and oxygenated species [53]. Thus, the OER on $SrRuO_3$ films must be based on a careful equilibrium between two opposing effects- the balance between active and unstable $Ru^{5+}$ sites along with stable and inactive $Ru^{4+}$ sites. An ordered ($Ru^{4+}$) to disordered ($Ru^{>4+}$) is strongly structure sensitive with onset potential increasing from more defective films with (111) orientation to highly ordered and less defective (001) films [54]. The structure of the composition plays a major role in the electronic structure manifested by the increased Jahn teller mediated exchange of the Ru –O-O-Ru and thus of the band edge position with respect to the redox chemical reaction at the interface of the electrolyte [54]. Due to greater disorder present in the system, it results in system to show greater defects and thus it changes the Ru coordination number. From Rietveld refinement it was observed that as the value of x increases the Sr-O1 bond decreases and the system shows greater octahedral distortion at higher values of x and particularly for x = 0.25 and 0.5, which results in change in Ru coordination number and results in system to show greater Jahn Teller distortion at Ru site. These combined factors result in change in different binding energy of the Ru at the surface with the electrolyte and which also changes the charge carrier density.

In the XPS measurement we have obtained the oxidation states of Ru and Ta ions, which were mainly found to be mainly in +5 and +4 oxidation states. It was observed that as the value of x increases the $Ru^{5+}$ concentration was found to increase. This also suggests greater ferromagnetic interactions arising from $Ru^{5+}$- O – O – $Ru^{4+}$ interactions at higher value of x [59]. The spectral lines were fit using Peakfit software. The data analysis involved spectral normalization, background subtraction and Voigt-area curve fitting. Best fits were evaluated using root mean square measure, with line shape, peak width (FWHM) and binding energies as adjustable parameters. The survey spectrum was standardized to adventitious $C_{1s}$ peak at *284.26 eV* (fig. 6.15 and 6.16). The total negative charge present in the composition $Sr_2YbRu_{1-x}Ta_xO_6$ per unit cell originating from Ru and Ta ions were found to be 0.79 (x = 0), 0.74 (x = 0.25), 1.68 (x = 0.5), 1.11 (x = 0.75) and 0.45 (x = 1) respectively. Quantification of area under the peak of Ru and Ta ions to find the correct stoichiometric ratios of Ru and Ta ions were found to be similar to their stoichiometric ratios. The ratios of Ru: Ta in $Sr_2YbRu_{1-x}Ta_xO_6$ (x = 0.25, 0.5 and 0.75) were found out to be 2.701, 1.065 and 0.485 which were found to be similar to their formula ratios (Ru/Ta) to be as: 3 (0.75/0.25) for x = 0.25, 1 (0.5/0.5) for x = 0.5 and 0.33 (0.25/0.75) for x = 0.75 respectively. The ratios of the stoichiometry of Ru and Ta were found by taking into account their spin multiplicity, photoionization cross section for that particular orbital, and the inelastic mean free path of the element present in a particular orbital as found from these reference papers [54-56]. For obtaining stoichiometric ratios of Ru and Ta, firstly the area under the peak of an orbital having a particular value was taken for each elements and were divided by the photoionization cross-section and inelastic mean free path length. The final ratios of the Ru and Ta for each element were taken in the stoichiometric ratios. The peaks for Ru and Ta were found to be similar in their binding energies as obtained from reference [57-58] after taking into account their peak shape for both Ru and Ta elements by same magnitude, (C corrected).

On correlation of structural with magnetic properties and optical properties it can be said from the Rietveld refinement that as the value of x increases, the distance between the Sr-O bond distance increases and the overall octahedral distortion between the Yb – O – Ru bond angle decreases. It can be said that the increased octahedral distortion leads to disruption in the magnetic spin ordering, taking place due to the ordered Yb – O – Ru atoms. It can be said that the these factors results in the increased spin canting taking place which as a result leads to increased ferromagnetic moments generated from the canted antiferromagnetic spins of the Ru ions. Furthermore, it can be said that this increase in tolerance factor as the value of x increases is responsible for increased disordered arrangement of oxidation states of Ru and Ta ions, which disrupts the A-type antiferromagnetic arrangement. As the value of x increases the Yb – O – Ru long range ordering decreases, which is responsible for the long range ordering of the A-type antiferromagnetic structure. The ferromagnetic moments generated from the Ru – O – O – Ru interaction is responsible for the increased ferromagnetic interaction. In the G and C type antiferromagnetic arrangement of the ions has been reported to show increased Ru – O – O – Ru ordering which can be correlated in our case for higher values of x.

On correlation of the optical bandgap with the magnetic ordering and structural ordering it can be said that due to the increased long range ordering of the orbitals in $Sr_2YbRuO_6$, with no Ta doping at the B-site, it results in decrease in the bandgap by 0.04 eV, as expected from linearity on plotting the bandgap values for $Sr_2YbRu_{1-x}Ta_xO_6$ (x = 0, 0.25,0.5 and 0.75).

## 6.4 Conclusions

$Sr_2YbRu_{1-x}Ta_xO_6$ (x = 0, 0.25, 0.5, 0.75 and 1) were obtained as pure phase with monoclinic spacegroup, $P2_1/n$. All the compositions showed antiferromagnetic behaviour with the decrease in the Weiss temperature as the Ta concentration increased with maximum qW in the case of $Sr_2YbRuO_6$. The bandgap increased linearly with an increase in Ta concentration except in the case of $Sr_2YbRuO_6$, probably due to increased magnetic ordering. Mott-Schottky plots indicated n-type behaviour for all the compositions. These plots further yielded the flatband potentials that correspond to the conduction band potentials for n-type semiconductors. Using the value of bandgaps and the positions of the conduction bands, a band energy diagram was constructed and it was seen that the system is suitable for the OER reaction. Photo electrocatalytic OER was carried out and it was observed that the photocurrent as well as the dark current decrease with x in the compositions $Sr_2YbRu_{1-x}Ta_xO_6$ increases from 0 to 1. The decrease in Ru content was attributed to the decrease in the dark current, while the positions of the valence band of the different compositions w.r.t the oxidation potential of water were attributed to the trends seen in the photocurrent.

# Chapter VII

# Conclusions and Future Prospects

# Chapter VII

## Conclusions and Future Prospects

Structural properties of several perovskites based on their structural distortions and magnetic properties were investigated. They were refined in orthorhombic *Pnma* and monoclinic $P2_1/n$ spacegroup which showed that the octahedral distortion increases and lattice volume decrease as the A-site radii decrease as mentioned in Table 7.1. From the DC magnetic susceptibility studies, a bifurcation between ZFC and FC suggest spin-glass behaviour as was observed in several of these new compositions.

In $Ho_{1-x}Pr_x(Mn_{0.5}Cr_{0.5})O_3$ (x = 0, 0.3, 0.5 and 0.7) the magnetic susceptibility was found to increase while it decreases in $Ho_{1-x}Pr_x(Fe_{0.25}Mn_{0.25}Cr_{0.5})O_3$ (x = 0, 0.5 and 1) as the value of x increases which could be attributed to Ho-Mn interaction and competing anisotropy of orbitals of Pr and Fe atoms. The Dzyaloshinskii-Moriya interaction plays a major role in the spin interactions of the transition metal ions. It was observed that the superexchange interaction between Pr and Mn ions depends on the charge, crystalline structure and the relative orientation of the corresponding orbitals. It was observed that incorporating Fe at Mn site results in the system to show greater anisotropy because as a result of competing anisotropy with Pr orbitals and thus results in a decrease in magnetic moment. The octahedral distortion with the change in A and B-site substitution has been mentioned in Table 7.1. An increased octahedral distortion and decreased B-O bond distances increase the Dzyaloshinskii Moriya interaction. This increased exchange interaction with the easy orientation of the spins results in a more negative value of Curie Weiss temperature but a lower value of the magnetic transition temperature has been compared in Table 7.1. It was also observed that the decreased anisotropy of rare-earth from Pr to La, further results in a decrease in spin glass temperature from 245 K to 101 K in $PrFe_{0.25}Mn_{0.25}Cr_{0.5}O_3$ and $LaFe_{0.25}Mn_{0.25}Cr_{0.5}O_3$ compositions respectively. A rapid change in magnetization at low temperatures was observed in $PrFe_{0.25}Mn_{0.25}Cr_{0.5}O_3$ which suggests a possibility of spin reorientation of rare-earth ions taking place.

In studies on a Ruthenium-deficient perovskite, $PrMn_{0.5}Ru_{0.35}O_3$ a Ru vacancy of 0.15(2) per formula unit was determined. The increased octahedral distortion and compression of the *b* lattice parameter further suggests the increased structural distortion in this structure. From the DC magnetic susceptibility data, the Ru-deficient perovskite was found to show ferromagnetic transition at temperatures below 90 K (Table 7.1) and a magnetic anomaly was observed at 18 K which was attributed to disorder-induced ferromagnetic ordering. Furthermore, the increased magnetic moment in ZFC at lower temperatures could be attributed to the magnetic ordering of the 'Pr' spins and Mn – O – Ru ordering. The lack of saturation magnetization even at high fields could be attributed to competing for ferromagnetic as well as antiferromagnetic interaction or absence of any long-range ordering in the system. A deviation from the Curie Weiss law at temperatures below 182 K suggests weak ferromagnetic behaviour arising from the canted antiferromagnetic alignment of the spins of the Mn ions. Greater octahedral distortion along the axial axis with Mn/Ru – $O_1$ – Mn/Ru bond angle is 157.15(1)$^{\circ}$ while the equatorial bond angle is 168.88(1)$^{\circ}$, (Table 7.1) and decreased *b* lattice parameter suggests greater B-site vacancy particularly along the axial direction. The core-level XPS studies confirmed that the majority of the Mn and Ru ions are in +3 and +5 oxidation states respectively.

New single phases of $La_{0.5-x}Sm_xPr_{0.5}CrO_3$ (x = 0, 0.1, 0.2, 0.3, 0.4 and 0.5) were successfully synthesized. Detailed magnetic studies were carried out, which showed that the Néel temperature was found to decrease with an increase in x, probably due to the increased structural distortion. Increased spin reorientation, as a result of reduced anisotropy, manifests in ordered antiferromagnetic alignment of rare-earth ions with respect to transition metal ions thus resulting in negative magnetization for x = 0.4 and 0.5 values. Evidence from FCC and FCW for x= 0 and 0.1, compositions suggest the possibility of the spin reorientation transition and magnetic transition as well. The ZFC and ZFW show a negative exchange bias for $La_{0.1}Sm_{0.4}Pr_{0.5}CrO_3$, at temperatures below 230 K, and temperatures closer to the Neel temperature (Table 7.1) and compensation temperature. The decrease in spin anisotropy plays a major role in an increased degree of spin canting and thus as a result leads to easy reorientation of canted spins which matches with the spin orientation of the rare-earth spins. A rapid increase in the moment at temperatures below 50 K was observed which could be attributed to rare-earth ordering and a bifurcation in FCC (field cooled cooling) and FCW (field cooled warming) for x = 0 and 0.1 in the temperature range 25-160 K further suggests magnetic phase transition. This degree of anisotropy present in the crystal system depending on the octahedral distortion leads to a correlation between the Néel temperature and spin reorientation temperature which depends on unidirectional anisotropy, structural distortion and Dzyaloshinskii Moriya interaction (Table 7.1).

New series of perovskite oxides $R_{0.5}Sr_{0.5}Fe_{0.5}Mn_{0.5}O_3$ (R: Gd, Nd, and Pr) were synthesized in pure phase and characterized in orthorhombic *Pnma* space group. The structural analysis from Rietveld refinement shows that the structures were found to show octahedral distortion primarily from octahedral tilting and Jahn-Teller distortion which was found to be maximum for $Gd_{0.5}Sr_{0.5}Fe_{0.5}Mn_{0.5}O_3$ followed by $Nd_{0.5}Sr_{0.5}Fe_{0.5}Mn_{0.5}O_3$ and $Pr_{0.5}Sr_{0.5}Fe_{0.5}Mn_{0.5}O_3$ compositions. The elongation of the *b* lattice parameter and the increased difference between the B-$O_{axial}$ and B-$O_{equatorial}$ confirmed maximum Jahn-Teller distortion in $Gd_{0.5}Sr_{0.5}Fe_{0.5}Mn_{0.5}O_3$ (Table 7.1). A canted antiferromagnetic nature was observed for $Gd_{0.5}Sr_{0.5}Fe_{0.5}Mn_{0.5}O_3$ at temperatures below 49 K. A spin-glass like behaviour for $Nd_{0.5}Sr_{0.5}Fe_{0.5}Mn_{0.5}O_3$ and $Pr_{0.5}Sr_{0.5}Fe_{0.5}Mn_{0.5}O_3$ at temperatures 47 K and 58 K respectively was observed. The non-saturating nature of the hysteresis loop at 2 K for $Nd_{0.5}Sr_{0.5}Fe_{0.5}Mn_{0.5}O_3$ and $Pr_{0.5}Sr_{0.5}Fe_{0.5}Mn_{0.5}O_3$ suggests the existence of antiferromagnetic and ferromagnetism or short-range ordering of the spins. The change octahedral distortion on substitution Sr at A-site leads to change in bond distances B-site octahedra and it was observed that $Pr_{0.5}Sr_{0.5}Fe_{0.5}Mn_{0.5}O_3$ shows a shorter B-O bond length than $PrFe_{0.25}Mn_{0.25}Cr_{0.5}O_3$ as mentioned in Table 7.1 XPS studies confirmed the presence of varied oxidation states of $Mn^{2+}$, $Mn^{3+}$, $Mn^{4+}$ and $Fe^{2+}$, $Fe^{3+}$ and $Fe^{4+}$ with major oxidation state in +3 and +4 states for all the three compositions.

$Sr_2YbRu_{1-x}Ta_xO_6$ (x = 0, 0.25, 0.5, 0.75 and 1) were obtained as pure phases with monoclinic spacegroup, $P2_1/n$. All the compositions showed antiferromagnetic behaviour with a decrease in the Weiss temperature as the Ta concentration increased with maximum $\theta_{CW}$ in the case of $Sr_2YbRuO_6$. (Table 7.1). The orbital ordering plays a major role in magnetic behaviour based on double exchange interaction between the B-site ions. The bandgap increased linearly with an increase in Ta concentration except in the case of $Sr_2YbRuO_6$, probably due to increased magnetic ordering. Mott-Schottky plots indicated n-type behaviour for all the compositions. These plots further yielded the flatband potentials that corresponded to the conduction band potentials for n-type semiconductors. Using the value of bandgaps and the positions of the conduction bands, a band energy diagram was constructed and it was seen that the system is suitable for the OER reaction. Photoelectrocatalytic OER was carried out and it was observed that the photocurrent as well as the dark current decrease with x in the compositions $Sr_2YbRu_{1-x}Ta_xO_6$ as x increases from 0 to 1. The decrease in Ru content was attributed to the decrease in the dark current, while the positions of the valence band of the different compositions w.r.t the oxidation potential of water were attributed to the trends seen in the photocurrent.

The above studies on these complex substituted perovskite structures have a great potential for future studies and applications. However, more detailed studies in obtaining the exact spin structure through neutron diffraction, and anisotropic magnetic data on single crystals could allow physicists to explore magnetic devices applications. Further studying thin films of these complex perovskites would be essential for taking them for applications. We predict that the spin reorientation transition of the rare-earth with respect to the transition metal ion as observed in the compositions could be changed by applying an electric field which can have its implementation in multiferroic device-based application. Also tuning the rare-earth composition based on its size and degree of anisotropy could be optimized to get the appropriate magnetic spin structure and the transition temperature. The B-site magnetic exchange interaction which determines the transition temperature could be adjusted based on varying transition metal oxidation states which thus could depend on the degree of magnetic exchange interaction. The structural properties based on the magnetic orbital interaction in turn shows dependence on the structural distortion and thus result in a change in defect induced transition, taking place as has been observed in $PrMn_{0.5}Ru_{0.35}O_3$. The short-range and long-range ordering of the perovskite structure needs to be optimized based on the structural properties and the magnetic exchange interaction. This spin reorientation depends on the orbital exchange interaction and in turn on the structural distortion and could be tuned and easily optimized. The complex spin structure induced by an increased disorder of the B-site and A-site atoms in the systems leading to magnetic transition can have its application in magnetocaloric materials. The charge and structural distortion of the B-site cations should be tuned to optimize the electronic and photoelectrocatalytic properties, which in turn depends on the size and the orbital interaction.

It was observed majorly in all the compositions that with increase in Curie temperature the ZFC and FC bifurcation temperature was also found to increase (Table 7.1). The rare-earth ordering temperature was also related with the increase in Curie temperature. The temperature at which the rare-earth and transition metal ion interaction takes place was found to be directly proportional with the magnetic transition temperature for all these compositions. It was also observed that lower value of magnetic moment (observed) from the inverse magnetic susceptibility in comparison with the calculated magnetic moment was found, particularly in compositions containing non-magnetic atoms at A-site, such as La (Table 7.1), and thus suggesting greater degree of spin alignment taking place, majorly offered by the increased spin alignment of the spins by the transition metal ions in these compositions. The decreased B-O bond distance and decreased B-O-B bond angle suggests, increased Dzyaloshinskii Moriya interaction which results in decrease in transition metal ion interaction as well as a decrease in temperature at which ZFC and FC bifurcation takes place. In order for the materials to show correlation between magnetic and structural properties arising from increased interaction between rare-earth and transition metal ions for its application in exchange bias and in multiferroic materials, it can be concluded that the magnetic exchange interaction of the transition metal ion should not be very strongly correlated, and the rare-earth should have moderate anisotropy value for the appropriate orbital interaction between the transition metal ions, to take place, for example in $PrFe_{0.25}Mn_{0.25}Cr_{0.5}O_3$. The increased orbital interaction could have its application in magnetically induced photoelectrocatalytic activities. The displacement of positively O-p band and Ru-d band centres thereby weakens the Ru adsorbate bonds, along with the change in oxidation states of Ru whereby a low spin configuration ($t_2g^3$ $t_2g^1$) of $Ru^{4+}$ favours Jahn-Teller distortion and long range ordering of Yb-O-Ru results in the system to show higher photoelectrocatalytic activity at lower values of x in $Sr_2YbRu_{1-x}Ta_xO_6$ (x = 0, 0.25, 0.5, 0.75 and 1). A similar correlation of structure based orbital interaction was found in $Gd_{0.5}Sr_{0.5}Fe_{0.5}Mn_{0.5}O_3$, where an increased octahedral distortion and the size mismatch between the A site and B site compositions results in chemical pressure in the system which probably results in disorder induced absence of long range magnetic ordering. The chemical pressure has been reported to favour the lattice stress, which generates Fermi dirac points as a result of absence of this symmetry [1]. The disordered arrangement of the electron density around the fermi energy level results in creation of these type of topological insulators [2]. If there is arrangement of long-range ordering taking place at the B-site ions, coupled with the anisotropic orbital interaction the rare earth orbitals, coupled along with the lattice stain, then it might result in creation of these topological kind of material. Another important observation made was that if the incommensurate arrangement of these B-site ions takes place which results in creation of spiral spin order and thus resulting in the creation of magnetic symmetry induced creation of polarization stress in the material [3]. On similar lines we predict that this B-site arrangement should be maintained in such arrangement by varying of the oxidation states of the ions which are Jahn Teller active. This Jahn Teller distortion induces anisotropy in the system. Another correlation that could be made about the easy orientation the spins with respect to the paramagnetic to antiferromagnetic or antiferromagnetic to ferromagnetic transition, is that with decrease in temperature it could be emphasized that increased spin glass transition (bifurcation temperature between ZFC and FC) temperature is directly proportional to the magnetic transition taking place, which also suggests that the short range and long range interactions are directly related with each other. The dominance of a particular magnetic ordering that is being favoured by charge ordering results in higher magnetic transitions, for example in $La_{0.5}Pr_{0.5}Fe_{0.25}Mn_{0.25}Cr_{0.5}O_3$. Another interesting observation made from Table 7.1 was that in La containing perovskites there is an increased difference between the observed and calculated magnetic moment values obtained from the Curie Weiss law which suggests the short range spin ordering or covalent interactions taking place. For application of these materials in spin hall effect, these materials should have greater spin orbit coupling, along with greater octahedral distortion, along with the rare earth and transition metal ion orbitals interacting together and thus inducing greater anisotropy to the system, which can have its application in spin hall effect materials, for example in $HoFe_{0.25}Mn_{0.25}Cr_{0.5}O_3$.

For materials in magneto-optical coupling it could have its application in optical data storage devices, because the bandgap is also related to the octahedral distortion [4-6] as we observed in our case in $Sr_2YbRuO_6$, where the bandgap showed a decrease by 0.04 eV as expected from the linearity. Greater the octahedral distortion results in more splitting of the orbitals taking place, resulting in smaller bandgap [4-6]. $Sr_2YbRuO_6$ shows a smaller band gap than expected due to the coupling of the orbitals taking place also suggesting long range antiferromagnetic arrangement of Yb-O-Ru. Detailed studies in magneto-optic coupling needs to be studied in this composition for its application in data storage and in catalysis. Studies where the spin ordering of electrons due to magnetic ordering resulting in enhanced photoelectrocatalytic activies needs to be investigated as future studies in $Sr_2YbRu_{1-x}Ta_xO_6$ (x = 0, 0.25, 0.5, 0.75 and 1) compositions [7]. Furthermore, studies where thin films showing Jahn Teller active ions results in increased magnetic exchange interaction between the thin films also needs to be investigated in these compositions [8].

Manganites are famous mostly for colossal magnetoresistance effects which involves phase separation between ferromagnetic phase and charge ordered type antiferromagnetic phase.

However, it has been predicted that charge domains walls may be stabilized by charge defects or free charge carriers. In $RMnO_3$ showing ferroelectricity is topologically inevitable because of highly curved vortex cores [9-10]. Measurements such as conductive AFM, piezoresponse force microscopy and Kelvin probe force microscopy at low temperatures needs to be investigated for finding applications of materials in topological insulators [9-10]. Since the band gap of $RMnO_3$ (R: Sc, Y, Ho or Lu) is relatively small that is 1.7 eV, so the conduction property of ferroelectric vortices may show interesting behavior [9-10]. The spin arrangement depends on exchange interaction depends on exchange strength and interatomic distances between the B-B atoms, for example in $Sr_2YbRu_{1-x}Ta_xO_6$ (x = 0, 0.25, 0.5, 0.75 and 1). Considering valence of $Mn^{(3+x)}$ or equivalent configuration of Fe or Cr ions, it can be said that the spin factors depends on strong Hund coupling, where the spin up channel of $t_2g$ orbitals are fully occupied with the spin up channel of $e_g$ which are partially filled (average $e_g$ electron density is 1-x). Due to high lattice symmetry due to exchange interaction the system has a generalized time reversal symmetry by the exchange interaction between the two sublattices [9-12]. These systems have $t_2g$ spin background which are controlled by the zig zag like quasi one dimensional chains with ferromagnetic coupling within each chain and antiferromagnetic coupling between nearest neighbouring chains [10]. Interplay between two types of spin orbit interaction results in anisotropic spin split Fermi contours which lead to anisotropic magnetoresistance, which needs to be investigated in $HoFe_{0.25}Mn_{0.25}Cr_{0.5}O_3$, which showed highest of remanent magnetization at -16.1 T amongst $RTO_3$ (R: La, Ho and Pr: T: Fe, Cr and Mn) perovskites. In topological insulators an enhancement of electron propagation along a narrow range of real space angle resulting in change of geometry and breaking central symmetry is required for a good topological insulating material. For example, in systems containing small rare-earth ions for example a composition containing Ho-Mn-O phase. In the literature it has been reported that in $TbMnO_3$ polarization is parallel to *a* axis with commensurate ordering taking place by ¼ along the *b* axis [3]. Thus in order to find applications of such materials as topological insulators, as we decrease the size of the rare-earth ion, the spin canting and magneto- structural changes, should not be very strong enough, such that it makes the material an insulator completely. Such spin frustration induces the lattice distortion that break the inversion symmetry at high magnetic field. By sandwiching bilayer between two different substrates. allows experimental control of band topology and decreases the symmetry from octahedral to trigonal, resulting in additional splitting of d-orbital which is crucial for topological insulator phase [9-12]. It has been reported that $t_2g^1$, $t_2g^2$, $t_2g^3$ and $t_2g^5$ transition metal oxides are candidate for topological insulators in strong spin orbit coupling limit and $t_2g^2$, $t_2g^4$ and $t_2g^5$ in weak spin orbit coupling limit, as found in our compositions [12]. The spin orbit coupling is found to decrease as the energy difference between eg and $t_2g$ levels increases. In our case we observed maximum Jahn-Teller distortion in case of R = Gd amongst $R_{0.5}Sr_{0.5}Fe_{0.5}Mn_{0.5}O_3$ (R = Gd, Nd and Pr) in comparison of their lattice parameters. Greater disorder results in less splitting of orbitals due to Jahn Teller distortion which results in higher energy difference required for ferromagnetic and antiferromagnetic transition. It has been reported that if a potential comparable to $\lambda$ is approximately of the same value as the gap at Dirac point, then it turns ions due to presence of $eg^2$ electrons into an insulator. However, it has been known that topological insulator works only in thin film system, because the suppressed trigonal field inside such thick island is expected to create gapless bulk modes [2]. If symmetry breaking by magnetic ordering is strong as in the case of $TbMnO_3$, then it would become a trivial insulator [3]. Anisotropic Hall effect is the antisymmetric transverse magnetoresistance coefficient, obeying where the magnetic vector is perpendicular to x, y plane [10]. The spin texture of Dresselhaus model should not be circularly symmetric however symmetries involving radial translational symmetry along [100] or[010] axes has been reported, which makes it unfavourable for small rare earth containing transition metal ions [9-12]. On incorporation of non-magnetic impurity, for example Ta substitution in our case in composition $Sr_2YbRu_{1-x}Ta_xO_6$ (x = 0, 0.25, 0.5, 0.75 and 1), it might result in polarization of carrier band with spin orbit coupling and without not much scattering from the impurity for magnetoresistance application.

To obtain the phenomenon of Hall effect, where a potential transvers an electric current in an electric conductor, which is being placed in magnetic field for it to show good magnetotransport mechanisms [12]. Thus, for planer Hall effect measurement, the magnetic anisotropy should decrease. Interplay of two spin orbit interactions results in anisotropic spin split Fermi contours, which leads to anisotropic magnetoresistance, and an enhancement of electron propagation along a narrow range of real space angles from isotropic source. For example on substitution at rare-earth site, Pr by Ho results in decrease in magnetic moment due to competeing Pr orbital with Fe orbital in case of $Ho_{1-x}Pr_xFe_{0.25}Mn_{0.25}Cr_{0.5}O_3$ (x = 0, 0.5 and 1). In presence of spin orbit interaction parallel to magnetic field, resulting in appearance of Zeeman energy which also affects dispersion of charge carriers, by changing geometry and breaking central symmetry of two dimensional fermi contours [9-12]. From Table 7.1 it was evident that in general for all the compositions as the magnetic transition temperature increases the bifurcation temperature also gets increased. This observation could be attributed to the competition between Zeeman energy due to electric field and strong magnetocrystalline anisotropy energy around the metamagnetic transitions. It was also observed in our case that in $La_{0.5-x}Sm_xPr_{0.5}CrO_3$ (x = 0 and 0.1) there is a magnetic hysteresis between FCC and FCW taking place due to magnetic transition in the temperature range 25-160 K. It has been known that there are varying magnetic spin reorientation taking place from GzAyFz to FxCyGz, where Gx, Ay, and Fz stand for spin component along x, y, and z axis in terms of Bertaut's notation [13], respectively, which becomes favourable on substitution of 50% Fe at Cr site in $HoCrO_3$ due to interaction of Ho and Fe taking place along with the Ho ordering at 13 K [14]. It has been reported in literature in case of $La_2CrMnO_6$ that there is a bifurcation between FC and ZFC which takes place at a higher temperature that is 170 K, as compared to 140 K for case of others [15]. This increase could be attributed to absence of Jahn Teller inactive ions which would have created magnetic inhomogeneities at low temperature. For spin glass signature this Jahn Teller active $Mn^{3+}$ ion in high spin state along with presence of $Mn^{4+}$ and $Mn^{3+}$ facilitate double exchange interaction along with super exchange interaction at 180˚ bond angle. This short range ordering of charges and higher octahedral distortion primarily from Jahn Teller distortion is postulated to result in charge ordering in $Gd_{0.5}Sr_{0.5}Fe_{0.5}Mn_{0.5}O_3$. We have observed Jahn Teller distortion by $Ru^{5+}/Ru^{4+}$ ion in case of $PrMn_{0.5}Ru_{0.35}O_3$ with short range ordering taking place at B-site while $Sr_2YbRuO_6$ is predicted to show long range ordering. A greater disorder resulting in less splitting of the orbital (Jahn Teller distortion), which in case of cubic symmetry shows uniform arrangement of orbitals and thus it results in higher energy difference between antiferromagnetic and ferromagnetic transitions taking place in case of small rare-earth substitution. The band gap is also found to be higher for small rare-earths [4-6]. Also due to less oriented orbitals the interaction of rare-earth with B - O - B takes place and thus it results in canting as observed in our case for $La_{0.5-x}Sm_xPr_{0.5}CrO_3$ (x = 0, 0.1, 0.2, 0.3, 0.4 and 0.5). Thus the orbital ordering gives rise not only to magnetic anisotropy but also strong localization effect. The random distribution of $Mn^{3+}$ and $Mn^{4+}$ double exchange interaction operates in spatially distinct disordered regions.

Thus more orthorhombic distortion along a particular axis suggests stronger anisotropy along that axis, especially in small rare-earth containing perovskites [3]. Magnetic spin reorientation taking place in $TbMnO_3$, induced by magnetic spin reorientation takes place for application in multiferroic behavior [3]. Rotation of $Cr^{3+}$ occurs more easily at temperatures below 30 K, because of reduction of anisotropy which suggests possible change of crystallographic or electronic structure at temperatures below T-minimum. In our case we obtained negative exchange bias of Cr ions taking place, due to this spin reorientation in $La_{0.5-x}Sm_xPr_{0.5}CrO_3$ (x = 0-0.5). The effective Cr anisotropy leads to preference for GxFz over GzFx [16-17]. On application of external magnetic field along *a* axis, it results in rotation of Cr moments in *ac* plane, until about 2 KOe when the GzFx mode is induced [16-17]. We predict that on studying magnetic exchange interaction in Ho – Cr – O phase, the rare-earth transition orbital interaction would induce easy rotation of spins. The Dzyaloshinskii field responsible for canting of Cr system causes net moment generated from canted spins of Cr ion ($M_{Cr}$) to increase as temperature decreases. Here $M_{Cr}$ is independent of temperature and the isotropic exchange does not contribute to internal magnetic field ($H_I$). In our case in the $La_{0.5-x}Sm_xPr_{0.5}CrO_3$ (x = 0.2 and 0.4) with the increase in octahedral distortion from x = 0.2 to 0.4, the $M_{Cr}$ and $H_I$ (obtained by extrapolation of isothermal magnetization curve [17]), is found to increase from to 474 emu/mol to 720 emu/mol and 2278 Oe to 12574 Oe respectively. We tried comparing the remanent magnetization value amongst all the compositions of formula $RTO_3$ (R: La, Ho and Pr and T: Fe, Cr and Mn) on the basis of the increase in octahedral distortion, where the remanent magnetization value, as obtained from the hysteresis loop at 5 K which was found to decrease with increased octahedral distortion, with minimum value obtained in case of $LaFe_{0.25}Mn_{0.25}Cr_{0.5}O_3$ (1.989 emu/mol) and maximum in case of $HoFe_{0.25}Mn_{0.25}Cr_{0.5}O_3$ (16.882 emu/mol). In the literature it has been reported that the effective field in $GdFeO_3$ has been reported at 310 Oe while in case of $GdCrO_3$ it is at 5500 Oe which has been attributed to increased pseudodipolar coupling [17]. Due to Gd-Cr interaction, GxFz to FxGz transition takes place in the temperature range 4.2-14 K, where $GdCrO_3$ favours GxFz configuration more [13,17]. As the type of spin configuration in $ABO_3$ perovskite depends on strength of B-B exchange interaction and their distances. In the composition $Sr_2YbRu_{1-x}Ta_xO_6$ (x = 0, 0.25, 0.5, 0.75 and 1) we observed two types of exchange interactions such as B- O –B' and B – O - O – B, where the relatively weaker $Ru^{5+}$ - O - O – $Ru^{4+}$ and $Ru^{4+}$ - O - O – $Ru^{4+}$ interaction is responsible for ferromagnetic interaction, which was found to maximum for higher values of x and $Ru^{5+}$ - O – O - $Ru^{5+}$ is responsible for antiferromagnetic interaction [18]. Due to less oriented orbitals the spins interact easily and easy canting of the spins takes place, as was observed also in $La_{0.5-x}Sm_xPr_{0.5}CrO_3$ (x = 0, 0.1, 0.2, 0.3, 0.4 and 0.5) at higher values of x. In the literature it has been reported that in $La_{1-x}Sr_xMnO_3$ that spin component along *a* axis and an ordering of spin component amongst all the four ions takes place. It has been reported in $GdCrO_3$ that antiferromagnetic transition takes place at temperatures below 170 K, where the weak ferromagnetism is observed due to Cr moment, which are canted from the true antiferromagnetic alignment [17].

In this thesis, we have studied the oxides of general formula $ABO_3$, having perovskite structure, show varying physical properties such as structural, magnetic and photoelectric properties and that may arise due the nature of rare earth metals(s) and transition metal(s) and their oxidation states. Manganite and ferrites are important in condensed matter as the correlation between their structural and magnetic properties lead to various phenomena like multiferroicity, colossal magnetoresistance etc. Thus, it is imperative to study both the electronic and magnetic properties of a library of these species to gain insights into the contribution of various factors (composition, temperature, magnetic field etc.) on the observed properties. In the course of this thesis, an attempt was made in trying to analyse various magnetic and structural interactions which may be favourable for various applications such as magnetoresistance, spin hall effect, topological insulators and multiferroics. The interplay of magnetic anisotropy, magnetic spin polarization and the effect of doping on octahedral distortion are inter-related and has been investigated. We have synthesized many rare earth based perovskites which will be a potential candidate for dielectric studies. The discussed systems in this thesis may also response as a multiferroic systems. Thus, this work paves the way for further studies in the hunt for exotic properties.

**Table 7.1: Lattice parameters and magnetic transition temperature for all the compositions.**

| Composition | a (Å) | b (Å) | c (Å) | β (˚) | B – O1(Å)<br>B – O2(Å) | B-$O_1$-B(˚)<br>B-$O_2$-B(˚) | $\theta_{CW}$ (K) | $T_C$ (K) | $T_N$ (K) | $T_{SR}$ (K) | $\mu_{obs}$<br>$\mu_{calc}$<br>$\mu_B$/ f.u. |
|---|---|---|---|---|---|---|---|---|---|---|---|
| **Chapter II** (Orthorhombic *Pnma*) | | | | | | | | | | | |
| $HoCr_{0.5}Mn_{0.5}O_3$ | 5.6174(4) | 7.4735(2) | 5.2435(3) | | 2.030(1)<br>1.980(1) | 133.61(1)<br>157.93(1) | -14 | 140 | | 54 | 11.345<br>11.284 |
| $Ho_{0.7}Pr_{0.3}Mn_{0.5}Cr_{0.5}O_3$ | 5.5975(6) | 7.5347(2) | 5.3000(6) | | 2.040(1)<br>1.910(1) | 134.35(1)<br>160.52(1) | -49 | 47 | | 37 | 9.703<br>9.928 |
| $Ho_{0.5}Pr_{0.5}Mn_{0.5}Cr_{0.5}O_3$ | 5.5854(6) | 7.5979(8) | 5.3539(6) | | 2.040(1)<br>2.030(1) | 137.50(1)<br>156.38(2) | -21 | 58 | | 32 | 8.357<br>8.909 |
| $Ho_{0.3}Pr_{0.7}Mn_{0.5}Cr_{0.5}O_3$ | 5.5589(6) | 7.6439(7) | 5.3938(2) | | 2.010(1)<br>2.050(1) | 144.49(1)<br>157.71(2) | -14 | 71 | | 39 | 7.539<br>7.759 |
| $HoFe_{0.25}Mn_{0.25}Cr_{0.5}O_3$ | 5.5603(4) | 7.5181(5) | 5.2475(3) | | 1.900(1)<br>1.970(1) | 162.33(1)<br>141.46(1) | -2 | 130 | | 12 | 11.216<br>11.407 |
| $Ho_{0.5}Pr_{0.5}Fe_{0.25}Mn_{0.25}Cr_{0.5}O_3$ | 5.5550(6) | 7.6317(6) | 5.3620(2) | | 2.070(1)<br>1.990(1) | 134.18(1)<br>173.16(2) | 14 | 137 | | 118 | 8.150<br>8.953 |
| $PrFe_{0.25}Mn_{0.25}Cr_{0.5}O_3$ | 5.5854(6) | 7.5979(8) | 5.3539(6) | | 2.040(1)<br>2.030(1) | 137.50(1)<br>156.38(1) | -58 | 245 | | 153 | 5.167<br>5.849 |
| $LaFe_{0.25}Mn_{0.25}Cr_{0.5}O_3$ | 5.5087(3) | 7.7917(6) | 5.4652(8) | | 1.950(1)<br>2.290(1) | 172.78(1)<br>159.37(1) | -25 | 101 | | 29 | 3.814<br>4.696 |
| $La_{0.5}Pr_{0.5}Fe_{0.25}Mn_{0.25}Cr_{0.5}O_3$ | 5.5005(2) | 7.7779(2) | 5.4711(2) | | 1.980(1)<br>2.100(1) | 158.59(1)<br>171.37(1) | 30 | 108 | | 38 | 4.162<br>5.299 |
| **Chapter III** (Orthorhombic *Pnma*) | | | | | | | | | | | |
| $PrMn_{0.5}Ru_{0.35}O_3$ | 5.6777(4) | 7.7026(3) | 5.4415(3) | | 1.970(1)<br>1.680(1) | 157.15(1)<br>168.88(1) | 21 | 90 | | | 5.104<br>5.248 |
| **Chapter IV** (Orthorhombic *Pnma*) | | | | | | | | | | | |
| $La_{0.5}Pr_{0.5}CrO_3$ | 5.4633(1) | 7.7465(2) | 5.4739(1) | | 1.942(1)<br>1.967(1) | 171.46(1)<br>155.76(1) | 60 | | 246 | | 2.857<br>4.624 |
| $La_{0.4}Sm_{0.1}Pr_{0.5}CrO_3$ | 5.4699(2) | 7.7392(1) | 5.45564(2) | | 1.949(1)<br>2.050(1) | 166.18(1)<br>154.14(1) | -27 | | 252 | | 2.957<br>4.630 |
| $La_{0.3}Sm_{0.2}Pr_{0.5}CrO_3$ | 5.4731(2) | 7.7147(2) | 5.4527(2) | | 1.978(1)<br>2.014(1) | 154.20(1)<br>155.77(1) | -163 | | 253 | | 3.427<br>4.640 |
| $La_{0.2}Sm_{0.3}Pr_{0.5}CrO_3$ | 5.4727(2) | 7.7001(2) | 5.4368(1) | | 1.987(1)<br>1.939(1) | 151.39(1)<br>172.77(1) | -166 | | 263 | | 3.100<br>4.647 |
| $La_{0.1}Sm_{0.4}Pr_{0.5}CrO_3$ | 5.4739(1) | 7.6848(2) | 5.4226(2) | | 1.937(1)<br>2.082(1) | 165.50(1)<br>147.83(1) | -167 | | 267 | | 3.087<br>4.655 |
| $Sm_{0.5}Pr_{0.5}CrO_3$ | 5.4748(1) | 7.6698(2) | 5.4048(1) | | 1.994(1)<br>2.001(1) | 148.14(1)<br>153.59(1) | -171 | | 275 | | 4.864<br>4.662 |
| **Chapter V** (Orthorhombic *Pnma*) | | | | | | | | | | | |
| $Gd_{0.5}Sr_{0.5}Fe_{0.5}Mn_{0.5}O_3$ | 5.423(1) | 7.649(1) | 5.417(1) | | 1.928(4)<br>2.187(3) | 162.69(1)<br>162.96(1) | 3 | | | 49 | 7.223<br>7.355 |
| $Nd_{0.5}Sr_{0.5}Fe_{0.5}Mn_{0.5}O_3$ | 5.4412(2) | 7.7214(1) | 5.4151(1) | | 1.942(1)<br>1.997(1) | 167.39(1)<br>158.46(1) | 55 | | | 47 | 5.022<br>5.478 |
| $Pr_{0.5}Sr_{0.5}Fe_{0.5}Mn_{0.5}O_3$ | 5.4695(1) | 7.7340(1) | 5.4287(1) | | 1.940(1)<br>2.006(1) | 170.66(1)<br>173.50(1) | 65 | | | 58 | 5.112<br>5.699 |
| **Chapter VI** (Monoclinic $P2_1/n$) | | | | | | | | | | | |
| $Sr_2YbRuO_6$ | 5.7926(1) | 5.7326(1) | 8.1138(1) | 90.24(1) | 1.871(1)<br>2.282(1)<br>2.049(1) | 161.19(1)<br>147.63(1)<br>163.25(1) | -148 | | 51 | | 5.966<br>5.837 |
| $Sr_2YbRu_{0.75}Ta_{0.25}O_6$ | 5.7929(4) | 5.7472(3) | 8.1321(3) | 90.157(7) | 2.078(1)<br>2.629(1)<br>1.915(1) | 168.51(1)<br>131.51(1)<br>168.27(1) | -125 | | 40 | | 5.643<br>5.560 |
| $Sr_2YbRu_{0.5}Ta_{0.5}O_6$ | 5.7963(4) | 5.4648(4) | 8.1539(5) | 90.213(1) | 2.269(1)<br>2.531(1)<br>2.496(1) | 162.12(1)<br>149.38(1)<br>151.62(1) | -118 | | 49 | | 5.301<br>5.320 |
| $Sr_2YbRu_{0.25}Ta_{0.75}O_6$ | 5.8001(1) | 5.7727(2) | 8.1701(2) | 90.136(3) | 2.057(1)<br>2.258(1)<br>2.007(1) | 167.43(1)<br>161.75(1)<br>158.29(1) | -102 | | 43 | | 4.935<br>4.797 |

| | | | | | | | | | | | | |
|---|---|---|---|---|---|---|---|---|---|---|---|---|
| **$Sr_2YbTaO_6$** | 5.8149(3) | 5.7839(3) | 8.1921(3) | 89.905(5) | 1.877(1)<br>2.326(1)<br>2.071(1) | 161.08(1)<br>149.31(1)<br>163.28(1) | | | | | | |

# Thank you